\documentclass{jfm}

\usepackage{subfiles}
\usepackage{graphicx}
\usepackage{epstopdf,epsfig,subfloat,subfig,floatrow}
\usepackage{newtxtext}
\usepackage{newtxmath}
\usepackage{natbib}
\usepackage{amsfonts}
\usepackage{amsmath}
\usepackage{amssymb,bm,mathtools}
\usepackage{color}
\usepackage{hyperref}
\usepackage{multirow,makecell,array,booktabs}
\usepackage{soul,cancel}
\usepackage{ulem}
\usepackage{lineno}

\hypersetup{
    colorlinks = true,
    urlcolor   = blue,
    citecolor  = blue,
}

\newcommand{\RomanNumeralCaps}[1]

\allowdisplaybreaks[4]

\definecolor{orange}{RGB}{255,100,0}

\title{Kinetic modelling of rarefied gas mixtures with disparate mass}

\author{Qi Li,
		Jianan Zeng
        \and Lei Wu
        \corresp{\email{wul@sustech.edu.cn}}
        }

\affiliation{
  Department of Mechanics and Aerospace Engineering, Southern University of Science and Technology, Shenzhen 518055, China
  }

\begin{document}
\maketitle


\begin{abstract}

The simulation of rarefied gas flow based on the Boltzmann equation is challenging, especially when the gas mixtures have disparate molecular masses. In this paper, a computationally tractable kinetic model is proposed for monatomic gas mixtures, to mimic the Boltzmann collision operator as closely as possible. The intra- and inter-collisions are modelled separately using relaxation approximations, to correctly recover the relaxation timescales that could span several orders of magnitude. The proposed kinetic model preserves the accuracy of the Boltzmann equation in the continuum regime by recovering the four critical transport properties of a gas mixture: the shear viscosity, the thermal conductivity, the coefficients of diffusion and the thermal diffusion. While in the rarefied flow regimes, the kinetic model is found to be accurate when comparing its solutions with those from the direct simulation Monte Carlo method in several representative cases (e.g. one-dimensional normal shock wave, Fourier flow and Couette flow, two-dimensional supersonic flow passing a cylinder and nozzle flow into a vacuum), for binary mixtures with a wide range of mass ratios (up to 1000), species concentrations, and different intermolecular potentials. Pronounced separations in species properties have been observed, and the flow characteristics of gas mixtures in shock waves are found to change as the mass difference increases from moderate to substantial. 
\end{abstract}


\section{Introduction}\label{sec:introduction}

The dynamics of rarefied gas mixtures have long been an important issue, and one of the particular interests lies in the non-equilibrium phenomena of disparate-mass mixtures widely encountered in plasma physics, aerospace engineering and chip industry. For example, during the re-entry of a vehicle into the planetary atmosphere at a significantly high Mach number, plasma comprising ions, electrons, as well as neutral species, is formed between the shock and the vehicle surface, where the mass ratio between the mixture components can be as large as $10^3$ to $10^5$ \citep{Brun2012High}; In the low-pressure environment of an extreme ultraviolet (EUV) lithography system, hydrogen is commonly employed as a clean gas to effectively inhibit the diffusion of pollutant gas molecules (e.g. hydrocarbon) generated by photoresists, which possess a mass several tens or hundreds times greater than that of clean gases \citep{Teng2023JCProd}; Similar situations are also encountered in the design of particle exhaust system in nuclear fusion device \citep{Tantos2024NuclFusion}.

In addition to the well-known gas rarefaction effects occurring when the mean free path of gas molecules is comparable to the characteristic flow length, in the rarefied gas mixtures with disparate mass, multiscale non-equilibrium is present not only spatially but also temporally. From the mesoscopic perspective, the light gas molecules have a larger thermal velocity than the heavy ones, thus leading to dispersed relaxation time scales of molecular collisions. According to \cite{Grad1960}, the light gas molecules will get the equilibrium among themselves first through intra-species collisions; then the heavy molecules reach their own equilibrium; and finally, all the species approach the common state through inter-species collisions. The large difference in relaxation time generates difficulties in the simulation of gas mixture flows. The Navier–Stokes-type equations involving a common flow temperature and approximated diffusion velocities of each species are adequate only when all these relaxation times are much smaller than the characteristic time of gas flow \citep{Nagnibeda2009}. Otherwise, the gas kinetic theory needs to be adopted to capture the non-equilibrium behaviours \citep{Sawant2020JFM,Agrawal2020JFM}. Although the Boltzmann equation is rigorously established for gas mixtures consisting of monatomic molecules at a mesoscopic level, it is practically difficult to apply in realistic applications, due to the numerical complexity of the high-dimensional nonlinear integral collision operator, especially for the mixtures with mass disparity. Even using the fast spectral method, the computational cost for each binary collision operator will be increased by the square root of the mass ratio, and hence the numerical simulation is stopped at the molecular mass ratio less than 36 \citep{Wu2015JCP}.

A well-acknowledged method for the simulations of rarefied gas flows is the Direct Simulation Monte Carlo (DSMC) method \citep{Bird1994}, which models the kinetic processes of a collection of simulated particles. It has been proven by \cite{wagner_consist} that DSMC is equivalent to the Boltzmann equation for monatomic gas, as the number of simulation particles approaches infinity. Although applicable in all flow regimes, DSMC is computationally costly in the simulation of flows with low Knudsen numbers ($\text{Kn}$), since the cell size and time step should be smaller than the mean free path and mean collision times of gas molecules, respectively. In a gas mixture with disparate mass, the mean collision time of different types of collisions spans multiple scales, thereby restricting the time step in DSMC to be smaller than the fastest relaxation time and significantly decelerating the numerical evolution of the system. In some simulation cases of collisional plasma, the electron mass is increased by three orders of magnitude to achieve an acceptable simulation time but sacrifice the accuracy \citep{Farbar2010PoF}. In addition, the concentration of components in the gas mixture varies significantly in many of the realistic applications. For instance, the ultraviolet radiation from nitric oxide is of particular interest in the flow fields surrounding hypersonic re-entry vehicles, while the mole fraction of nitric oxide is typically less than $10^{-5}$ \citep{Erdman1993JTHT}. Thus, the conventional DSMC method with equally weighted particles has enormous difficulties in terms of either huge computational costs or significant statistical noise. The differentially weighted schemes, although solve the disparate mole fraction problem, face conservation issues during each collision \citep{Boyd1996JTHT}, and involve complicated splitting/merging treatment for colliding particles \citep{Alves2018PSST}. Therefore, the multiscale feature in gas mixtures with disparate mass and concentrations makes the DSMC method time-consuming and even intractable in some cases.

Therefore, it is important to develop kinetic model equations with much-simplified collision operators to imitate as closely as possible the behaviour of the Boltzmann equation, and multiscale numerical methods to solve those kinetic models. For a single-species monatomic gas, the Bhatnagar-Gross-Krook (BGK) model equation replaces the Boltzmann collision operator with a relaxation approximation \citep{Bhatnagar1954}, thus achieves high computational efficiency and lays the foundation for more sophisticated models after that. However, the main drawback of the BGK model is the incorrect Prandtl number produced by its single relaxation rates for both stress and heat flux. This issue has been addressed by the modified versions of the original BGK model, that is, the ellipsoidal-statistical BGK (ES-BGK) model \citep{Holway1966,andries2000gaussian,Mathiaud2022EjM} and the Shakhov model \citep{Shakhov1968,Shakhov_S}, both of which can reproduce correct shear viscosity and thermal conductivity simultaneously in the continuum flow regime. Together with the multiscale numerical methods, these kinetic models have found applications in many engineering problems \citep{Su2020JCP,Liu2020JCP,Pfeiffer2022PRE,Zeng2023CiCP,Liu2024JCP}.

However, the extension of the single-species kinetic model equations to gas mixtures is a nontrivial task with substantial difficulties. The first fundamental reason for this is that collisions between different species of molecules lead to exchanges of momentum and energy between mixture components, exhibiting notable disparities in relaxation times due to variations in species properties. Second, apart from the components-specific shear viscosities and thermal conductivities, a multi-species gas also possesses effective mixture viscosity and thermal conductivity, as well as diffusion and thermal diffusion coefficients that correspond to the Fick and Soret effects, respectively \citep{CE}, which have to be recovered by the kinetic models in the continuum limit. Previously proposed kinetic models using BGK approaches can be classified into two types \citep{Pirner2021Fluids}. One uses single-relaxation term involving both inter- and intra-species collisions, and the other one has a sum of collision terms modelling each type of binary collision individually. For the single-relaxation models, The model of \cite{Andries2002JSP} is the most widely applied one, which reduces to the single-species BGK model for mechanically identical components that cannot recover shear viscosity and thermal conductivity of the mixture simultaneously. Besides, the diffusion coefficient is not correctly captured by this model. Later, more adjustable parameters are introduced into the single-relaxation model by adopting ES-BGK and Shakhov-type operators, thus thermal conductivity \citep{Brull2015CMS} and diffusion coefficients \citep{Groppi2011EPL,Todorova2019EJMB} can be recovered. It should be noted that, since the thermal diffusion effect is not reproduced, the proposed models are incapable of modelling the thermally induced flow of non-Maxwellian molecules, which is an important transport phenomenon in gas mixtures \citep{Sharipov2024IJHMT}. Although the single-relaxation models are found to be accurate enough for mixtures with small mass ratio \citep{Pfeiffer2021PoF} and also easy to be extended to polyatomic gases with internal energy \citep{Bisi2020PhysicaA,Todorova2020AIP} and reactions \citep{Groppi2004PoF,Bisi2018JPA}, they are however not able to distinguish the multiple scales of relaxation times and different types of interactions. On the other hand, several multi-relaxation models have been proposed \citep{Morse1964PoF,Hamel1965PoF,Greene1973PoF,Haack2017JSP,Klingenberg2017KRM,Bobylev2018KRM,Bisi2022AMC}, which mainly differ in the construction of the auxiliary properties in the inter-species collision terms. Because of the complexity of the collision operators, these models were basically derived mathematically but rarely applied to a realistic problem \citep{Tantos2021PoF,Bisi2022AMC}. More importantly, the recovery of transport properties from the multi-relaxation models in the continuum limit is usually overlooked, and the determination of the adjustable parameters (e.g. the multiple relaxation times) is still an open question. 
It is also noted that, the models proposed by \cite{McCormack1973PoF} and \cite{Kosuge2009EJMB} replace the Boltzmann collision operator by the use of polynomial expansions in the molecular velocity, where the model coefficients are determined by matching the moments of the model collision operator to the Boltzmann collision operator. The polynomial models correctly recover all the transport coefficients of a gas mixture, and show excellent accuracy in slightly to moderately non-equilibrium flows \citep{Ho2016IJHMT,Tantos2018IJHMT}. However, tremendously complicated model coefficients and collision integrals based on molecular interactions are involved in the model equations, which brings difficulties in realistic applications due to the significant computational cost and the lack of accurate intermolecular potentials.

Despite the great effort made in the past decades, establishing accurate and computationally tractable kinetic models for gas mixtures with disparate mass is still of significant challenge. Hence, the present work is dedicated to developing a kinetic model based on the idea of multi-relaxation models for rarefied monotonic gas mixtures with disparate molecular mass, which not only recoveries the transport properties of a gas mixture including shear viscosity, thermal conductivity, diffusion coefficient and thermal diffusion coefficients, but also correctly captures the multiscale relaxation rates of different collision processes. More importantly, with the deterministic numerical methods and multiscale schemes, the proposed kinetic model can be used to solve gas mixture flows with disparate mass and mole fractions efficiently and accurately in all flow regimes.

The rest of the paper is organized as follows: in \S \ref{sec:kinetic_model}, the kinetic model is proposed with all the adjustable parameters determined by the transport properties of the gas, and the transport coefficients and hydrodynamic equations in the continuum limit are given; in \S \ref{sec:parameters}, to make a consistent comparison to DSMC method as validation, the kinetic model parameters are obtained based on the DSMC collision model; the accuracy of the proposed kinetic model is assessed by the DSMC method in several one-dimensional and two-dimensional problems in \S \ref{sec:1D_cases} and \S \ref{sec:2D_cases}, respectively, and the flow characteristics of gas mixtures with disparate mass are discussed. Finally, conclusions are presented in \S \ref{sec:conclusion}.


\section{Kinetic model equation}\label{sec:kinetic_model}

The gas kinetic theory describes the status of a gaseous system in the phase space using the velocity distribution functions, whose evolution is governed by the Boltzmann equations when only binary collisions are considered. Because of the unaffordable computational cost of solving the $N^2$ number of Boltzmann operators for a $N$-components gas mixture, kinetic model equations with simplified collision operators are highly demanded. 
Theoretically, several requirements need to be followed by a kinetic model for monatomic gas mixtures: (i) the collision terms satisfy the conservations of mass, momentum and energy, and restore the equilibrium state for an isolated system; (ii) the relaxation rates for different types of intermolecular interaction can be correctly captured; (iii) all the transport properties given by the model equation are consistent with those obtained from the Boltzmann equation in the hydrodynamic limit; (iv) the momentum and energy exchange during inter-species collisions are close to those obtained from the Boltzmann equation; (v) the kinetic model complies with the indifferentiability principle and H-theorem. 

However, it is impractical to build such an ideal kinetic model to satisfy all the requirements with affordable computational cost. Even for a single-species gas, though the ES-BGK model has been proven to keep the non-negativity of the velocity distribution functions and satisfy the H-theorem \citep{andries2000gaussian}, the Shakhov model is found to be more accurate for many strong non-equilibrium problems due to its better approximations of high-order moments of molecular velocity distributions \citep{ChenAAMM2015,Fei2020AIAA,Yuan2022JFM, Park2024PoF}. Moreover, despite the fact that the model of \cite{Andries2002JSP} complies with the indifferentiability principle unconditionally, it recovers only one transport coefficient and ignores the multiple relaxation time scales.

Therefore, to achieve a balance between the accuracy and computational burden for a gas mixture with disparate mass, we build our kinetic model using multi-relaxation operators to distinguish different types of binary interactions. The tunable parameters in the kinetic model are determined by recovering the transport properties in the continuum flow regime, including viscosity, thermal conductivity, diffusion coefficients and thermal diffusion coefficients. Additionally, the conditional indifferentiability principle of the proposed model equation can be satisfied in the near-equilibrium condition.

\subsection{Kinetic description of monatomic gas mixture}

We consider a $N$-components mixture of monatomic gases with the velocity distribution functions $f_s(\bm{x},\bm{v},t)$ describing their mesoscopic states, where $s$ indicates the species, $t$ is the time, $\bm{x}\in \mathbb{R}^3$ is the spatial coordinates, $\bm{v}\in \mathbb{R}^3$ is the molecular velocity. Since all the collisions, either intra-species or inter-species, are binary, the distribution function for species $s$ under external body accelerations $\bm{a}_s$ evolve according to the Boltzmann equation,
\begin{equation}\label{eq:Boltzmann_equation}
	\begin{aligned}[b]
		\underbrace{\frac{\partial{f_{s}}}{\partial{t}}+\bm{v}\cdot\frac{\partial{f_{s}}}{\partial{\bm{x}}}+ \bm{a}_s\cdot\frac{\partial{f_{s}}}{\partial{\bm{v}}}}_{\mathcal{D}f_{s}} &= \sum_{r=1}^N Q_{sr}\left(f_s,f_r\right), \quad s=1,2,\cdots,N,
	\end{aligned}
\end{equation}
where the left-hand side is known as the streaming term $\mathcal{D}f_{s}$, and the right-hand side is a sum over all binary Boltzmann collision operator $Q_{sr}$ between molecules of species $s$ and $r$,
\begin{equation}\label{eq:Boltzmann_operator}
	\begin{aligned}[b]
		Q_{sr}\left(f_s,f_r\right) = \int_{\mathbb{R}^3}\int_{4\pi} \sigma_{sr}\left(|\bm{v}_{*}|,\bm{\varOmega}\cdot\frac{\bm{v}_{*}}{|\bm{v}_{*}|}\right)\left[f_{s}(\bm{v}')f_{r}(\bm{w}')-f_{s}(\bm{v})f_{r}(\bm{w})\right]|\bm{v}_{*}|\mathrm{d}\bm{\varOmega}\mathrm{d}\bm{w}.
	\end{aligned}
\end{equation}
Here, $\sigma_{sr}$ is the scattering cross-section depending on the intermolecular potential between the two species, and $\bm{\varOmega}$ is the unit vector of the solid angle; $\bm{v}$ and $\bm{w}$ are the pre-collision velocities of the two molecules of species $s$ and $r$, respective. Hence $\bm{v}_{*}=\bm{v}-\bm{w}$ is the relative velocity, which determine the post-collision velocities $\bm{v}'$ and $\bm{w}'$ of the collision pair as,
\begin{equation}\label{eq:post_collision_velocities}
	\begin{aligned}[b]
		\bm{v}' = \bm{v} - \frac{2m_r}{m_s+m_r}\left(\bm{v}_{*}\cdot\bm{\varOmega}\right)\bm{\varOmega}, \quad \bm{w}' = \bm{w} + \frac{2m_s}{m_s+m_r}\left(\bm{v}_{*}\cdot\bm{\varOmega}\right)\bm{\varOmega}.
	\end{aligned}
\end{equation}

The macroscopic variables of each species $s$, namely, the number density $n_s$, mass density $\rho_s$, flow velocity $\bm{u_s}$, temperatures $T_s$, pressure tensor $\bm{P}_s$, and heat flux $\bm{q}_s$, are obtained by taking the moments of the respective velocity distribution function $f_s$,
\begin{equation}\label{eq:species_macroscopic_variables_f}
	\begin{aligned}[b]
		n_s &= \left<1,f_s\right>, \quad
        \rho_s = \left<m_s,f_s\right>, \quad
        \rho_s\bm{u}_s = \left<m_s\bm{v},f_s\right>, \quad
        \frac{3}{2}n_sk_BT_s = \left<\frac{1}{2}m_s(\bm{v}-\bm{u}_s)^2,f_s\right>, \notag \\
        \bm{P}_s &= \left<m_s(\bm{v}-\bm{u}_s)(\bm{v}-\bm{u}_s)f_s,f_s\right>, \quad
        \bm{q}_s = \left<\frac{1}{2}m_s(\bm{v}-\bm{u}_s)^2(\bm{v}-\bm{u}_s),f_s\right>, 
	\end{aligned}
\end{equation}
where $m_s$ is the molecular mass, $k_B$ is the Boltzmann constant; and the operator $\left<h,\psi\right>$ is defined as an integral of $h\psi$ over the velocity space,
\begin{equation}\label{eq:velocity_integral}
	\begin{aligned}[b]
		\left<h,\psi\right> \equiv \int_{\mathbb{R}^3}{h\psi}\mathrm{d}\bm{v}.
	\end{aligned}
\end{equation}
Then, the corresponding macroscopic quantities for the mixture, the number density $n$, mass $\rho$, flow velocity $\bm{u}$, temperatures $T$, pressure tensor $\bm{P}$, and heat flux $\bm{q}$, are given by
\begin{align}\label{eq:mixture_macroscopic_variables_f}
		n &= \sum_s\left<1,f_s\right> = \sum_{s}n_s, \notag \\
        \rho &= \sum_s\left<m_s,f_s\right> = \sum_{s}\rho_s, \notag \\
        \rho\bm{u} &= \sum_s\left<m_s\bm{v},f_s\right> = \sum_{s}\rho_s\bm{u}_s, \notag \\
        \frac{3}{2}nk_BT &= \sum_{s}\left<\frac{1}{2}m_sc^2,f_s\right> = \sum_{s}\frac{3}{2}n_sk_BT_s + \frac{1}{2}\sum_{s}\rho_s\left(\bm{u}_s-\bm{u}\right)^2, \notag \\
        \bm{P} &= \sum_{s}\left<m_s\bm{c}\bm{c},f_s\right> = \sum_{s}\bm{P}_s + \sum_{s}\rho_s\left(\bm{u}_s-\bm{u}\right)\left(\bm{u}_s-\bm{u}\right), \notag \\
        \bm{q} &= \sum_{s}\left<\frac{1}{2}m_sc^2\bm{c},f_s\right> \notag \\
		&= \sum_{s}\bm{q}_s + \sum_{s}\frac{3}{2}n_sk_BT_s\left(\bm{u}_s-\bm{u}\right) +  \frac{1}{2}\sum_{s}\rho_s\left(\bm{u}_s-\bm{u}\right)^2\left(\bm{u}_s-\bm{u}\right) +  \sum_{s}\bm{P}_s\cdot\left(\bm{u}_s-\bm{u}\right),
\end{align}
where $\bm{c}=\bm{v}-\bm{u}$ is the peculiar velocity with respect to the mixture velocity $\bm{u}$, and therefore the diffusion velocities $\bm{u}_s-\bm{u}$ contribute to the mixture temperature, stress and heat flux.

\subsection{Kinetic model with multi-relaxation collision operators}

The proposed kinetic model for gas mixtures adopts relaxation time approximations for each pair of gas components individually to simplify the Boltzmann collision operator \eqref{eq:Boltzmann_operator}, and thus the evolution of distribution functions $f_{s}$ can be written as,
\begin{equation}\label{eq:kinetic_equation}
	\begin{aligned}[b]
		{\mathcal{D}f_{s}} &= 
		{\sum_{r=1}^N\frac{1}{\tau_{sr}}\left(g_{sr}-f_s\right)}, \quad s=1,2,\cdots,N,
	\end{aligned}
\end{equation}
where $s=r$ indicates a intra-species collision operator, $s\neq r$ are inter-species collisions; $\tau_{sr}$ is the corresponding relaxation time; $g_{sr}$ is the reference distribution function constructed in the form,
\begin{equation}\label{eq:g_sr}
	\begin{aligned}[b]
		g_{sr}=~&\hat{n}_{sr}\left(\frac{m_s}{2\pi k_B{T}_{s}}\right)^{3/2}\exp\left(-\frac{m_s(\bm{v}-\hat{\bm{u}}_{sr})^2}{2k_B{T}_{s}}\right) \times \\
		&\left[1+\frac{\hat{T}_{sr}-T_s}{T_s}\left(\frac{m_s(\bm{v}-\hat{\bm{u}}_{sr})^2}{2k_B{T}_{s}}-\frac{3}{2}\right)+\frac{2m_s\hat{\bm{q}}_{sr}\cdot(\bm{v}-\hat{\bm{u}}_{sr})}{5\hat{n}_{sr}k_B^2\hat{T}_{sr}^2}\left(\frac{m_s(\bm{v}-\hat{\bm{u}}_{sr})^2}{2k_B\hat{T}_{sr}}-\frac{5}{2}\right)\right],
	\end{aligned}
\end{equation}
with $\hat{n}_{sr},~\hat{T}_{sr},~\hat{\bm{u}}_{sr},~\hat{\bm{q}}_{sr}$ being the auxiliary parameters.

Construction of the auxiliary parameters is the most crucial task in building the kinetic model. As the essential constraints, the conservations of mass, momentum and energy have to be guaranteed during any binary collisions. For the intra-species collision operators, the conservations are the same as those for a single-species gas,
\begin{equation}\label{eq:conservation_intra}
	\begin{aligned}[b]
		\left<1,\frac{1}{\tau_{ss}}(g_{ss}-f_{s})\right> = 0, \quad
        \left<m_s\bm{v},\frac{1}{\tau_{ss}}(g_{ss}-f_{s})\right> = 0, \quad
        \left<\frac{1}{2}m_sv^2,\frac{1}{\tau_{ss}}(g_{ss}-f_{s})\right> = 0.
	\end{aligned}
\end{equation}
Therefore, the auxiliary number density, flow velocity and temperature in intra-species collision operators can be simply determined by the macroscopic properties of each species as $\hat{n}_{ss}=n_s$, $\hat{\bm{u}}_{ss}=\bm{u}_s$, $\hat{T}_{ss}=T_s$. Thus, the reference distribution function $g_{ss}$ in the intra-species collision term reduces to the Shakhov model.

For the inter-species collision operators, the number density of each inert species is still unchanged, while only the total momentum and energy of the collision pairs of species $s$ and $r$ meet the conservations, 
\begin{equation}\label{eq:conservation_inter}
	\begin{aligned}[b]
		\left<1,\frac{1}{\tau_{sr}}(g_{sr}-f_{s})\right> = 0, &\quad
        \left<1,\frac{1}{\tau_{rs}}(g_{rs}-f_{r})\right> = 0, \\
		\left<m_s\bm{v},\frac{1}{\tau_{sr}}(g_{sr}-f_{s})\right> &+ \left<m_r\bm{v},\frac{1}{\tau_{rs}}(g_{rs}-f_{r})\right> = 0, \\
		\left< \frac{1}{2}m_sv^2,\frac{1}{\tau_{sr}}(g_{sr}-f_{s})\right> &+ \left<\frac{1}{2}m_rv^2,\frac{1}{\tau_{rs}}(g_{rs}-f_{r})\right> = 0.
	\end{aligned}
\end{equation}
Then, the auxiliary number density is obtained as $\hat{n}_{sr} = n_s$ and $\hat{n}_{rs} = n_r$, while the auxiliary velocities and temperatures $\hat{\bm{u}}_{sr},~\hat{\bm{u}}_{rs},~\hat{T}_{sr},~\hat{T}_{rs}$ cannot be uniquely determined, but yield the constraints,
\begin{equation}\label{eq:conservation_inter_2}
	\begin{aligned}[b]
		\frac{\rho_s}{\tau_{sr}}\hat{\bm{u}}_{sr} +\frac{\rho_r}{\tau_{rs}}\hat{\bm{u}}_{rs} &= \frac{\rho_s}{\tau_{sr}}{\bm{u}}_{s} +\frac{\rho_r}{\tau_{rs}}{\bm{u}}_{r}, \\
		\frac{1}{\tau_{sr}}\left[\frac{3}{2}n_sk_B(\hat{T}_{sr}-T_s)+\frac{1}{2}\rho_s\left(\hat{u}^2_{sr}-u^2_s\right)\right] &+ \frac{1}{\tau_{rs}}\left[\frac{3}{2}n_rk_B(\hat{T}_{rs}-T_r)+\frac{1}{2}\rho_r\left(\hat{u}^2_{rs}-u^2_r\right)\right] =0. 
	\end{aligned}
\end{equation}

Therefore, we further impose the following assumptions \eqref{eq:assumptions_auxiliary} to determine the auxiliary velocities and temperatures in inter-species collision operators,
\begin{equation}\label{eq:assumptions_auxiliary}
	\begin{aligned}[b]
		\hat{\bm{u}}_{sr}-\hat{\bm{u}}_{rs} &= (1-a_{sr})\left({\bm{u}}_{s}-{\bm{u}}_{r}\right)-b_{sr}\left(\nabla{\ln T_s}+\nabla{\ln T_r}\right), \\
		\hat{T}_{sr}-\hat{T}_{rs} &= (1-c_{sr})\left(T_{s}-T_{r}\right)-d_{sr}\left({\bm{u}}_{s}-{\bm{u}}_{r}\right)^2,
	\end{aligned}
\end{equation}
where $a_{sr}=a_{rs},~b_{sr}=-b_{rs},~c_{sr}=c_{rs},~d_{sr}=-d_{rs}$ are the adjustable parameters, which describe how rapidly the equilibrium among different gas components can be achieved through inter-species collisions. Note that, compared with those in literature (such as \cite{Haack2017JSP,Bobylev2018KRM}), \eqref{eq:assumptions_auxiliary} gives a more general expression for the relations between the auxiliary and macroscopic properties, by adding an additional term of the temperature gradient to phenomenologically model the possible thermally induced flow. Then, combined with \eqref{eq:conservation_inter_2}, the auxiliary velocities and temperatures are given by
\begin{equation}\label{eq:auxiliary_u_T}
	\begin{aligned}[b]
		\hat{\bm{u}}_{sr} &= {\bm{u}}_{s}-\frac{\rho_r\tau_{sr}}{\rho_s\tau_{rs}+\rho_r\tau_{sr}}\bm{{X}}_{sr}, \\
		\hat{T}_{sr} &= T_s-\frac{n_r\tau_{sr}}{n_s\tau_{rs}+n_r\tau_{sr}}{Y}_{sr}-\frac{\rho_s\rho_r\tau_{sr}\tau_{rs}\bm{{X}}_{sr}\cdot\left[\bm{{X}}_{sr}-2\left({\bm{u}}_{s}-{\bm{u}}_{r}\right)\right]}{3k_B\left(n_s\tau_{rs}+n_r\tau_{sr}\right)\left(\rho_s\tau_{rs}+\rho_r\tau_{sr}\right)},
	\end{aligned}
\end{equation}
with
\begin{equation}\label{eq:auxiliary_X_Y}
	\begin{aligned}[b]
		\bm{{X}}_{sr} &= a_{sr}\left({\bm{u}}_{s}-{\bm{u}}_{r}\right)+b_{sr}\left(\nabla{\ln T_s}+\nabla{\ln T_r}\right), \\
		{Y}_{sr} &= c_{sr}\left(T_{s}-T_{r}\right)+d_{sr}\left({\bm{u}}_{s}-{\bm{u}}_{r}\right)^2.
	\end{aligned}
\end{equation}

In addition, the auxiliary properties $\hat{\bm{q}}_{sr}$ are constructed to adjust the relaxation rates of heat fluxes as
\begin{equation}\label{eq:auxiliary_q}
	\begin{aligned}[b]
		\hat{\bm{q}}_{sr} &= \left(1-\text{Pr}_{sr}\right){\bm{q}}_{s}+\gamma_{sr}\left({\bm{q}}_{sr}-{\bm{q}}_{s}\right),
	\end{aligned}
\end{equation}
where $\text{Pr}_{sr}$ is an effective Prandtl number giving the thermal relaxation of species $s$ due to collisions with species $r$, which reduces to $\text{Pr}_{ss}=2/3$ for the intra-species collisions of a monatomic gas; ${\bm{q}}_{sr}$ is defined as the heat flux of species $s$ measured relative to auxiliary velocity $\hat{\bm{u}}_{sr}$,
\begin{equation}\label{eq:qsr}
	\begin{aligned}[b]
		{\bm{q}}_{sr} &= \left<\frac{1}{2}m_s(\bm{v}-\hat{\bm{u}}_{sr})^2(\bm{v}-\hat{\bm{u}}_{sr}),f_s\right>,
	\end{aligned}
\end{equation}
and thus \eqref{eq:auxiliary_q} yields $\hat{\bm{q}}_{ss} = \left(1-\text{Pr}_{ss}\right){\bm{q}}_{s}$ for the intra-species collision term; $\gamma_{sr}=-\gamma_{rs}$ is a dimensionless coefficient taking into account the Dufour effects caused by the diffusive thermal conductivity. Based on the asymptotic analysis of the Boltzmann equation \citep{CE}, it is an effect inverse to thermal diffusion and hence $\gamma_{sr}$ is not an independent parameter, which can be determined by the other adjustable parameters as shown in the following section.

\subsection{Continuum limit}

The multi-relaxation kinetic model equations with the auxiliary properties have been constructed for a monatomic gas mixture with disparate mass, which involves several adjustable parameters: relaxation rates $\tau_{sr}$, effective Prandtl number $\text{Pr}_{sr}$, and the set of coefficients $a_{sr}$, $b_{sr}$, $c_{sr}$, $d_{sr}$, $\gamma_{sr}$ used for calculating auxiliary properties. Here, we perform the Chapman–Enskog analysis to get the asymptotic limit of the proposed model equation and determine the relevant parameters to recover the transport coefficients, so that the kinetic model and macroscopic fluid dynamics are consistent in the continuum flow regime, which is the basic requirement to achieve accurate kinetic modelling.

Without loss of generality, we consider a binary mixture of monatomic gases, and the extension to a multi-species mixture can be straightforward. In the continuum limit, when all the relaxation times are considerably smaller than the characteristic time of gas flow, only the species number density $n_s$, total momentum $\rho\bm{u}$ and energy $\frac{3}{2}nk_BT+\frac{1}{2}\rho u^2$ are the collisional invariants. Then, the set of equations for the conserved macroscopic variables $n_s,~\bm{u},~T$ can be obtained by taking momentums of the kinetic equations \eqref{eq:kinetic_equation}, and summing over the species ($s=1,2$) for momentum and energy equations
\begin{equation}\label{eq:macroscopic_equation}
	\begin{aligned}[b]
		\frac{\partial{n_s}}{\partial{t}} + \nabla\cdot\left(n_s\bm{u}\right) + \nabla\cdot\left(n_s(\bm{u}_s-\bm{u})\right) &= 0, \\
		\frac{\partial}{\partial{t}}\left(\rho\bm{u}\right) + \nabla\cdot\left(\rho\bm{u}\bm{u}\right) + \nabla\cdot\bm{P} &= \rho_1\bm{a}_1+\rho_2\bm{a}_2, \\
		\frac{\partial}{\partial{t}}\left(\frac{3}{2}nk_BT+\frac{1}{2}\rho u^2\right) + \nabla\cdot\left(\left(\frac{3}{2}nk_BT+\frac{1}{2}\rho u^2\right)\bm{u}\right) + \nabla\cdot\left(\bm{P}\cdot\bm{u}\right) + \nabla\cdot\bm{q} &= \rho_1\bm{a}_1\cdot\bm{u}_1+\rho_2\bm{a}_2\cdot\bm{u}_2,
	\end{aligned}
\end{equation} 
where $\bm{u}_s$ is the species macroscopic velocity defined in \eqref{eq:species_macroscopic_variables_f}, $\bm{P}$ and $\bm{q}$ are the mixture pressure tensor and heat flux, respectively, given in \eqref{eq:mixture_macroscopic_variables_f}. To close the above set of equations, it is necessary to find the approximations to distribution functions $f_s$, and hence the properties $\bm{u}_s,~\bm{P},~\bm{q}$ can be expressed as functions of the macroscopic variables $n_s,~\bm{u},~T$, giving the constitutive relations and transport properties of the gas mixture (see the details in Appendix \ref{app:A}).

The Navier–Stokes-type equation for gas mixture is then obtained by the second approximation of $f_s$, and the species velocities $\bm{u}_s^{\text{NS}}$, stress tensor $\bm{P}^{\text{NS}}$ and heat flux $\bm{q}^{\text{NS}}$ are given by,
\begin{equation}\label{eq:u_P_q_NS}
	\begin{aligned}[b]
		\bm{u}_1^{\text{NS}} =&~ \bm{u} -\frac{\rho_1\tau_{21}+\rho_2\tau_{12}}{a_{12}}\frac{p}{\rho\rho_1}\bm{d}_{12}-\frac{2b_{12}\rho_2}{a_{12}\rho}\nabla{\ln T}, \\
		\bm{u}_2^{\text{NS}} =&~ \bm{u}+ \frac{\rho_1\tau_{21}+\rho_2\tau_{12}}{a_{12}}\frac{p}{\rho\rho_2}\bm{d}_{12}+\frac{2b_{12}\rho_1}{a_{12}\rho}\nabla{\ln T}, \\
		\bm{P}^{\text{NS}} =&~ nk_BT\bm{\mathrm{I}} - k_BT\left(\frac{\tau_{11}\tau_{12}}{\tau_{11}+\tau_{12}}n_1+\frac{\tau_{22}\tau_{21}}{\tau_{22}+\tau_{21}}n_2\right)\left(\nabla\bm{u}+\nabla\bm{u}^{\mathrm{T}}-\frac{2}{3}\nabla\cdot\bm{u}\bm{\mathrm{I}}\right) , \\
		\bm{q}^{\text{NS}} =&~ \frac{5}{2}k_BT\left(n_1\bm{u}_1^{\text{NS}}+n_2\bm{u}_2^{\text{NS}}-n\bm{u}\right) + \frac{2b_{12}\rho_1\rho_2}{\rho_1\tau_{21}+\rho_2\tau_{12}}\left({\bm{u}}_{1}^{\text{NS}}-{\bm{u}}_{2}^{\text{NS}}\right) \\
		&-\left[\left(\frac{n_1}{m_1}\frac{\tau_{11}\tau_{12}}{\text{Pr}_{12}\tau_{11}+\text{Pr}_{11}\tau_{12}}+\frac{n_2}{m_2}\frac{\tau_{22}\tau_{21}}{\text{Pr}_{21}\tau_{22}+\text{Pr}_{22}\tau_{21}}\right)\frac{5k_B^2T}{2} - \frac{4b_{12}^2\rho_1\rho_2}{a_{12}\left(\rho_1\tau_{21}+\rho_2\tau_{12}\right)T}\right]\nabla T.
	\end{aligned}
\end{equation}   
where $p=nk_BT$ is the pressure, $\bm{\mathrm{I}}$ is a $3\times3$ identity matrix, and $\bm{d}_{12}$ is given as
\begin{equation}\label{eq:d12}
	\begin{aligned}[b]
		\bm{d}_{12} = \nabla \left(\frac{n_1}{n}\right) +\frac{n_1n_2(m_2-m_1)}{n\rho}\nabla{\ln p} -\frac{\rho_1\rho_2}{\rho p}\left(\bm{a}_1-\bm{a}_2\right),
	\end{aligned}
\end{equation}
which yields $\bm{d}_{12}=-\bm{d}_{21}$, since $\nabla \left({n_1}/{n}\right) =-\nabla \left({n_2}/{n}\right)$. 

\subsubsection{Viscosity}

The shear viscosity $\mu$ of a binary mixture can be obtained from the off-diagonal components of the non-equilibrium stress tensor $\bm{P}$ in \eqref{eq:u_P_q_NS}, which leads to
\begin{equation}\label{eq:mu_mixture}
	\begin{aligned}[b]
		\mu &= k_BT\left(\frac{\tau_{11}\tau_{12}}{\tau_{11}+\tau_{12}}n_1+\frac{\tau_{22}\tau_{21}}{\tau_{22}+\tau_{21}}n_2\right).
	\end{aligned}
\end{equation} 
Clearly, the mixture shear stress depends on all the relaxation times including both intra- and inter-species interactions. It is known that the dependence of species shear viscosity $\mu_s$ on mean collision time $\tau_{ss}$ of intra-species collisions is $\mu_s=n_sk_BT\tau_{ss}$. However, the relaxation time $\tau_{sr}~(s\neq r)$ in the kinetic model is no longer a mean molecular collision time of inter-species interactions, but measures the time scale approaching the reference states $g_{sr}$ for the component $s$ due to the collisions with component $r$. In other words, $n_r\tau_{sr}$ should be smaller than $n_s\tau_{rs}$ when $m_s<m_r$, because of the fact that collisions between molecules with mass disparity have a more significant influence on the light one than on the heavy one. Particularly, by considering a mixture consisting of ions and electrons with mass $m_i$ and $m_e$, respectively, the approximated relation between the inter-species relaxation times given in the literature is $n_i\tau_{ei}/n_e\tau_{ie}=m_e/m_i\ll 1$ \citep{Bellan2006}.

Therefore, here we define 
\begin{equation}
	\begin{aligned}
	 \phi_{sr}=\frac{n_s\tau_{ss}}{n_r\tau_{sr}}
	\end{aligned}
\end{equation}
to quantify the ratio between the relaxation times of one single intra- and inter-species collision for species $s$, where $\phi_{sr}$ is only determined by the intermolecular interactions and temperature, but independent of the concentrations of the components. Then the mixture viscosity \eqref{eq:mu_mixture} can be rewritten in the form
\begin{equation}\label{eq:mu_mixture_s}
	\begin{aligned}[b]
		\mu &= \frac{\mu_1}{1+\frac{n_2}{n_1}\phi_{12}} + \frac{\mu_2}{1+\frac{n_1}{n_2}\phi_{21}},
	\end{aligned}
\end{equation} 
which is a linear combination of the species viscosities of the mixture components and shares the exact same form as that given by Wilke’s mixture rule \citep{Wilke1950JCP}. Nevertheless, the more accurate values of $\phi_{sr}$ can be obtained by fitting the mixture viscosity measured experimentally. 

\subsubsection{Diffusion}

In a gas mixture, the ordinary diffusion coefficient $D_{sr}$ is usually measured when the gas mixture is uniform in temperature and pressure and without external forces acting on the gas molecules; while the thermal diffusion with coefficient $D_{T,sr}$ also contributes to the diffusion velocity when there is a temperature gradient present. In general, the mass flux $J_s$ of the species $s$ in a binary mixture caused by gradients of concentration, pressure and temperature is given by \citep{Hirschfelder1954},
\begin{equation}\label{eq:mass_flux}
	\begin{aligned}[b]
		J_s = \rho_s\left(\bm{u}_s-\bm{u}\right) &= -\frac{n^2}{\rho}m_sm_r\left[D_{sr}\bm{d}_{sr}-D_{T,sr}\nabla{\ln T}\right].
	\end{aligned}
\end{equation} 
Meanwhile, the diffusion velocity in the proposed kinetic model is obtained as \eqref{eq:u_P_q_NS} from the Chapman-Enskog method,
\begin{equation}\label{eq:diffusion_velocity}
	\begin{aligned}[b]
		\bm{u}_s-\bm{u} &= -\frac{\rho_1\tau_{21}+\rho_2\tau_{12}}{a_{sr}}\frac{p}{\rho_1\rho_2}\bm{d}_{sr}-\frac{2b_{sr}}{a_{sr}}\nabla{\ln T}.
	\end{aligned}
\end{equation} 
Therefore, the binary diffusion coefficient $D_{12}$ yields
\begin{equation}\label{eq:D12}
	\begin{aligned}[b]
		D_{12} & = \frac{k_BT}{m_1m_2n}\frac{\rho_1\tau_{21}+\rho_2\tau_{12}}{a_{12}},
	\end{aligned}
\end{equation} 
which determines the parameter $a_{12}=a_{21}$ when the relaxation times are known from viscosity. Also, the thermal diffusion coefficient $D_{T,12}$, as well as the thermal-diffusion ratio $k_{T,12}=D_{T,12}/D_{12}$ are obtained,
\begin{equation}\label{eq:DT_kT}
	\begin{aligned}[b]
		D_{T,12} & = \frac{2b_{12}n_1n_2}{a_{12}n^2}, \quad
		k_{T,12} = \frac{2b_{12}\rho_1\rho_2}{p\left(\rho_1\tau_{21}+\rho_2\tau_{12}\right)}.
	\end{aligned}
\end{equation} 
Thus, the parameter $b_{12}=-b_{21}$ is adjusted to match the thermal-diffusion ratio, and recover the Soret effect in the continuum limit. The Soret effect describes mass separation due to a temperature gradient, and it may be positive or negative depending on the mass difference and intermolecular potentials. In general, thermal diffusion becomes stronger in molecules with a larger mass difference, and is thus crucial in the modelling of gas mixtures with significant mass disparity. Meanwhile, though the absolute value of thermal-diffusion ratio $k_{T,12}$ usually has an order of magnitude less than $10^{-1}$ for neutral gas mixture, it can be greatly increased in ionized gases \citep{Chapman1958PPS}.

\subsubsection{Thermal conductivity}

In addition to the direct transfer of kinetic energy during collisions, the diffusional migration of molecules also carries thermal energy and contributes to the heat transport in a gas mixture, which is known as the Dufour effect. Based on the Chapman-Enskog method, the mixture heat flux $\bm{q}$ is obtained as,
\begin{equation}\label{eq:heat_flux}
	\begin{aligned}[b]
		\bm{q} =&~ \frac{5}{2}k_BT\left(n_1\left(\bm{u}_1-\bm{u}\right)+n_2\left(\bm{u}_2-\bm{u}\right)\right) + \frac{5}{2}k_BTA\left({\bm{u}}_{1}-{\bm{u}}_{2}\right)\\
		& - \frac{5}{2}k_BT\left(\frac{n_1}{m_1}\frac{\tau_{11}\tau_{12}}{\text{Pr}_{12}\tau_{11}+\text{Pr}_{11}\tau_{12}}+\frac{n_2}{m_2}\frac{\tau_{22}\tau_{21}}{\text{Pr}_{21}\tau_{22}+\text{Pr}_{22}\tau_{21}} - A\frac{2b_{12}}{a_{12}k_BT}\right)k_B\nabla T.
	\end{aligned}
\end{equation} 
with
\begin{equation}\label{eq:heat_flux_A}
	\begin{aligned}[b]
		A =&~ \gamma_{12} \left(\frac{1}{m_1}\frac{\tau_{11}\tau_{12}}{\text{Pr}_{12}\tau_{11}+\text{Pr}_{11}\tau_{12}}+\frac{1}{m_2}\frac{\tau_{22}\tau_{21}}{\text{Pr}_{21}\tau_{22}+\text{Pr}_{22}\tau_{21}}\right)\frac{a_{12}\rho_1\rho_2}{\rho_1\tau_{21}+\rho_2\tau_{12}}.
	\end{aligned}
\end{equation} 
It can be seen that there are three parts contributing to the total heat flux: (i) The first term occurs since the heat flux is measured relative to the mixture flow velocity, instead of the species flow velocity, and thus represents energy carried by the molecular flux in the presence of diffusion $\bm{u}_s-\bm{u}$. (ii) The second term with binary diffusion velocity $\bm{u}_1-\bm{u}_2$ arises as an inverse process to thermal diffusion, and has a coefficient $\frac{5}{2}k_BTA=nk_BTk_{T,12}$ based on the asymptotic analysis of the original Boltzmann equation \citep{CE}, leading to
\begin{equation}\label{eq:gamma}
	\begin{aligned}[b]
		\gamma_{12} =\left(\frac{1}{m_1}\frac{\tau_{11}\tau_{12}}{\text{Pr}_{12}\tau_{11}+\text{Pr}_{11}\tau_{12}}+\frac{1}{m_2}\frac{\tau_{22}\tau_{21}}{\text{Pr}_{21}\tau_{22}+\text{Pr}_{22}\tau_{21}}\right)^{-1}\frac{4b_{12}}{5a_{12}k_BT}.
	\end{aligned}
\end{equation} 
(iii) The third term $-\kappa\nabla T$ is generated by a temperature gradient, where $\kappa$ is the ordinary thermal conductivity of the mixture that is usually measured experimentally in the absence of any diffusion velocity
\begin{equation}\label{eq:kappa_mixture}
	\begin{aligned}[b]
		\kappa = \left(\frac{n_1}{m_1}\frac{\tau_{11}\tau_{12}}{\text{Pr}_{12}\tau_{11}+\text{Pr}_{11}\tau_{12}}+\frac{n_2}{m_2}\frac{\tau_{22}\tau_{21}}{\text{Pr}_{21}\tau_{22}+\text{Pr}_{22}\tau_{21}}\right)\frac{5k_B^2T}{2} - \frac{4b_{12}^2\rho_1\rho_2}{a_{12}\left(\rho_1\tau_{21}+\rho_2\tau_{12}\right)T}.
	\end{aligned}
\end{equation} 
In analogy with the parameter $\phi_{sr}$ used to measure the ratio between the relaxation times of intra- and inter-species collisions, we also define
\begin{equation}
	\begin{aligned}
		\varphi_{sr}=\frac{\text{Pr}_{sr}}{\text{Pr}_{ss}}
	\end{aligned}
\end{equation}
to represent the ratio of the thermal relaxation rates. Given the species thermal conductivity $\kappa_s ={5n_sk_B^2T\tau_{ss}}/{2m_s\text{Pr}_{ss}}$, the mixture thermal conductivity \eqref{eq:kappa_mixture} can be rewritten in the form,
\begin{equation}\label{eq:kappa_mixture_s}
	\begin{aligned}[b]
		\kappa = \frac{\kappa_1}{1+\frac{n_2}{n_1}\phi_{12}\varphi_{12}} + \frac{\kappa_2}{1+\frac{n_1}{n_2}\phi_{21}\varphi_{21}} - \frac{D_{12}k_T^2n^3k_B}{n_1n_2},
	\end{aligned}
\end{equation} 
where the last term indicates the effect of thermal diffusion on the thermal conductance of a gas mixture. Similar to the viscosity, the values of $\varphi_{sr}$ can be determined by matching the mixture thermal conductivity measured at different proportions of gas components. 

\subsection{Inter-species energy relaxation}

We have shown that the temperatures of different components stay the same up to the Navier–Stokes approximation of the proposed model in the continuum limit, when all the relaxation times are considerably smaller than the characteristic time of gas flow. Thus the transport coefficients in the continuum limit are not affected by the parameters $c_{sr}$ and $d_{sr}$. However, these parameters determine the auxiliary temperatures and hence the energy relaxation rates between different species, which may have a significant impact in strong non-equilibrium situations. It implies that in a gas mixture, having all the transport coefficients is not enough to exactly describe the energy relaxation during inter-species collisions. Recovery of underlying relaxation processes is crucial to give accurate kinetic modelling of rarefied gas flow, which is systematically analysed in our previous work on single-species polyatomic gas flows \citep{Li2021JFM,Li2023JFM,Zeng2022AAS}.

Therefore, we determine the parameters $c_{sr}$ and $d_{sr}$ for calculating auxiliary temperature by imposing that the energy exchange rates of the inter-species collision operator coincide with that of the Boltzmann collision operator,
\begin{equation}\label{eq:energy_exchange_condition}
	\begin{aligned}[b]
		\left<\frac{1}{2}m_sv^2,\frac{1}{\tau_{sr}}(g_{sr}-f_{s})\right> &= \left<\frac{1}{2}m_sv^2,Q_{sr}\right>. 
	\end{aligned}
\end{equation}
The exchange rates of the kinetic model (left-hand side) can be calculated straightforwardly, while that of the Boltzmann collision operator may only be explicitly evaluated for the Maxwellian intermolecular potential,
\begin{equation}\label{eq:energy_exchange_Maxwell}
	\begin{aligned}[b]
		\frac{n_s}{\tau_{sr}}\left[\frac{3}{2}k_B\left(\hat{T}_{sr}-T_s\right)+\frac{m_s}{2}\left(\hat{\bm{u}}_{sr}-\bm{u}_s\right)^2\right] &= \lambda_{sr}\frac{m_sm_r}{(m_s+m_r)^2}n_sn_r\left[3k_B(T_r-T_s)+m_r(\bm{u}_r-\bm{u}_s)^2\right],
	\end{aligned}
\end{equation}
where $\lambda_{sr}=\lambda_{rs}$ are constants related to the integral of collision cross-sections. By substituting the auxiliary velocity and temperature \eqref{eq:auxiliary_u_T} into \eqref{eq:energy_exchange_Maxwell}, we immediately get
\begin{equation}\label{eq:c_d_Maxwell}
	\begin{aligned}[b]
		c_{sr} &= \frac{2\lambda_{sr}m_sm_r}{(m_s+m_r)^2}\left(n_s\tau_{rs}+n_r\tau_{sr}\right), \\
		d_{sr} &= \frac{\lambda_{sr}m_sm_r}{3k_B\left(m_s+m_r\right)^2}\left[{\lambda_{sr}\left(n_r\rho_r\tau_{sr}^2-n_s\rho_s\tau_{rs}^2\right)} - {2\left(\rho_r\tau_{sr}-\rho_s\tau_{rs}\right)}\right]. 
	\end{aligned}
\end{equation}
Note that the same form of \eqref{eq:energy_exchange_Maxwell} can be obtained for non-Maxwellian intermolecular potentials with non-constant $\lambda_{sr}$, when it is subject to the restriction that the distribution functions are Maxwellians at different temperatures but with small diffusion velocities \citep{Morse1964PoF}. Therefore, we calculate the parameters $c_{sr}$ and $d_{sr}$ using \eqref{eq:c_d_Maxwell} for any type of intermolecular potential. 

Furthermore, the collision cross-sections related variables $\lambda_{sr}$ can be approximated by matching the momentum exchange rates of the inter-species collision operator with that of the Boltzmann collision operator for the Maxwellian intermolecular potential,
\begin{equation}\label{eq:momentum_exchange_condition}
	\begin{aligned}[b]
		\left<m_sv,\frac{1}{\tau_{sr}}(g_{sr}-f_{s})\right> &= \left<m_sv,Q_{sr}\right>, 
	\end{aligned}
\end{equation}
which leads to
\begin{equation}\label{eq:momentum_exchange_Maxwell}
	\begin{aligned}[b]
		\frac{n_sm_s}{\tau_{sr}}\left(\hat{\bm{u}}_{sr}-{\bm{u}}_{s}\right) &= \lambda_{sr}\frac{m_sm_r}{m_s+m_r}n_sn_r\left({\bm{u}}_{r}-{\bm{u}}_{s}\right).
	\end{aligned}
\end{equation}
Therefore, $\lambda_{sr}$ can be evaluated by parameter $a_{sr}$ as
\begin{equation}\label{eq:integral_constant}
	\begin{aligned}[b]
		\lambda_{sr} = \frac{a_{sr}(m_s+m_r)}{\rho_s\tau_{rs}+\rho_r\tau_{sr}},
	\end{aligned}
\end{equation}
which is thus determined by the binary diffusion coefficient as given by \eqref{eq:D12}.

\subsection{Indifferentiability principle}

The indifferentiability principle states that, when all the molecules are mechanically identical (e.g. they have the same mass and scattering cross-section), the model equation reduces to a single one by adding the distribution functions \citep{Garzo1989PoFA}. This property holds for the Boltzmann equation with only binary collisions because of the bilinearity of its operators. It is however nontrivial to inherit for a model equation since the operators constructed are usually highly nonlinear. Historically, several kinetic models using a single relaxation collision operator have been proven to fulfil the principle. It should be noted that some of them require a condition that the diffusion velocities vanish for the indifferentiable molecules, when the models contain parameters to recover the Fick law \citep{Brull2012EJMB,Todorova2019EJMB}.

Therefore, we also adopt the assumption that $\bm{u}_s=\bm{u}$ for all the indifferentiable species $s$, to demonstrate that our multi-relaxation model with linearised collision operator complies with the indifferentiability principle. Consider a system of gas mixture that slightly deviates from an equilibrium state with flow velocity $\bm{u}$, temperature $T$, and number density $n_s$ of each component, the reference distribution in the linearised collision operator is given as
\begin{equation}\label{eq:linear_g}
	\begin{aligned}[b]
		g_{sr}^{linear} &= f_{s}^{eq}\left[1 + \frac{m_s\left(\hat{\bm{u}}_{sr}-\bm{u}\right)\cdot\bm{c}}{k_BT} + \frac{\hat T_{sr}-T}{T}\left(\frac{m_sc^2}{2k_B{T}}-\frac{3}{2}\right) + \frac{2m_s\hat{\bm{q}}_{sr}\cdot\bm{c}}{5{n}_{s}k_B^2{T}^2}\left(\frac{m_sc^2}{2k_B{T}}-\frac{5}{2}\right)\right],
	\end{aligned}
\end{equation}
with
\begin{equation}\label{eq:fs_eq}
	\begin{aligned}[b]
		f_s^{eq} = n_s\left(\frac{m_s}{2\pi k_BT}\right)^{3/2}\exp{\left(-\frac{m_s\left(\bm{v}-{\bm{u}}\right)^2}{2k_BT}\right)}.
	\end{aligned}
\end{equation} 

For the indifferentiable molecules, we have (i) $m_s=m$, and $n_r\tau_{sr}$ is constant for any $s$ and $r$ due to the identical mass and scattering cross-sections, respectively; (ii) $\hat{\bm{u}}_{sr}=\bm{u}$ based on \eqref{eq:auxiliary_u_T} with the thermal diffusion coefficient vanishing; (iii) $\hat T_{sr}+\hat T_{rs}=T_s+T_r$ and $nT=\sum n_sT_s$ from total energy conservation \eqref{eq:conservation_inter_2} and calculation of mixture temperature \eqref{eq:mixture_macroscopic_variables_f}, respectively; (iv) $\hat{\bm{q}}_{sr}=(1-\text{Pr})\bm{q}_{s}$ from \eqref{eq:auxiliary_q}.
Then, the sum of the kinetic equations over all species yields,
\begin{equation}\label{eq:indifferentiability}
	\begin{aligned}[b]
		{\mathcal{D}\left(\sum_{s=1}^Nf_{s}\right)} &= 
		\sum_{s=1}^N{\sum_{r=1}^N\frac{1}{\tau_{sr}}\left(g_{sr}^{linear}-f_s\right)} \\
		&= \frac{1}{\tau}\left(\sum_sf_s^{eq}\left(1+\frac{2m(1-\text{Pr}){\left(\sum_s\bm{q}_{s}\right)}\cdot\bm{c}}{5{n}_{s}k_B^2{T}^2}\left(\frac{mc^2}{2k_B{T}}-\frac{5}{2}\right)\right)-\sum_{s=1}^Nf_{s}\right),
	\end{aligned}
\end{equation}
where $\tau=n_r\tau_{sr}/n$ is the overall relaxation time. Clearly, the kinetic model equation reduces to the Shakhov model of single-species monatomic gas with distribution function $f=\sum f_{s}$.

\subsection{Dimensionless forms}

Let $L_0,~T_0,~n_0,~m_0$ be the reference length, temperature, number density and mass, respectively, then the most probable speed is $v_m=\sqrt{2k_BT_0/m_0}$ and reference pressure is $p_0=n_0k_BT_0$. The dimensionless variables are introduced as,
\begin{equation}\label{eq:dimensionless_variables}
	\begin{aligned}[b]
		&\tilde{\bm{x}}=\bm{x}/L_0, \quad\quad \tilde{n}=n/n_0, \quad\quad \tilde{m}=m/m_0, \quad\quad \tilde{T}=T/T_0, \\
		&\tilde{\bm{v}}=\bm{v}/v_m, \quad\quad \tilde{\bm{c}}=\bm{c}/v_m,  \quad\quad \tilde{t}=v_mt/L_0, \quad\quad \tilde{\tau}=v_m\tau/L_0, \\
		&\tilde{p}=p/p_0, \quad\quad \tilde{\bm{q}}=\bm{q}/(p_0v_m), \quad\quad \tilde{f}_s=v_m^{3}f_s/n_0.
	\end{aligned}
\end{equation}

The Knudsen numbers $\text{Kn}_{s}$ of each species $s$ is defined as
\begin{equation}\label{eq:Kn_s}
	\begin{aligned}[b]
		\text{Kn}_{s}&=\frac{\mu_s(T_0)}{n_0L_0}\sqrt{\frac{\pi}{2m_sk_BT_0}}.
	\end{aligned}
\end{equation}
It is noted that the species-specific Knudsen numbers are correlated as $\text{Kn}_{r}=\text{Kn}_{s}\beta^{\mu}_{rs}/\sqrt{\beta^m_{rs}}$, with $\beta^{\mu}_{rs}=\mu_r(T_0)/\mu_s(T_0)$ being the viscosity ratio at the reference temperature, and $\beta^m_{rs}=m_r/m_s$ the mass ratio.
Therefore, the dimensionless relaxation times can be written in terms of the Knudsen numbers,
\begin{equation}\label{eq:dimensionless_tau}
	\begin{aligned}[b]
		\tilde{\tau}_{ss} &= 2\text{Kn}_{s}\sqrt{\frac{\tilde{m}_s}{\pi}}\frac{\tilde{T}_s^{\omega_s-1}}{\tilde{n}_s}, \quad
		\tilde{\tau}_{sr} = \tilde{\tau}_{ss}\phi_{sr}^{-1}\frac{\tilde{n}_s}{\tilde{n}_r},
	\end{aligned}
\end{equation}
where $\omega_s$ is the viscosity index of species $s$ in
\begin{equation}\label{eq:viscosity_temperature}
	\mu_s(T)=\mu_s(T_0)\left(\frac{T}{T_0}\right)^\omega_s.
\end{equation}

Then, the kinetic model equations are non-dimensionalised as
\begin{equation}\label{eq:nondimensional_kinetic_equation}
	\begin{aligned}[b]
		\frac{\partial{\tilde{f}_{s}}}{\partial{\tilde{t}}}+\tilde{\bm{v}} \cdot \frac{\partial{\tilde{f}_{s}}}{\partial{\tilde{\bm{x}}}}+ \tilde{\bm{a}}_s\cdot\frac{\partial{\tilde{f}_{s}}}{\partial{\tilde{\bm{v}}}}&=\sqrt{\frac{\pi}{\tilde{m}_s}}\frac{\tilde{T}_s^{1-\omega_s}}{2\text{Kn}_{s}}\left[\tilde{n}_s\left(\tilde{g}_{ss}-\tilde{f}_s\right)+\sum_r\tilde{n}_r\phi_{sr}\left(\tilde{g}_{sr}-\tilde{f}_s\right)\right],
	\end{aligned}
\end{equation}
with the dimensionless reference velocity distribution function
\begin{equation}\label{eq:dimensionless_g}
	\begin{aligned}[b]
		\tilde{g}_{sr}=~&\tilde{{n}}_{sr}\left(\frac{\tilde{m}_s}{\pi\tilde{\hat{T}}_{sr}}\right)^{3/2}\exp\left(-\frac{\tilde{m}_s(\tilde{\bm{v}}-\tilde{\hat{\bm{u}}}_{sr})^2}{\tilde{\hat{T}}_{sr}}\right) \times \\
		&\left[1+\frac{\tilde{\hat{T}_{sr}}-\tilde{T_s}}{\tilde{T_s}}\left(\frac{\tilde{m}_s(\tilde{\bm{v}}-\tilde{\hat{\bm{u}}}_{sr})^2}{\tilde{\hat{T}}_{sr}}-\frac{3}{2}\right)+\frac{4\tilde{m}_s\tilde{\hat{\bm{q}}}_{sr}\cdot(\tilde{\bm{v}}-\tilde{\hat{\bm{u}}}_{sr})}{5\tilde{\hat n}_{sr}\tilde{\hat{T}}_{sr}^2}\left(\frac{\tilde{m}_s(\tilde{\bm{v}}-\tilde{\hat{\bm{u}}}_{sr})^2}{\tilde{\hat{T}}_{sr}}-\frac{5}{2}\right)\right].
	\end{aligned}
\end{equation}

It clearly shows that the strengths of intra- and inter-species collisions are indicated by the magnitudes of $\tilde{n}_s$ and $\tilde{n}_r\phi_{sr}$, respectively. Therefore, in a gas mixture with a large disparity in concentration or mass, the intra-species collisions become dominant for the component $s$ with a major proportion of number density ($\tilde{n}_s \gg \tilde{n}_r$) or significantly heavier mass ($\phi_{sr} \ll 1$ when $m_r\ll m_s$).

\section{Determination of parameters}\label{sec:parameters}

Given the relationship between the model parameters and transport coefficients of the mixtures as \eqref{eq:mu_mixture_s}, \eqref{eq:D12}, \eqref{eq:DT_kT} and \eqref{eq:kappa_mixture_s}, the adjustable parameters can be uniquely determined by the experimentally measured properties of gas mixtures directly, without the knowledge of any intermolecular potentials that is also constructed to approximate the real gas properties.

Nevertheless, to validate our kinetic model in various rarefied flow problems of gas mixtures with a wide range of mass ratios and different types of molecular interactions, we compare the solutions of the kinetic model with DSMC results for virtual gases with well-defined intermolecular potentials. The transport properties of a gas in DSMC simulations are the result of the transfer of mass, momentum, and energy through particle movement and collision dynamics of corresponding collision models. Given the information of any intermolecular potentials, the DSMC method with the variable-soft-sphere (VSS) collision model \citep{Koura1991PoFA} captures the viscosity and diffusion cross sections simultaneously, and thus provides reference solutions consistent with those from the Boltzmann equation for monatomic gases. Therefore, to make a consistent comparison between the results from the proposed kinetic model and the DSMC method, we determine the parameters in the model equation by matching the transport properties of a gas mixture from the VSS model applied in the DSMC simulations.

The transport coefficients for both simple gases and gas mixtures can be approximated by the Chapman–Enskog solutions of the Boltzmann equation. In a binary gas mixture of monatomic molecules, the first approximations of the transport coefficients, denoted by $[\cdot]_1$, are given by \citep{CE},
\begin{equation}\label{eq:CE_approximations_transport_coefficients}
	\begin{aligned}[b]
        \left[\mu_s\right]_1 &= \frac{5k_BT}{8\Omega_s^{(2)}(2)}, \\
        \left[\kappa_s\right]_1 &= \frac{75k_B^2T}{32m_s\Omega_s^{(2)}(2)}, \\
		\left[D_{12}\right]_1 &= \frac{3E}{2n(m_1+m_2)}, \\
        \left[k_T\right]_1 &= 5C\frac{n_1n_2}{n}\frac{n_1S_1-n_2S_2}{n_1^2Q_1+n_2^2Q_2+n_1n_2Q_{12}}, \\
        \left[\mu\right]_1 &= \frac{n_1^2R_1+n_2^2R_2+n_1n_2R'_{12}}{n_1^2R_1/\left[\mu_1\right]_1+n_2^2R_2/\left[\mu_2\right]_1+n_1n_2R_{12}}, \\
        \left[\kappa\right]_1 &= \frac{n_1^2Q_1\left[\kappa_1\right]_1+n_2^2Q_2\left[\kappa_2\right]_1+n_1n_2Q'_{12}}{n_1^2Q_1+n_2^2Q_2+n_1n_2Q_{12}}, 
	\end{aligned}
\end{equation} 
where $\Omega_s$ is the collision integral of intra-species interactions, the variables $S,~Q,~R$ can be expressed in terms of species viscosities $\left[\mu_s\right]_1$ and mass fraction $M_s=m_s/(m_1+m_2)$,
\begin{align}\label{eq:SQR}
		&S_1 = \frac{M_1^2E}{[\mu_1]_1}-M_2\left(3(M_2-M_1)+4M_1A\right), \notag \\
		&S_2  = \frac{M_2^2E}{[\mu_2]_1}-M_1\left(3(M_1-M_2)+4M_2A\right), \notag \\
		&Q_1 = \frac{M_1E}{[\mu_1]_1}\left(6M_2^2+5M_1^2-4M_1^2B+8M_1M_2A\right), \notag \\
		&Q_2  = \frac{M_2E}{[\mu_2]_1}\left(6M_1^2+5M_2^2-4M_2^2B+8M_2M_1A\right), \notag \\
		&Q_{12} = 3(M_1-M_2)^2(5-4B)+4M_1M_2A(11-4B)+\frac{2M_1M_2E^2}{[\mu_1]_1[\mu_2]_1}, \notag \\
		&Q'_{12} = \frac{15k_BE}{2(m_1+m_2)}\left(\frac{M_1E}{[\mu_1]_1}+\frac{M_2E}{[\mu_2]_1}+(11-4B-8A)M_1M_2\right), \\
		&R_1 = \frac{2}{3}+\frac{M_1A}{M_2}, \quad
		R_2  = \frac{2}{3}+\frac{M_2A}{M_1}, \notag \\
		&R_{12} = \frac{E}{2[\mu_1]_1[\mu_2]_1}+\frac{4A}{3EM_1M_2}, \notag \\
		&R'_{12} = \frac{E}{2[\mu_1]_1}+\frac{E}{2[\mu_2]_1}+2\left(\frac{2}{3}-A\right). 
\end{align} 
Here, $A,~B,~C,~E$ are functions of collision integrals $\Omega_{12}$ of inter-species interactions,
\begin{equation}\label{eq:collision_integrals_ABCE}
	\begin{aligned}[b]
        A &= \frac{\Omega_{12}^{(2)}(2)}{5\Omega_{12}^{(1)}(1)}, \quad 
        B  = \frac{5\Omega_{12}^{(1)}(2)-\Omega_{12}^{(1)}(3)}{5\Omega_{12}^{(1)}(1)}, \\
        C &= \frac{2\Omega_{12}^{(1)}(2)}{5\Omega_{12}^{(1)}(1)}-1, \quad
        E  = \frac{k_BT(m_1+m_2)^2}{8m_1m_2\Omega_{12}^{(1)}(1)},
	\end{aligned}
\end{equation} 

Based on the VSS collision model, the collision integrals can be calculated as given by \cite{Stephani2012PoF},
\begin{align}\label{eq:Omega_integrals_VSS}
        \Omega_{s}^{(2)}(2) &= \frac{4\alpha_s}{(\alpha_s+1)(\alpha_s+2)}\frac{\pi}{2}\left(\frac{k_BT}{\pi m_{s}}\right)^{1/2}\left(\frac{5}{2}-\omega_{s}\right)\left(\frac{7}{2}-\omega_{s}\right)\left(\frac{T_{0}}{T}\right)^{\omega_{s}-1/2}d_{s,ref}^2, \notag \\
        \Omega_{12}^{(1)}(1) &= \frac{2}{\alpha_{12}+1}\frac{\pi}{2}\left(\frac{k_BT}{2\pi m_{12}}\right)^{1/2}\left(\frac{5}{2}-\omega_{12}\right)\left(\frac{T_{0}}{T}\right)^{\omega_{12}-1/2}d_{12,ref}^2, \notag \\
        \Omega_{12}^{(1)}(2) &= \Omega_{12}^{(1)}(1)\left(\frac{7}{2}-\omega_{12}\right), \notag \\
        \Omega_{12}^{(1)}(3) &= \Omega_{12}^{(1)}(1)\left(\frac{7}{2}-\omega_{12}\right)\left(\frac{9}{2}-\omega_{12}\right), \notag \\
        \Omega_{12}^{(2)}(2) &= \frac{4\alpha_{12}}{(\alpha_{12}+1)(\alpha_{12}+2)}\frac{\pi}{2}\left(\frac{k_BT}{2\pi m_{12}}\right)^{1/2}\left(\frac{5}{2}-\omega_{12}\right)\left(\frac{7}{2}-\omega_{12}\right)\left(\frac{T_{0}}{T}\right)^{\omega_{12}-1/2}d_{12,ref}^2, 
\end{align} 
where $m_{12}=m_1m_2/(m_1+m_2)$ is the reduced mass; $d_{ref}$ is the reference collision diameter, $\omega$ is the viscosity index, $\alpha$ is the angular scattering parameter in VSS model, with subscripts $s$ and $12$ indicating the values for intra- and inter-species collisions, respectively. 

Therefore, given the collision parameters in the VSS model, the first approximations of the transport properties of a gas mixture can be explicitly evaluated based on equations \eqref{eq:CE_approximations_transport_coefficients} to \eqref{eq:Omega_integrals_VSS}. The parameters $\phi_{sr}$ measuring the ratio between relaxation times of intra- and inter-species collisions are obtained by fitting the expression \eqref{eq:mu_mixture_s} to the calculated mixture viscosities at different proportions of gas components. Similarly, the ratio of thermal relaxation rates $\varphi_{sr}$ can be fitted by matching the mixture thermal conductivity using \eqref{eq:kappa_mixture_s}, once $\phi_{sr}$ is determined. Also, the parameters $a_{sr}$ and $b_{sr}$ are calculated based on the diffusion coefficients \eqref{eq:D12} and thermal-diffusion ratio \eqref{eq:DT_kT}, respectively. It should be noted that, the first approximations of the transport coefficients for non-Maxwell gases may have differences compared to the exact values. Based on the Chapman–Enskog solutions to higher order in Sonine polynomials \citep{Tipton2009EJMB,Tipton2009EJMB_2}, for a gas mixture consisting of hard-sphere molecules, the discrepancies in mixture viscosity and diffusion coefficient are usually limited, namely, less than 3\% for a wide range of molecular mass, sizes and mole fractions. On the other hand, the discrepancies can be non-negligible in thermal conductivity, that is, around 10\% in some mixtures, and even significant for thermal diffusion coefficients (higher than 20\%).

\begin{figure}[t]
	\centering
	\sidesubfloat[]{\includegraphics[scale=0.3,clip=true]{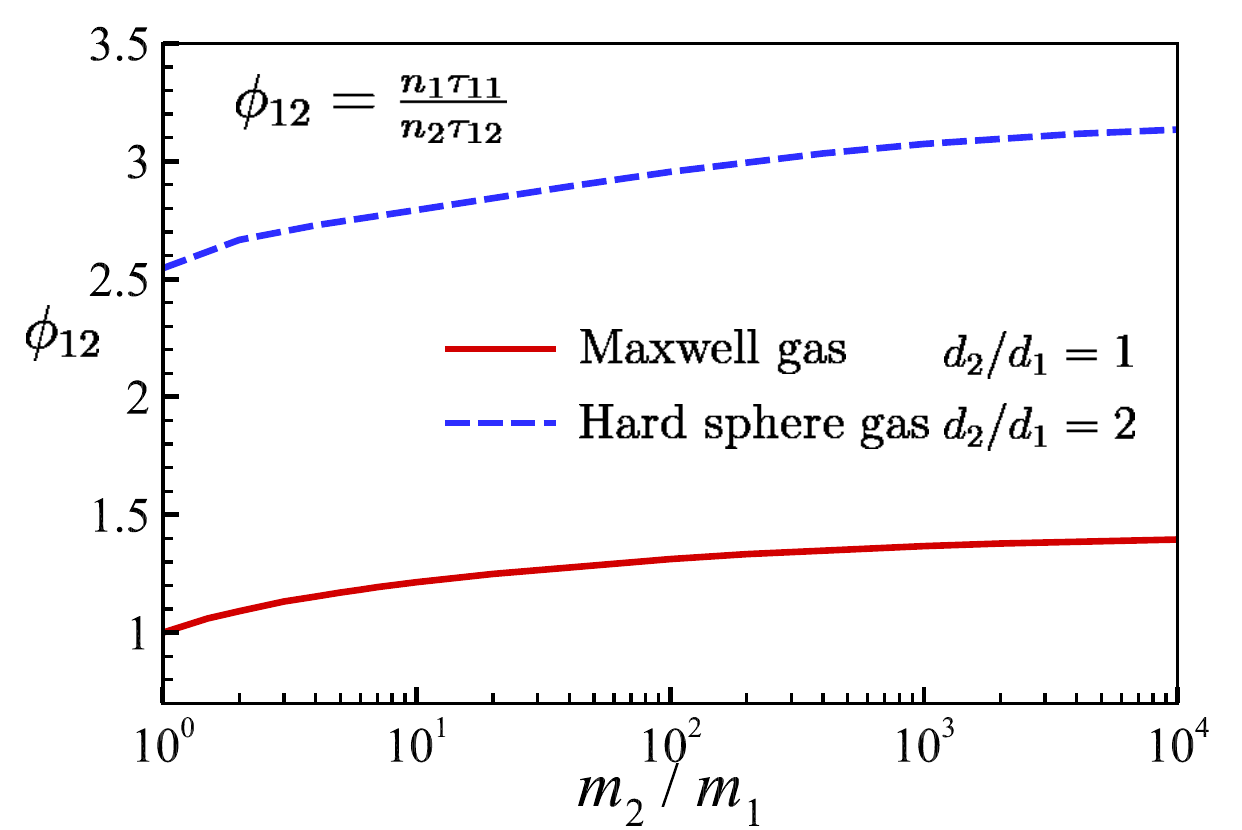}\label{fig:phi_VSS:phi12}}   
	\sidesubfloat[]{\includegraphics[scale=0.3,clip=true]{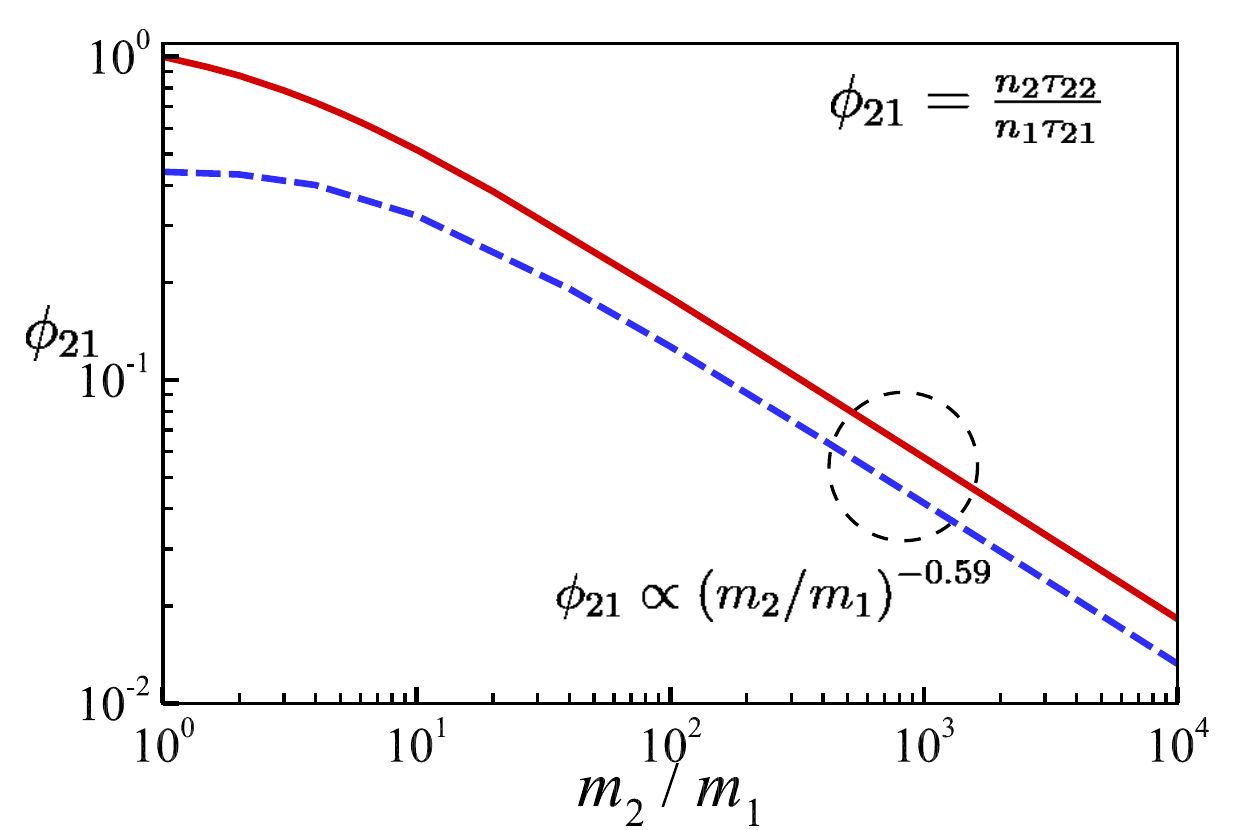}\label{fig:phi_VSS:phi21}}  
	\caption{The relaxation time ratios (a) $\phi_{12}$ and (b) $\phi_{21}$ fitted by the first approximations of the mixture viscosities with the mass ratio $m_2/m_1$ varies from 1 to $10^4$, for Maxwell gas mixtures with a fixed reference diameter ratio $d_2/d_1=1$ and hard-sphere gas mixtures with $d_2/d_1=2$.}
	\label{fig:phi_VSS}
\end{figure}

The value of $\phi_{sr}=n_s\tau_{ss}/n_r\tau_{sr}$ is not only essential to recover the shear viscosity of the gas mixture, but also reveals the relaxation timescales of different types of binary collisions and their corresponding significance in the kinetic model. Then, to explore the dependence of the relaxation time ratio on mass disparity, we obtain $\phi_{sr}$ across a wide range of mass ratios $m_2/m_1=1 \sim 10^4$ for Maxwell gas mixtures with a fixed reference diameter ratio $d_2/d_1=1$ and hard-sphere gas mixtures with $d_2/d_1=2$, as shown in figure \ref{fig:phi_VSS}. It is found that for lighter species (typically smaller in diameter), $\phi_{12}=n_1\tau_{11}/n_2\tau_{12}$ is greater than 1 and also increases gradually with mass ratio (figure \ref{fig:phi_VSS:phi12}), which implies an important role of inter-species collisions with heavy gas molecules. On the other hand, $\phi_{21}=n_2\tau_{22}/n_1\tau_{21}$ rapidly decreases as the mass disparity grows, and roughly scales as $\left({m_2}/{m_1}\right)^{-0.59}$ when the mass ratio exceeds 20 (figure \ref{fig:phi_VSS:phi21}). Consequently, in a gas mixture with a large mass difference, the inter-species collision term in the proposed kinetic model will have a negligible impact on the dynamics of the heavier species unless its mole fraction is significantly small; while it is usually important in determining the behaviors of the lighter species. Note that, knowing the relaxation timescale of each collision term is also crucial for developing a multi-temperature hydrodynamic equation for mixtures with disparate mass, since the corresponding constitutive relations depend on the orders of magnitude of the inter-species relaxation rates.

\begin{table}[ht]
    \begin{center}
	  \begin{tabular}{cccccccccc}
		\hline
		Mixture & constituents & $m_2/m_1$ & $d_2/d_1$ & $\omega_{12}$ & $\alpha_{12}$ & $\phi_{12}$  & $\phi_{21}$ & $\varphi_{12}$ & $\varphi_{21}$\\
		\specialrule{0em}{4pt}{4pt}
		 1 & \multirow{2}{*}{Maxwell gas} &  10 & 1 & 1.0 & 2.14 & 1.214 & 0.5154 & 1.035 & 1.779 \\
		 2 & ~&  1000 & 1 & 1.0 & 2.14 & 1.367 & 0.05754 & 0.999 & 2.259 \\
		 3 & hard-sphere gas &  100 & 2 & 0.5 & 1.0 & 2.955 & 0.1269 & 1.425 & 1.261 \\
        \hline
      \end{tabular}
      \caption{The constituents of the three binary mixtures considered in the present work, and the corresponding parameters $\phi_{sr}$ and $\varphi_{sr}$ in the kinetic model fitted by matching the mixture viscosity and thermal conductivity from VSS collision model, respectively.}
    \label{tab:parameter_gas}
    \end{center}
\end{table}

In the present work, three types of binary gas mixtures are considered in the following simulations: Mixtures 1 and 2 consist of Maxwell gas molecules possessing identical reference diameters and mass ratios of 10 and 1000, respectively; Mixture 3 composites hard-sphere molecules characterized by a mass ratio of 100 and a diameter ratio of 2. All the parameters associated with inter-species collisions in the VSS model (namely, $\omega_{12}$, $\alpha_{12}$ and $d_{12}$) are determined simply through the arithmetic averaging of the corresponding parameters of the individual components in each mixture. The fitting parameters $\phi_{sr}$ and $\varphi_{sr}$ for relaxation rate ratios are given in Table \ref{tab:parameter_gas}. Due to the discrepancy by the first approximation of mixture thermal conductivity for hard-sphere molecules as mentioned above, we determine the values of $\varphi_{sr}$ for Mixture 3 based on the thermal conductivities calculated from DSMC directly, to make a more consistent comparison for model validation. It is noteworthy that the fitted values of $\phi_{sr}$ and $\varphi_{sr}$ for the considered mixtures are independent of temperature, while this is generally not the case for arbitrary mixtures containing gas molecules with different interaction potentials.

\section{Numerical results of one-dimensional problems}\label{sec:1D_cases}

In this section, the accuracy of the proposed kinetic model is assessed by the DSMC method in one-dimensional normal shock wave, Fourier flow and Couette flow of the binary gas mixtures listed in Table \ref{tab:parameter_gas}. We compare not only the average properties of the mixture but also those of the individual components, which is crucial for accurately describing mixture flows, as the different species in the mixture can vary significantly in concentration, velocity, and temperature in non-equilibrium flows. 

The DSMC simulations are conducted using the open-source code SPARTA \citep{SPARTA}. On the other hand, to reduce the computational cost of solving kinetic model equations, the velocity distribution functions are dimensionally reduced to be quasi-one-dimensional in velocity space for the normal shock wave and Fourier flow, by introducing functions $f_{s,x1}$ and $f_{s,x2}$,
\begin{equation}\label{eq:quasi_one_dimensional}
	\begin{aligned}[b]
		f_{s,x1} = \int_{\mathbb{R}^2}{f_s}\mathrm{d}{v_y}\mathrm{d}{v_z}, \quad
		f_{s,x2} = \int_{\mathbb{R}^2}{\left(v_y^2+v_z^2\right)f_s}\mathrm{d}{v_y}\mathrm{d}{v_z}.
	\end{aligned}
\end{equation}
Similarly, $f_{s}$ can be reduced to be quasi-two-dimensional in velocity space for the Couette flow, by introducing functions $f_{s,xy1}$ and $f_{s,xy2}$,
\begin{equation}\label{eq:quasi_two_dimensional}
	\begin{aligned}[b]
		f_{s,xy1} = \int_{\mathbb{R}^1}{f_s}\mathrm{d}{v_z}, \quad
		f_{s,xy2} = \int_{\mathbb{R}^1}{v_z^2f_s}\mathrm{d}{v_z}.
	\end{aligned}
\end{equation}
The macroscopic variables are then calculated by taking moments of the reduced distribution functions. The kinetic model equations are solved by the discretized velocity method, and a finite difference method with a second-order upwind scheme is adopted in the numerical implementation. Since the characteristic molecular velocity of the lighter gas can be much higher than that of the heavier one by the square root of the mass ratio, the individual velocity space of each component is used with different truncations and discretization.

\subsection{Normal shock wave}

\begin{figure}[t]
	\centering
	\sidesubfloat[]{\includegraphics[scale=0.19,clip=true]{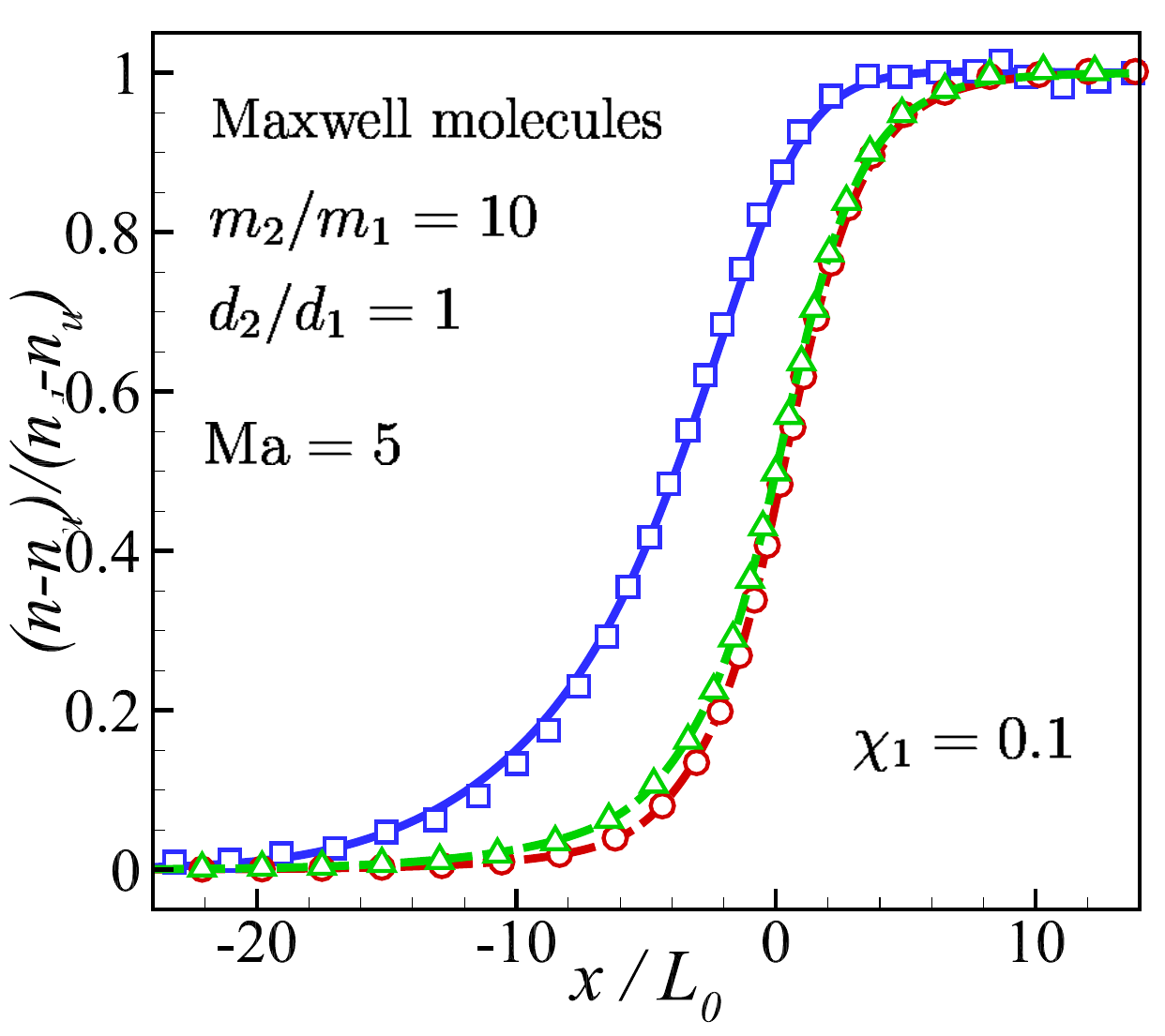}\label{fig:1DNormalShockWave_Mix1:X1_01_n}}   
	\sidesubfloat[]{\includegraphics[scale=0.19,clip=true]{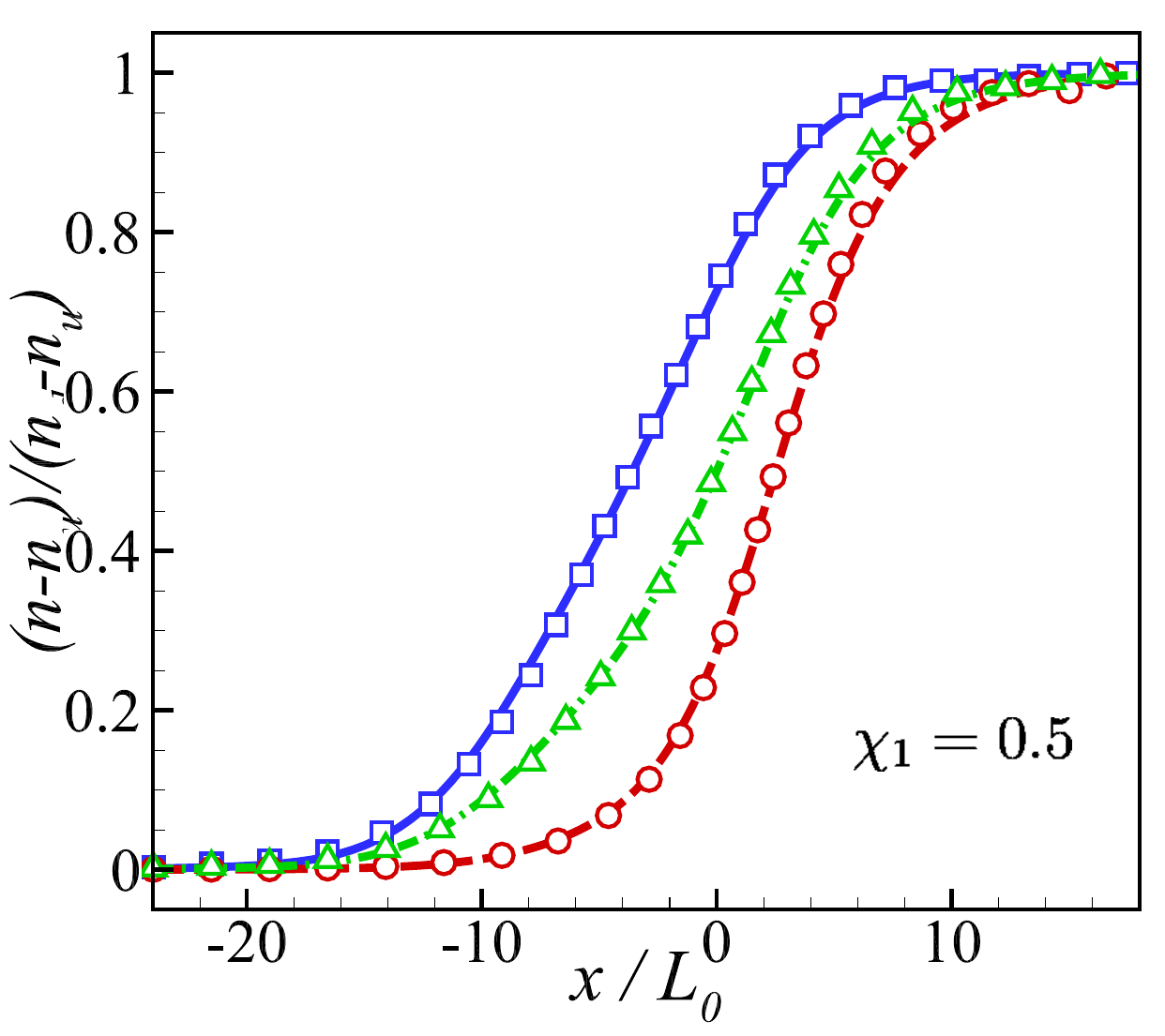}\label{fig:1DNormalShockWave_Mix1:X1_05_n}}  
    \sidesubfloat[]{\includegraphics[scale=0.19,clip=true]{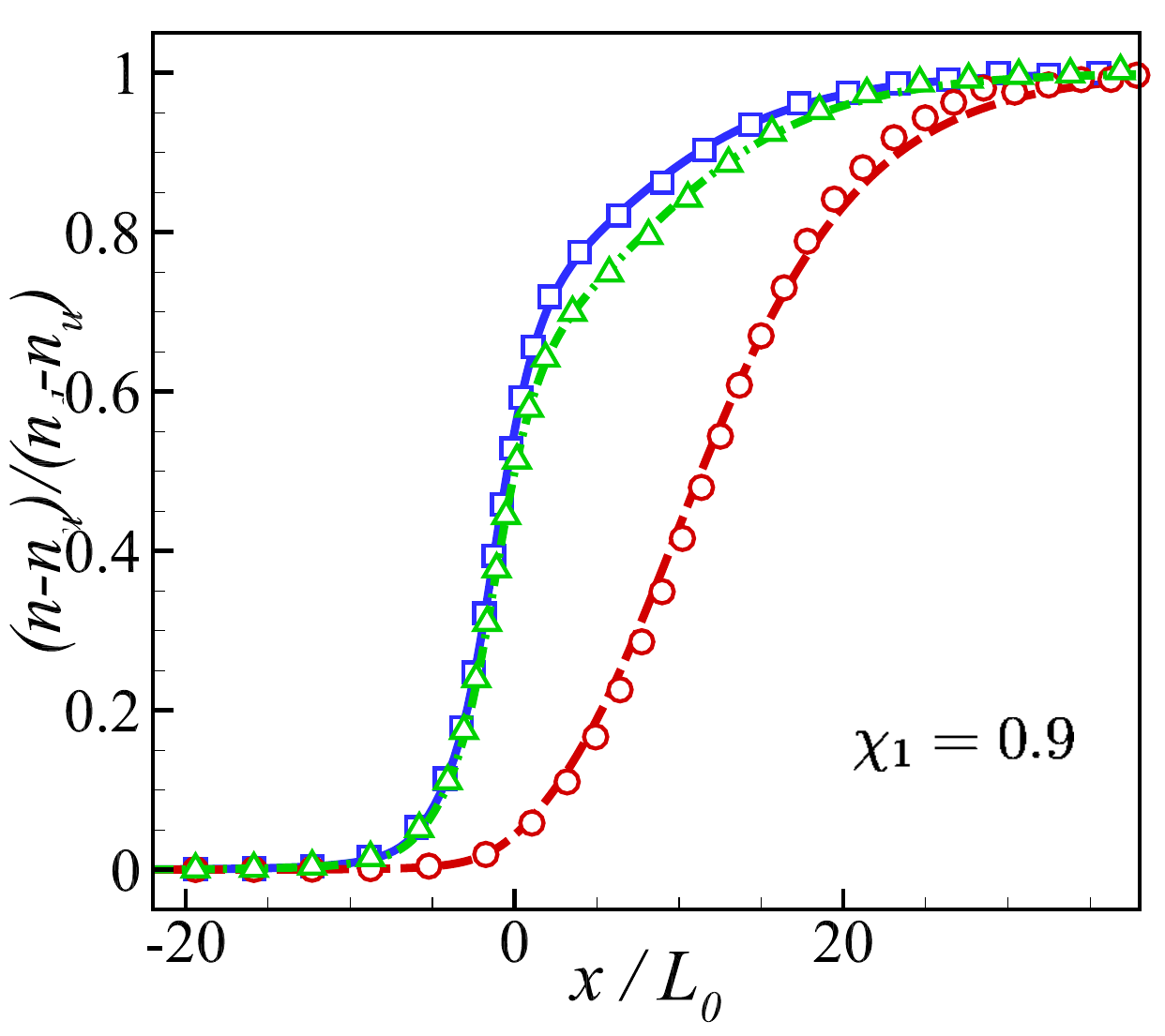}\label{fig:1DNormalShockWave_Mix1:X1_09_n}}  \\ 
    \sidesubfloat[]{\includegraphics[scale=0.19,clip=true]{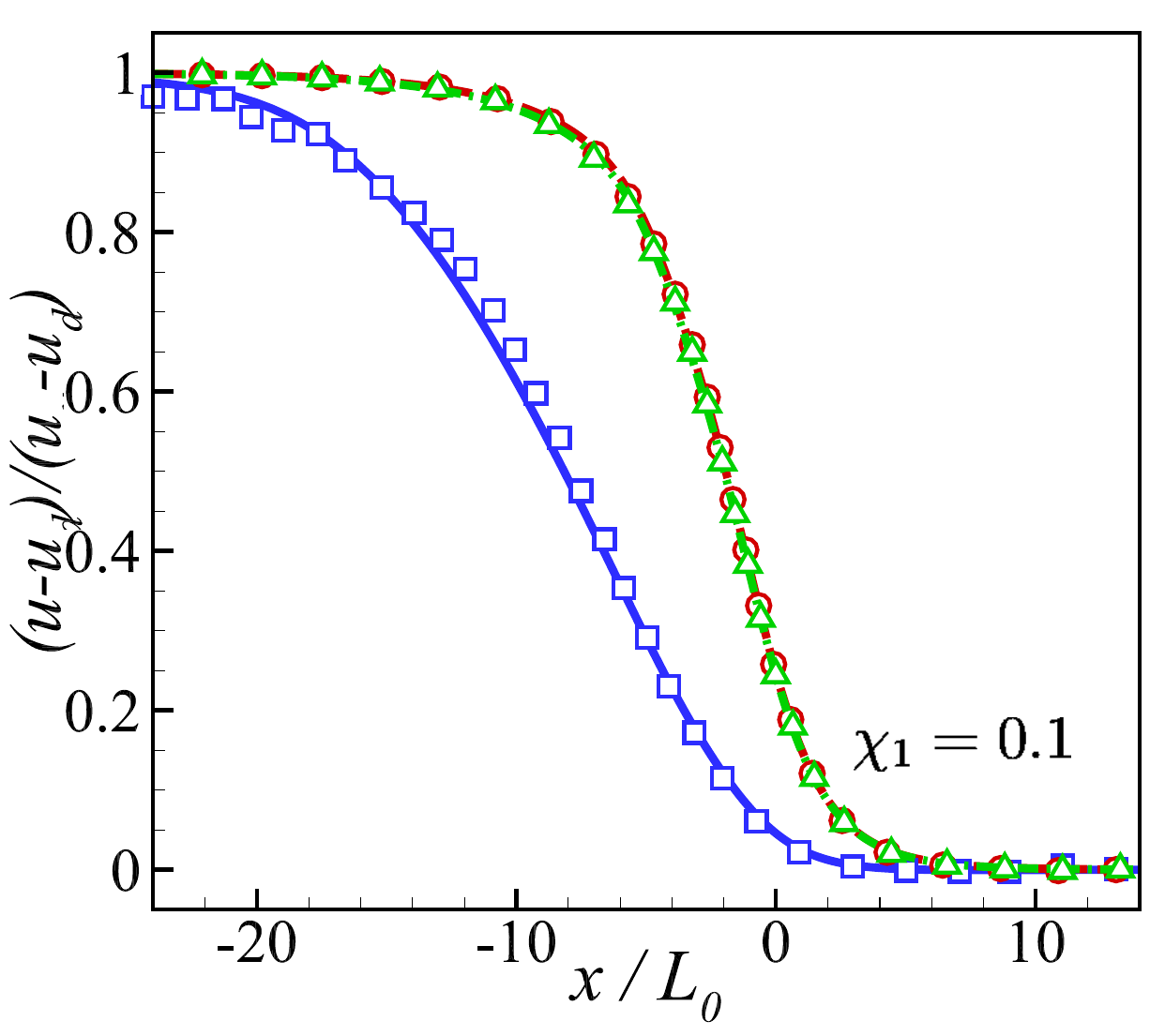}\label{fig:1DNormalShockWave_Mix1:X1_01_u}}   
	\sidesubfloat[]{\includegraphics[scale=0.19,clip=true]{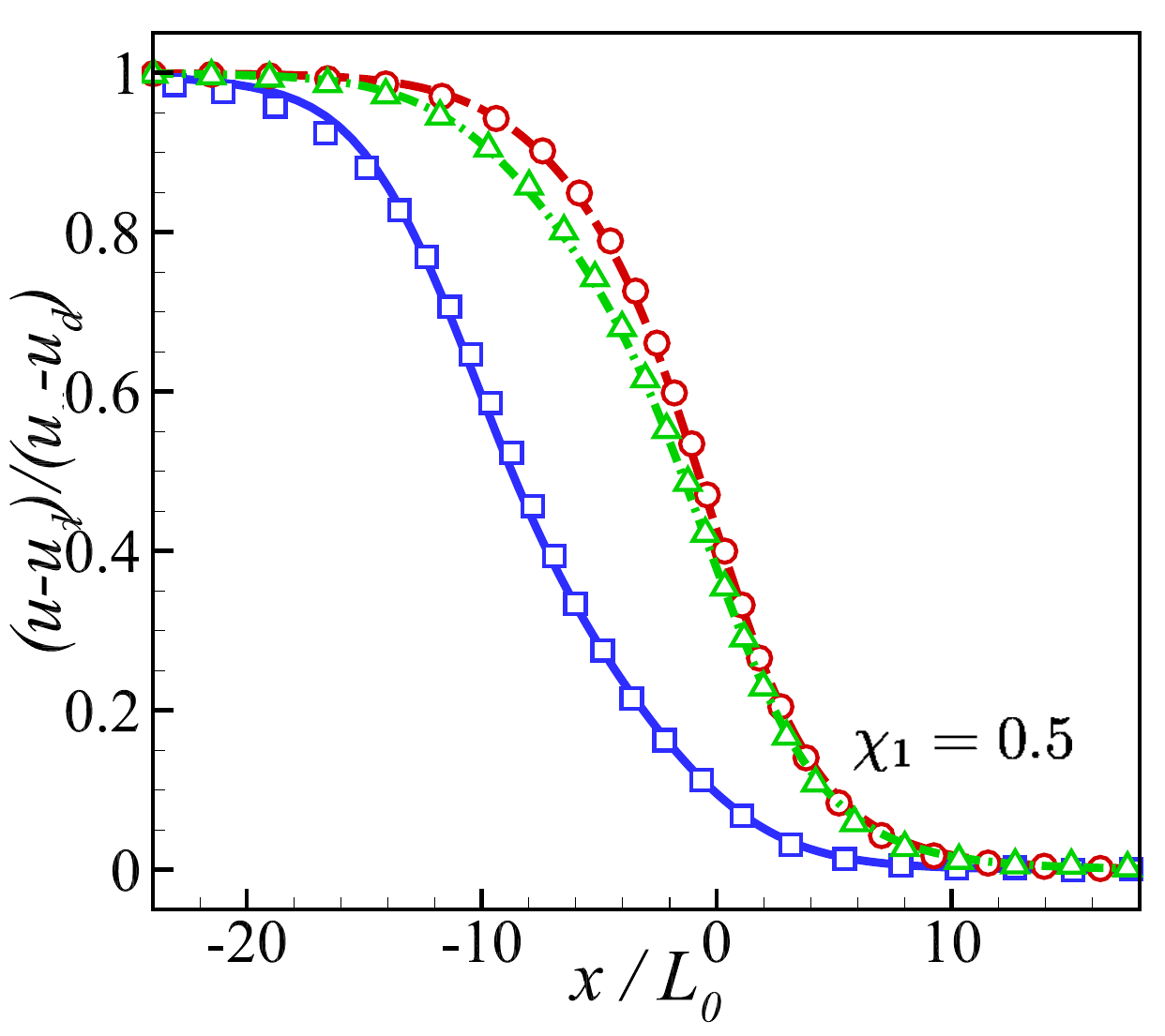}\label{fig:1DNormalShockWave_Mix1:X1_05_u}}  
    \sidesubfloat[]{\includegraphics[scale=0.19,clip=true]{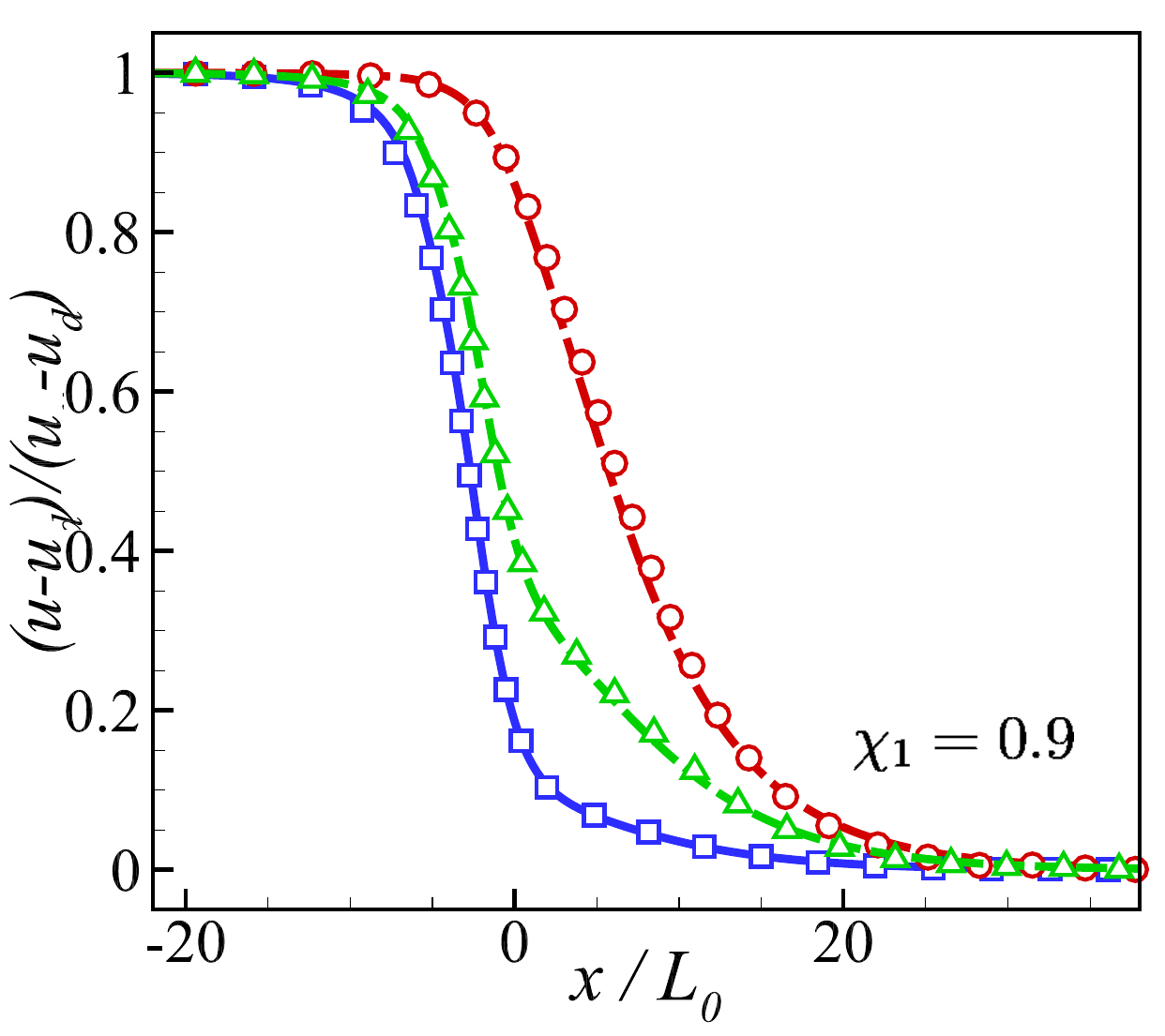}\label{fig:1DNormalShockWave_Mix1:X1_09_u}}  \\ 
    \sidesubfloat[]{\includegraphics[scale=0.19,clip=true]{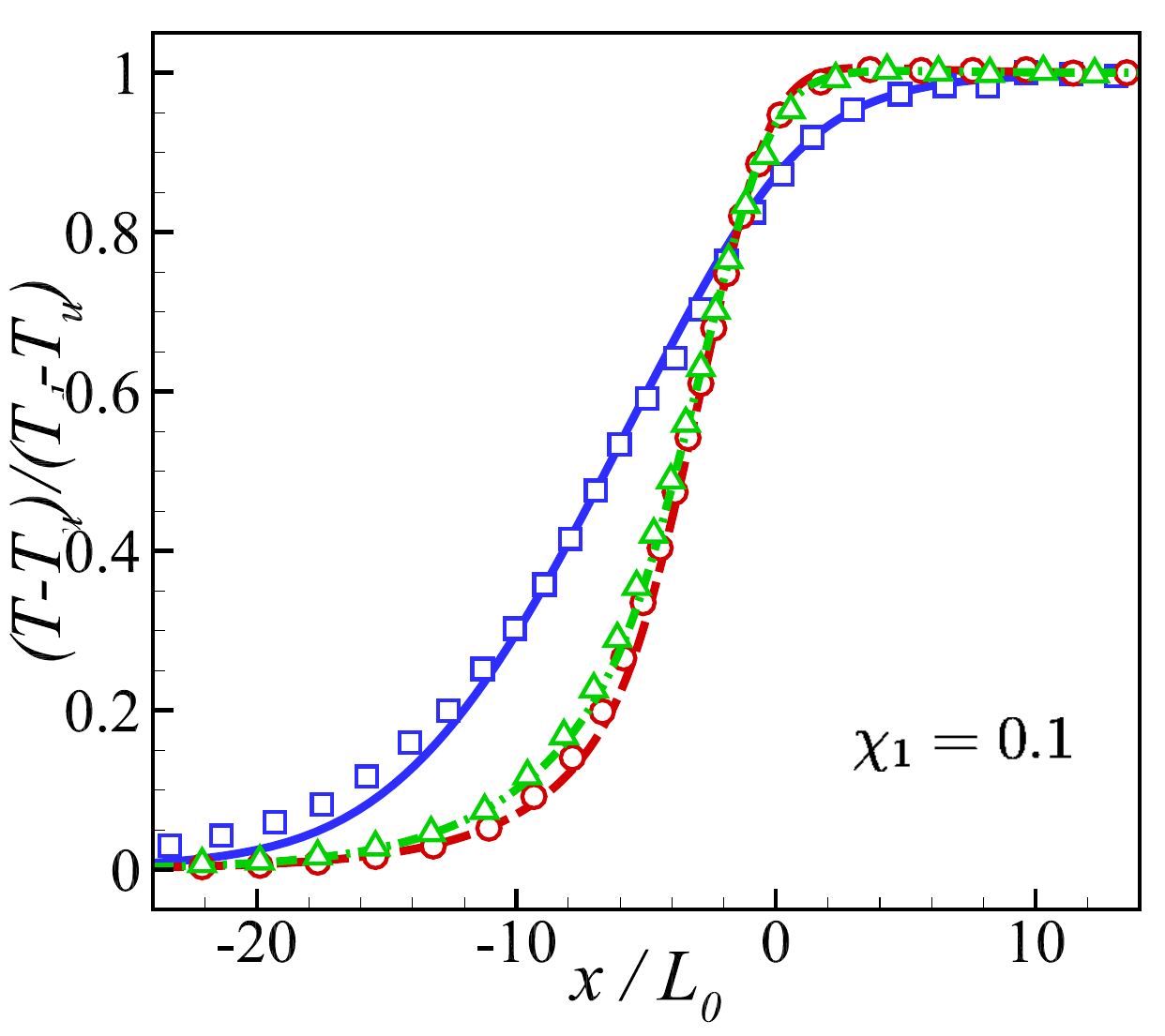}\label{fig:1DNormalShockWave_Mix1:X1_01_T}}   
	\sidesubfloat[]{\includegraphics[scale=0.19,clip=true]{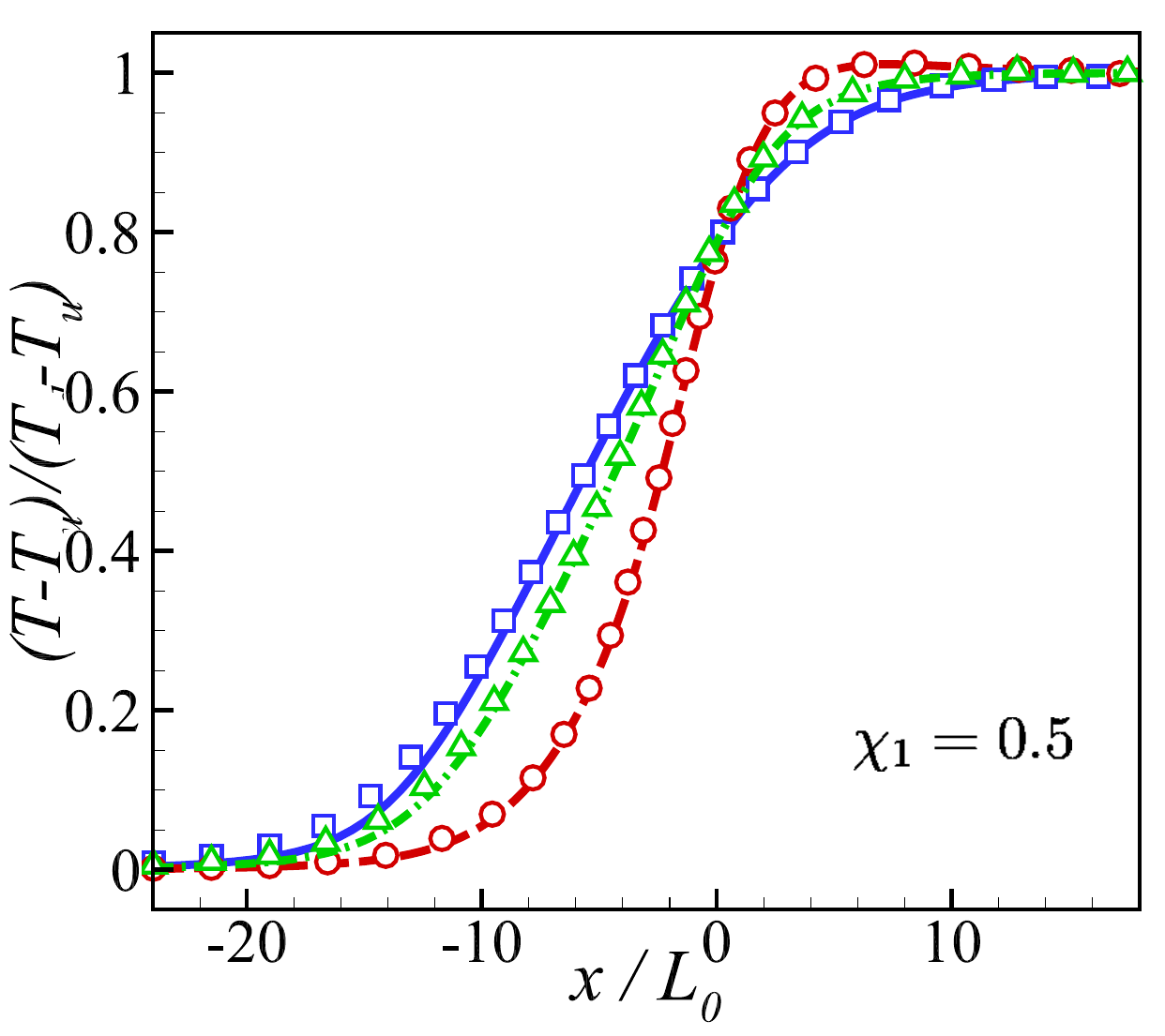}\label{fig:1DNormalShockWave_Mix1:X1_05_T}}  
    \sidesubfloat[]{\includegraphics[scale=0.19,clip=true]{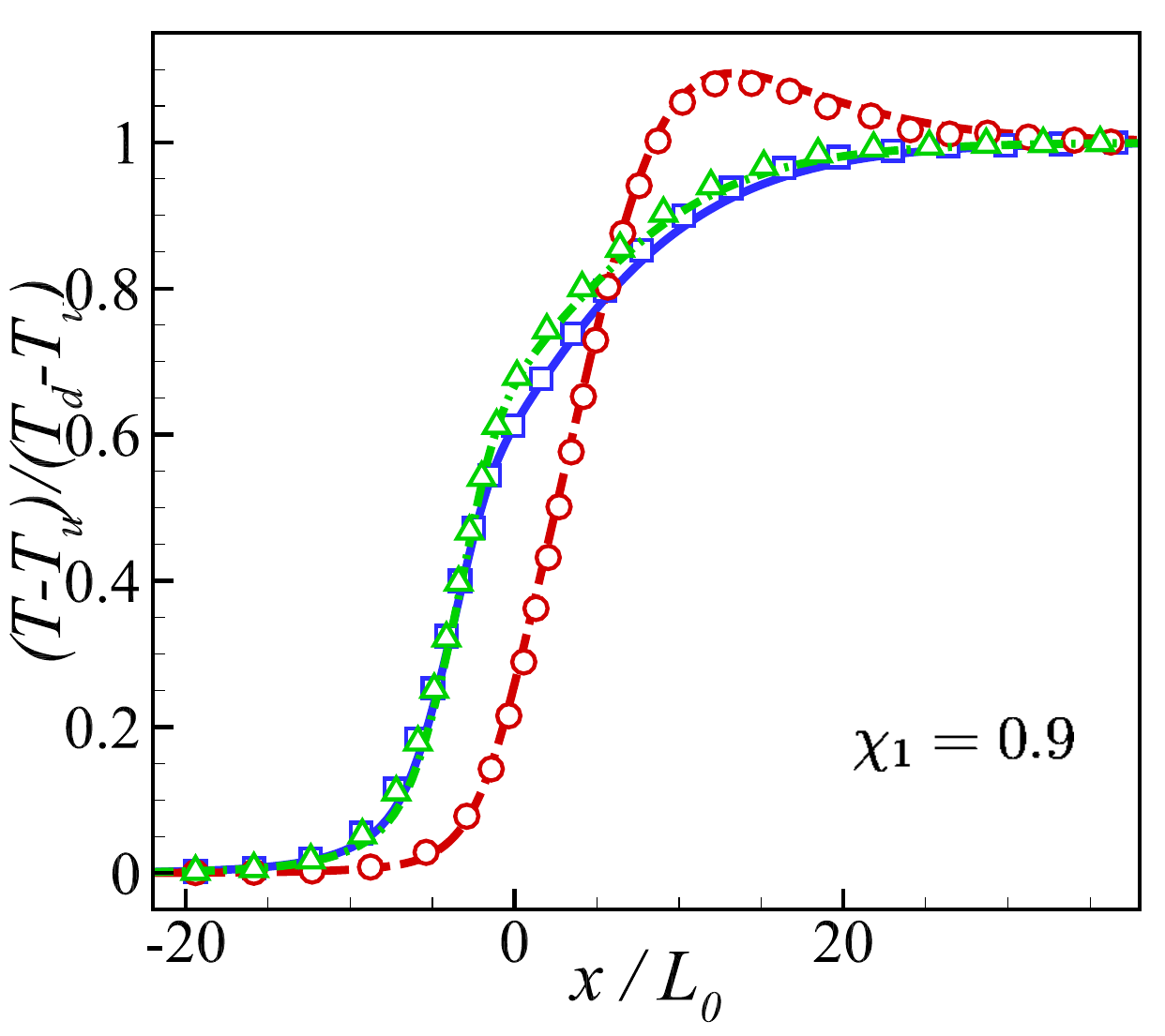}\label{fig:1DNormalShockWave_Mix1:X1_09_T}}  \\ 
    \sidesubfloat[]{\includegraphics[scale=0.19,clip=true]{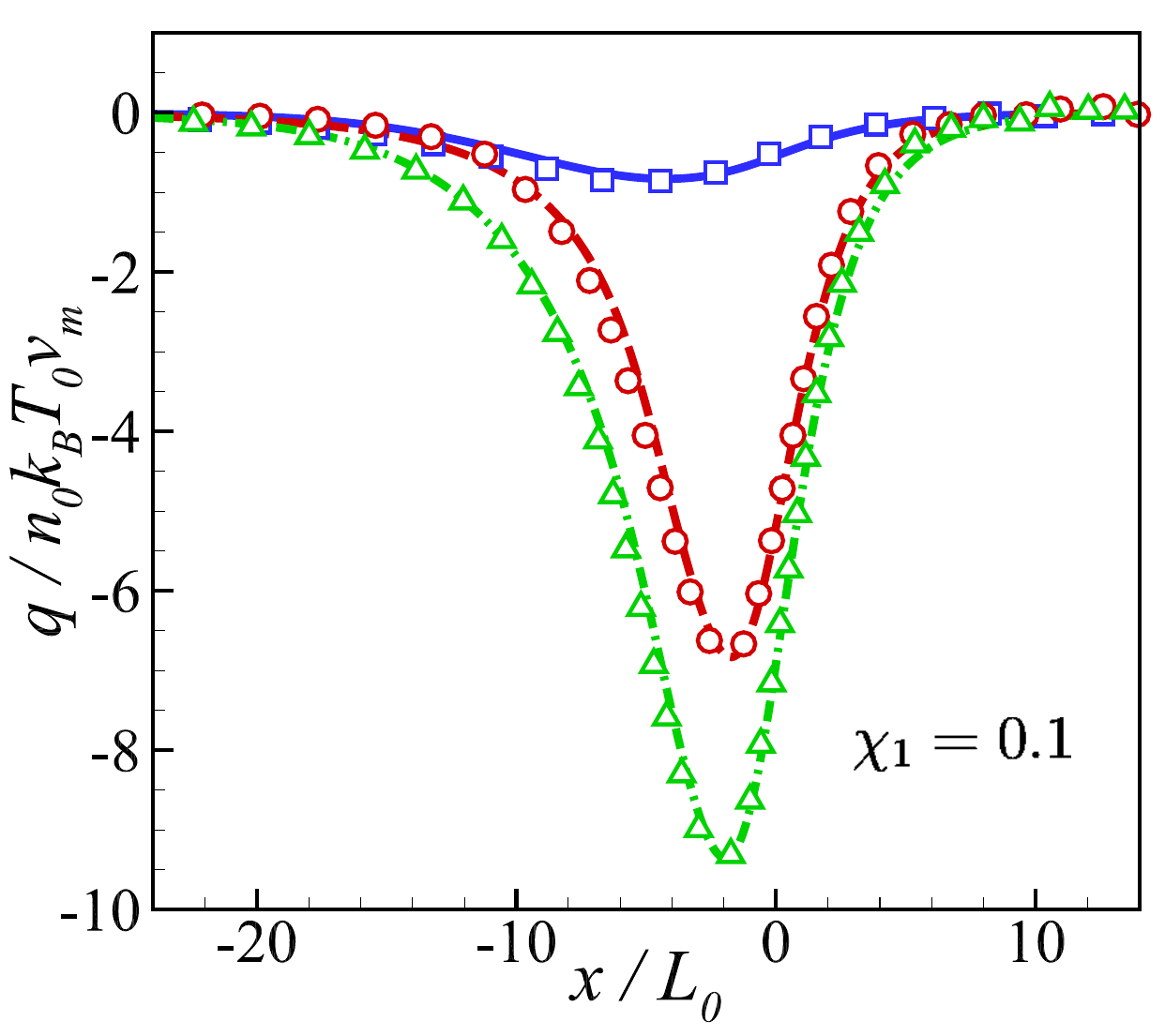}\label{fig:1DNormalShockWave_Mix1:X1_01_q}}   
	\sidesubfloat[]{\includegraphics[scale=0.19,clip=true]{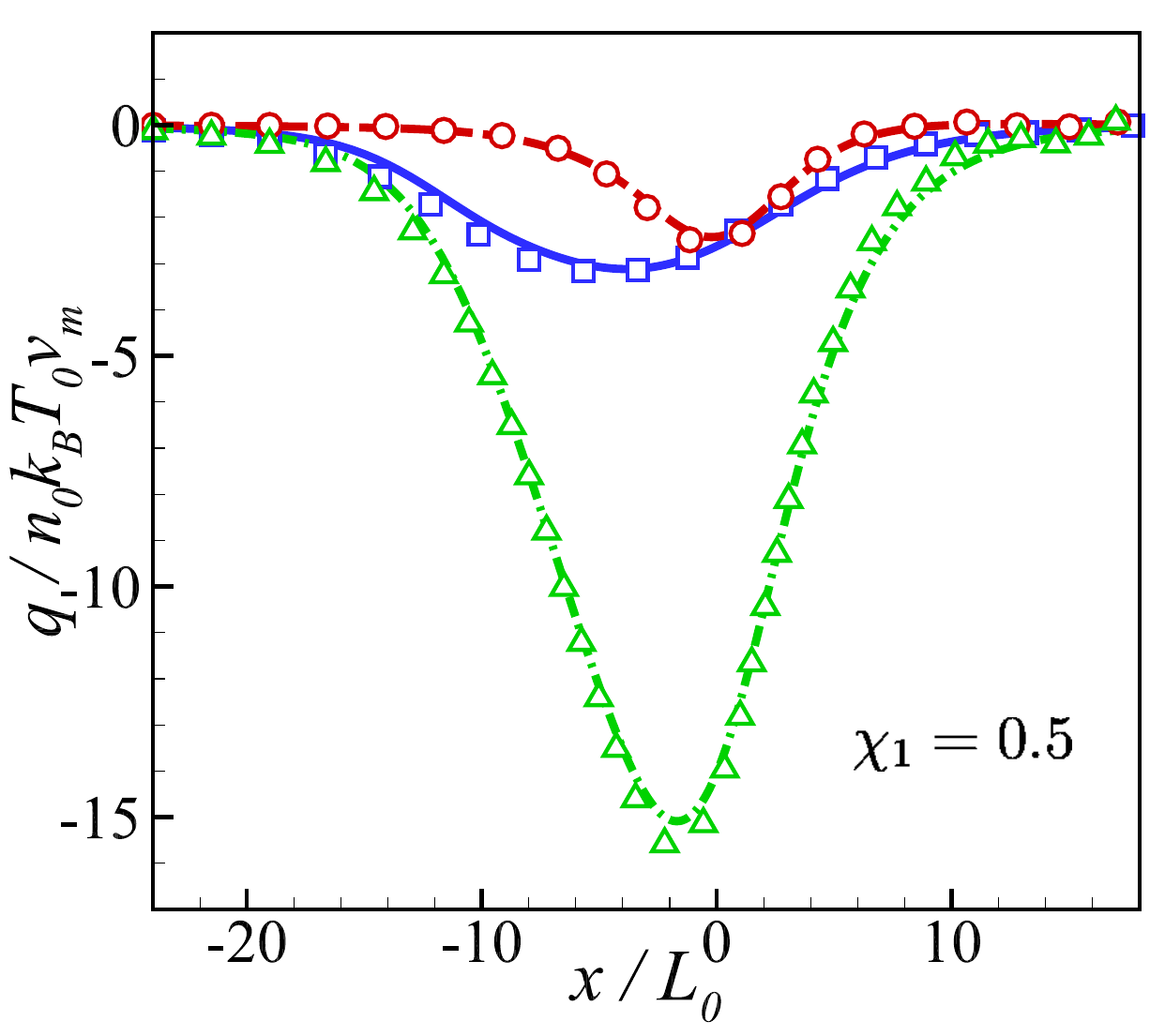}\label{fig:1DNormalShockWave_Mix1:X1_05_q}}  
    \sidesubfloat[]{\includegraphics[scale=0.19,clip=true]{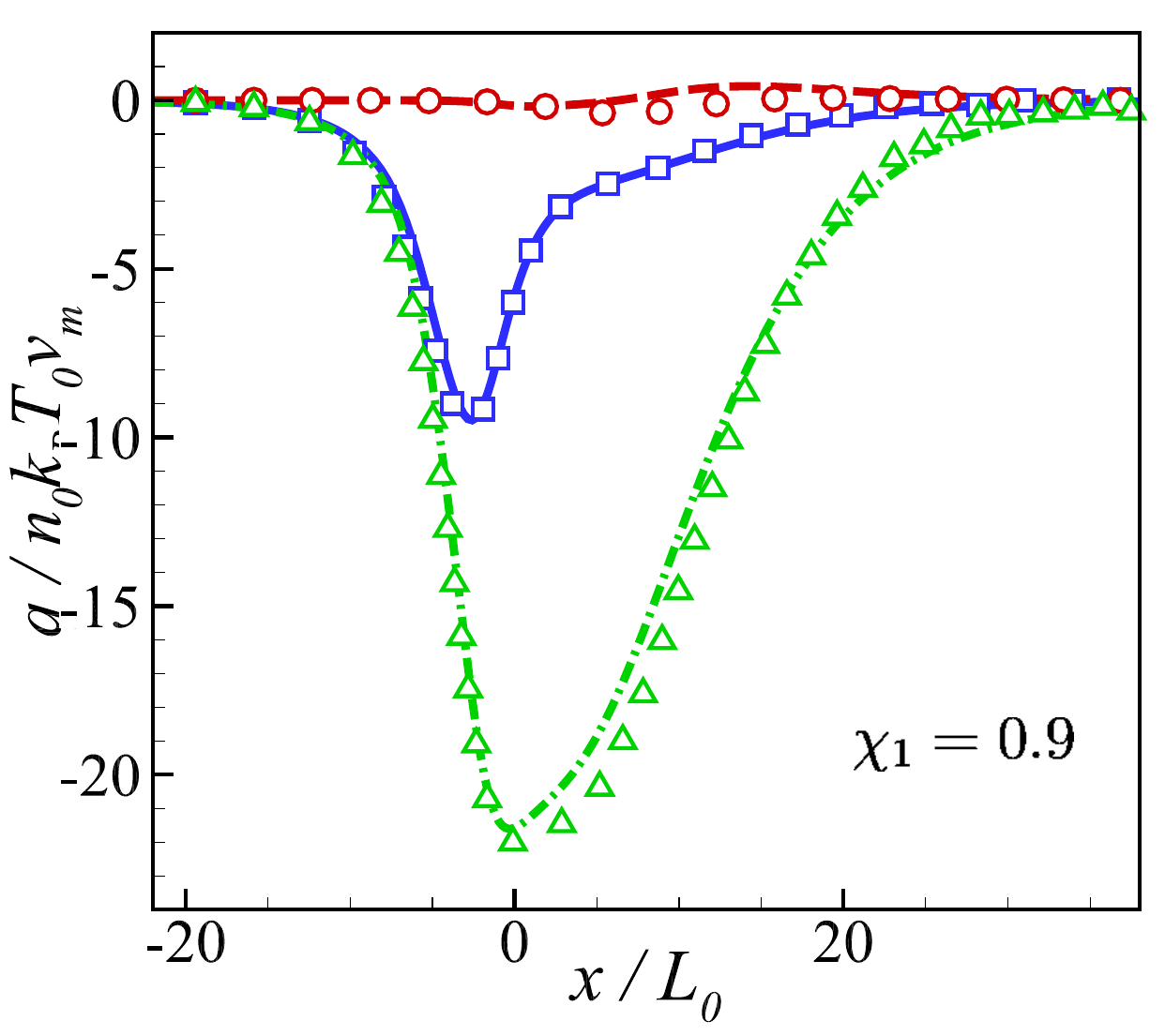}\label{fig:1DNormalShockWave_Mix1:X1_09_q}}  \\ 
    \includegraphics[scale=0.22,clip=true]{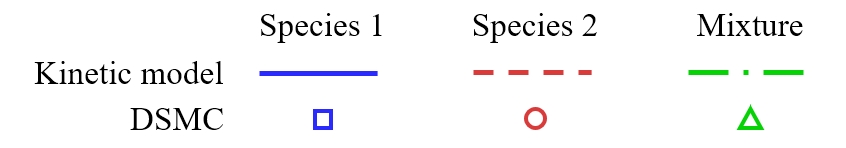}
	\caption{Comparisons of the normalized (a-c) number density, (d-f) flow velocity, (g-i) temperature, and dimensionless (j-l) heat flux of the gas mixture between kinetic model (lines) and DSMC (symbols) for the normal shock wave at $\text{Ma}=5$. The binary mixture consists of Maxwell molecules with a mass ratio $m_2/m_1=10$, diameter ratio $d_2/d_1=1$, and the mole fraction of light species $\chi_1=0.1,0.5,0.9$.}
	\label{fig:1DNormalShockWave_Mix1}
\end{figure}

\begin{figure}[t]
	\centering
	\sidesubfloat[]{\includegraphics[scale=0.19,clip=true]{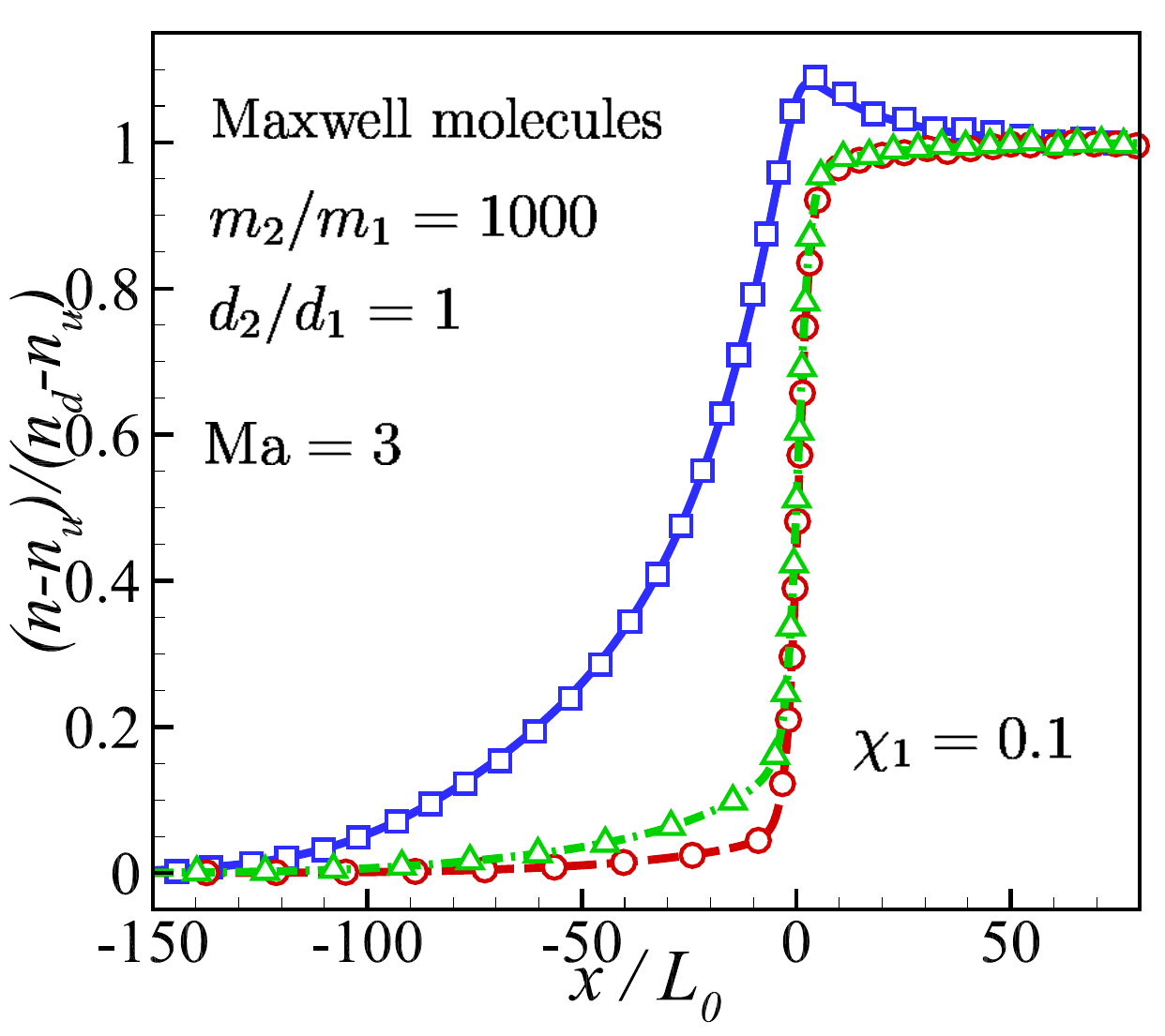}\label{fig:1DNormalShockWave_Mix2:X1_01_n}}   
	\sidesubfloat[]{\includegraphics[scale=0.19,clip=true]{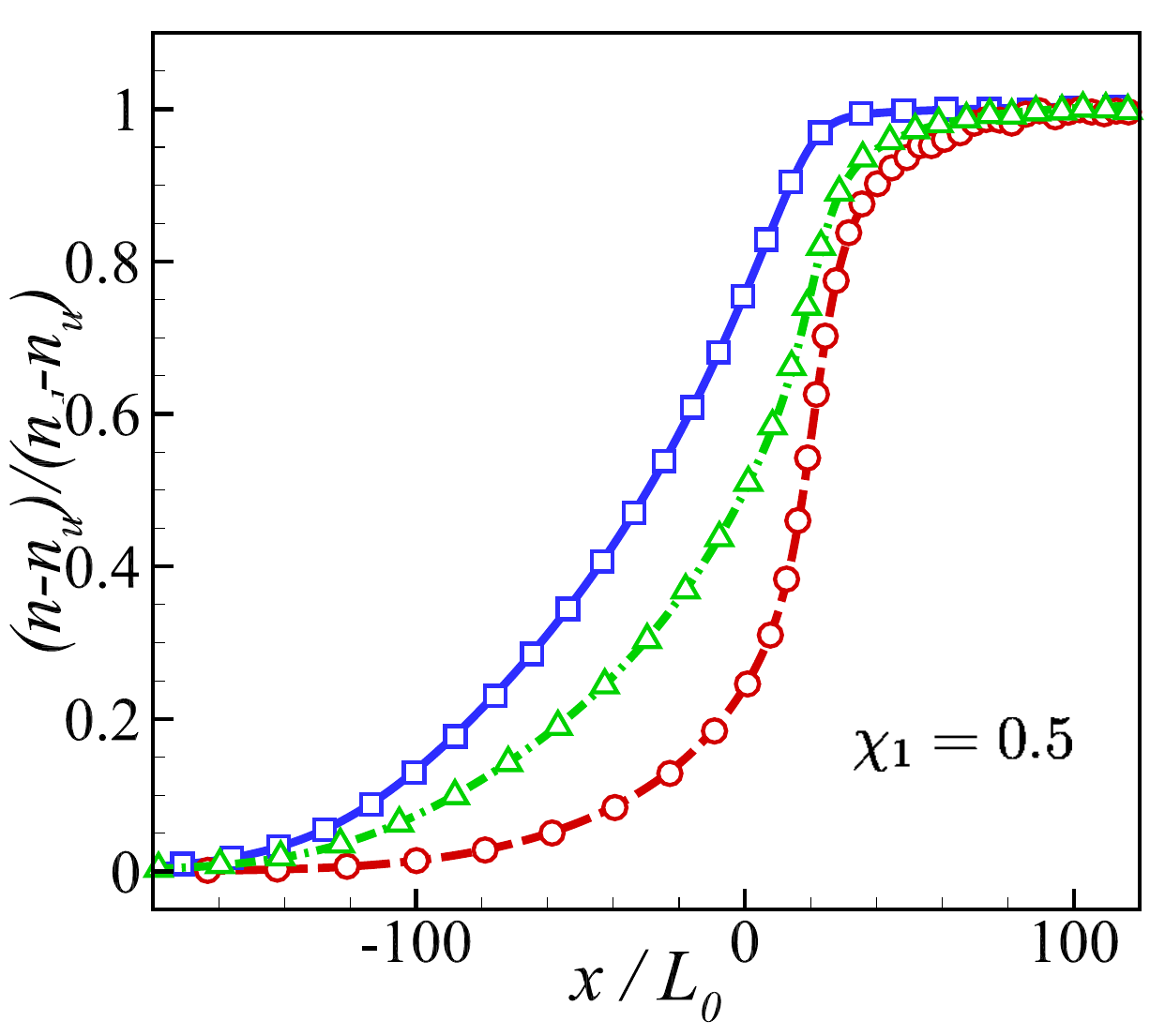}\label{fig:1DNormalShockWave_Mix2:X1_05_n}}  
    \sidesubfloat[]{\includegraphics[scale=0.19,clip=true]{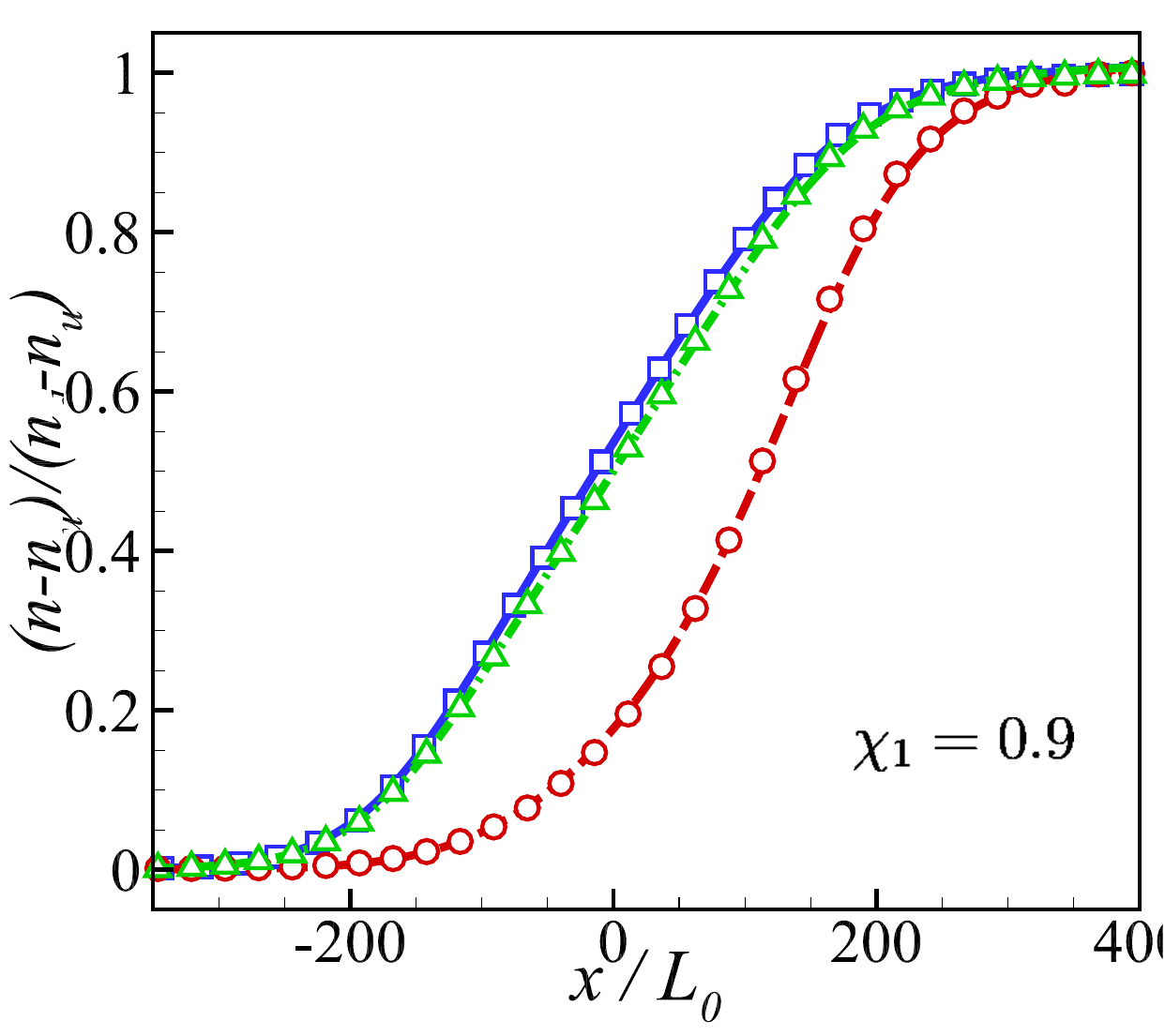}\label{fig:1DNormalShockWave_Mix2:X1_09_n}}  \\ 
    \sidesubfloat[]{\includegraphics[scale=0.19,clip=true]{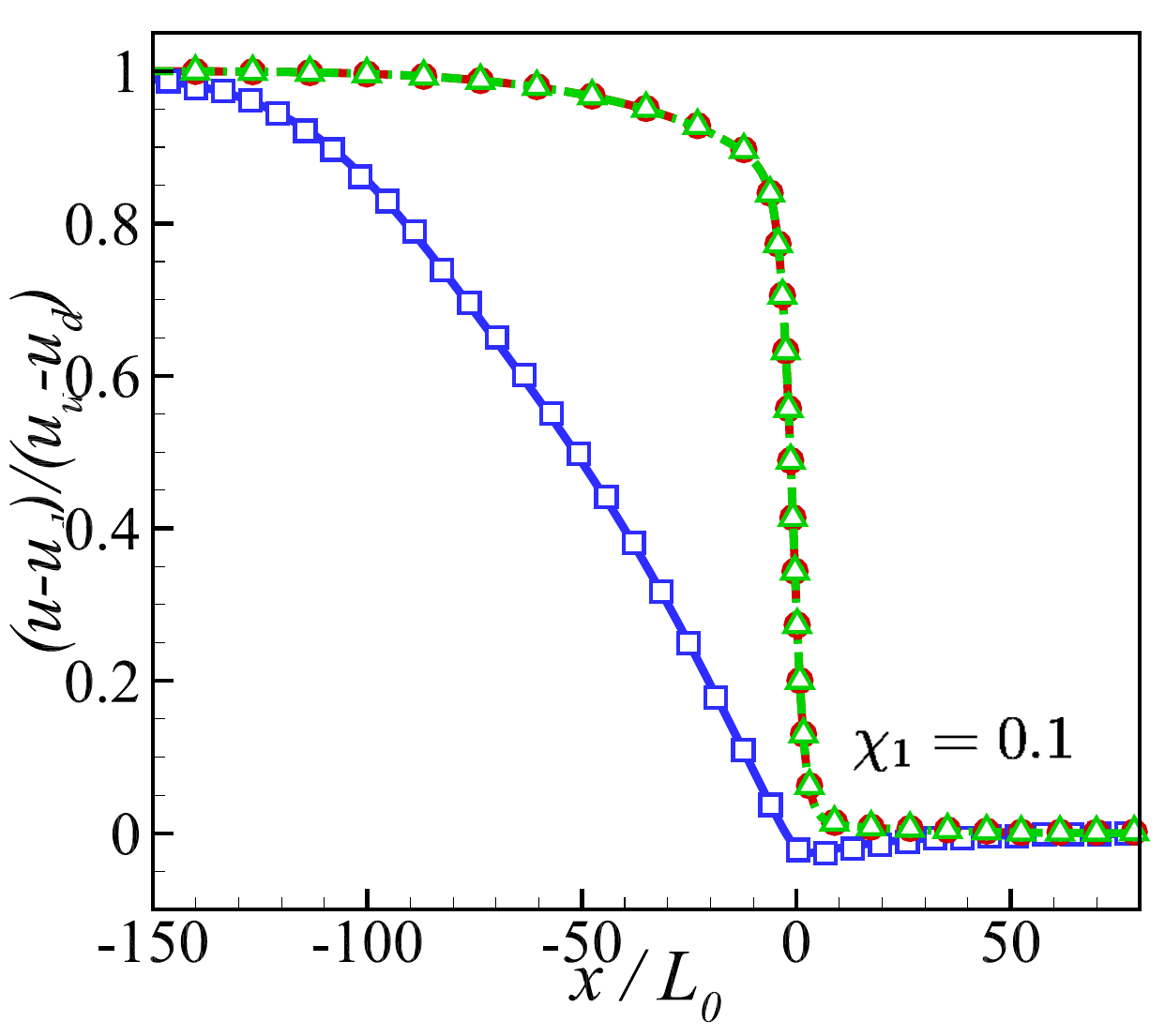}\label{fig:1DNormalShockWave_Mix2:X1_01_u}}   
	\sidesubfloat[]{\includegraphics[scale=0.19,clip=true]{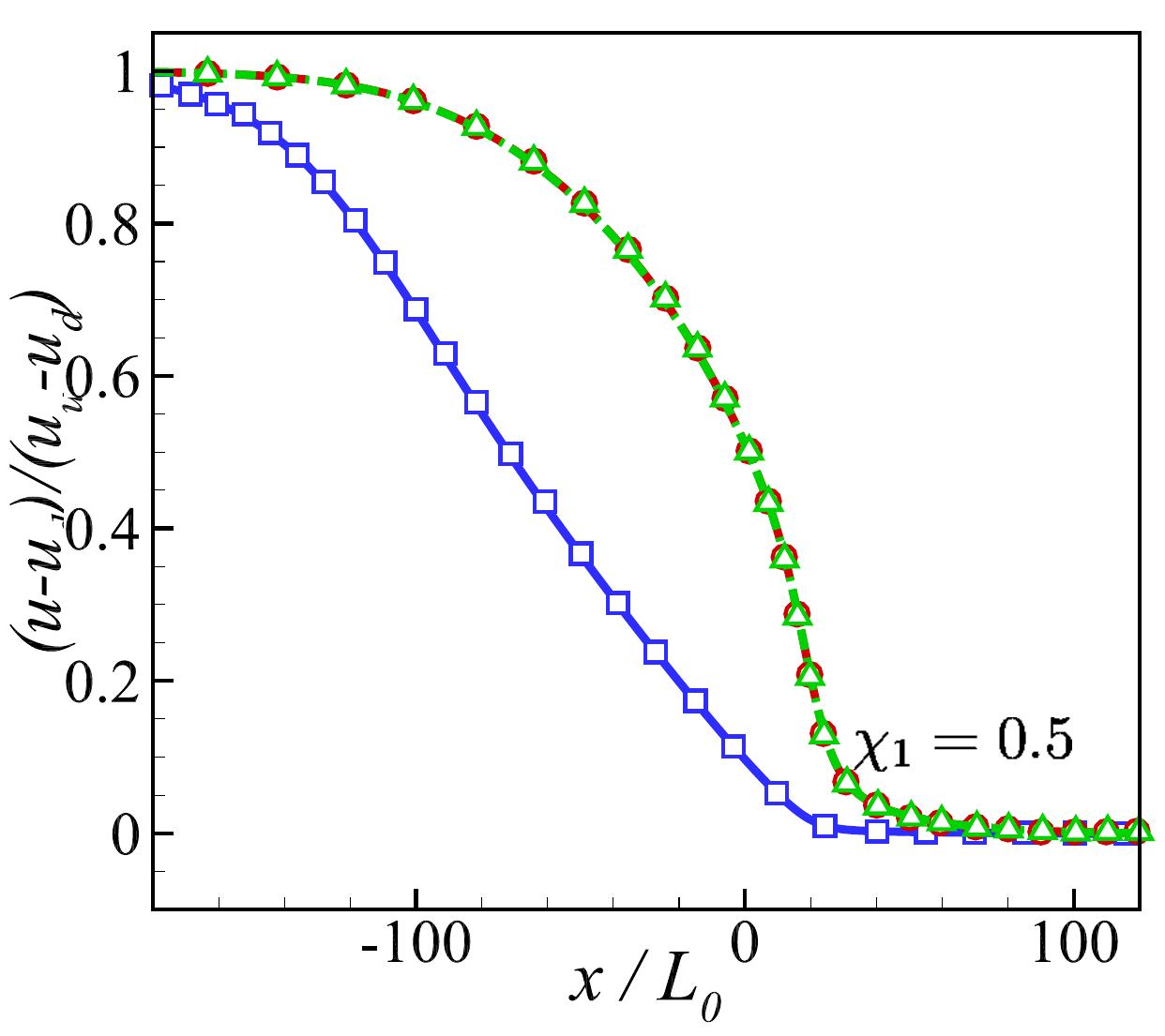}\label{fig:1DNormalShockWave_Mix2:X1_05_u}}  
    \sidesubfloat[]{\includegraphics[scale=0.19,clip=true]{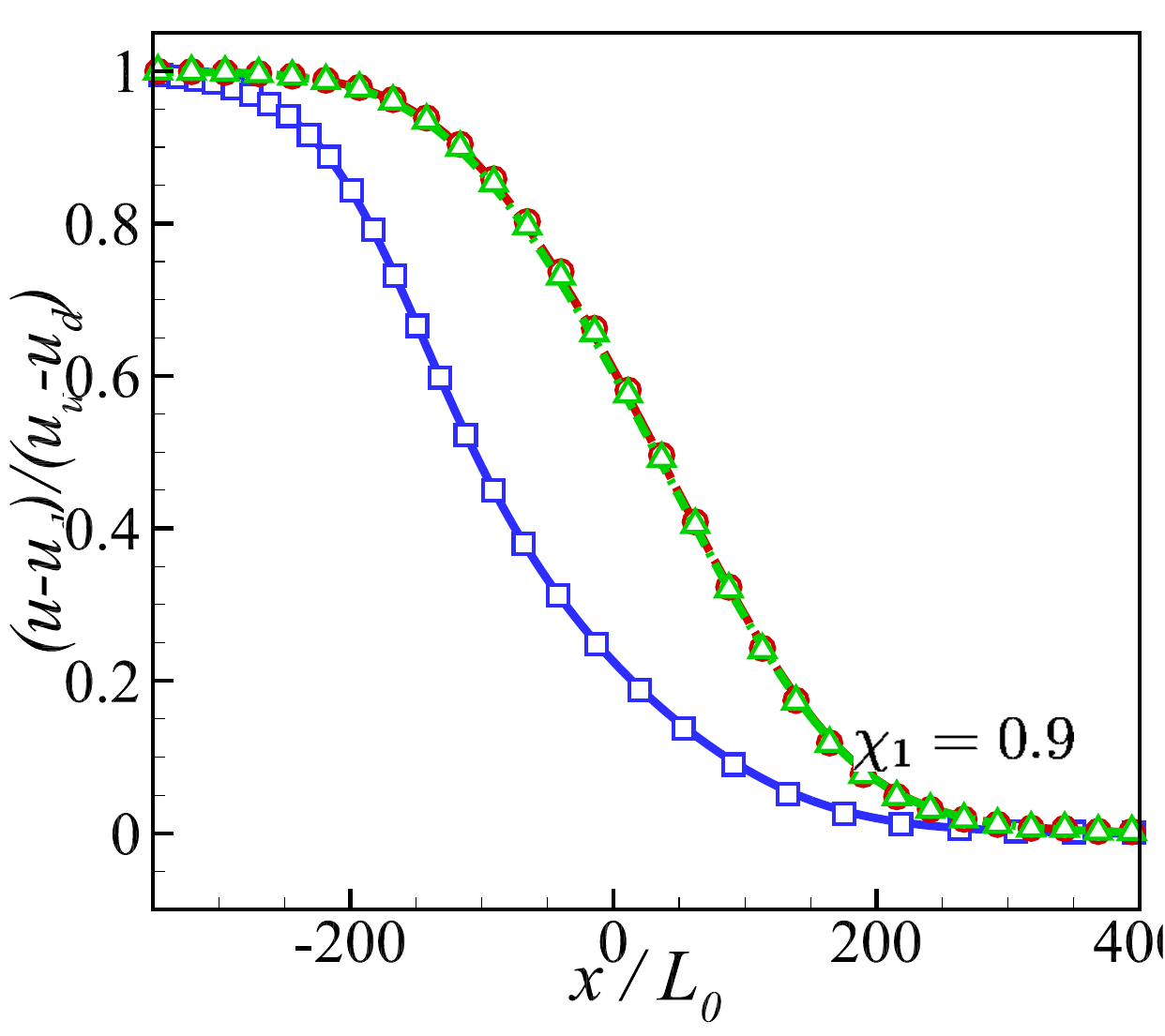}\label{fig:1DNormalShockWave_Mix2:X1_09_u}}  \\ 
    \sidesubfloat[]{\includegraphics[scale=0.19,clip=true]{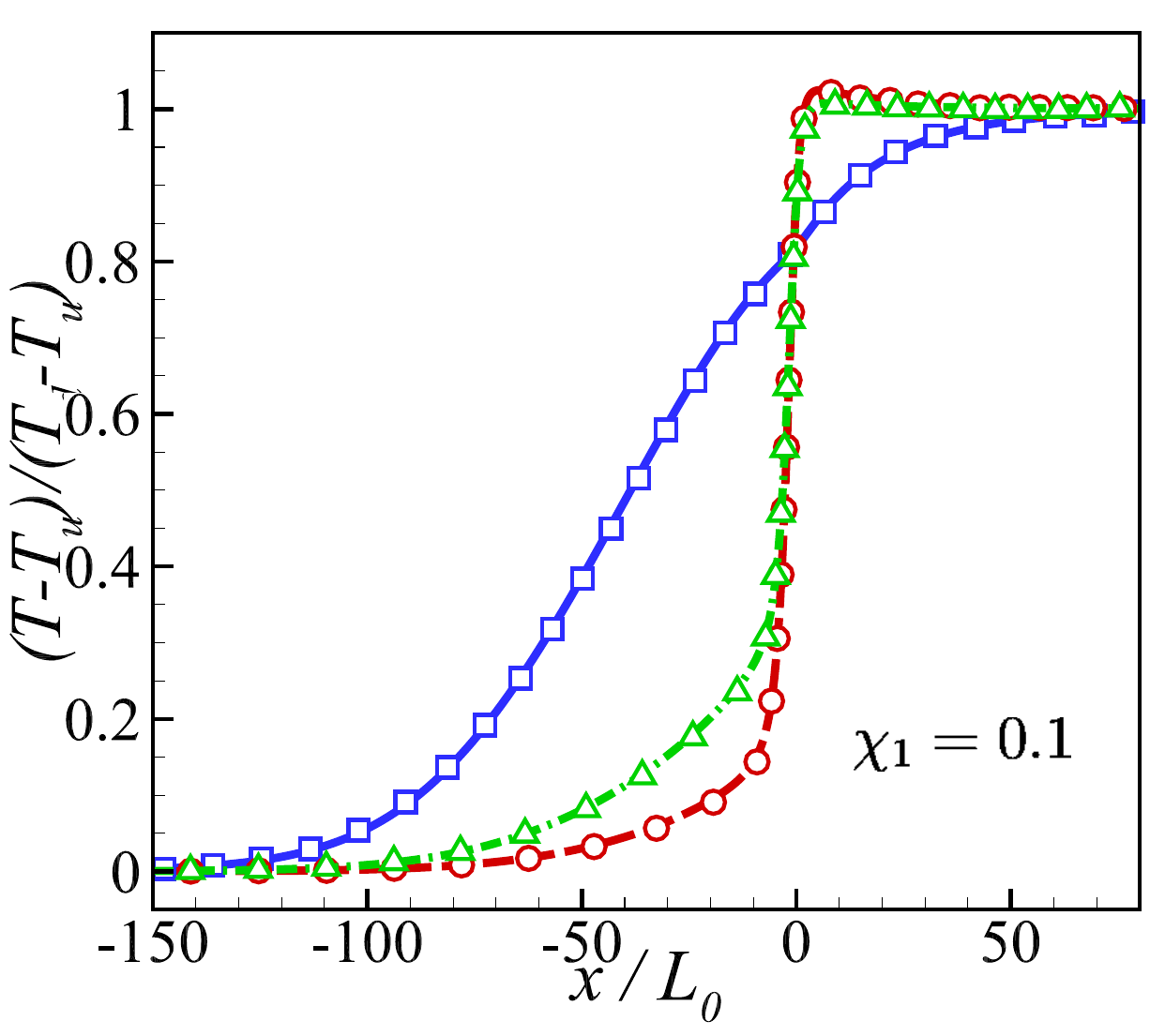}\label{fig:1DNormalShockWave_Mix2:X1_01_T}}   
	\sidesubfloat[]{\includegraphics[scale=0.19,clip=true]{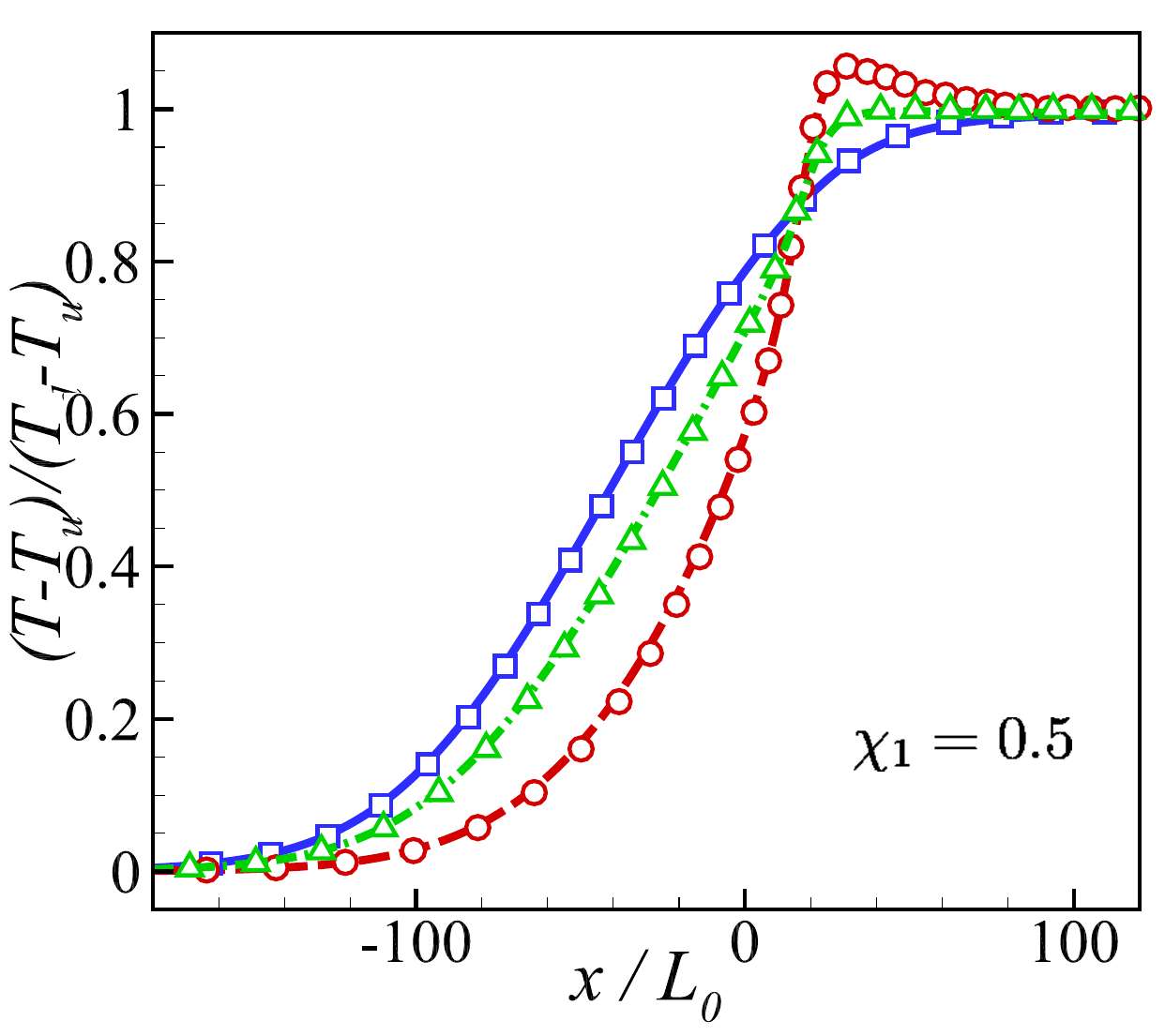}\label{fig:1DNormalShockWave_Mix2:X1_05_T}}  
    \sidesubfloat[]{\includegraphics[scale=0.19,clip=true]{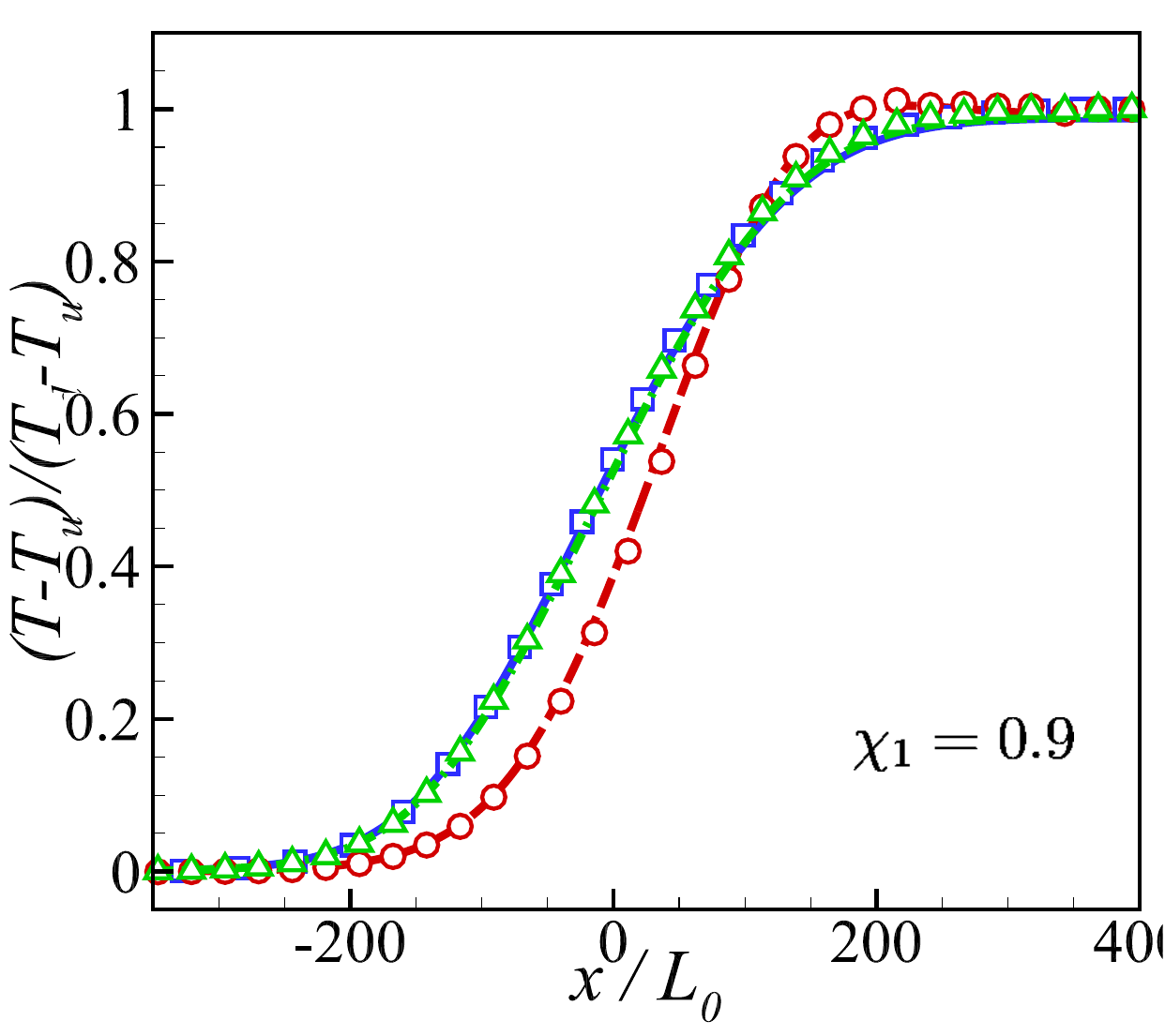}\label{fig:1DNormalShockWave_Mix2:X1_09_T}}  \\ 
    \sidesubfloat[]{\includegraphics[scale=0.19,clip=true]{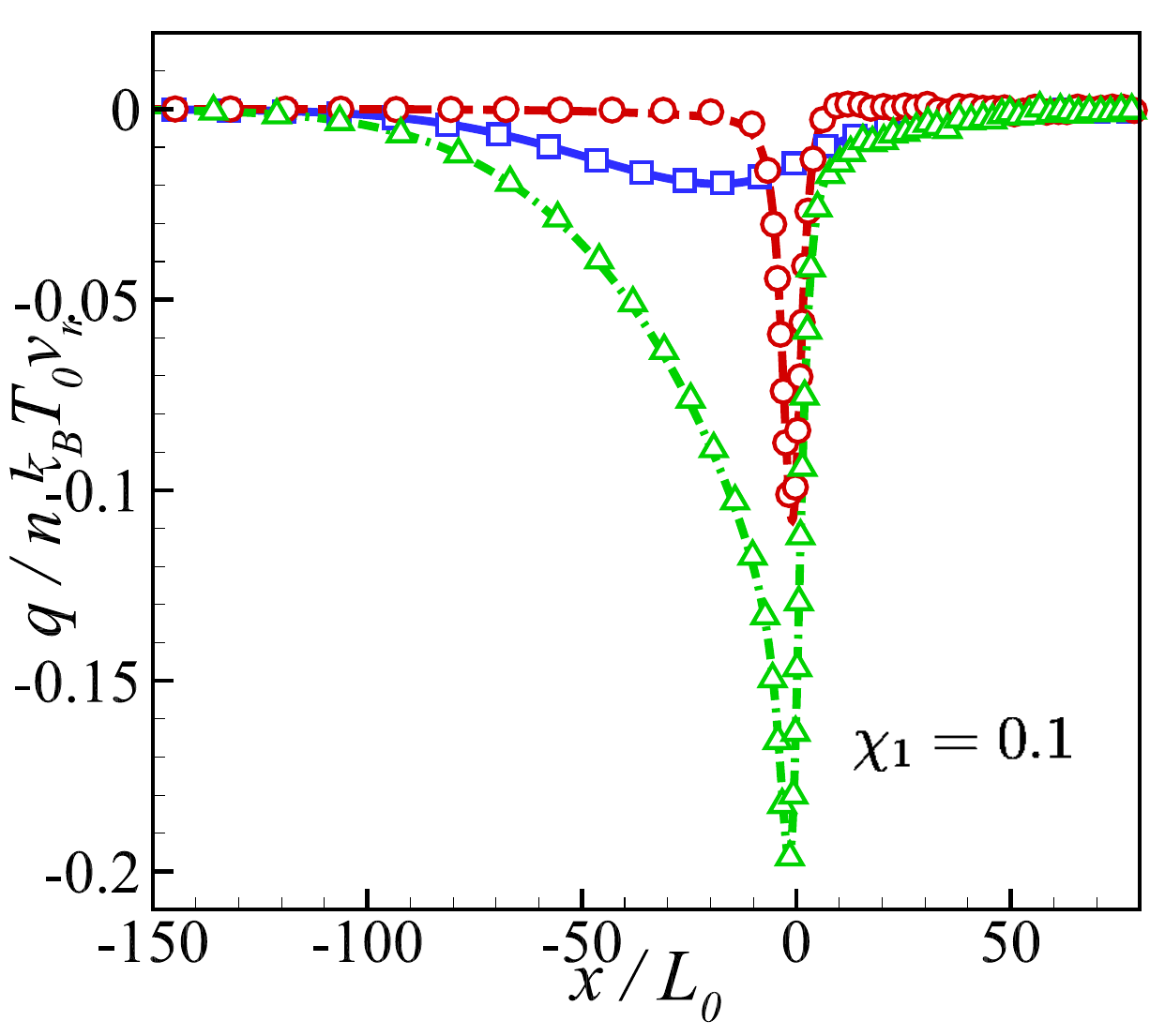}\label{fig:1DNormalShockWave_Mix2:X1_01_q}}   
	\sidesubfloat[]{\includegraphics[scale=0.19,clip=true]{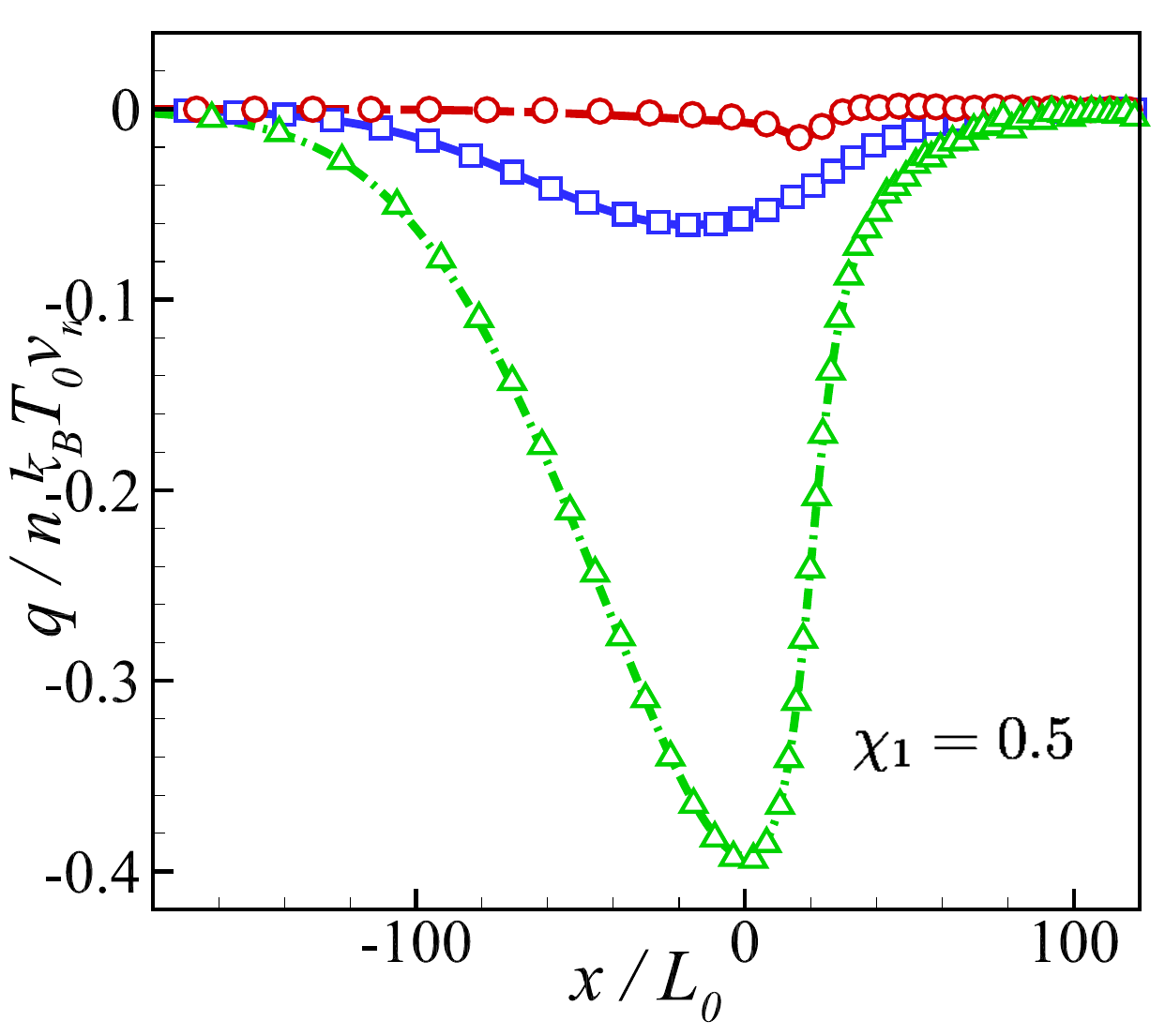}\label{fig:1DNormalShockWave_Mix2:X1_05_q}}  
    \sidesubfloat[]{\includegraphics[scale=0.19,clip=true]{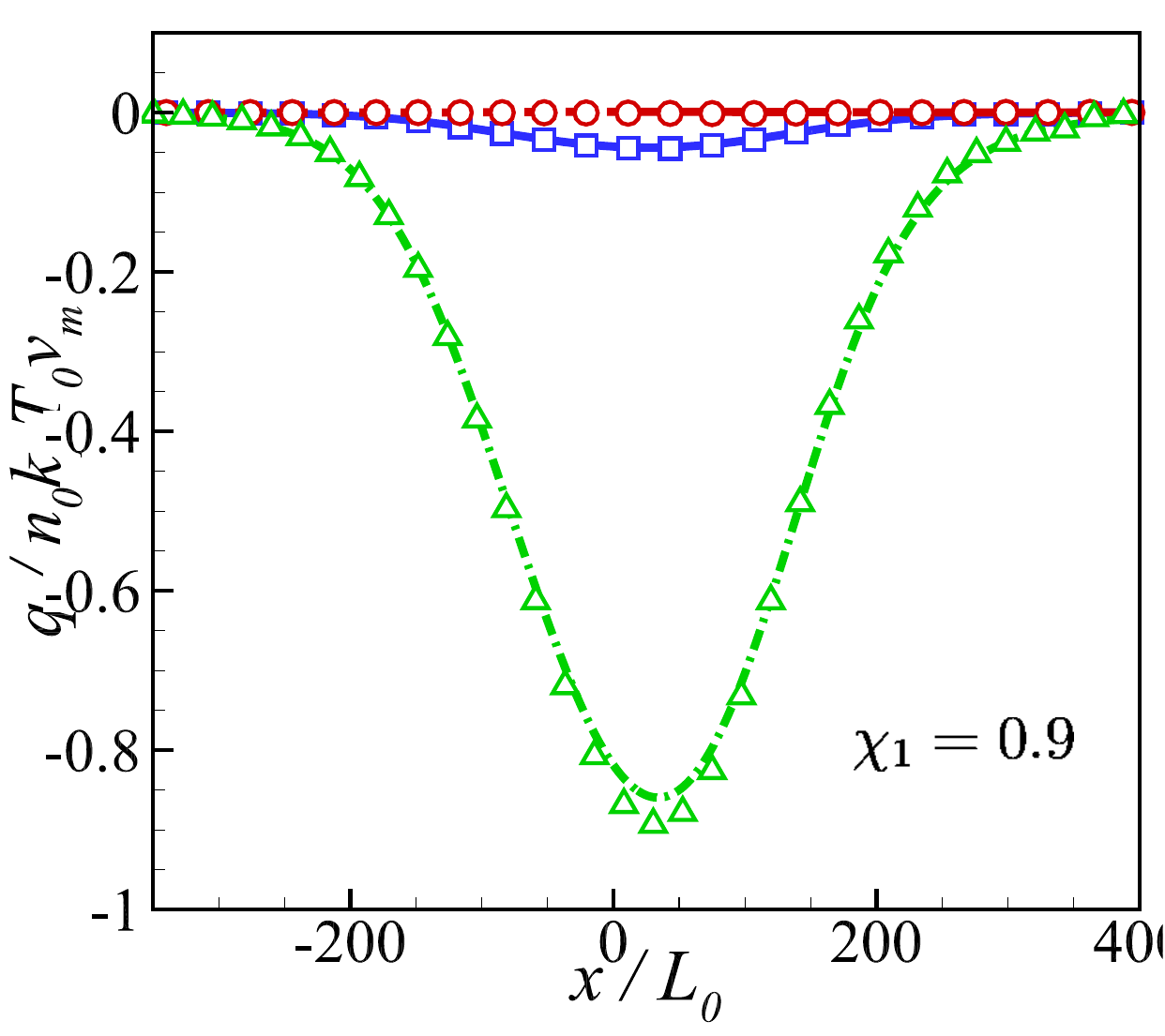}\label{fig:1DNormalShockWave_Mix2:X1_09_q}}  \\ 
    \includegraphics[scale=0.22,clip=true]{Figures/legend_1D.png}
	\caption{Comparisons of the normalized (a-c) number density, (d-f) flow velocity, (g-i) temperature, and dimensionless (j-l) heat flux of the gas mixture between kinetic model (lines) and DSMC (symbols) for the normal shock wave at $\text{Ma}=3$. The binary mixture consists of Maxwell molecules with a mass ratio $m_2/m_1=1000$, diameter ratio $d_2/d_1=1$, and the mole fraction of light species $\chi_1=0.1,0.5,0.9$.}
	\label{fig:1DNormalShockWave_Mix2}
\end{figure}

\begin{figure}[t]
	\centering
	\sidesubfloat[]{\includegraphics[scale=0.19,clip=true]{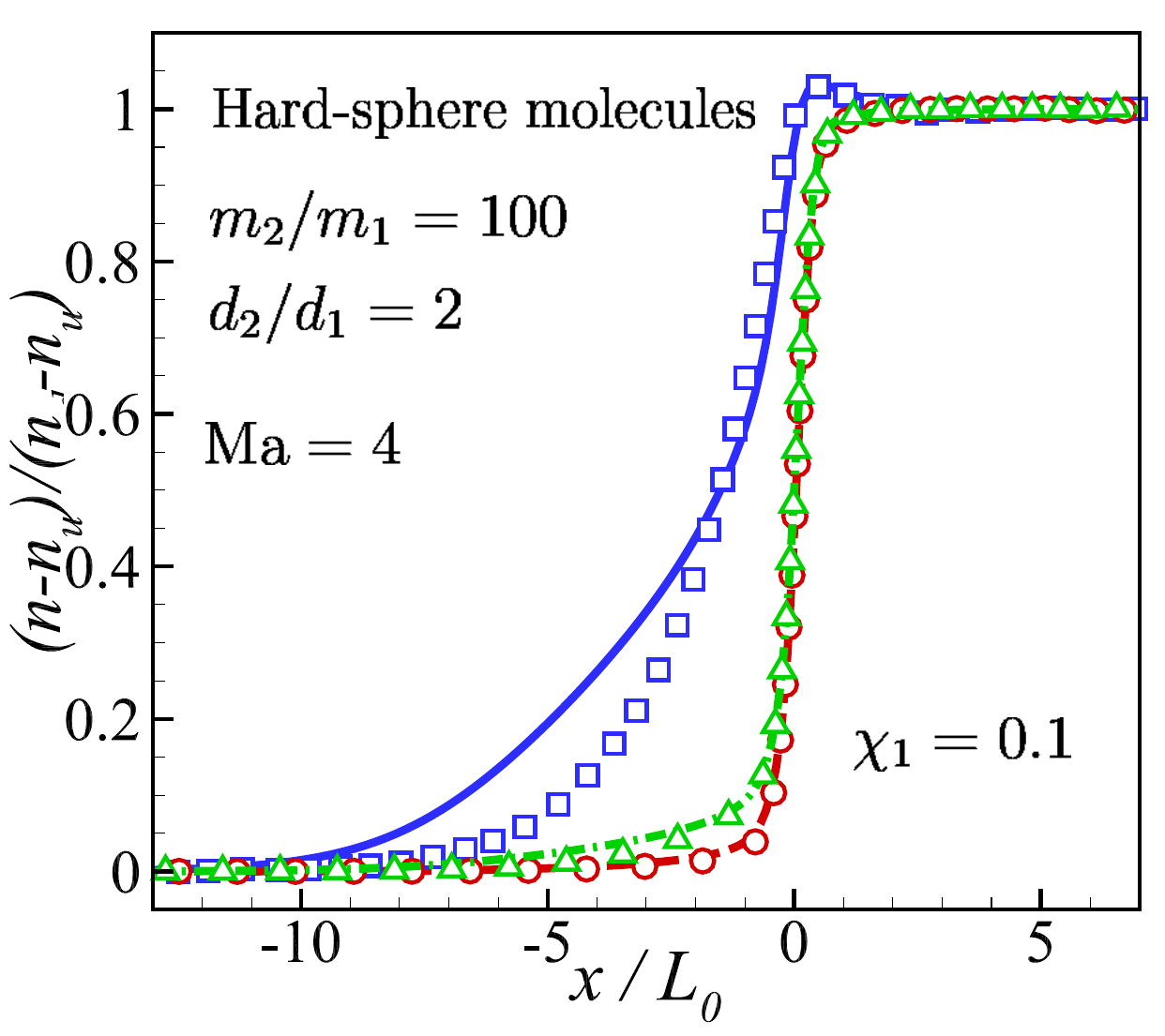}\label{fig:1DNormalShockWave_Mix3:X1_01_n}}   
	\sidesubfloat[]{\includegraphics[scale=0.19,clip=true]{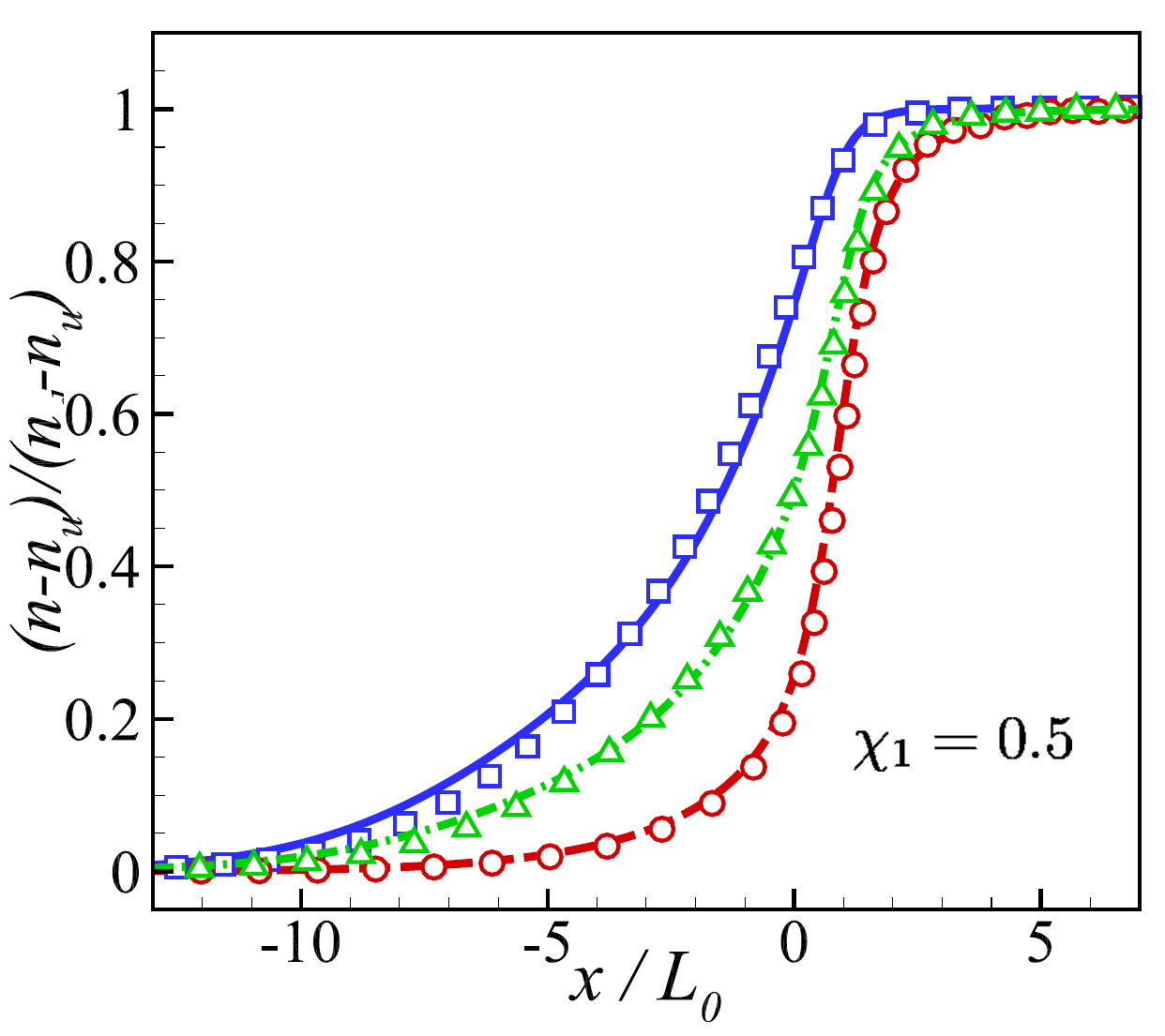}\label{fig:1DNormalShockWave_Mix3:X1_05_n}}  
    \sidesubfloat[]{\includegraphics[scale=0.19,clip=true]{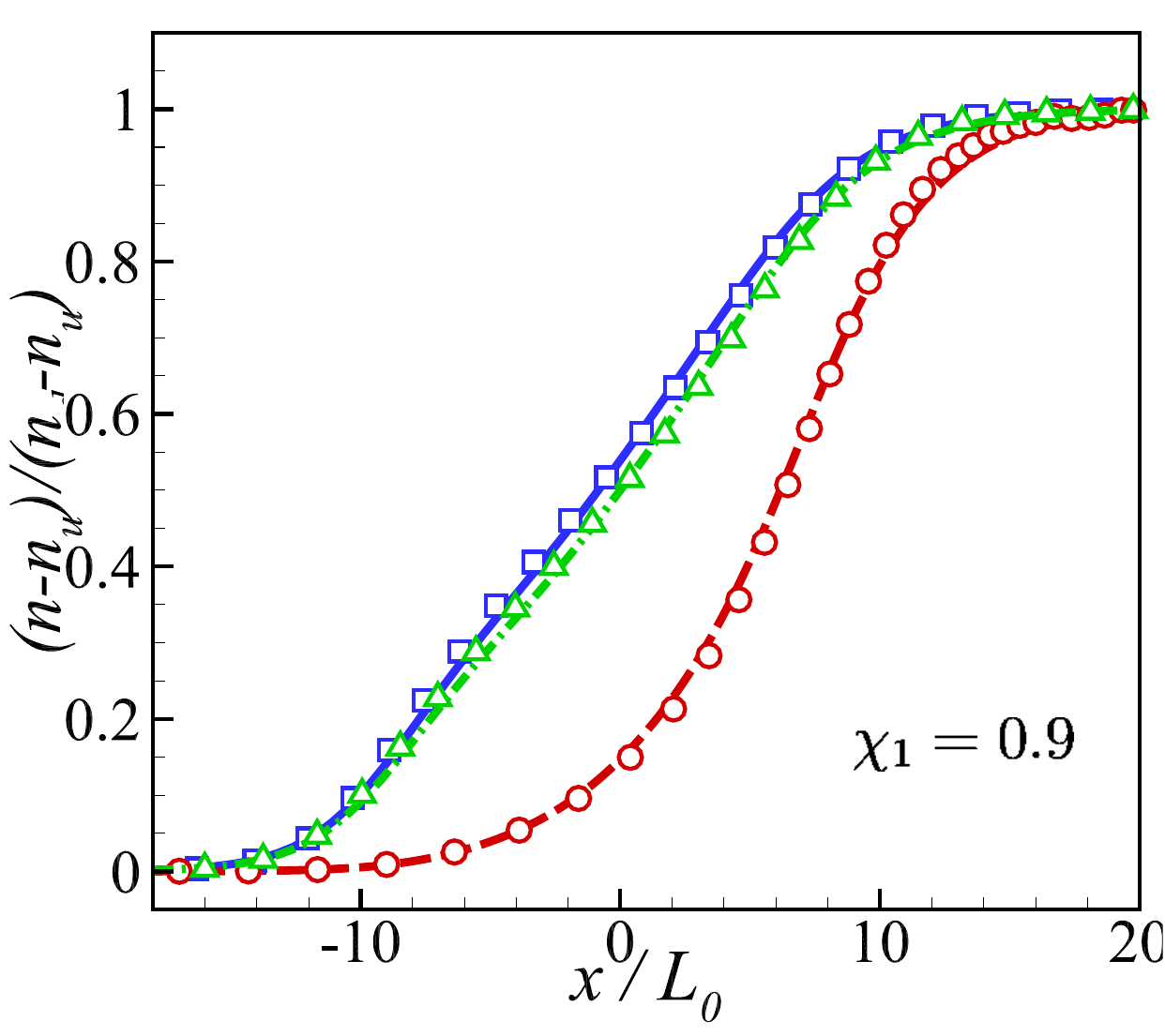}\label{fig:1DNormalShockWave_Mix3:X1_09_n}}  \\ 
    \sidesubfloat[]{\includegraphics[scale=0.19,clip=true]{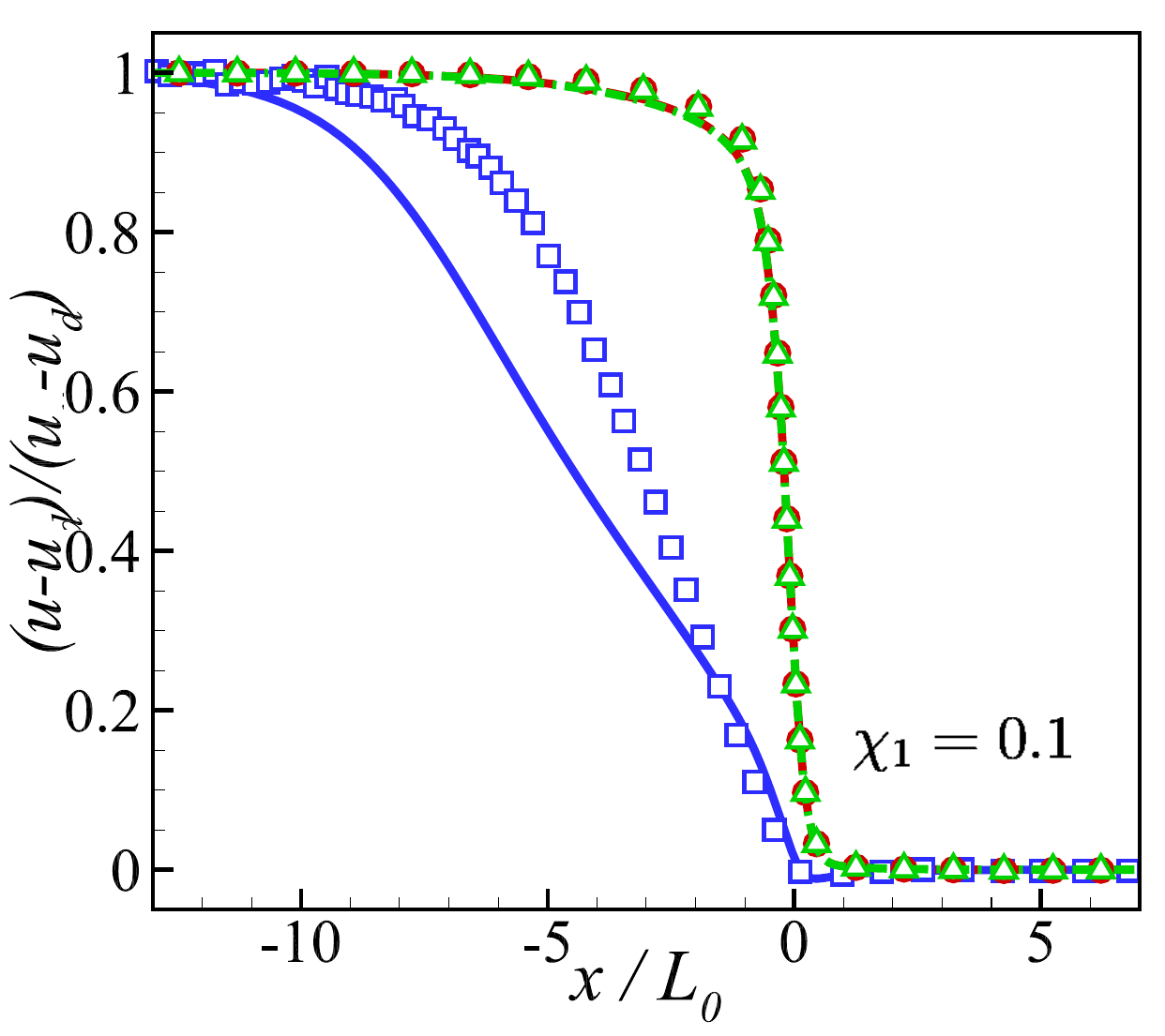}\label{fig:1DNormalShockWave_Mix3:X1_01_u}}   
	\sidesubfloat[]{\includegraphics[scale=0.19,clip=true]{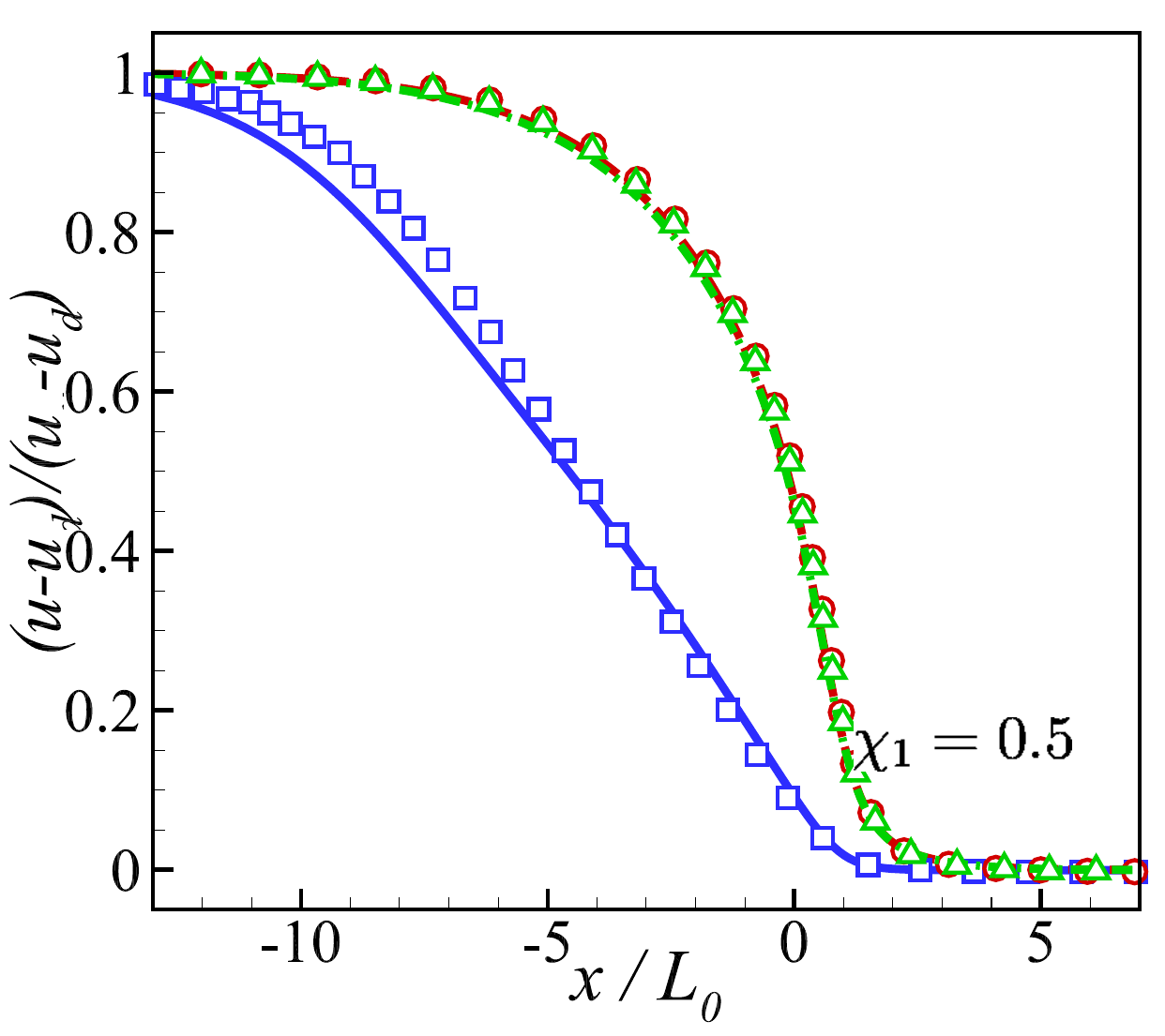}\label{fig:1DNormalShockWave_Mix3:X1_05_u}}  
    \sidesubfloat[]{\includegraphics[scale=0.19,clip=true]{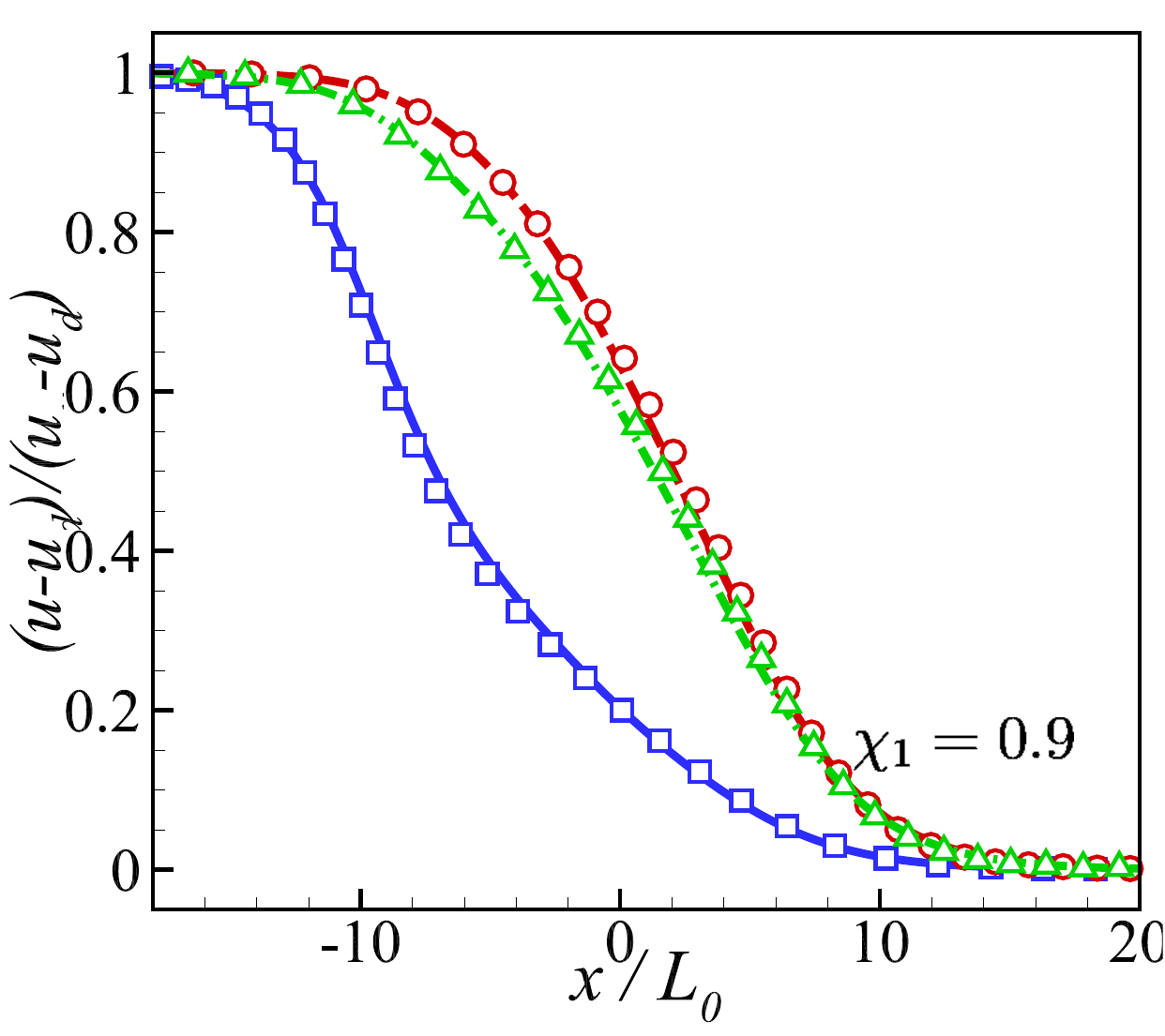}\label{fig:1DNormalShockWave_Mix3:X1_09_u}}  \\ 
    \sidesubfloat[]{\includegraphics[scale=0.19,clip=true]{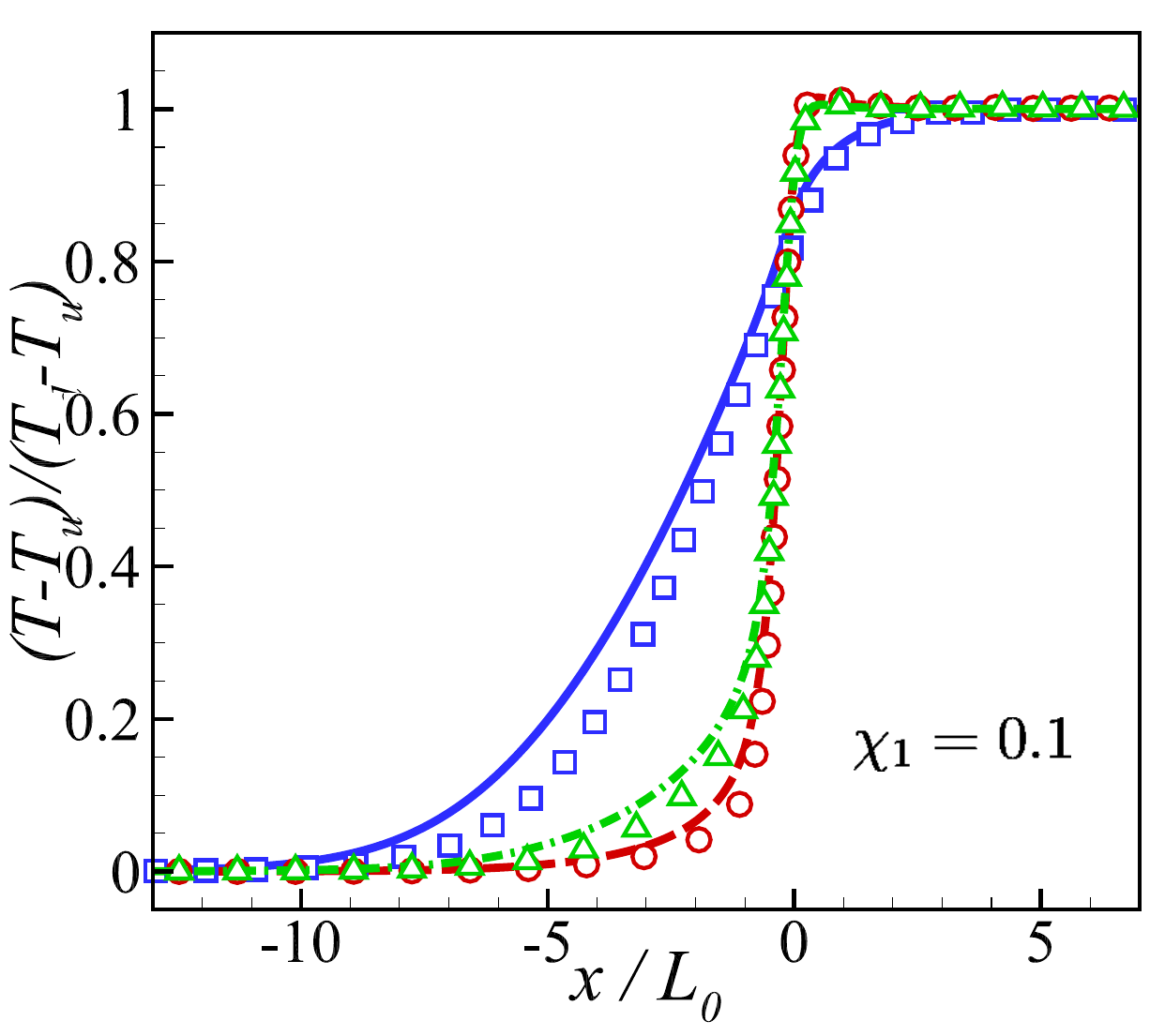}\label{fig:1DNormalShockWave_Mix3:X1_01_T}}   
	\sidesubfloat[]{\includegraphics[scale=0.19,clip=true]{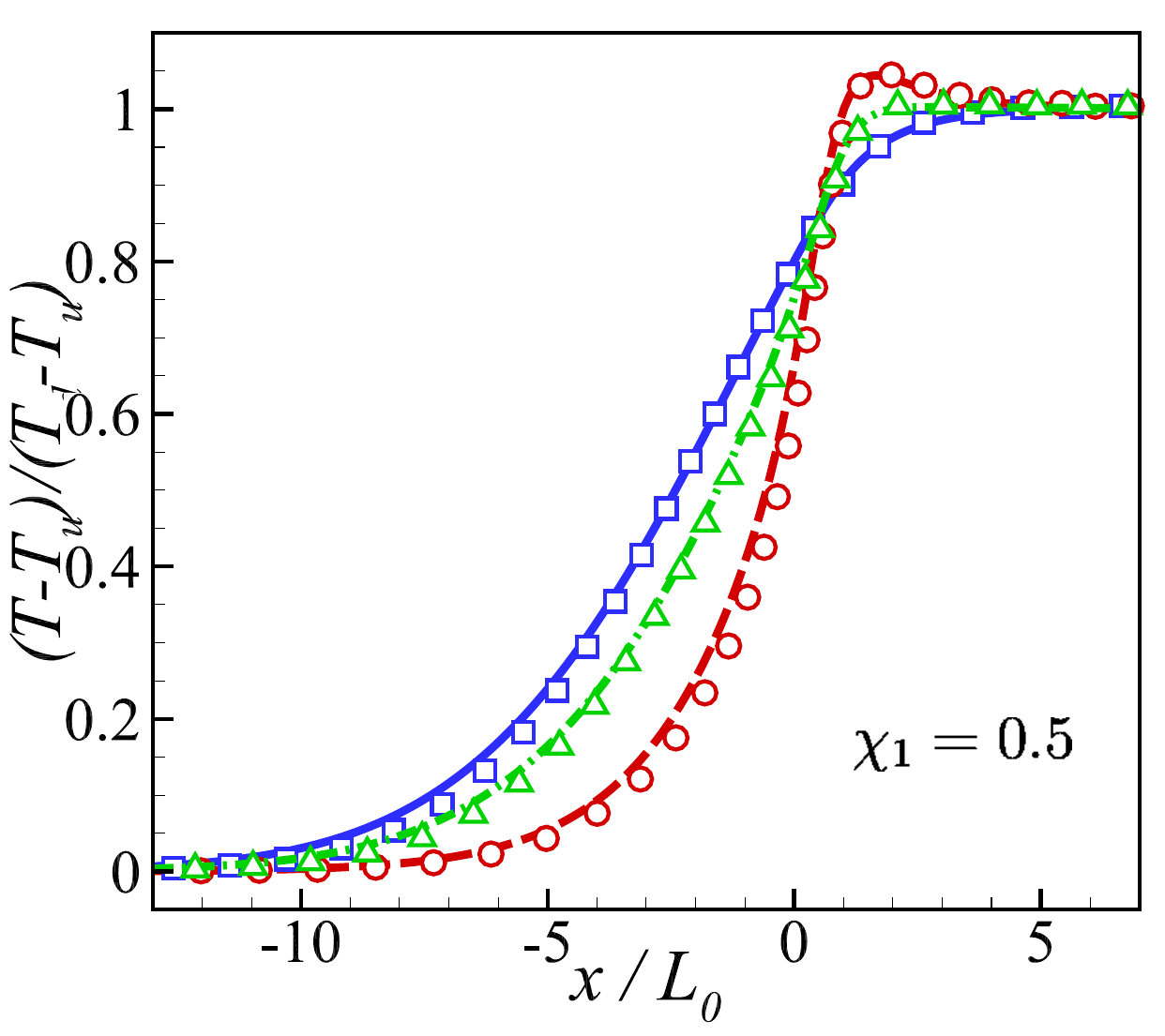}\label{fig:1DNormalShockWave_Mix3:X1_05_T}}  
    \sidesubfloat[]{\includegraphics[scale=0.19,clip=true]{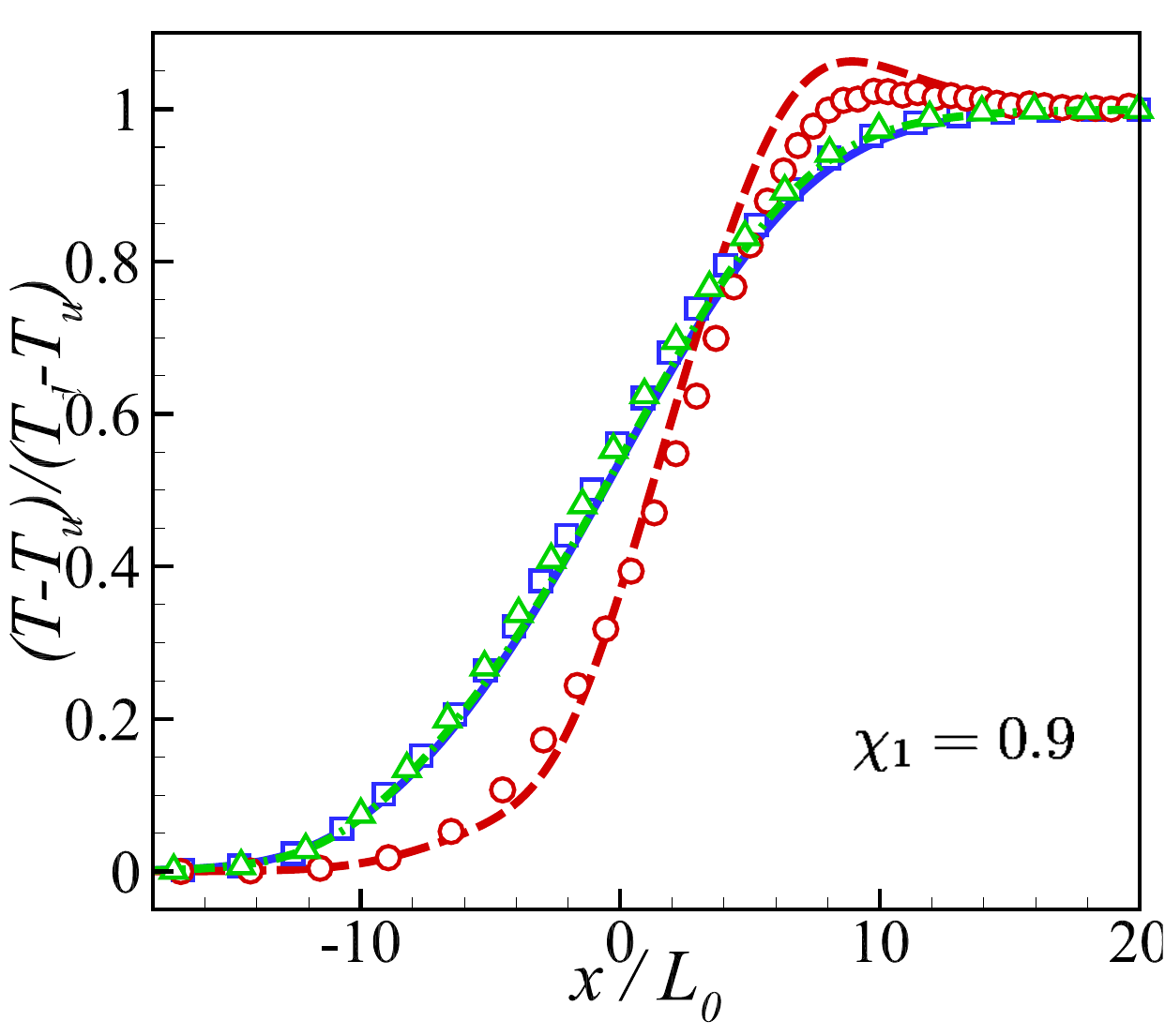}\label{fig:1DNormalShockWave_Mix3:X1_09_T}}  \\ 
    \sidesubfloat[]{\includegraphics[scale=0.19,clip=true]{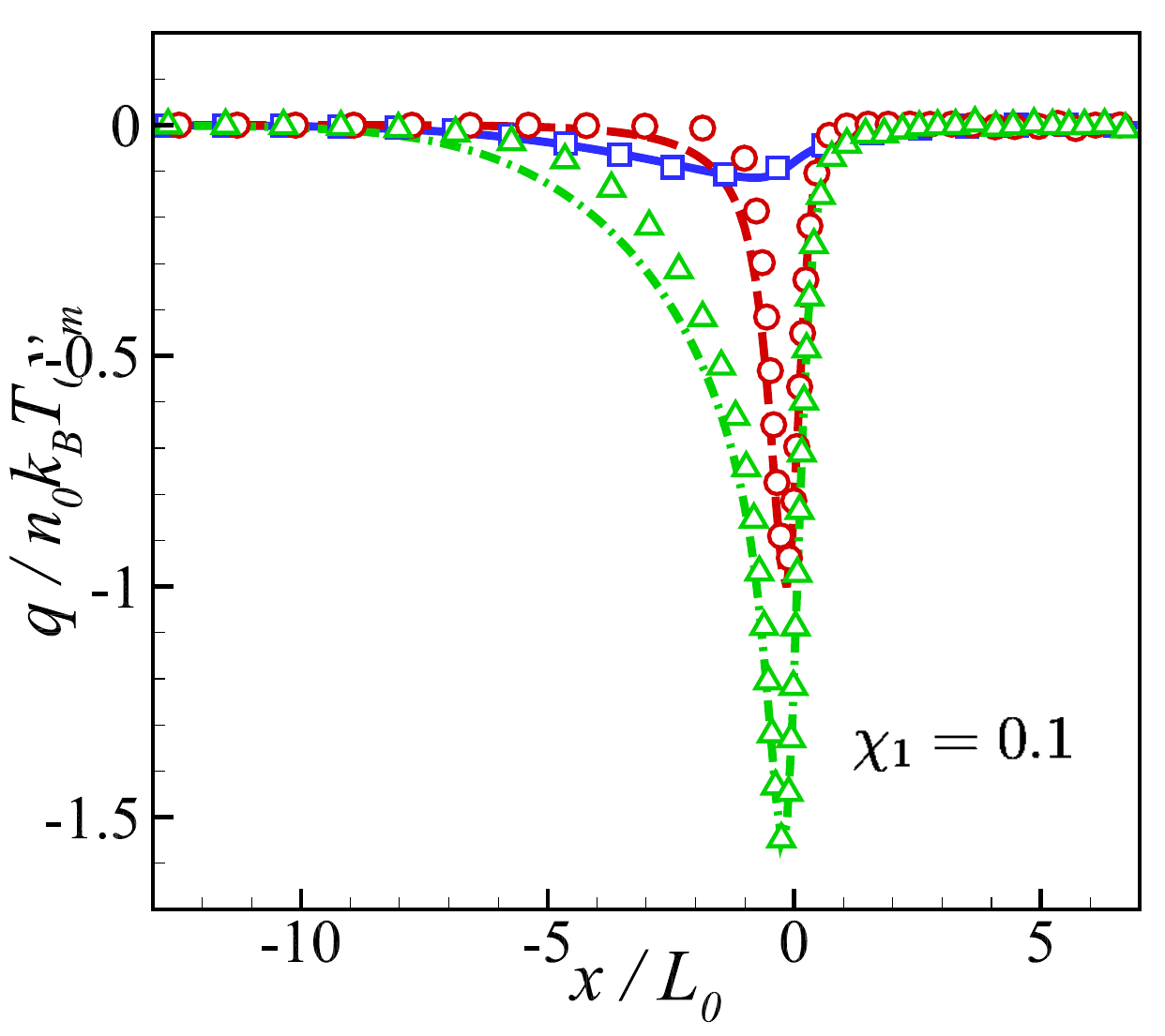}\label{fig:1DNormalShockWave_Mix3:X1_01_q}}   
	\sidesubfloat[]{\includegraphics[scale=0.19,clip=true]{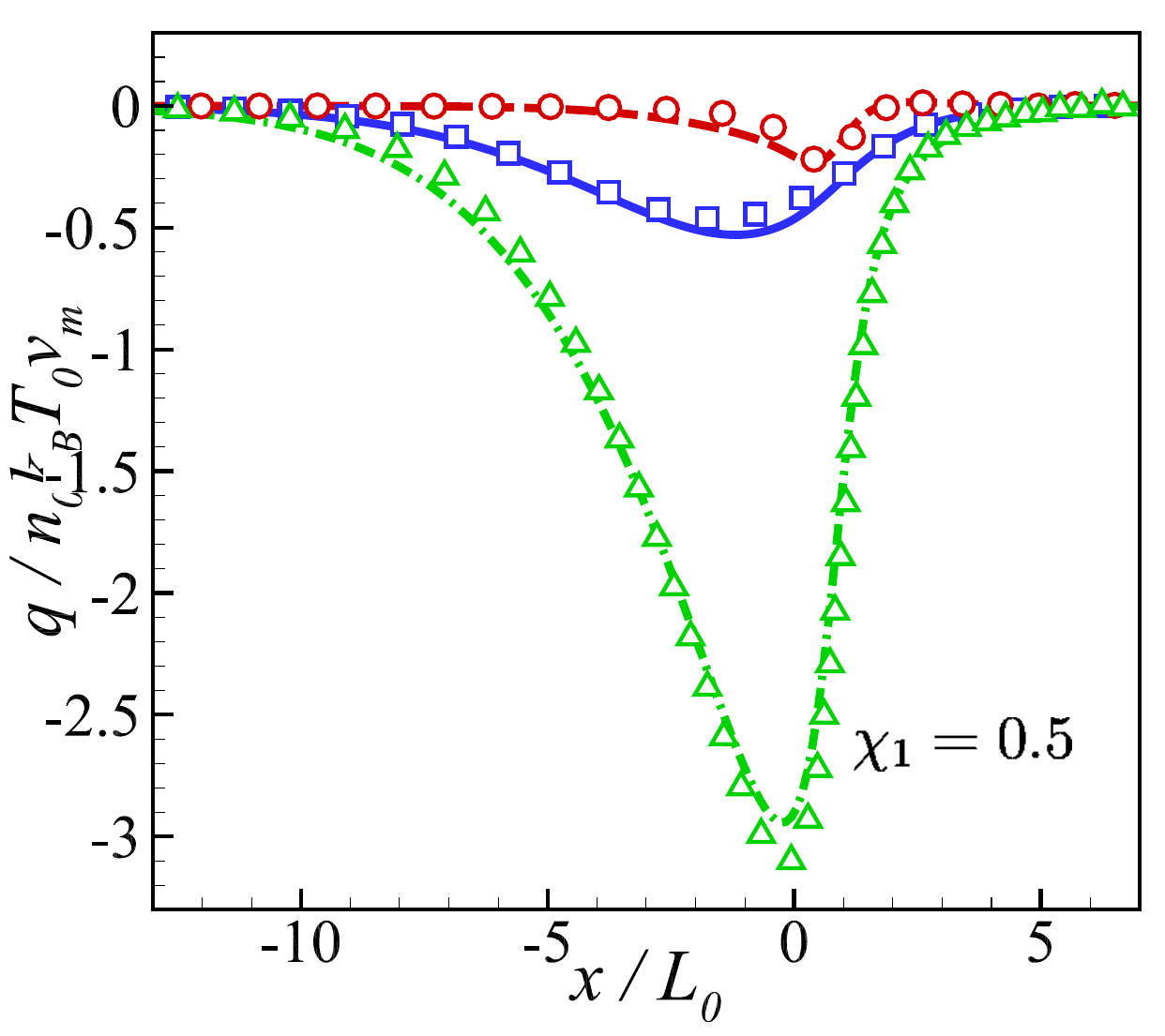}\label{fig:1DNormalShockWave_Mix3:X1_05_q}}  
    \sidesubfloat[]{\includegraphics[scale=0.19,clip=true]{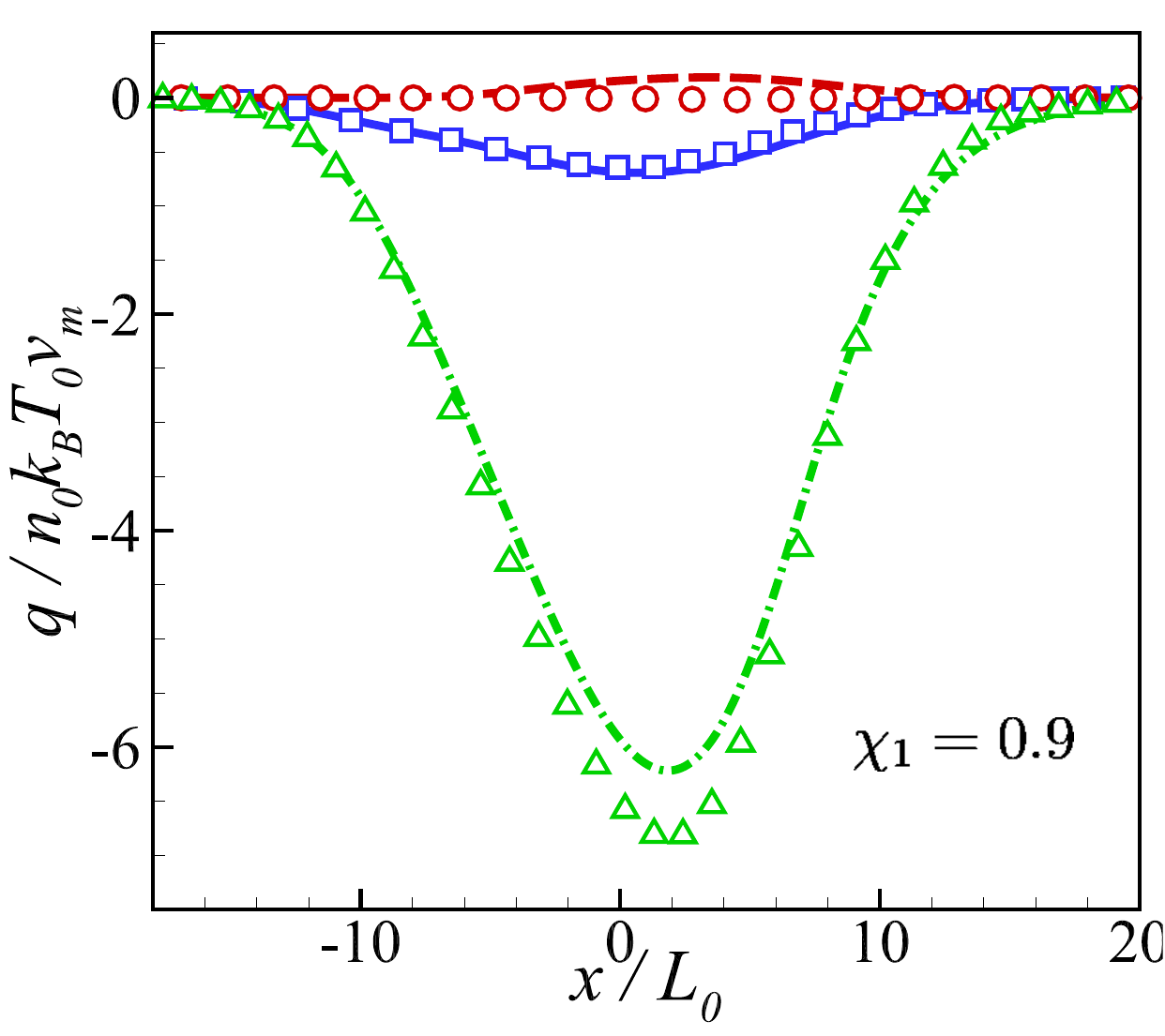}\label{fig:1DNormalShockWave_Mix3:X1_09_q}}  \\ 
    \includegraphics[scale=0.22,clip=true]{Figures/legend_1D.png}
	\caption{Comparisons of the normalized (a-c) number density, (d-f) flow velocity, (g-i) temperature, and dimensionless (j-l) heat flux of the gas mixture between kinetic model (lines) and DSMC (symbols) for the normal shock wave at $\text{Ma}=4$. The binary mixture consists of hard-sphere molecules with a mass ratio $m_2/m_1=100$, diameter ratio $d_2/d_1=2$, and the mole fraction of light species $\chi_1=0.1,0.5,0.9$.}
	\label{fig:1DNormalShockWave_Mix3}
\end{figure}

We investigate the structure of normal shock waves under various freestream conditions. The conditions are defined by the species mole fraction $\chi_s$ in the unperturbed freestream mixture, as well as the Mach number ($\text{Ma}$) calculated based on the speed of sound ${v}_{mix}=\sqrt{5k_BT_u/3m_{mix}}$ in the upstream, where $T_u$ is the upstream temperature and $m_{mix}=\sum{m_s\chi_s}$ is the averaged mass of the mixture. Given the freestream conditions, the macroscopic quantities at the downstream end are determined by the classical Rankine–Hugoniot relation.  

The mass of the lighter species (denoted as species 1), the mixture number density $n_u$ and the temperature $T_u$ of the upstream flow are taken as reference values, namely, $m_0=m_1$, $n_0=n_u$, $T_0=T_u$. The characteristic length $L_0$ is set equal to the mean free path of the lighter species in the upstream flow, thus leading to $\text{Kn}_1=1$. The simulation domain $[-L_x,L_x]$ is selected to ensure that the boundary conditions at the upstream and downstream ends can be approximated by equilibrium states, and then the specific values of $L_x$ are chosen as $40L_0,500L_0,30L_0$ for Mixtures 1, 2 and 3, respectively, due to the significant differences in the properties of these gas mixtures and the freestream conditions.

Numerical results of both the kinetic model and DSMC are compared in figures \ref{fig:1DNormalShockWave_Mix1}, \ref{fig:1DNormalShockWave_Mix2} and \ref{fig:1DNormalShockWave_Mix3} for Mixtures 1, 2 and 3, respectively. We present the normalized values for number density, flow velocity, temperature, and dimensionless heat flux for each species and also the mixture, under the specified conditions with different concentration $\chi_1=0.1,0.5,0.9$ and $\text{Ma}=3,4,5$. For Mixture 1 and 2 consisting of Maxwell gases, excellent agreements between the results of the kinetic model and DSMC are achieved (figure \ref{fig:1DNormalShockWave_Mix1} and \ref{fig:1DNormalShockWave_Mix2}), even when the mass ratio is as high as 1000. For Mixture 3 consisting of hard-sphere molecules, the number density and velocity of the lighter species predicted by the kinetic model deviate from DSMC results, when the lighter species only present in a small amount ($\chi_1=0.1$, see figure \ref{fig:1DNormalShockWave_Mix3:X1_01_n} and \ref{fig:1DNormalShockWave_Mix3:X1_01_u}). However, the average properties of the mixture remain highly accurate despite this discrepancy, since in this situation, the lighter species, due to its low concentration, has a minimal impact on the overall behavior of the mixture. Generally, three reasons may contribute to the possible deviation predicted by the kinetic model for mixtures consisting of non-Maxwell gases: (i) the parameters $c_{sr}$ and $d_{sr}$ accounting for the energy relaxation are derived in the sense of approximation for non-Maxwell molecules; (ii) the term with parameter $b_{sr}$ is designed to capture the correct thermal diffusion phenomena in the continuum limit, thus may have a discrepancy in strong non-equilibrium cases; (iii) the velocity-dependent collision frequency for non-Maxwell molecules is not recovered in the BGK-types operators. Nevertheless, as shown in figure \ref{fig:1DNormalShockWave_Mix3}, the kinetic model gives good overall agreement with DSMC simulations for hard-sphere gas mixtures. Therefore, the agreement suggests that the kinetic model can predict accurate results for real gases, whose behavior usually lies between that of hard-sphere and Maxwell molecules.

Although the shock waves in mixtures with large mass ratios (e.g. 32.8 for Helium-Xenon) have been studied in the literature, and have shown unique characteristics absent in those composed of similar gas molecules. The shock structures can be significantly altered by a further substantial mass disparity within the mixture. By comparing the shock structures in a wide range of mass ratios and species concentrations, the following features can be observed and correctly captured by our kinetic model:
\begin{enumerate}
	\item The shock wave thickness of a gas mixture is markedly greater than that of a pure gas, especially when the mass ratio is large, as the mixture viscosity and diffusivity become stronger and the relaxation between components gets slower. For example, as shown in figure \ref{fig:1DNormalShockWave_Mix2}, the Maxwell gases with a mass ratio of 1000 and diameter ratio of 1 form a significantly large transition zone from the upstream to the downstream, which spans several hundreds of the molecular mean free path.
	\item In Mixture 1 with a mass ratio of 10 (moderate mass difference), a pronounced temperature overshoot of the heavy species (higher than the downstream temperature) is observed when the heavier gas has only a small proportion in the mixture (figure \ref{fig:1DNormalShockWave_Mix1:X1_09_T}), which has been shown in the literature from kinetic modelling \citep{Bird1968JFM,Kosuge2001EJMB,Sharipov2018EJMB} and hydrodynamic equations \citep{Schmidt1984JFM}. However, when the mass ratio increases to 1000, the temperature overshoot gradually vanishes with the concentration unchanged. As illustrated in figure \ref{fig:1DNormalShockWave_Mix2}, a comparable or even larger proportion of the heavier gas is required for the temperature overshoot to occur.
	\item Flow velocity undershoot of the lighter gas happens in a mixture with a large mass ratio (100 and 1000) and a small proportion of lighter species ($\chi_1=0.1$) (figure \ref{fig:1DNormalShockWave_Mix2:X1_01_u} and \ref{fig:1DNormalShockWave_Mix3:X1_01_u}), which confirms the phenomena predicted by the multi-temperature hydrodynamic equations on weak shock \citep{Goldman1969JFM}. In these situations, the lighter gas decelerates to a velocity lower than that of the downstream flow, and thus is compressed to a density above the downstream one. 
	\item A two-stage shock structure with distinct gradients of the mixture properties can be observed. In Mixture 1 with a mass ratio of 10 and a large concentration of light species ($\chi_1=0.9$), the properties of the shock exhibit a very steep change on the upstream side, while followed by a sudden change in the form of a long tail downstream. This phenomenon happened in a mixture with a moderate mass difference has been reported in experiments of Helium-Xenon \citep{Gmurczyk1979RGD} and found by hydrodynamic equations \citep{Ruyev2002JAMTP}. However, when the mass ratio becomes significantly large in Mixtures 2 and 3 (figure \ref{fig:1DNormalShockWave_Mix2} and \ref{fig:1DNormalShockWave_Mix3}), this phenomenon disappears. On the contrary, the two-stage shock structure with an opposite trend occurs when there is only a small proportion of light species in the mixture ($\chi_1=0.1$). This structure consists of a smooth change in gas properties on the upstream side followed by a sudden and dramatic change on the downstream side.
\end{enumerate}

\subsection{Planar Fourier flow}

\begin{figure}[t]
	\centering
	\sidesubfloat[]{\includegraphics[scale=0.19,clip=true]{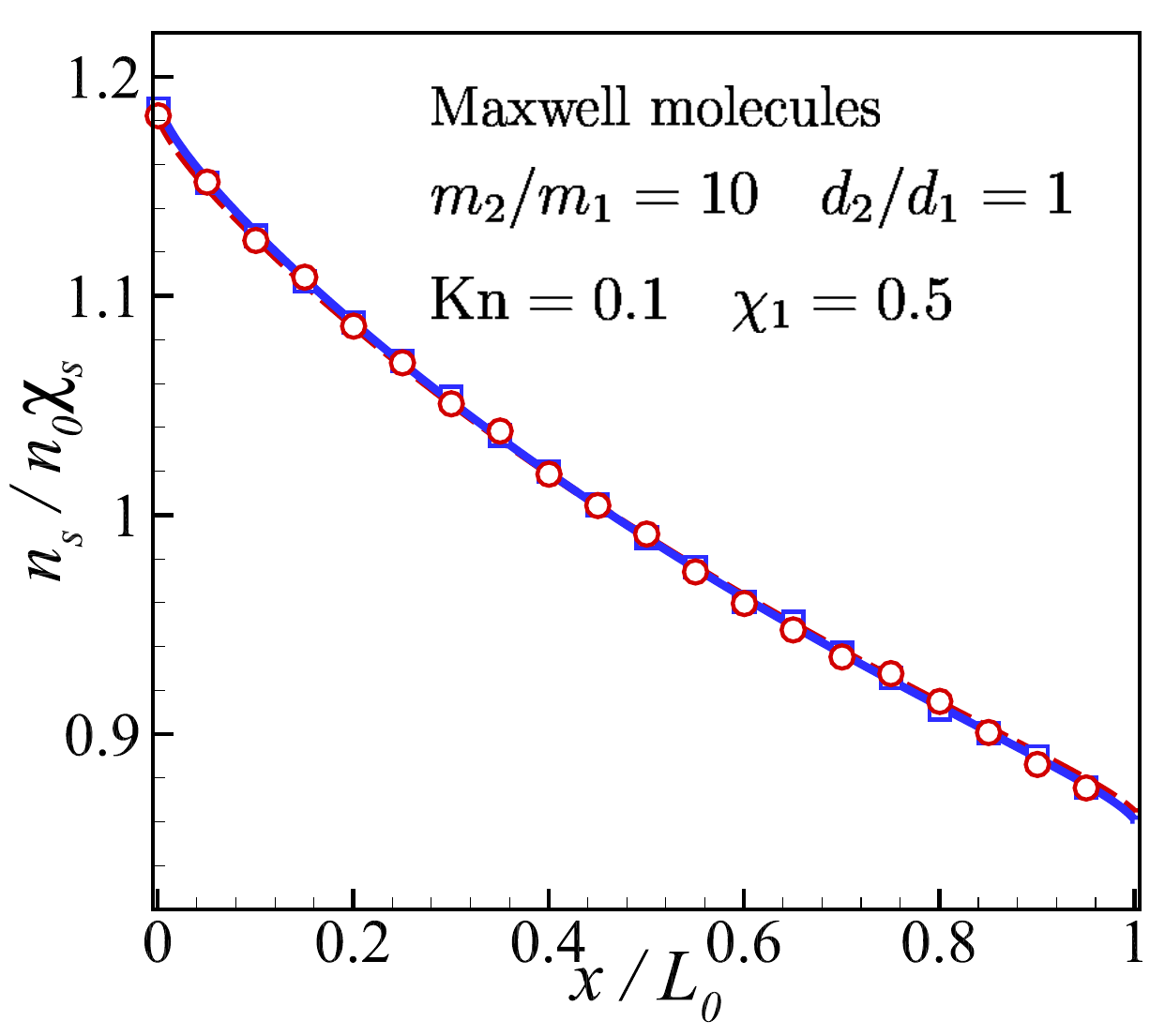}\label{fig:1DFourierFlow_Mix1_2:Mix1_Kn_01_X1_05_n}} 
    \sidesubfloat[]{\includegraphics[scale=0.19,clip=true]{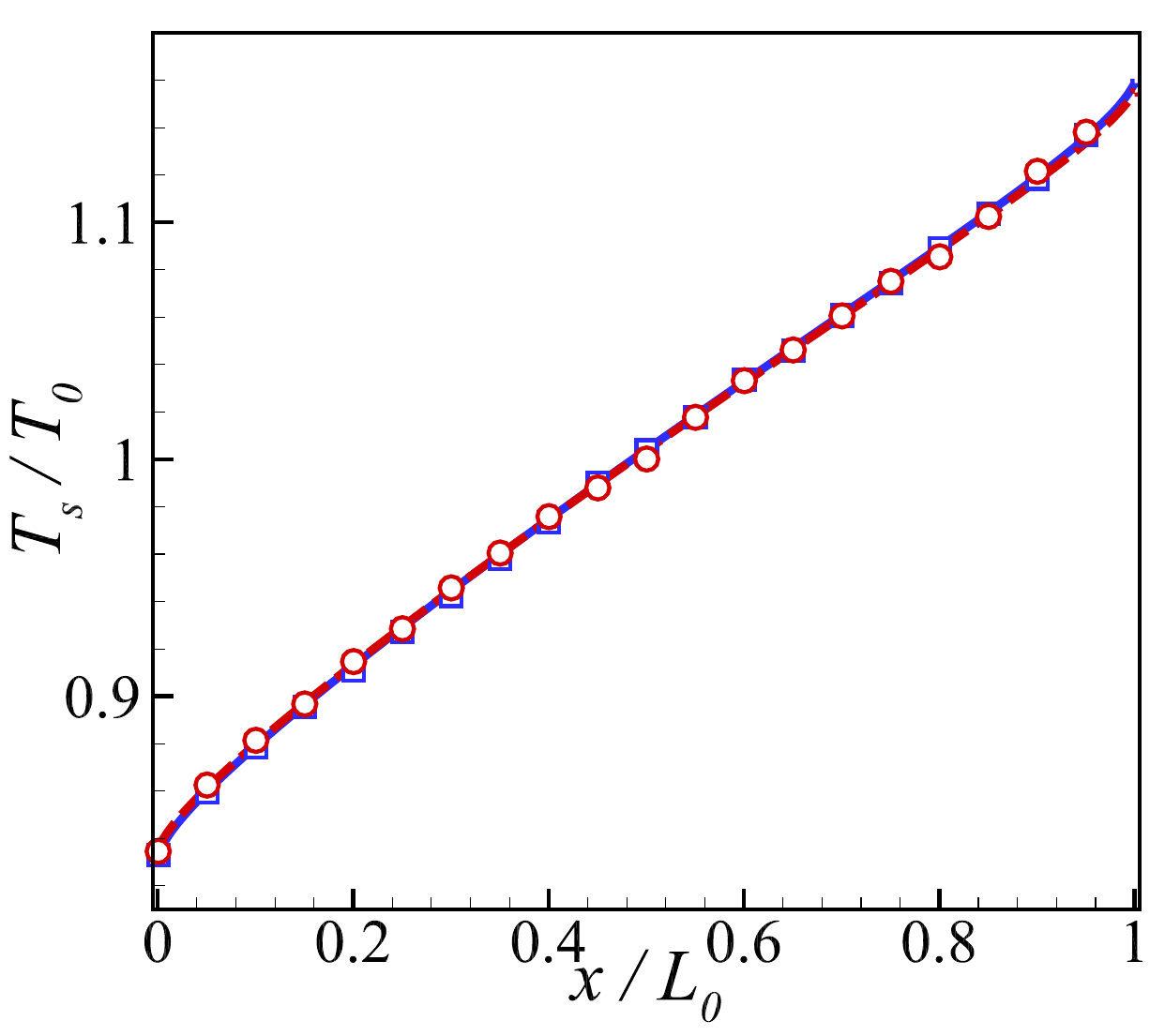}\label{fig:1DFourierFlow_Mix1_2:Mix1_Kn_01_X1_05_T}} 
    \sidesubfloat[]{\includegraphics[scale=0.19,clip=true]{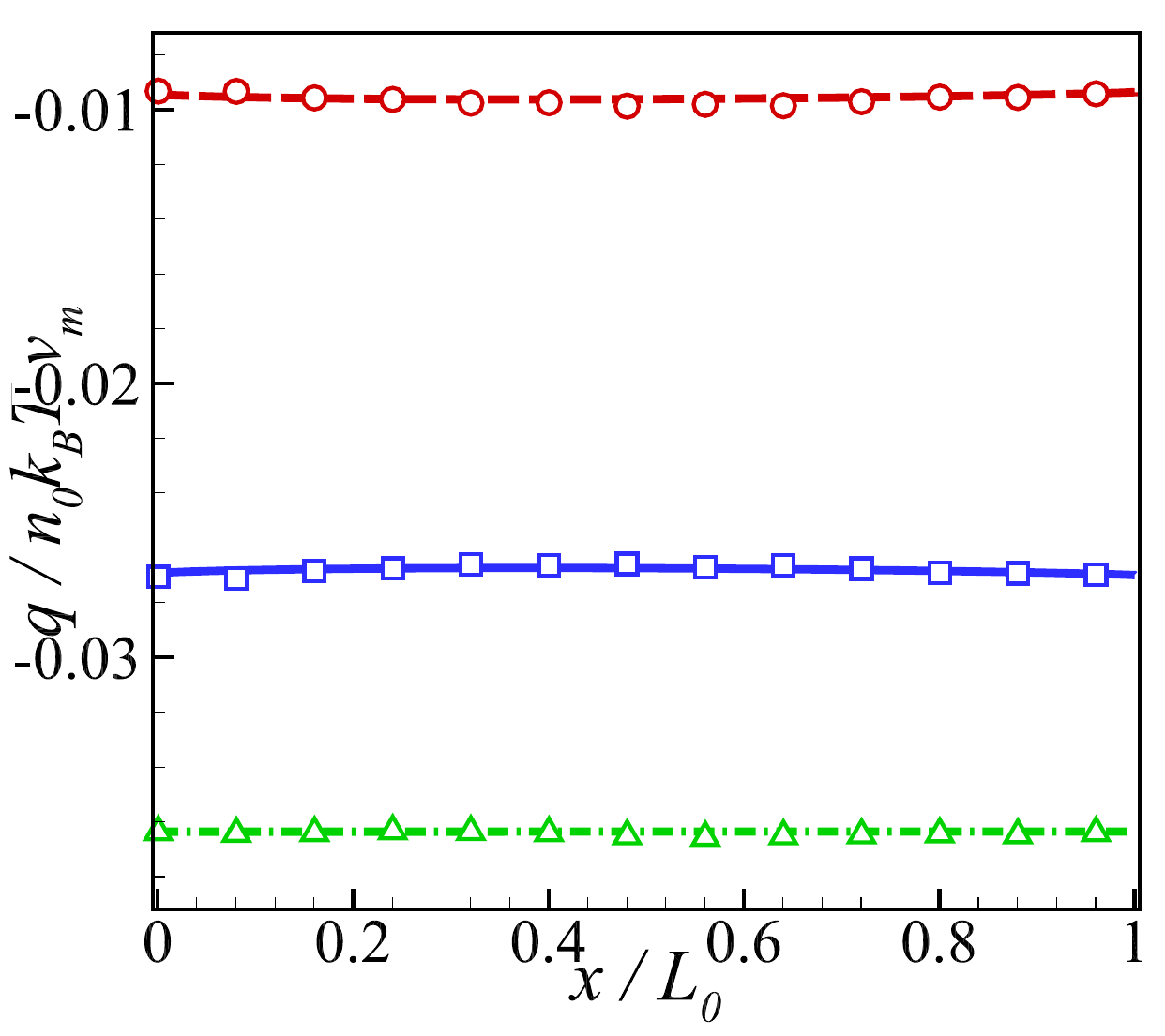}\label{fig:1DFourierFlow_Mix1_2:Mix1_Kn_01_X1_05_q}} \\ 
    \sidesubfloat[]{\includegraphics[scale=0.19,clip=true]{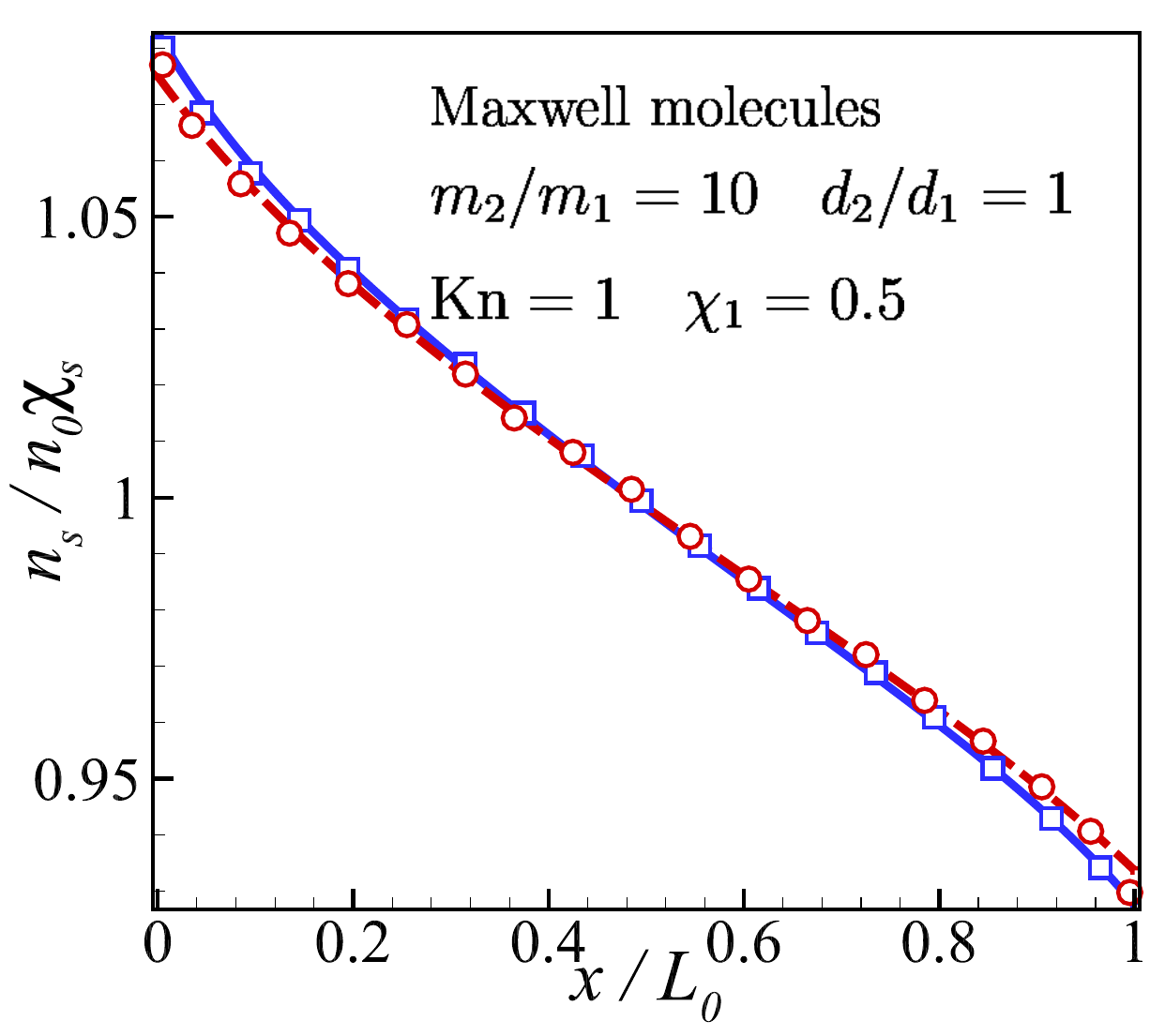}\label{fig:1DFourierFlow_Mix1_2:Mix1_Kn_1_X1_05_n}} 
    \sidesubfloat[]{\includegraphics[scale=0.19,clip=true]{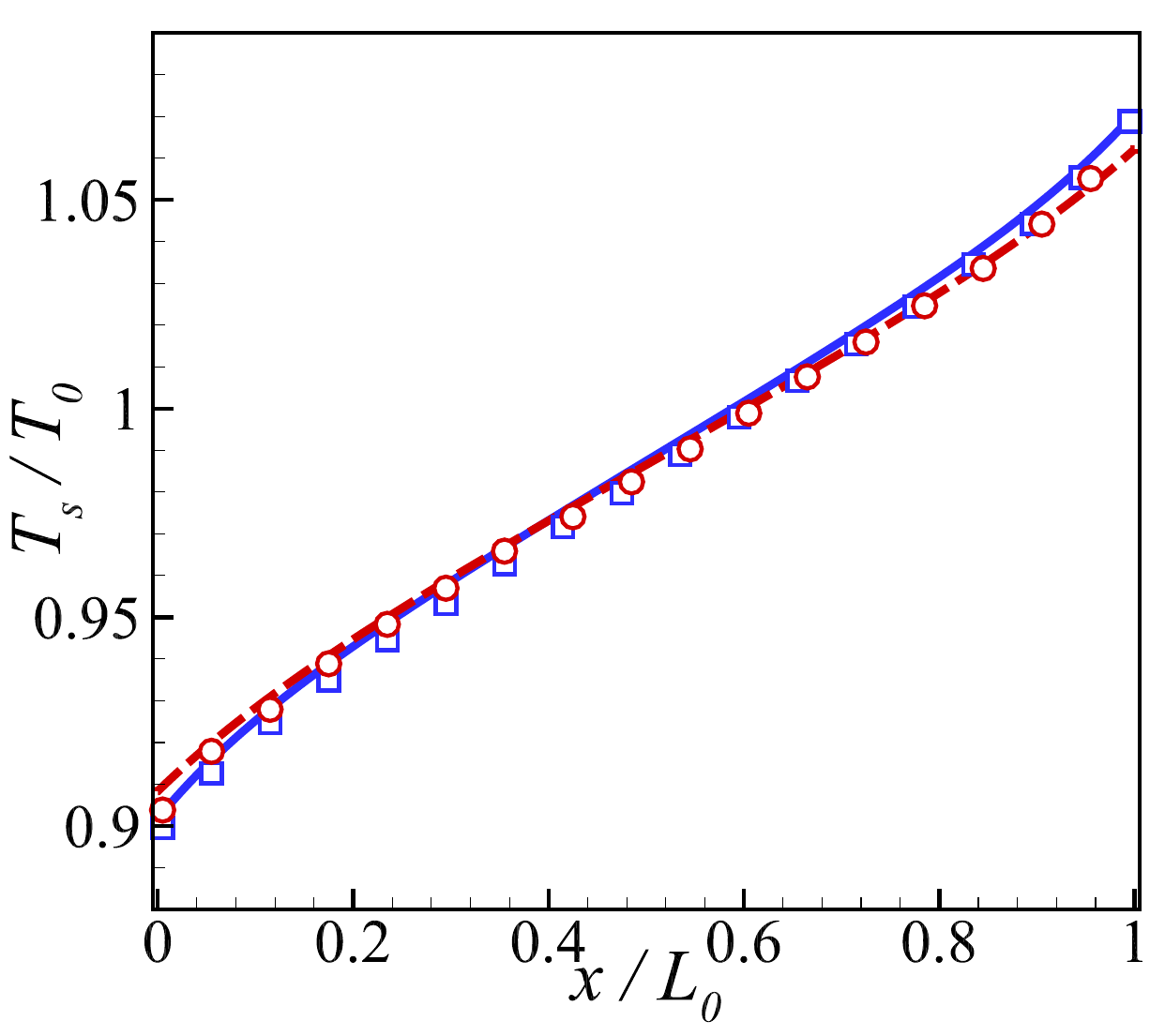}\label{fig:1DFourierFlow_Mix1_2:Mix1_Kn_1_X1_05_T}} 
    \sidesubfloat[]{\includegraphics[scale=0.19,clip=true]{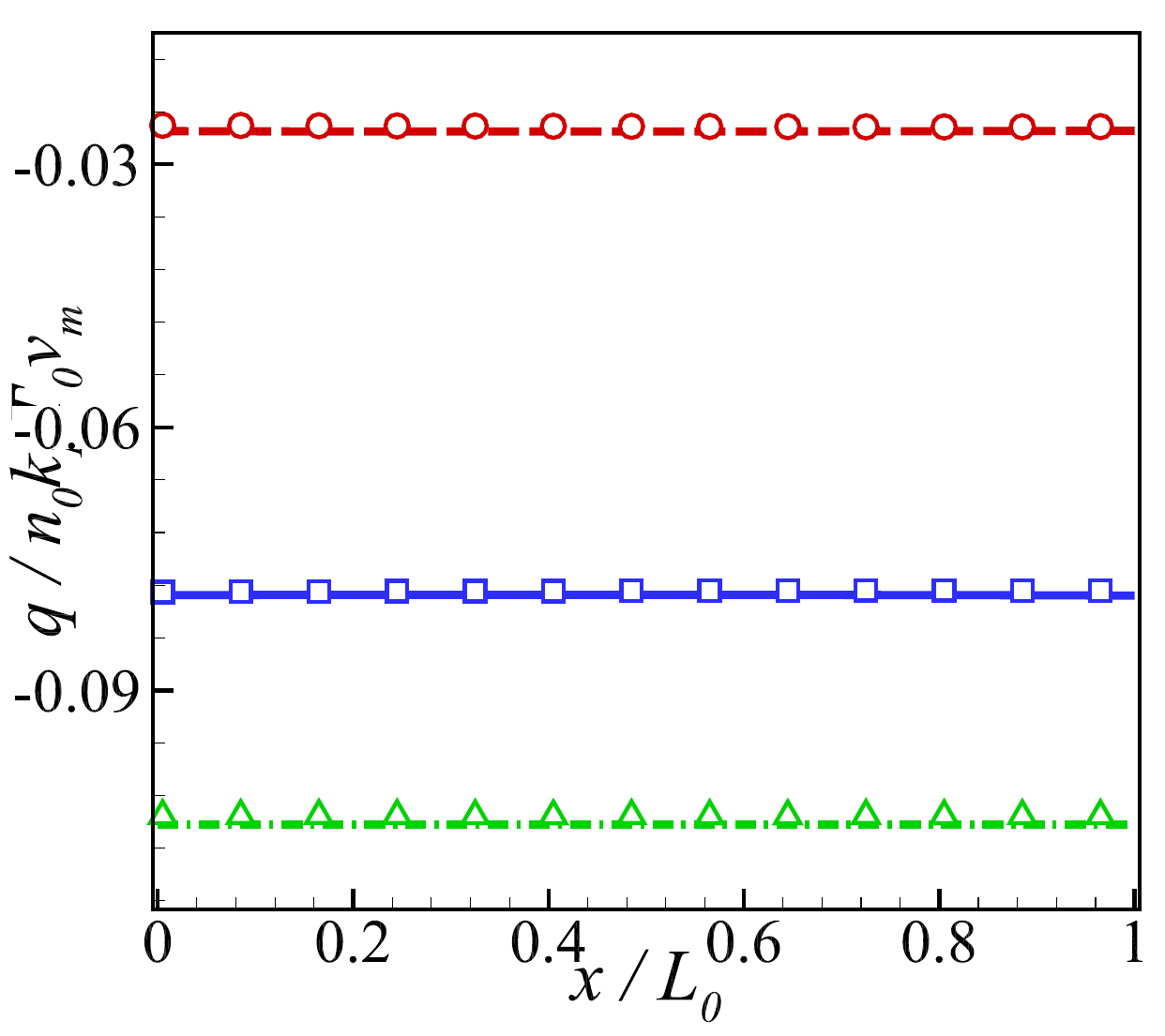}\label{fig:1DFourierFlow_Mix1_2:Mix1_Kn_1_X1_05_q}} \\
	\sidesubfloat[]{\includegraphics[scale=0.19,clip=true]{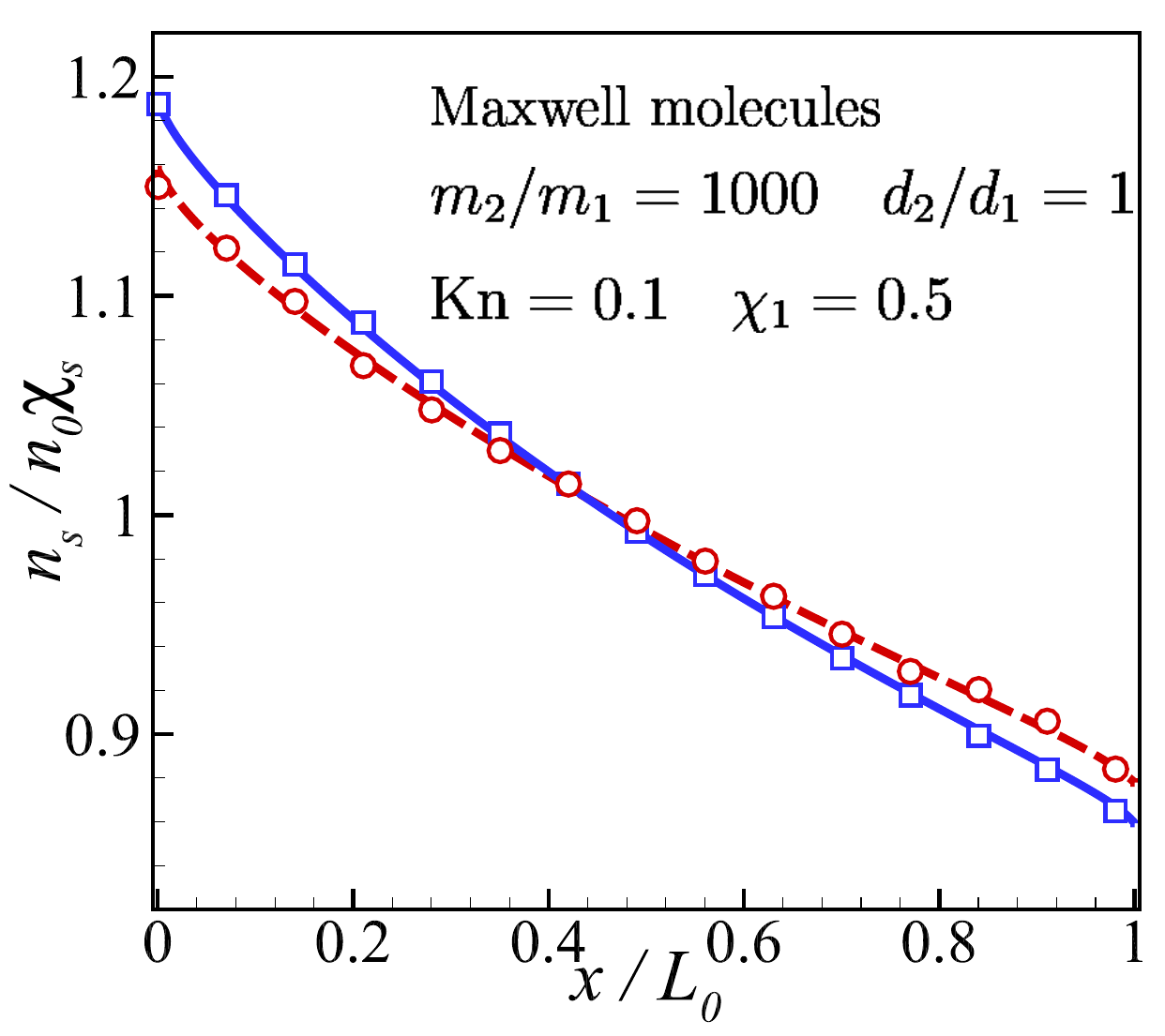}\label{fig:1DFourierFlow_Mix1_2:Mix2_Kn_01_X1_05_n}} 
    \sidesubfloat[]{\includegraphics[scale=0.19,clip=true]{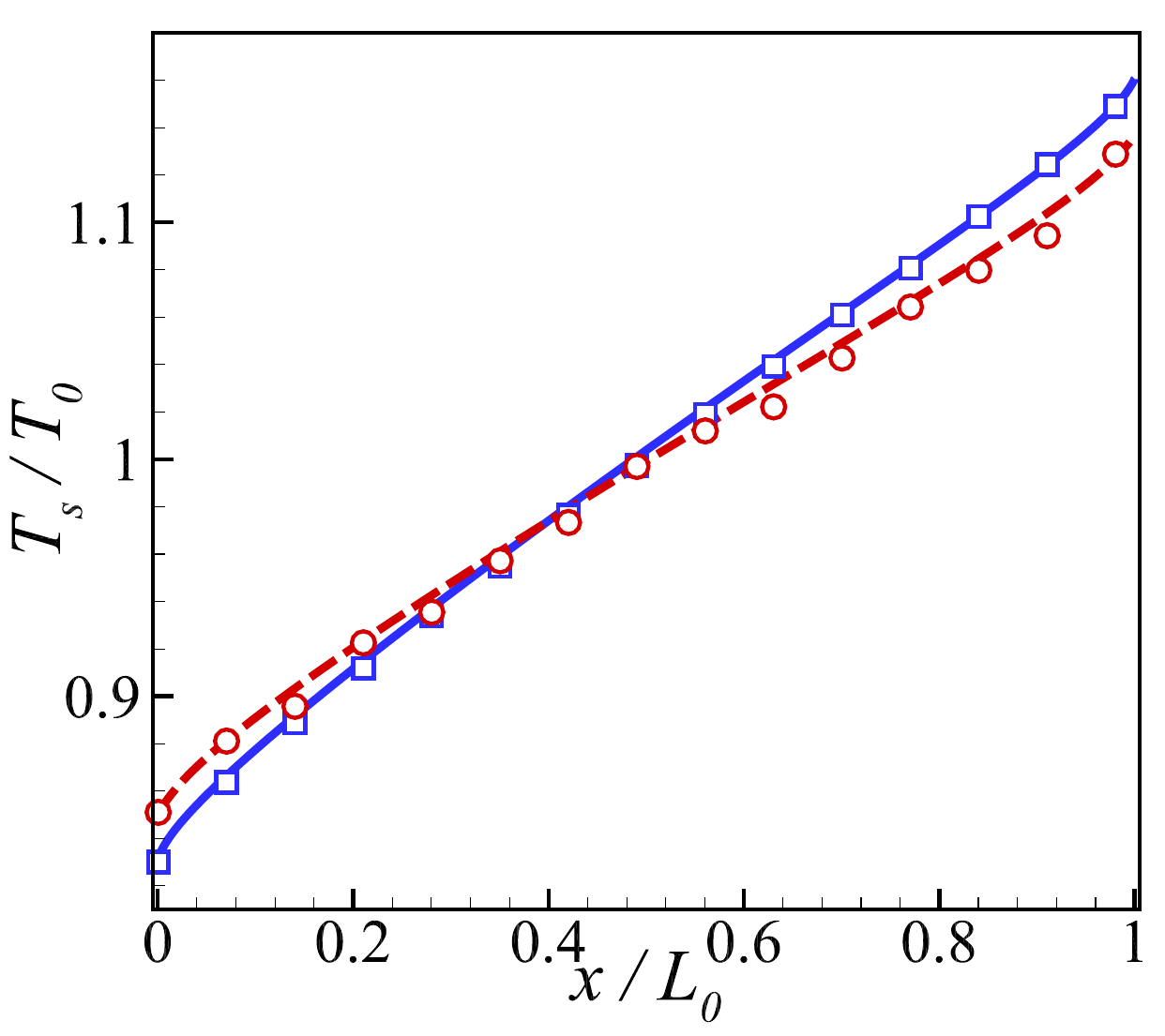}\label{fig:1DFourierFlow_Mix1_2:Mix2__Kn_01_X1_05_T}} 
    \sidesubfloat[]{\includegraphics[scale=0.19,clip=true]{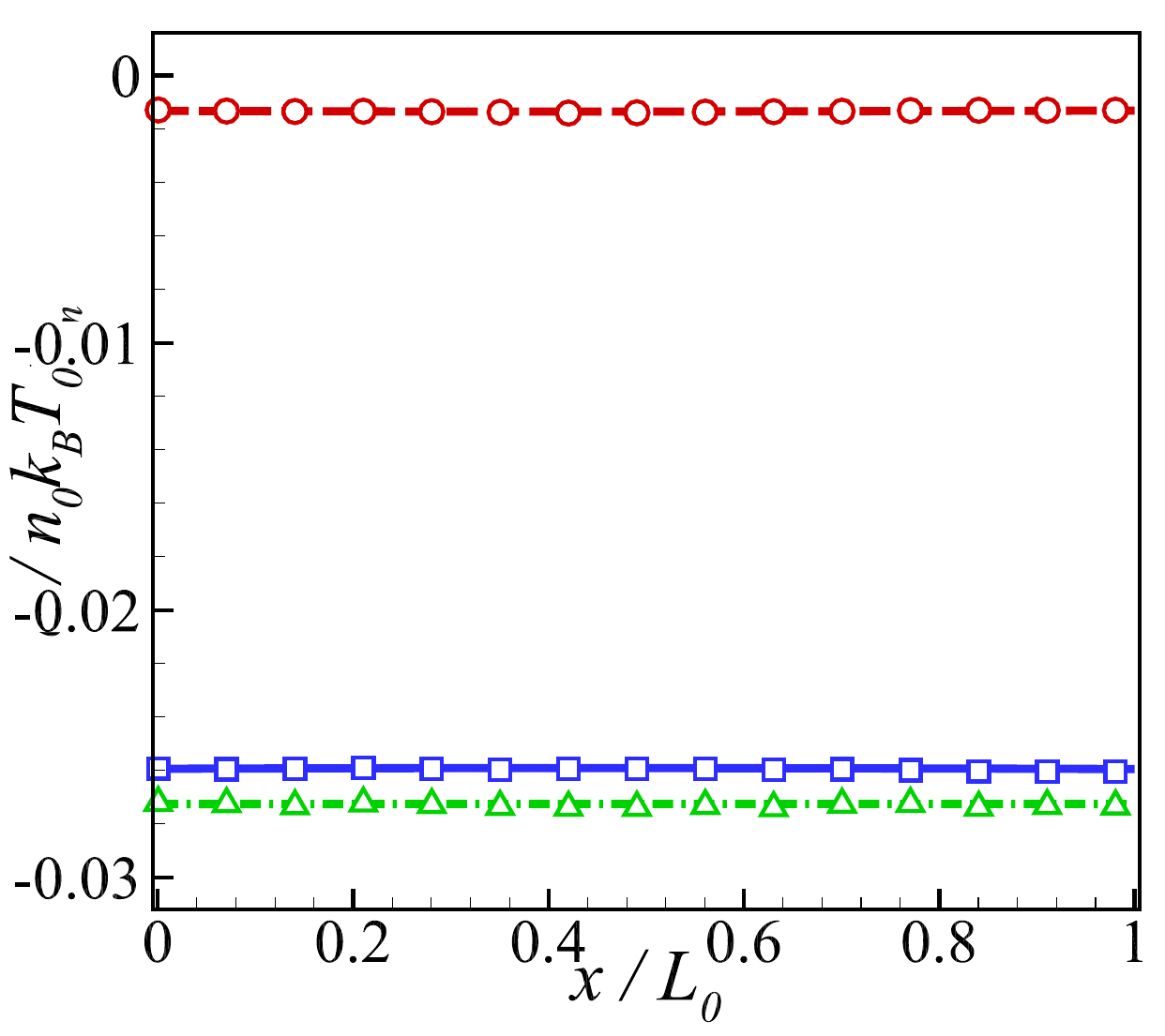}\label{fig:1DFourierFlow_Mix1_2:Mix2_Kn_01_X1_05_q}} \\ 
    \sidesubfloat[]{\includegraphics[scale=0.19,clip=true]{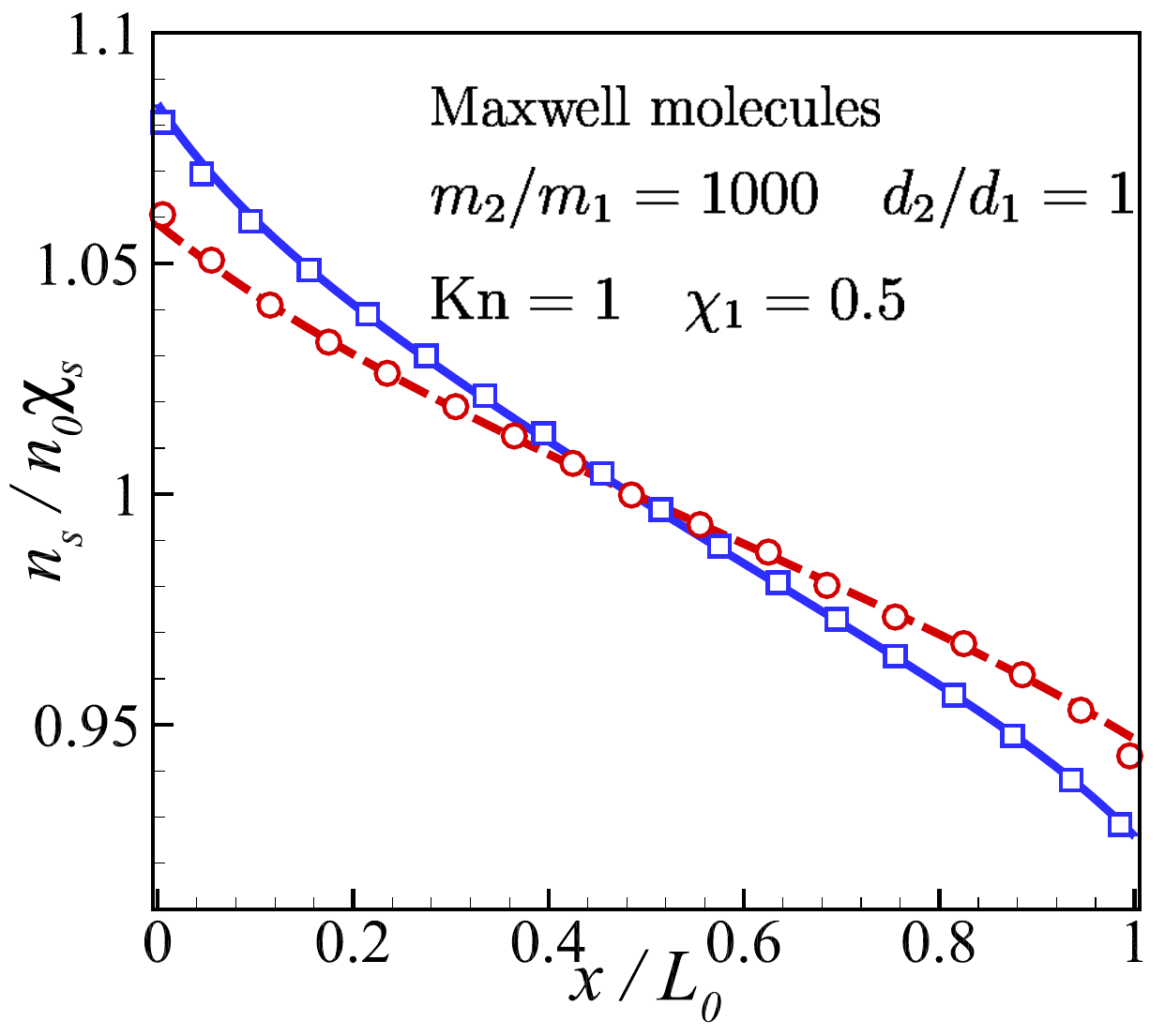}\label{fig:1DFourierFlow_Mix1_2:Mix2_Kn_1_X1_05_n}} 
    \sidesubfloat[]{\includegraphics[scale=0.19,clip=true]{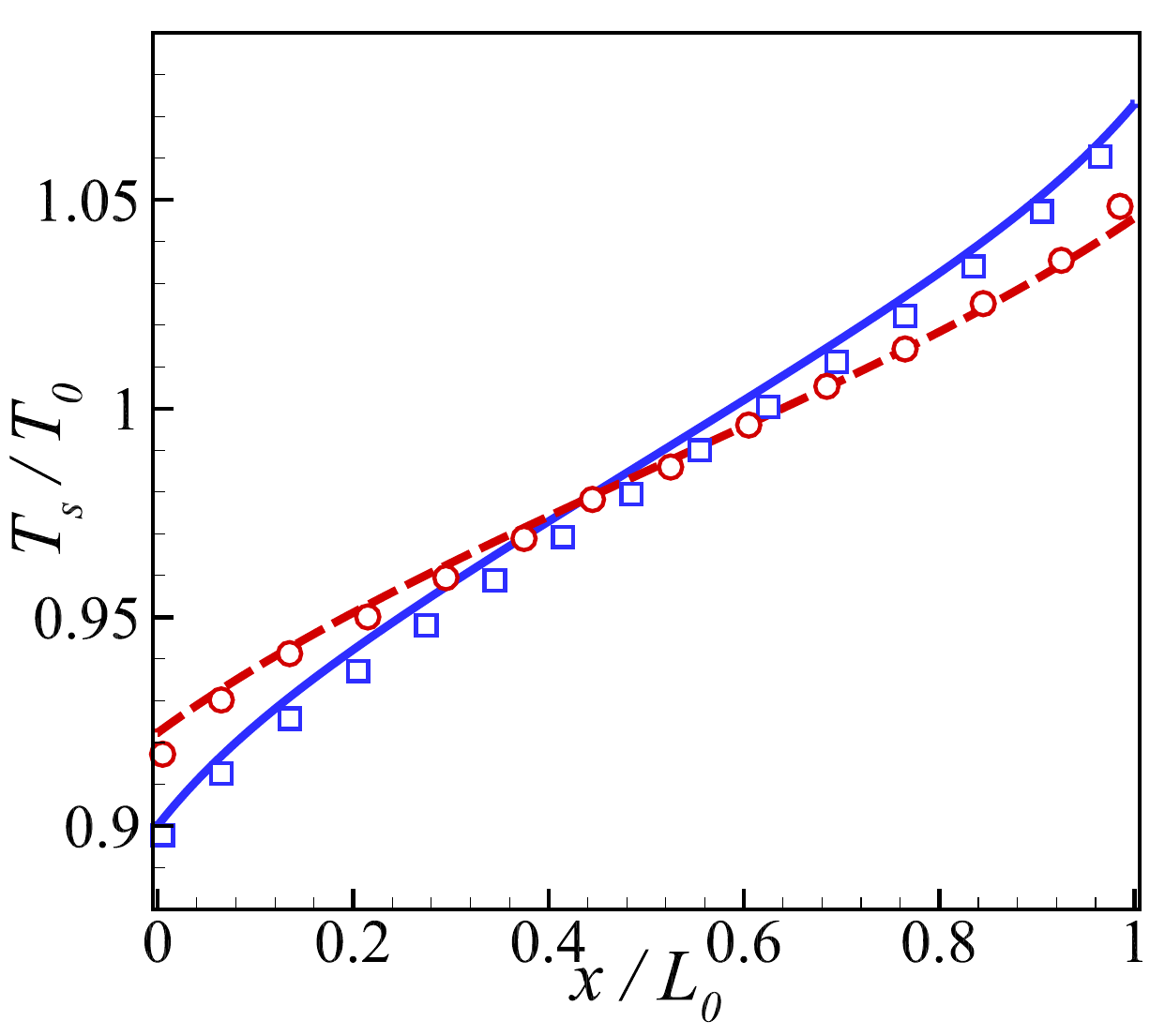}\label{fig:1DFourierFlow_Mix1_2:Mix2_Kn_1_X1_05_T}} 
    \sidesubfloat[]{\includegraphics[scale=0.19,clip=true]{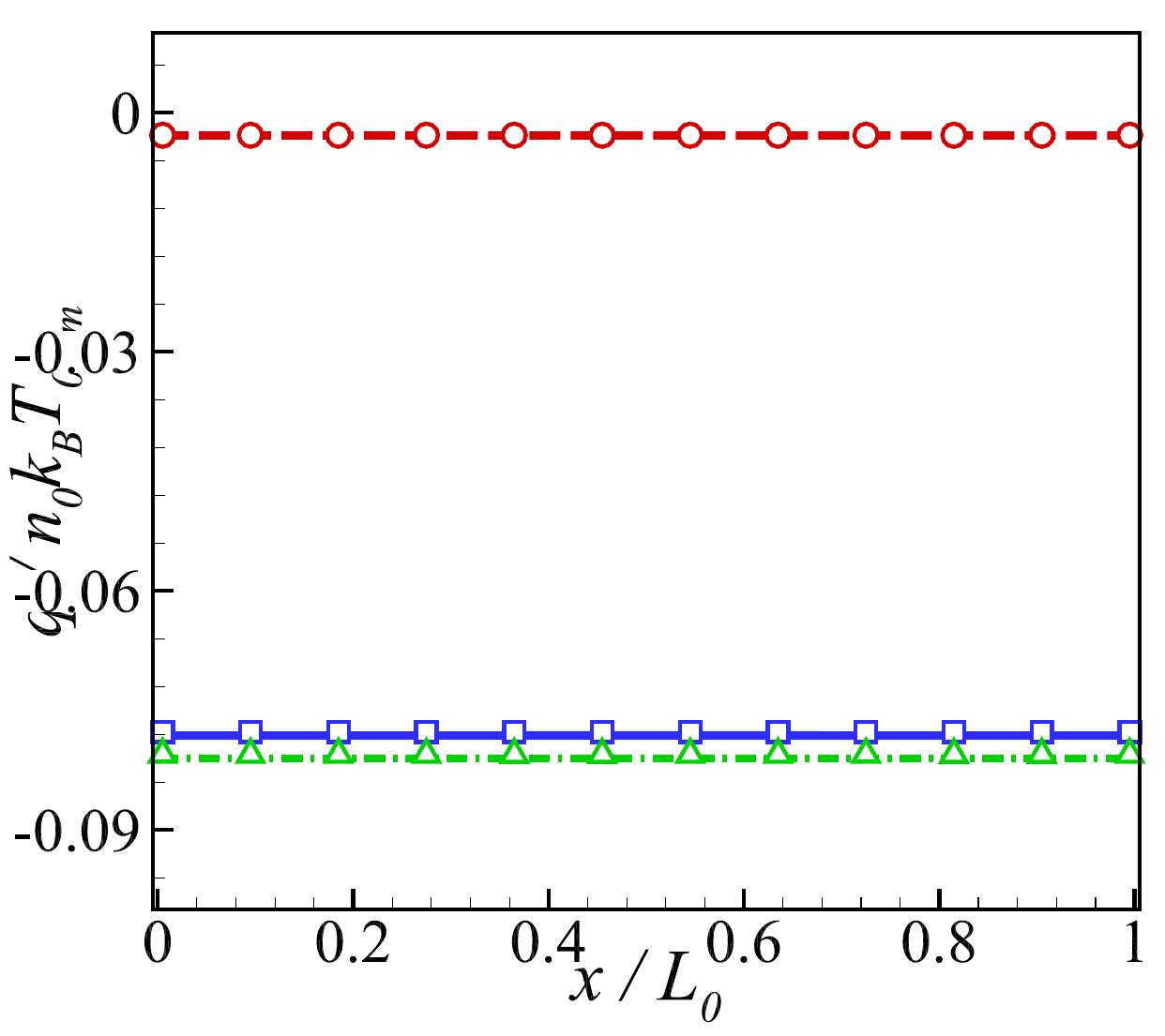}\label{fig:1DFourierFlow_Mix1_2:Mix2_Kn_1_X1_05_q}} \\
    \includegraphics[scale=0.22,clip=true]{Figures/legend_1D.png}
	\caption{Comparisons of the normalized number density (first column), dimensionless temperature (second column), and heat flux (third column) of the gas mixture between kinetic model (lines) and DSMC (symbols) for the planar Fourier flow with $\Delta T=0.2T_0$. The binary mixture consists of Maxwell molecules with a diameter ratio $d_2/d_1=1$, and the mole fraction of light species $\chi_1=0.5$. Each row belongs to a specific flow condition: (a-c) Mixture 1 ($m_2/m_1=10$), $\text{Kn}_1=0.1$, (d-f) Mixture 1, $\text{Kn}_1=1$, (g-i) Mixture 2 ($m_2/m_1=1000$), $\text{Kn}_1=0.1$, (j-l) Mixture 2, $\text{Kn}_1=1$.}
	\label{fig:1DFourierFlow_Mix1_2}
\end{figure}

\begin{figure}[t]
	\centering
	\sidesubfloat[]{\includegraphics[scale=0.19,clip=true]{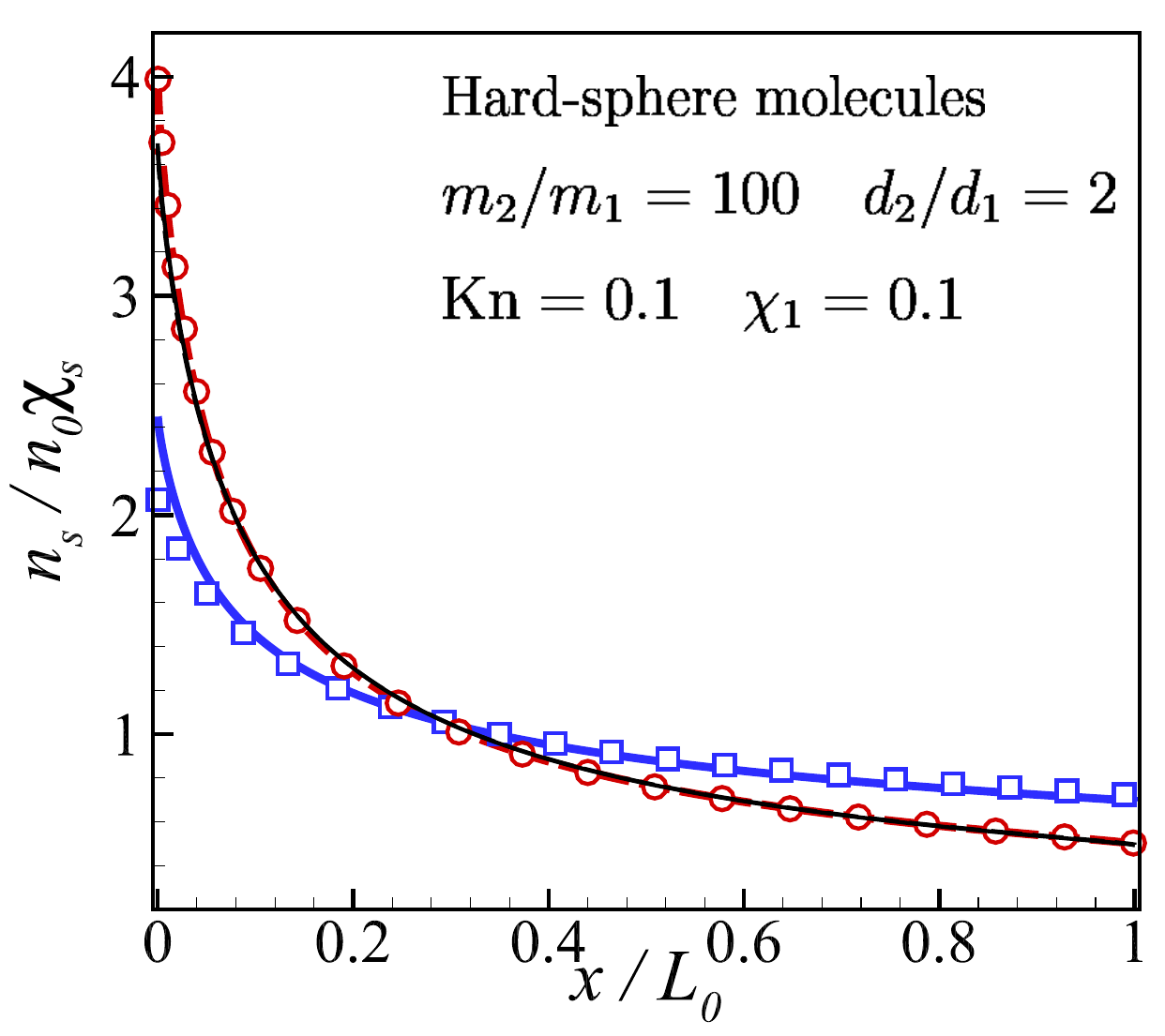}\label{fig:1DFourierFlow_Mix3:Kn_01_X1_01_n}} 
    \sidesubfloat[]{\includegraphics[scale=0.19,clip=true]{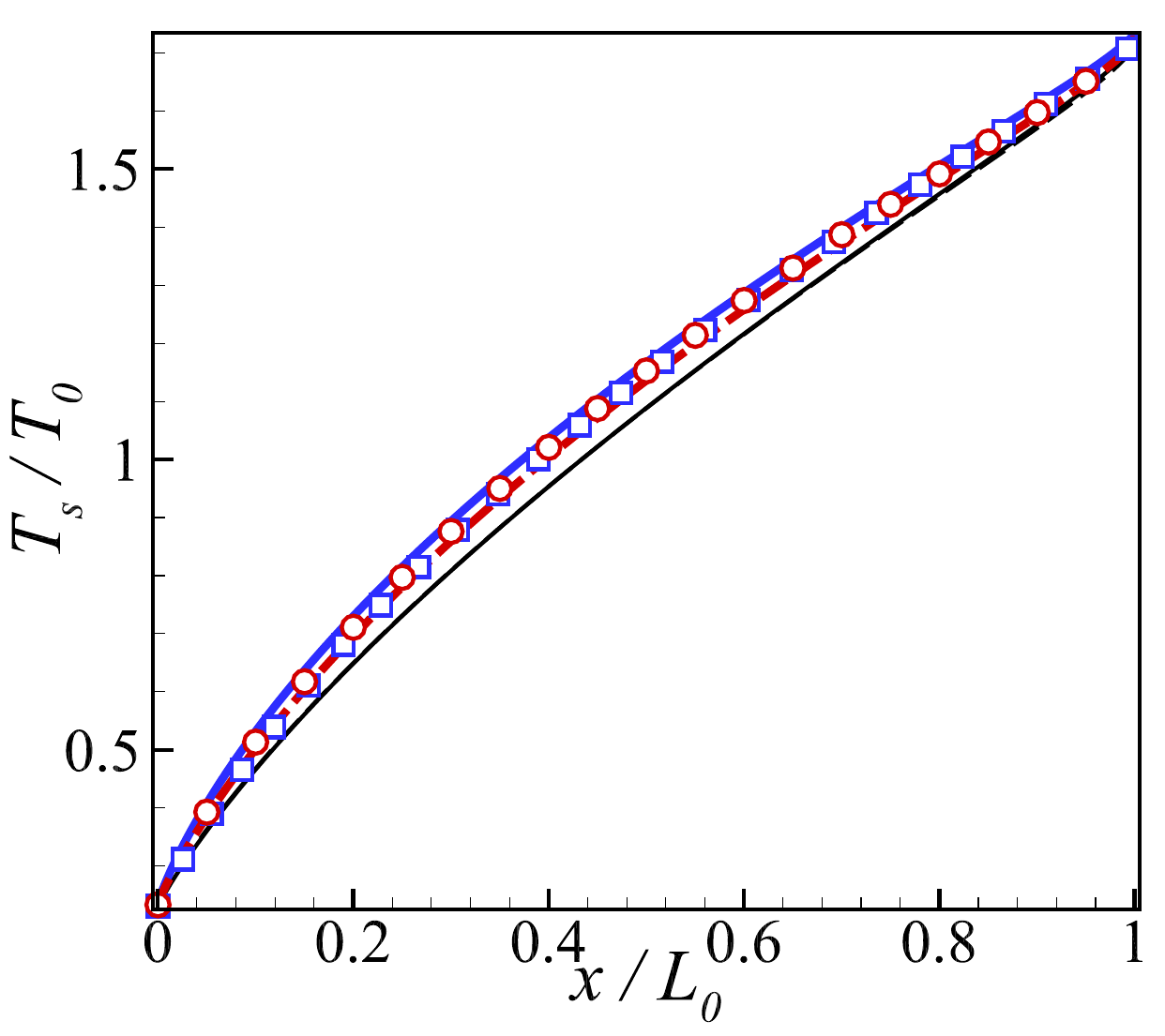}\label{fig:1DFourierFlow_Mix3:Kn_01_X1_01_T}} 
    \sidesubfloat[]{\includegraphics[scale=0.19,clip=true]{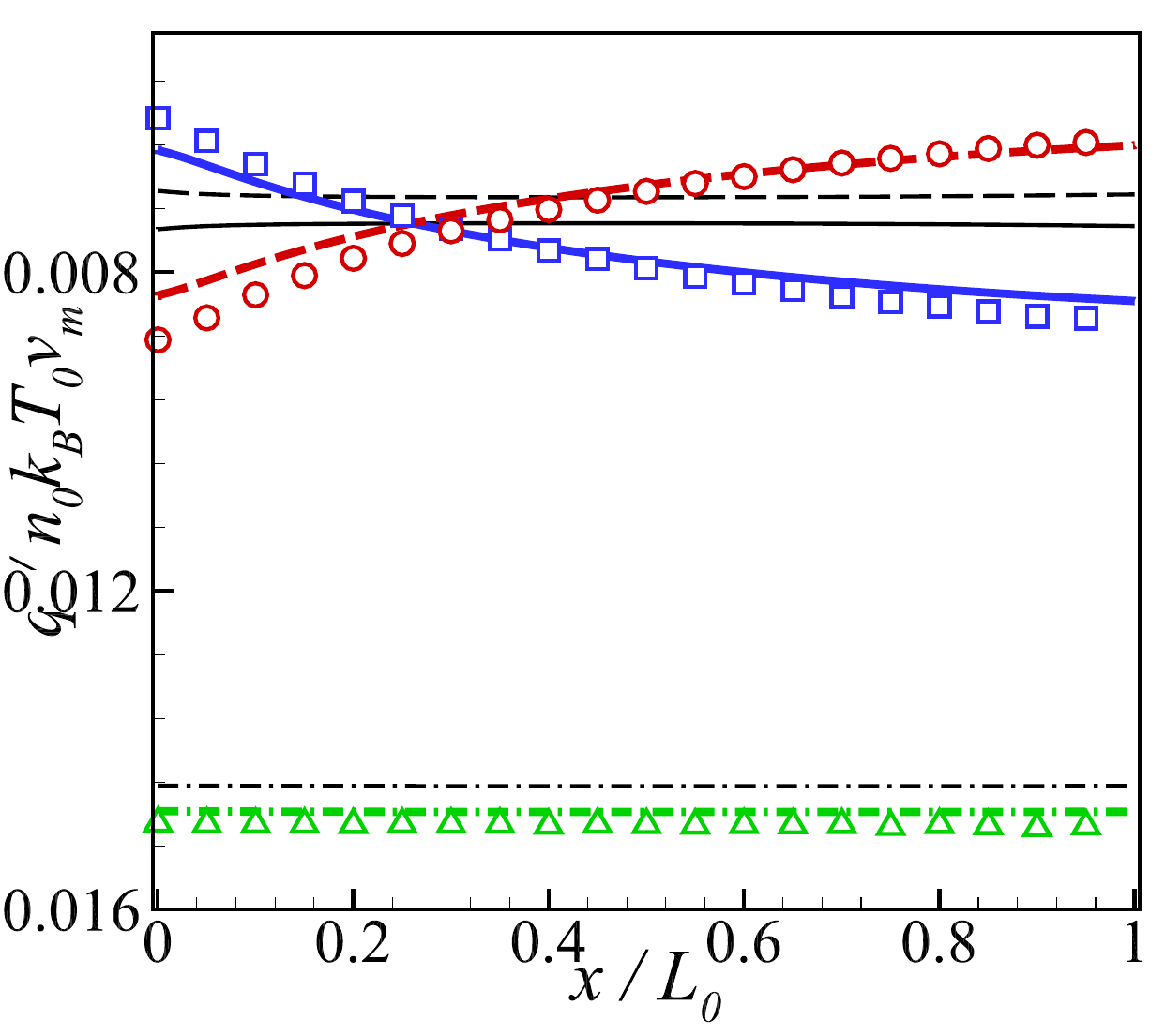}\label{fig:1DFourierFlow_Mix3:Kn_01_X1_01_q}} \\ 
    \sidesubfloat[]{\includegraphics[scale=0.19,clip=true]{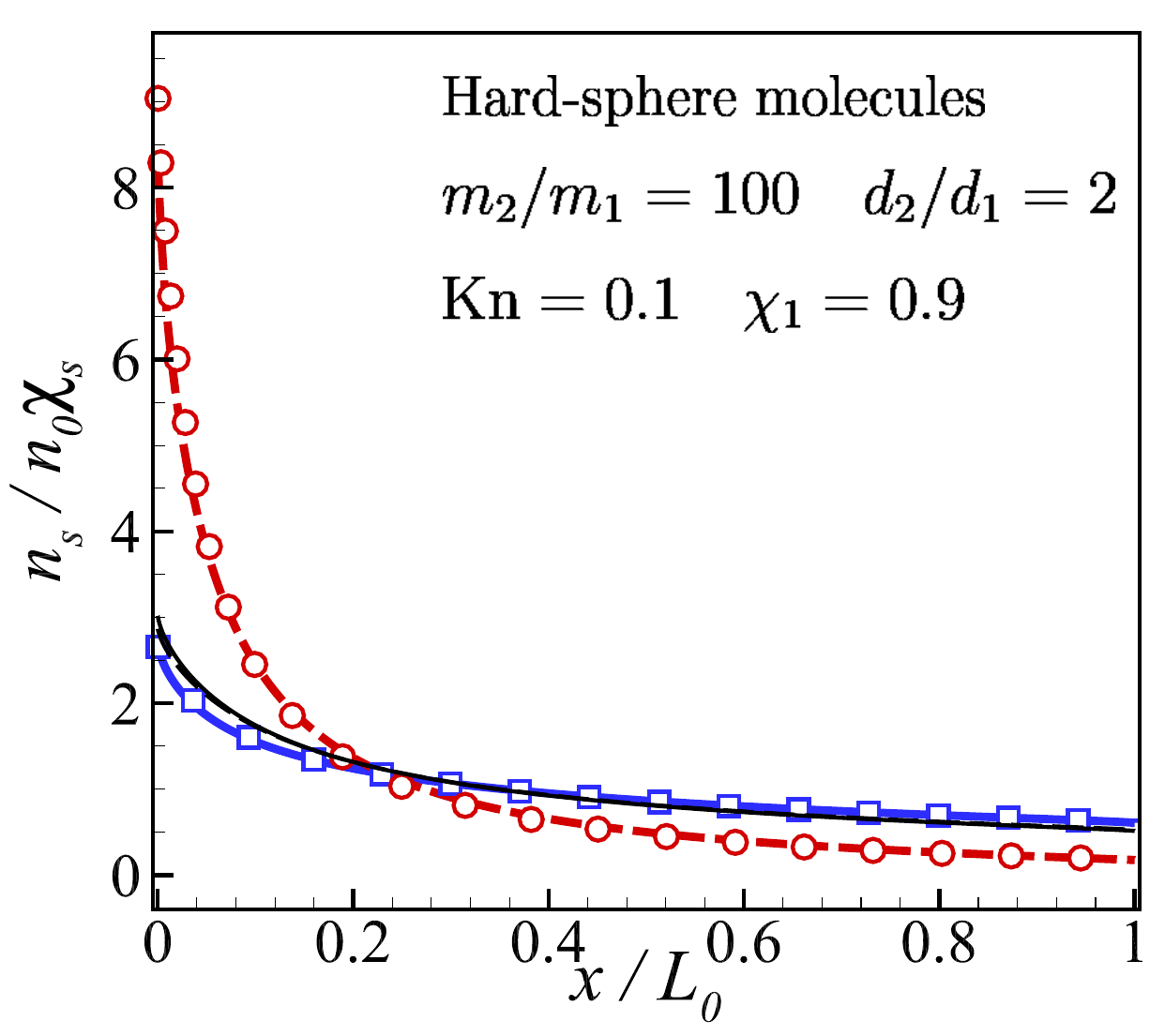}\label{fig:1DFourierFlow_Mix3:Kn_01_X1_09_n}} 
    \sidesubfloat[]{\includegraphics[scale=0.19,clip=true]{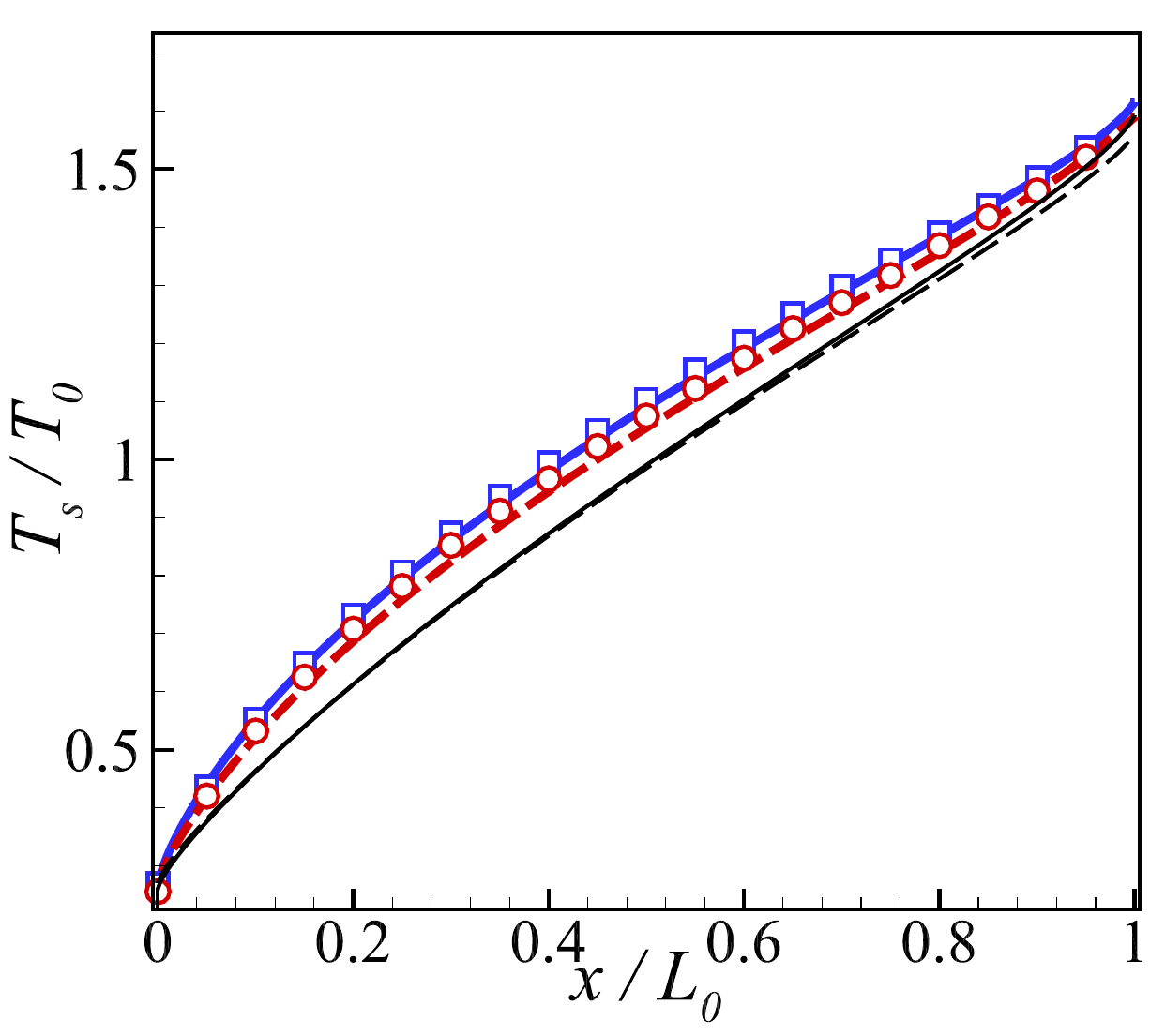}\label{fig:1DFourierFlow_Mix3:Kn_01_X1_09_T}} 
    \sidesubfloat[]{\includegraphics[scale=0.19,clip=true]{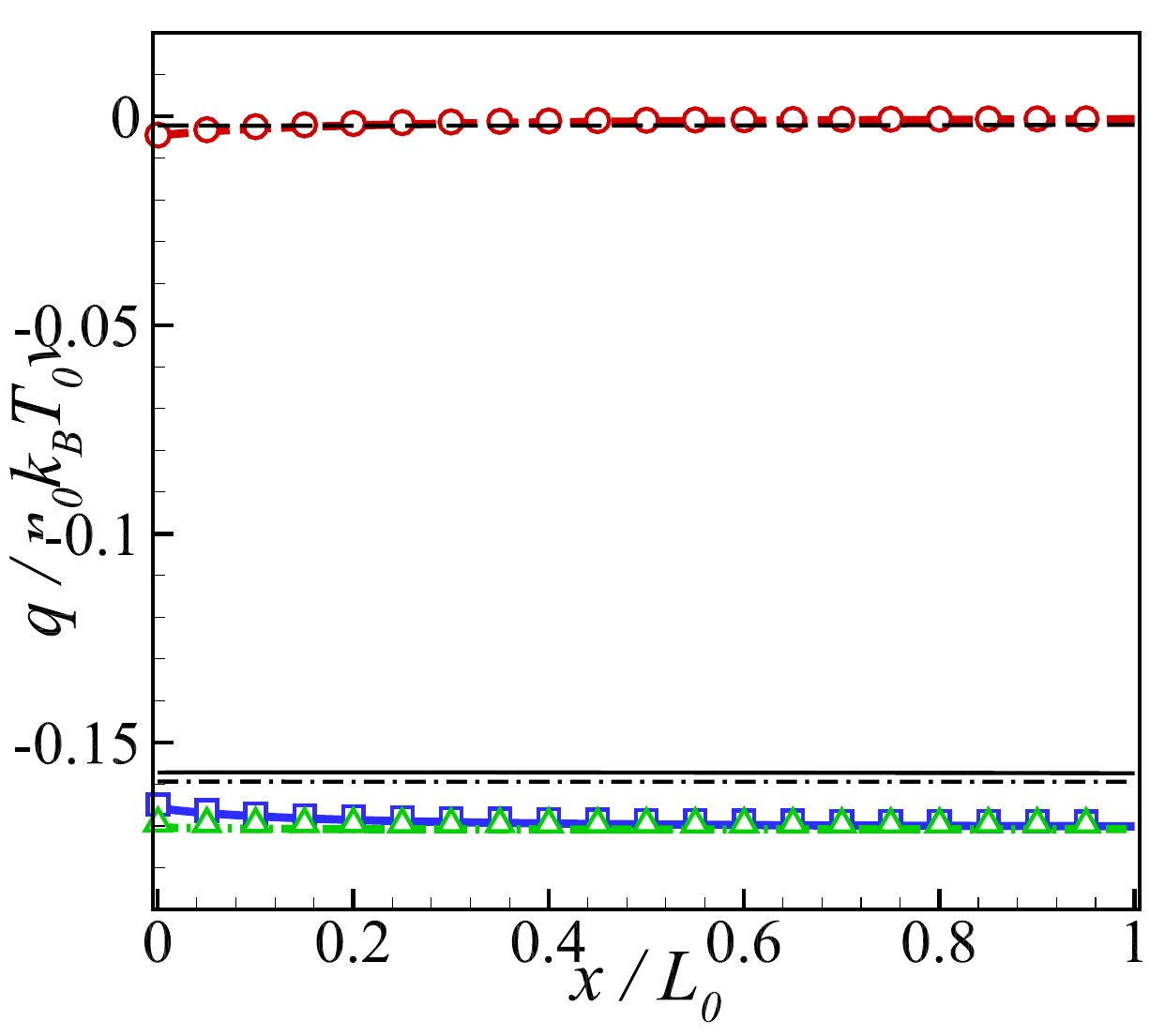}\label{fig:1DFourierFlow_Mix3:Kn_01_X1_09_q}} \\
    \includegraphics[scale=0.22,clip=true]{Figures/legend_1D.png}
	\caption{Comparisons of the normalized number density (first column), dimensionless temperature (second column), and heat flux (third column) of the gas mixture between kinetic model (lines) and DSMC (symbols) for the Fourier flow, when $\text{Kn}_1=0.1$ and $\Delta T=0.8T_0$. The binary mixture consists of hard-sphere molecules with a mass ratio $m_2/m_1=100$, diameter ratio $d_2/d_1=2$, and the mole fraction of light species $\chi_1=0.1$ (first row) and 0.9 (second row). The black lines are the kinetic model solutions without thermal diffusion effects ($b_{12}=0$).}
	\label{fig:1DFourierFlow_Mix3}
\end{figure}

A stationary rarefied mixture confined between two infinite parallel plates located at $x=0$ and $x=L_0$ is considered. The surfaces of the plates have different temperatures $T_w=T_0 \pm \Delta T$, and reflect the gas molecules in a fully diffuse way. The Knudsen number is determined in terms of the averaged number density of the mixture $n_0$, the temperature $T_0$, and the distance between the two plates $L_0$. A variety of cases were considered for the mixtures with different mole fractions $\chi_s$, temperature difference $\Delta T$ and Knudsen numbers, while a selection of the representative results is shown in figures \ref{fig:1DFourierFlow_Mix1_2} and \ref{fig:1DFourierFlow_Mix3}. It can be seen that the solutions given by the kinetic model agree well with the DSMC results.

The temperatures of the components in Mixture 1 ($m_2/m_1=10$) stay close when $\text{Kn}$ is up to 1 (figure \ref{fig:1DFourierFlow_Mix1_2:Mix1_Kn_01_X1_05_n} to \ref{fig:1DFourierFlow_Mix1_2:Mix1_Kn_1_X1_05_q}), while pronounced temperature separation and concentration variation can be observed for Mixture 2 ($m_2/m_1=1000$) when $\text{Kn}=0.1$ (figure \ref{fig:1DFourierFlow_Mix1_2:Mix2_Kn_01_X1_05_n} to \ref{fig:1DFourierFlow_Mix1_2:Mix2_Kn_1_X1_05_q}). It is noteworthy that all these Maxwell molecules under consideration have the same size of the mean free path, due to their identical diameter and interaction potential. However, the inter-species relaxation is much slower for the gases with disparate mass, although the spatial Knudsen number is the same. In other words, the mixtures with larger mass ratios may exhibit significant non-equilibrium phenomena even at small $\text{Kn}$, and hence shrink the applicable range of the hydrodynamic description of the mixtures. The same observation has been found in a previous work solving linearized Boltzmann equation and McCormack model \citep{Ho2016IJHMT}.

Unlike Maxwell gas mixtures, where the thermal diffusion effect is absent, mixtures of hard-sphere molecules have a significant thermal diffusion effect. As shown in figure \ref{fig:1DFourierFlow_Mix3} ($\text{Kn}_1=0.1$ and $\Delta T=0.8T_0$), although the temperature of the two species remains the same, the concentration ratio between components varies across the domain due to the temperature gradient, and the heavy gas molecules tend to concentrate in the cold region. Meanwhile, the highly nonlinear feature arising from the pronounced temperature difference between the two plates is accurately captured. We also solve the kinetic model without the thermal diffusion effect for these cases by setting parameter $b_{12}=0$ (all the other transport properties and relaxation rates remain unchanged), and the corresponding results are shown in figure \ref{fig:1DFourierFlow_Mix3} by black lines. Clearly, the species separation phenomenon cannot be reproduced, thus leading to an incorrect prediction of concentration and hence the heat flux of each species.

\subsection{Planar Couette flow}

\begin{figure}[t]
	\centering
	\sidesubfloat[]{\includegraphics[scale=0.19,clip=true]{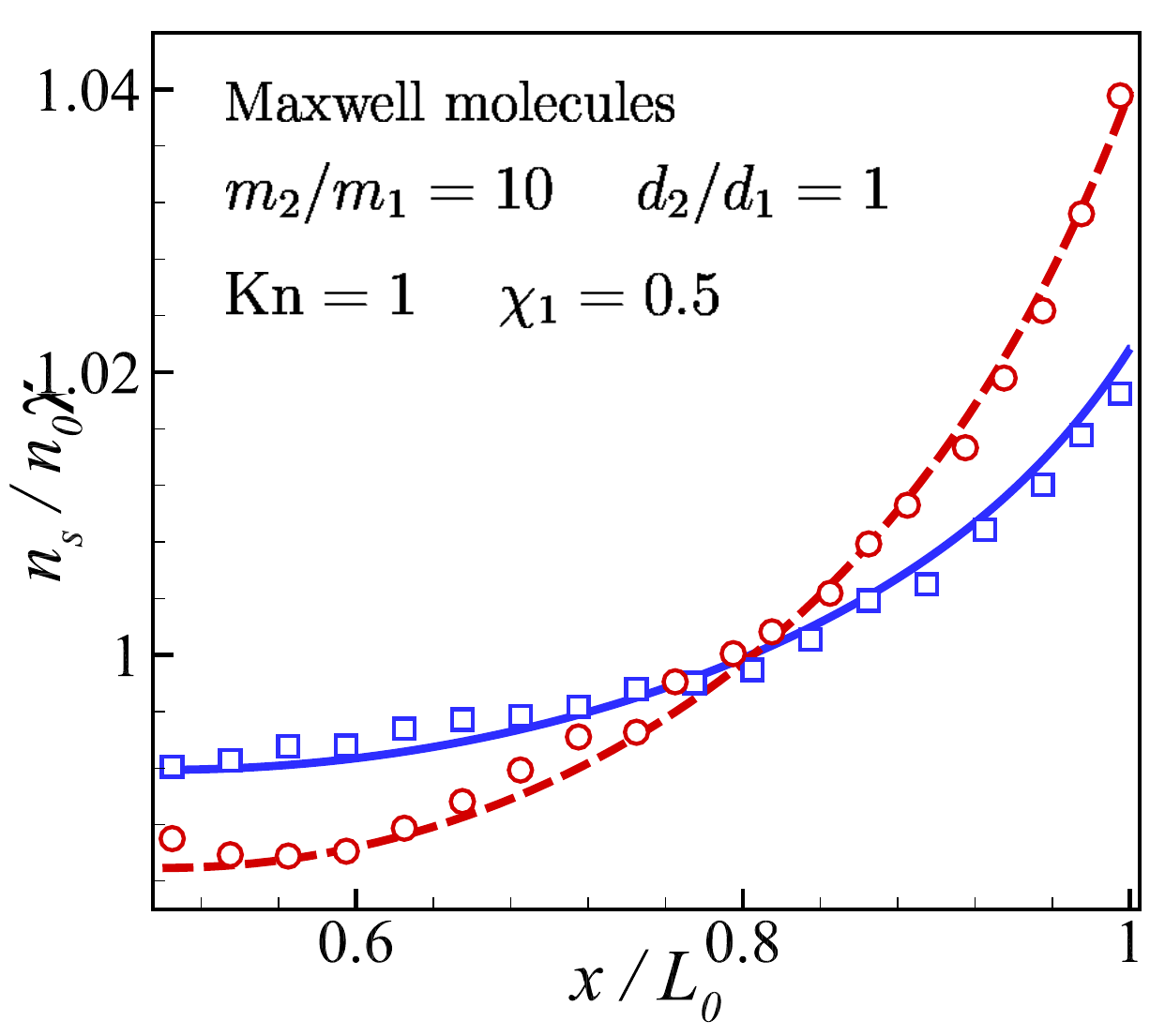}\label{fig:1DCouetteFlow_Mix1:Kn_1_X1_05_n}} 
	\sidesubfloat[]{\includegraphics[scale=0.19,clip=true]{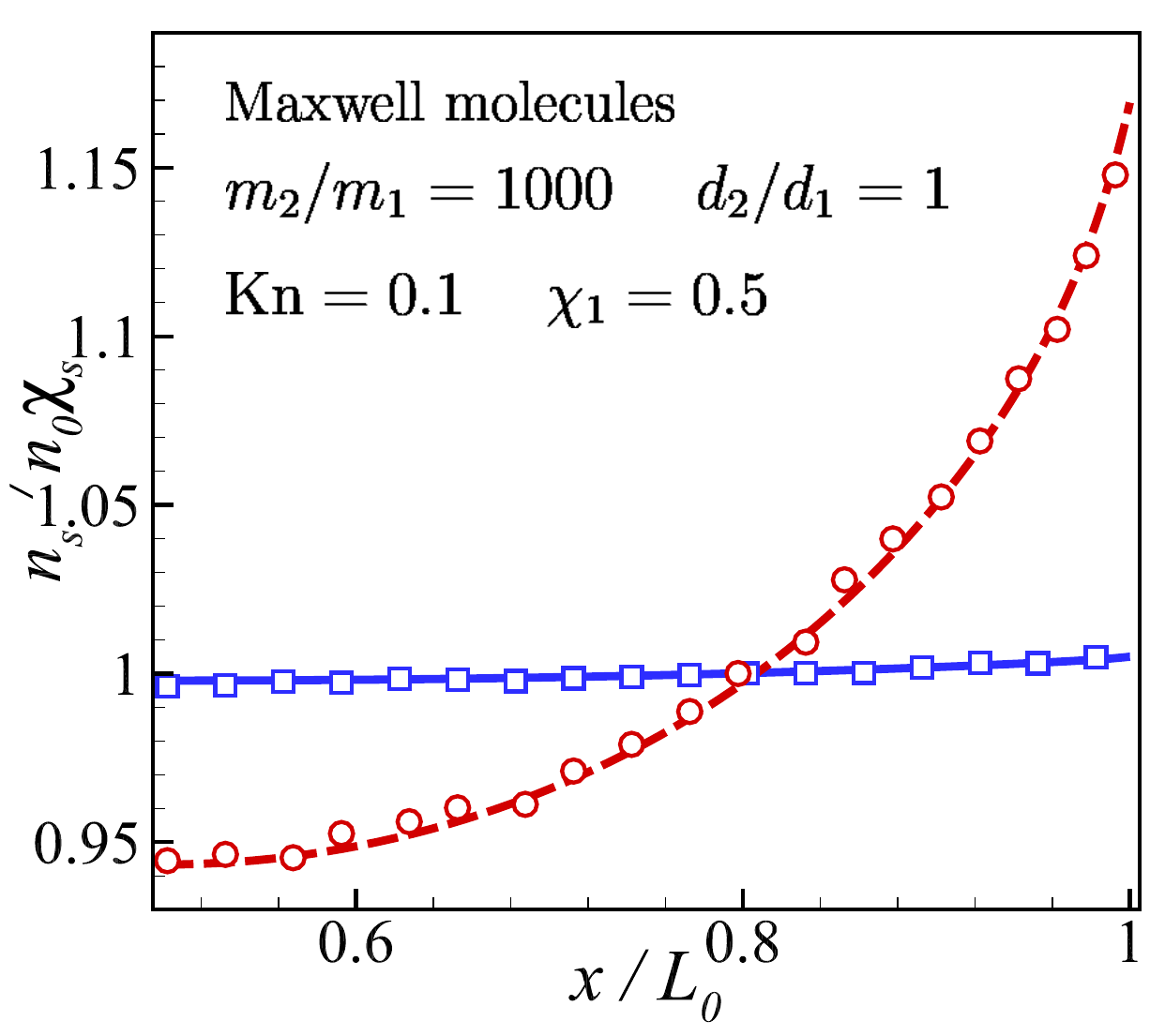}\label{fig:1DCouetteFlow_Mix2:Kn_01_X1_05_n}} 
	\sidesubfloat[]{\includegraphics[scale=0.19,clip=true]{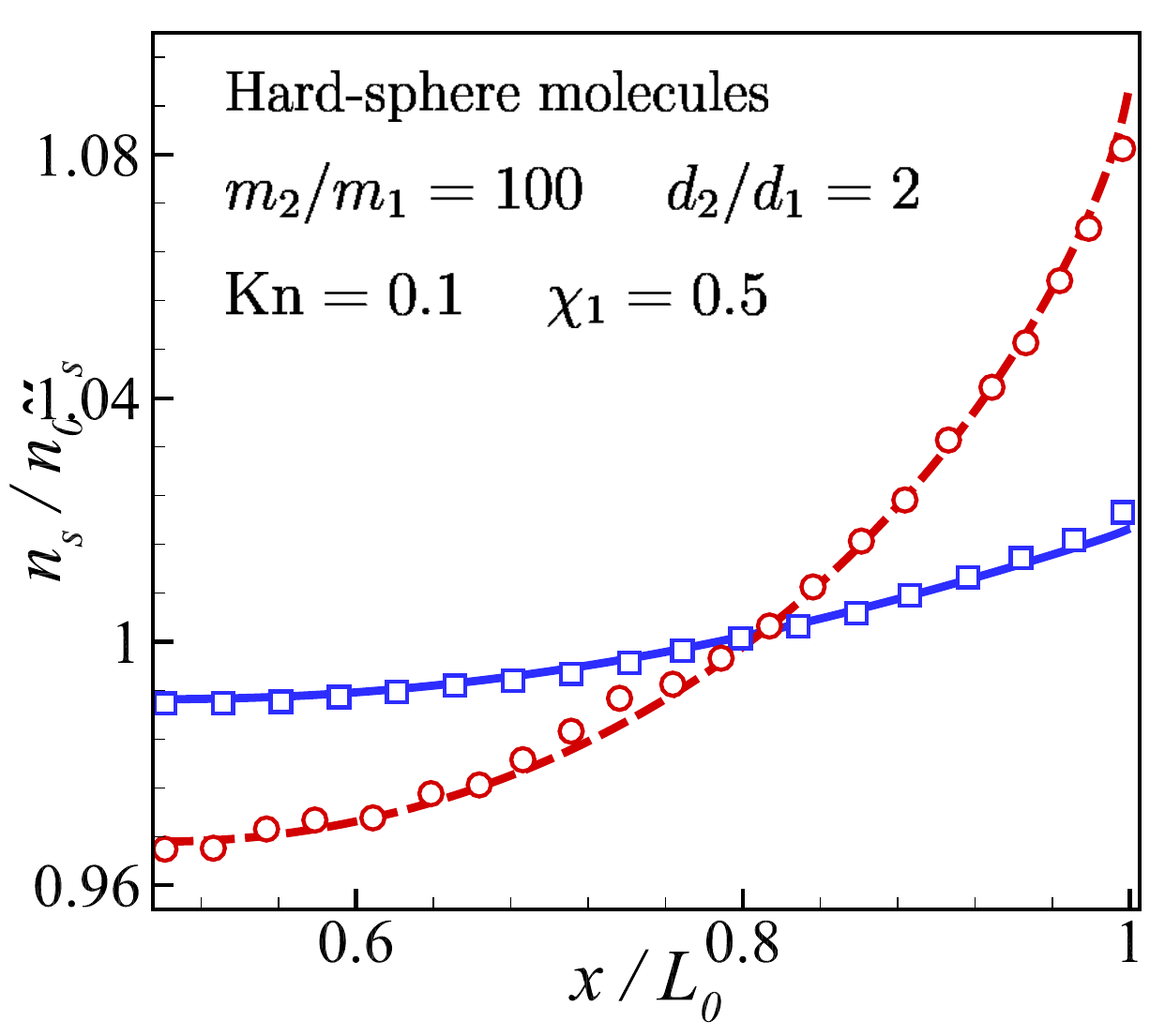}\label{fig:1DCouetteFlow_Mix3:Kn_01_X1_05_n}} \\
    \sidesubfloat[]{\includegraphics[scale=0.19,clip=true]{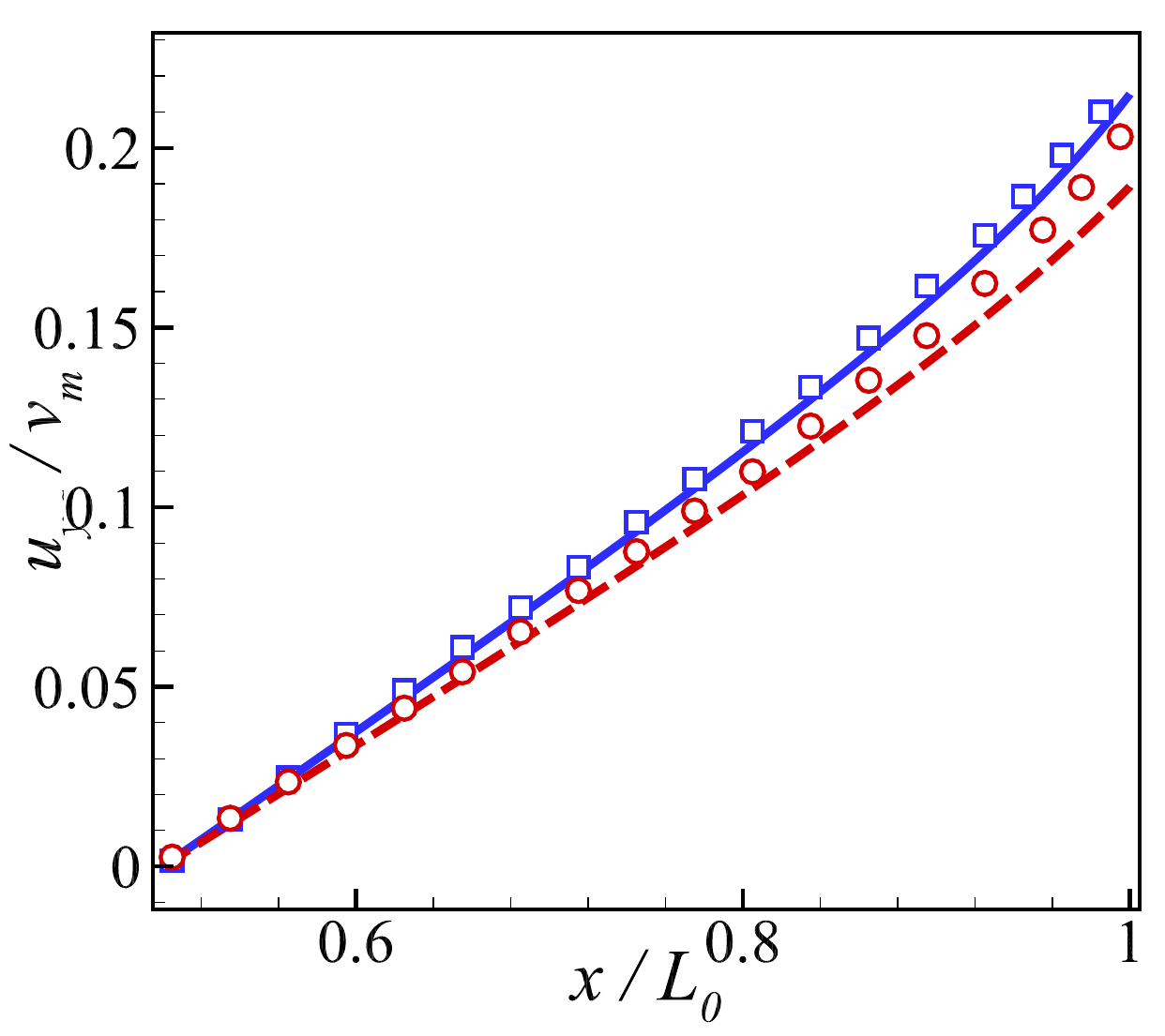}\label{fig:1DCouetteFlow_Mix1:Kn_1_X1_05_uy}} 
	\sidesubfloat[]{\includegraphics[scale=0.19,clip=true]{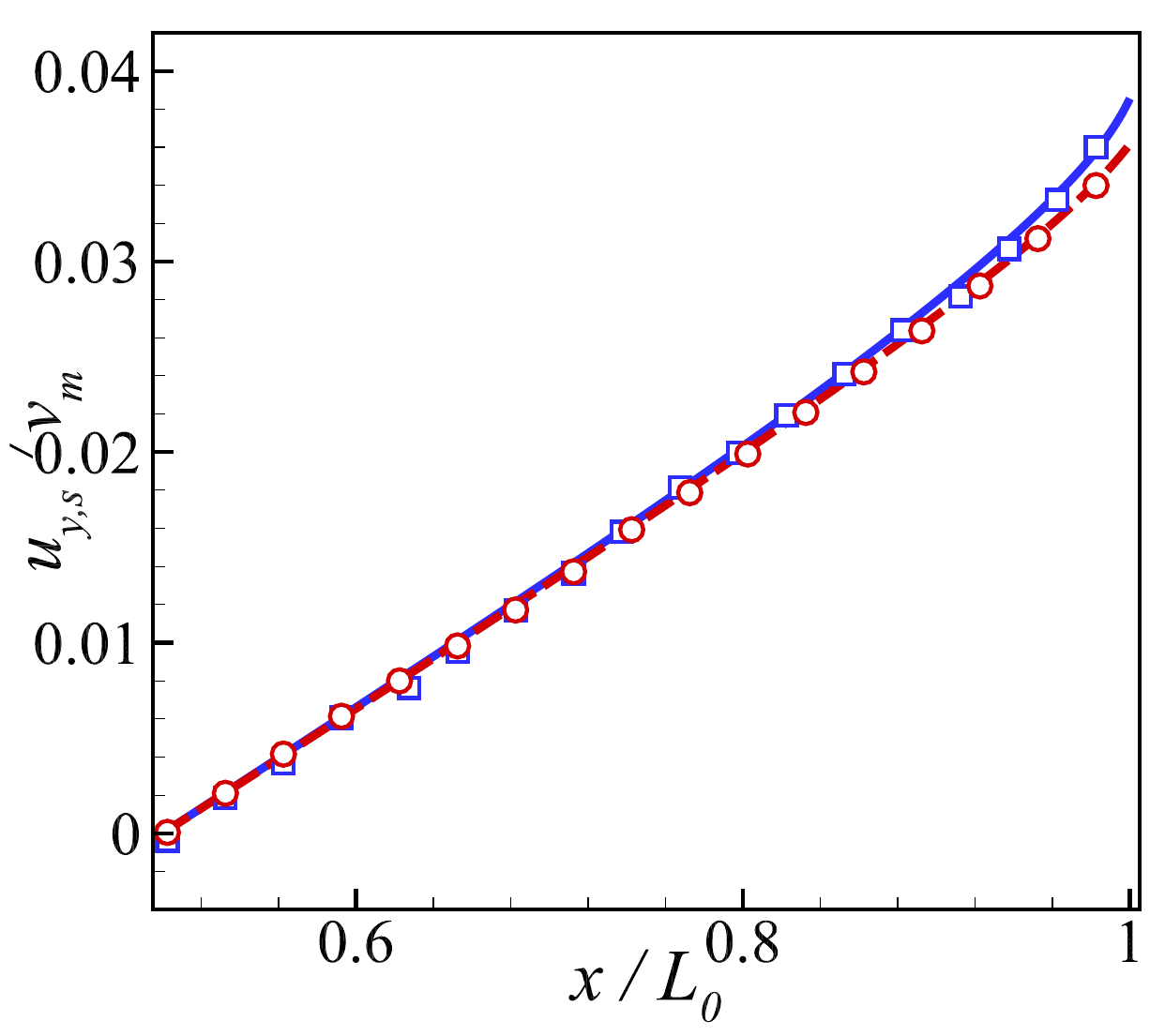}\label{fig:1DCouetteFlow_Mix2:Kn_01_X1_05_uy}} 
	\sidesubfloat[]{\includegraphics[scale=0.19,clip=true]{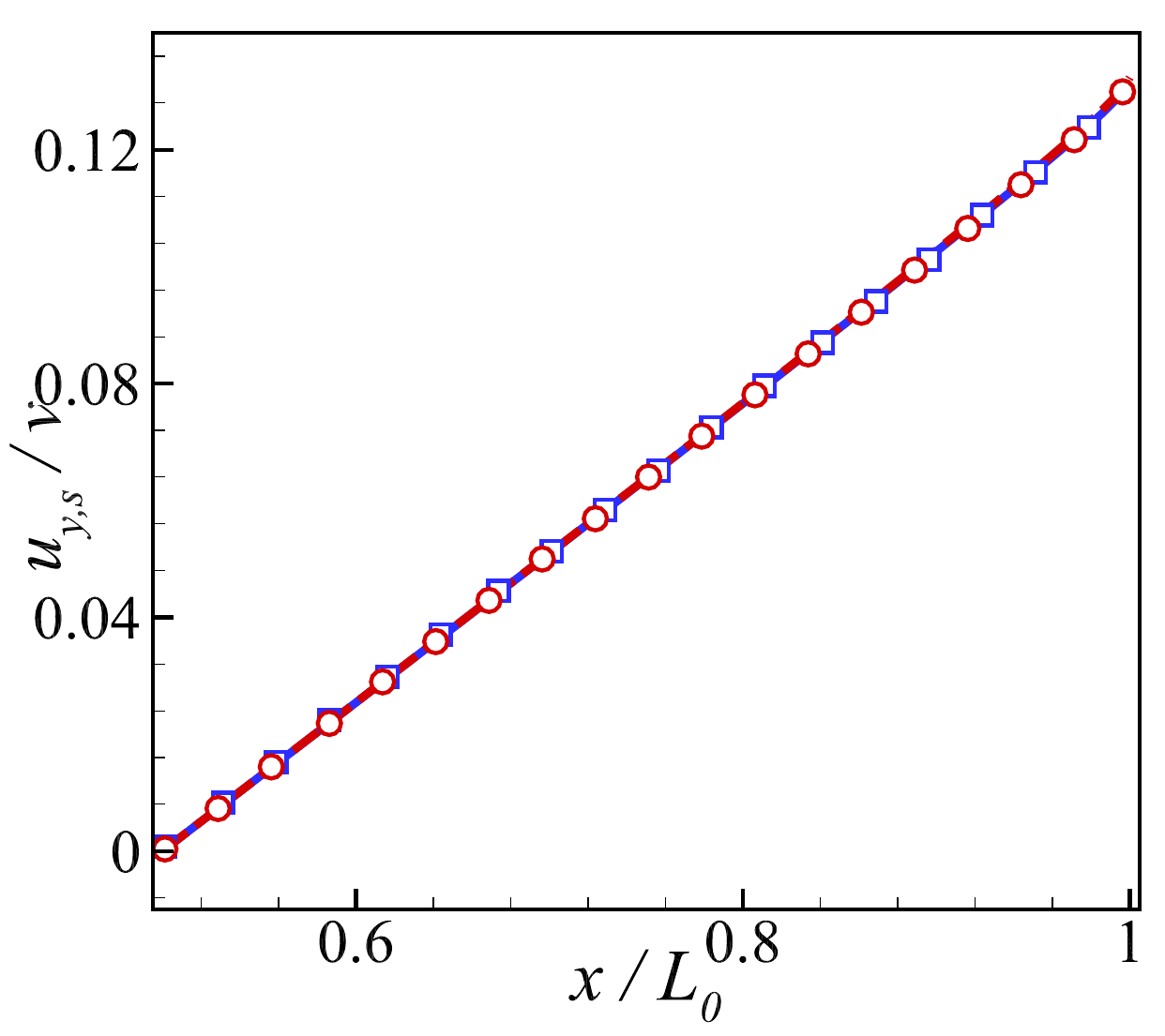}\label{fig:1DCouetteFlow_Mix3:Kn_01_X1_05_uy}} \\
    \sidesubfloat[]{\includegraphics[scale=0.19,clip=true]{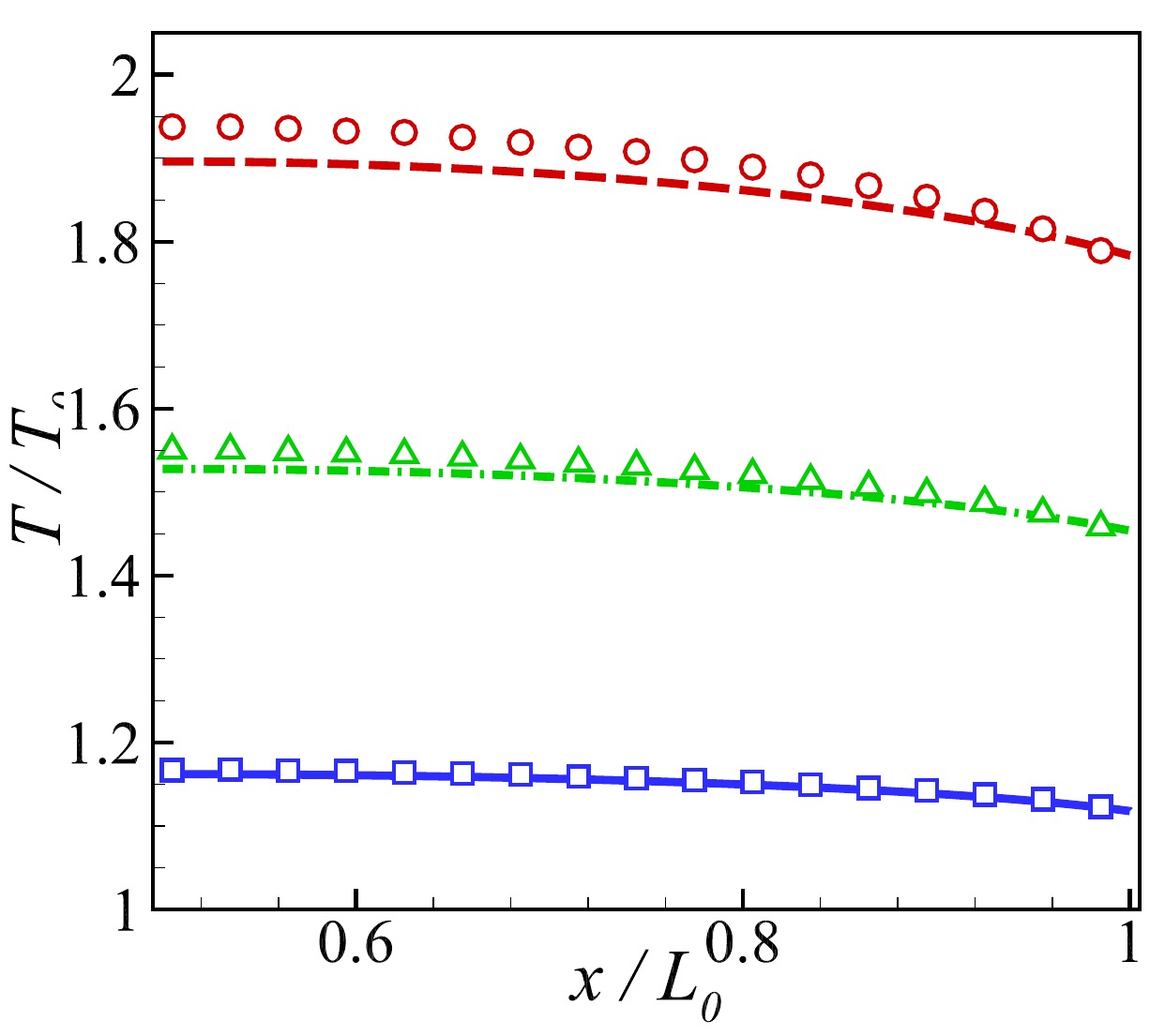}\label{fig:1DCouetteFlow_Mix1:Kn_1_X1_05_T}} 
	\sidesubfloat[]{\includegraphics[scale=0.19,clip=true]{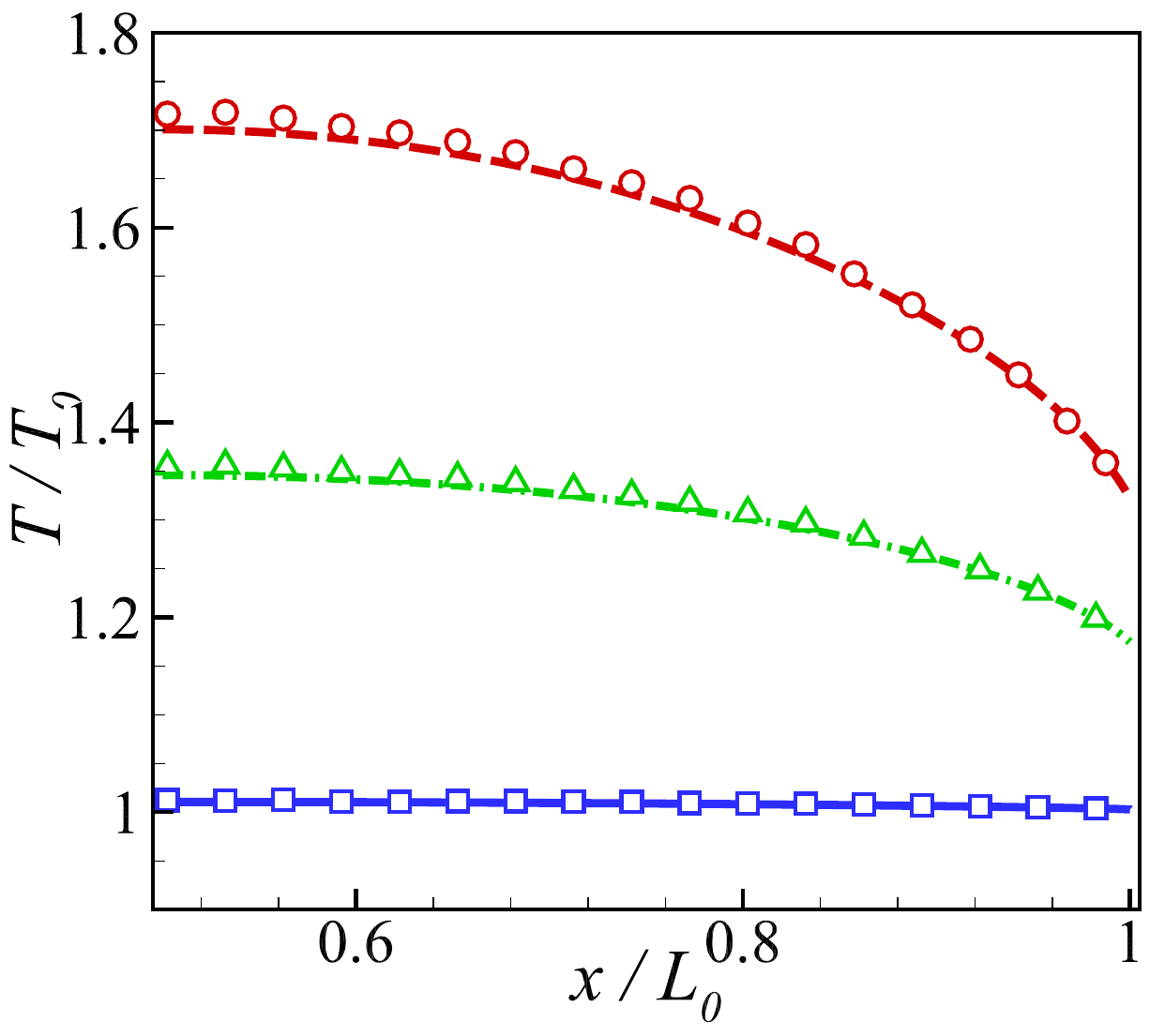}\label{fig:1DCouetteFlow_Mix2:Kn_01_X1_05_T}} 
	\sidesubfloat[]{\includegraphics[scale=0.19,clip=true]{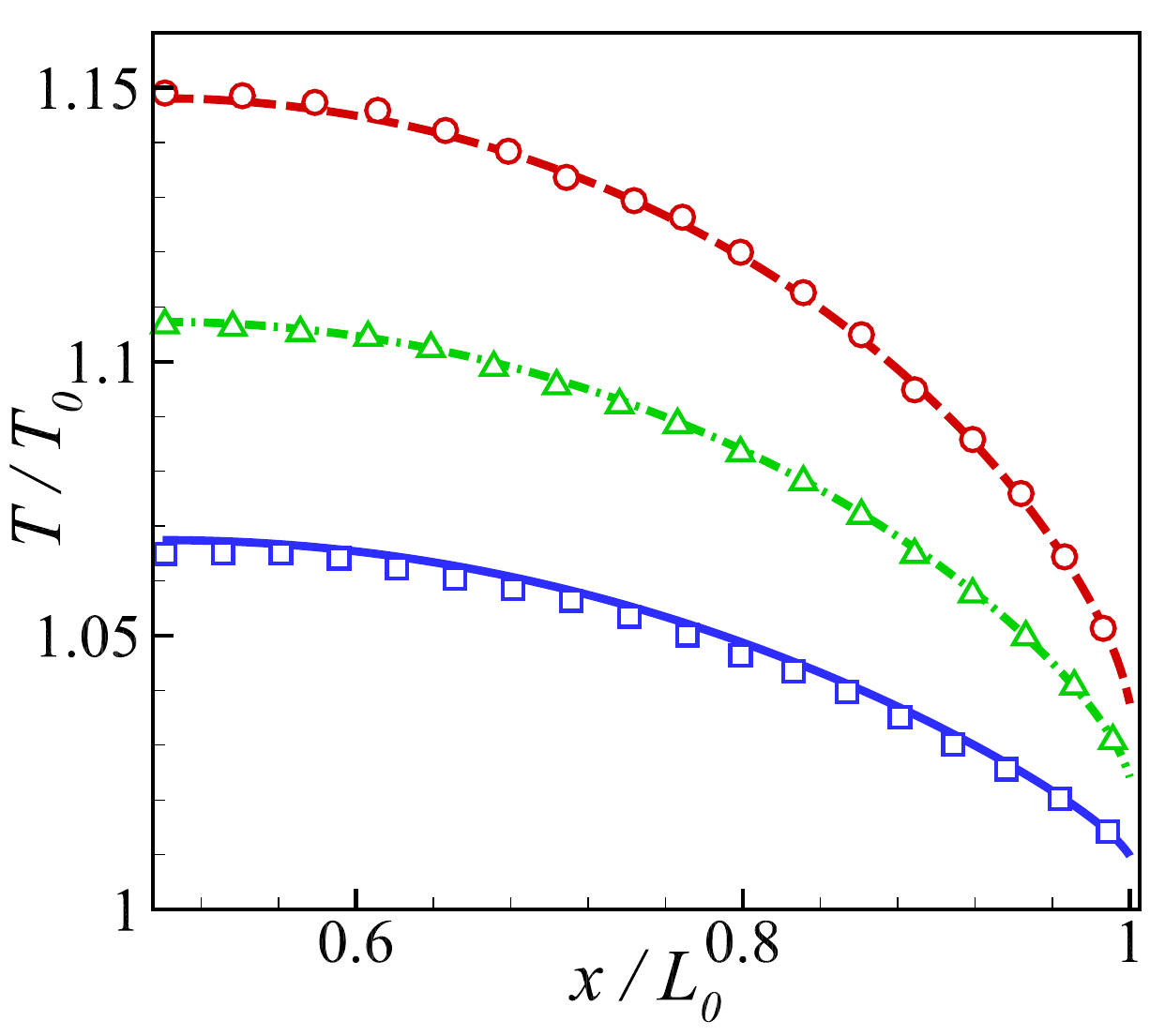}\label{fig:1DCouetteFlow_Mix3:Kn_01_X1_05_T}} \\ 
    \sidesubfloat[]{\includegraphics[scale=0.19,clip=true]{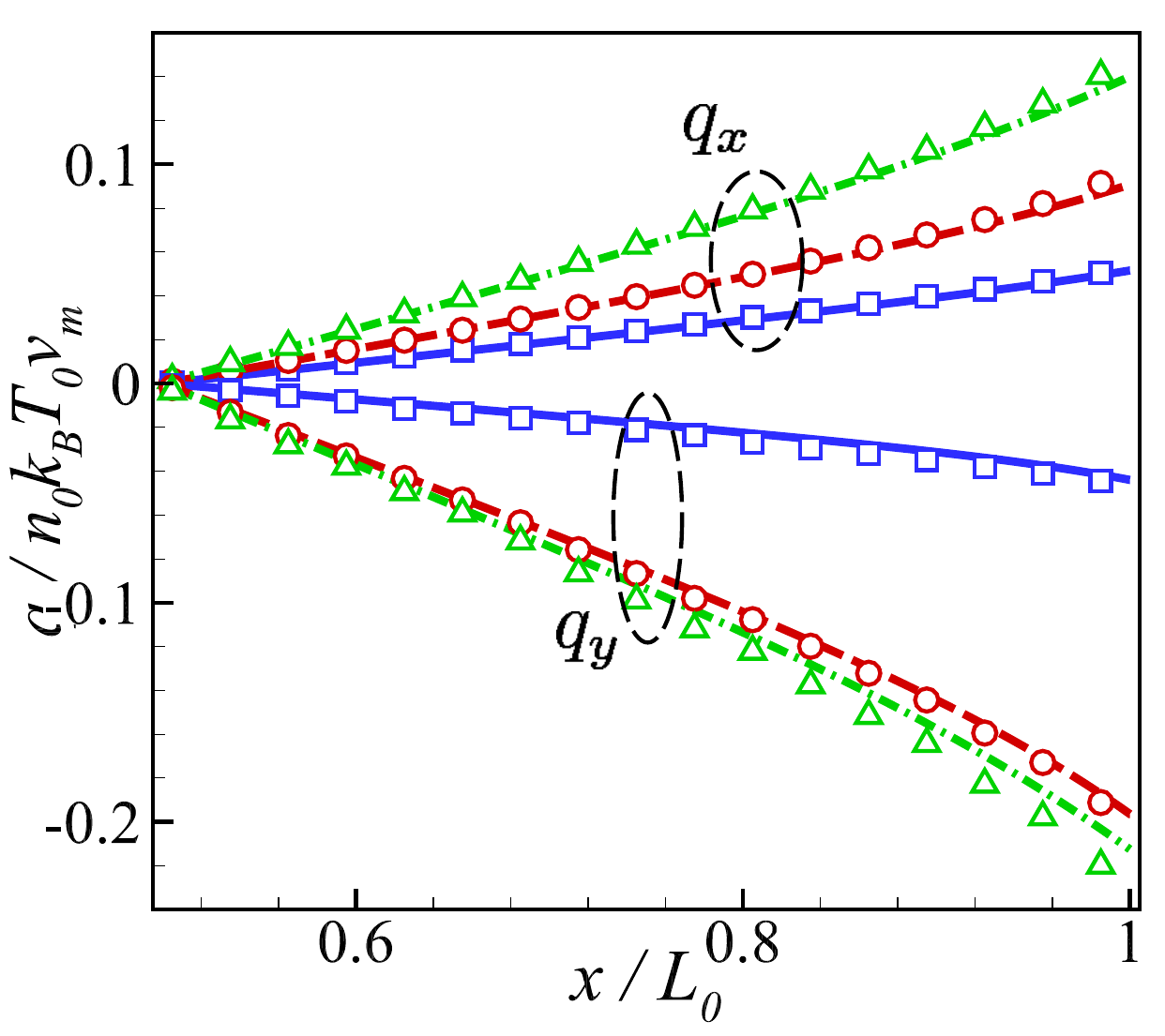}\label{fig:1DCouetteFlow_Mix1:Kn_1_X1_05_q}} 
	\sidesubfloat[]{\includegraphics[scale=0.19,clip=true]{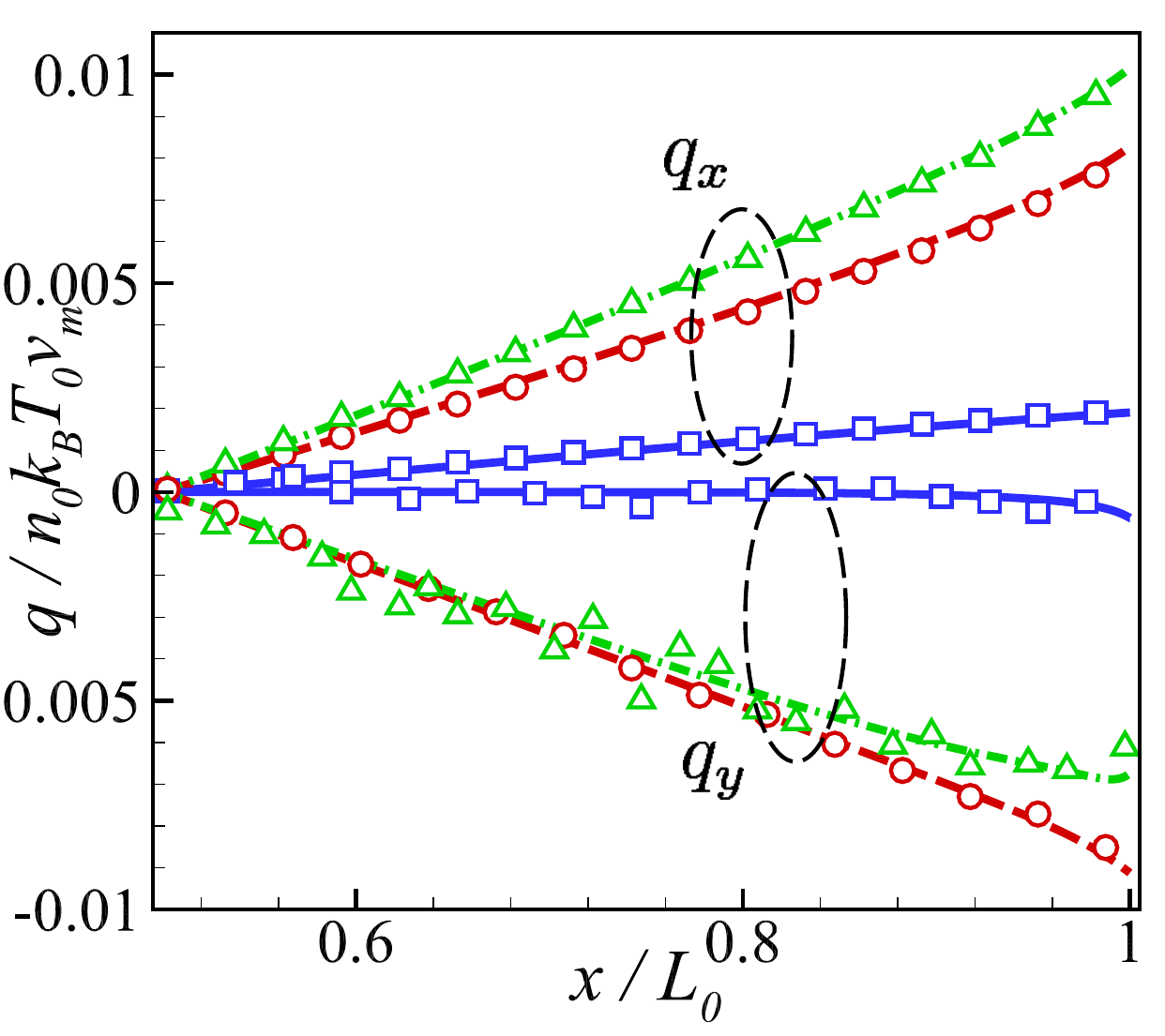}\label{fig:1DCouetteFlow_Mix2:Kn_01_X1_05_q}} 
	\sidesubfloat[]{\includegraphics[scale=0.19,clip=true]{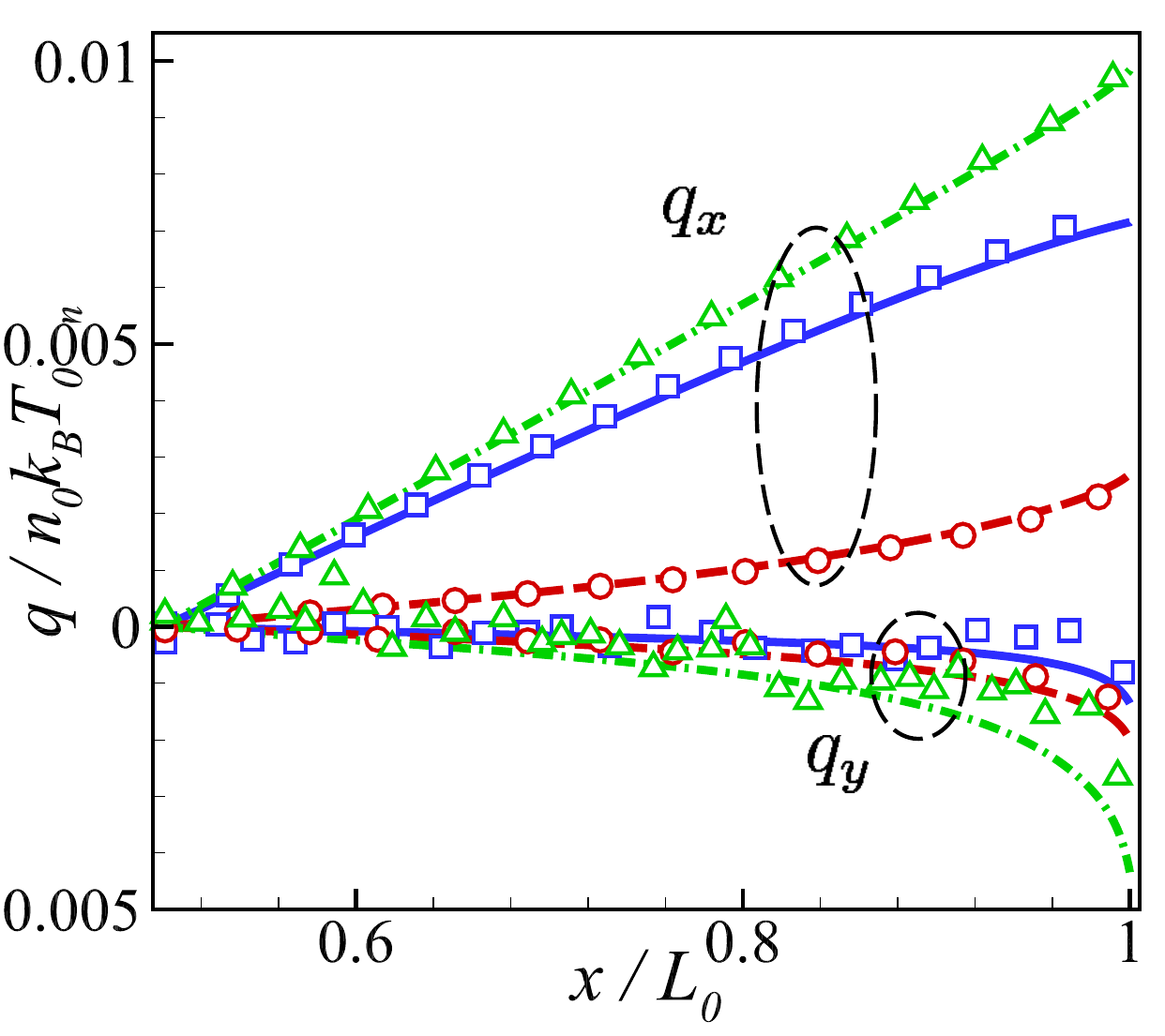}\label{fig:1DCouetteFlow_Mix3:Kn_01_X1_05_q}} \\
    \includegraphics[scale=0.22,clip=true]{Figures/legend_1D.png}
	\caption{Comparisons of the normalized (a-c) number density, dimensionless (d-f) flow velocity, (g-i) temperature, and (j-l) heat flux between kinetic model (lines) and DSMC (symbols) for the Couette flow with the mole fraction of light species $\chi_1=0.5$, when $\text{Kn}_1=0.1$ (Mixture 2 and 3 in the first and second column, respectively) and 1 (Mixture 1 in the third column).}
	\label{fig:1DCouetteFlow}
\end{figure}

The Couette flow shares the same configuration as that of the Fourier flow, but the temperatures of both plates are kept the same at $T_0$, and the lower and upper plates move along y direction with velocity $u_{1,y}=-v_w$ and $u_{2,y}=v_w$, respectively, where $v_w=\sqrt{2k_BT_0/m_{mix}}$. Due to the symmetry, only half of the domain $(L_0/2\le{}x_2\le{}L_0)$ is used in the simulations.

\begin{table}[ht]
    \begin{center}
	  \begin{tabular}{cccccccccc}
		\hline
		~ & \multicolumn{6}{c}{Maxwell gas}   & \multicolumn{3}{c}{Hard-sphere gas}  \\
		\specialrule{0em}{4pt}{4pt}
		$m_2/m_1$ & \multicolumn{3}{c}{10}  & \multicolumn{3}{c}{1000} & \multicolumn{3}{c}{100}  \\
		$\chi_1$ & 0.1 & 0.5 & 0.9 & 0.1 & 0.5 & 0.9 & 0.1 & 0.5 & 0.9  \\
        \specialrule{0em}{4pt}{4pt}
		\multirow{2}{*}{$\text{Kn}_1=0.1$} & 0.224 & 0.223 & 0.216 & 0.233 & 0.281 & 0.289 & 0.600 &  0.074 & 0.110 \\
		 ~ & (0.1\%) & (0.6\%) & (1.7\%) & (0.2\%) & (0.9\%) &  (3.2\%) & (1.4\%) & (1.7\%) & (2.9\%) \\
		 \multirow{2}{*}{1} & 0.782 & 0.731 & 0.706 & 0.781 & 0.692 & 0.411 & 0.397 & 0.425 & 0.449  \\
		 ~ & (1.7\%) & (2.3\%) & (3.0\%) & (1.5\%) & (1.1\%) & (3.2\%) & (1.3\%) & (1.6\%) & (3.2\%)  \\
		 \multirow{2}{*}{10} & 1.058 & 0.963 & 0.952 & 1.035 & 0.809 & 0.450 & 0.904 & 0.787 & 0.626  \\
		 ~ & (0.2\%) & (0.7\%) & (0.3\%) & (0.2\%) & (0.1\%) & (0.0\%) & (0.0\%) & (0.1\%) & (2.8\%)  \\
        \hline
      \end{tabular}
      \caption{The dimensionless shear stress $P_{xy}/n_0k_BT_0$ of the mixture calculated by the kinetic model equation for the Couette flow, with the mole fraction $\chi_1=0.1,0.5,0.9$ and $\text{Kn}_1=0.1,1,10$. The values in parentheses are the relative errors between the results given by the kinetic model and the DSMC method.}
    \label{tab:1DCouetteFlow_Pxy}
    \end{center}
\end{table}

The simulation results of our kinetic model and DSMC method are shown in figure \ref{fig:1DCouetteFlow} for the three mixtures with $\chi_1=0.5$, when $\text{Kn}_1=1$ for Mixture 1 and $\text{Kn}_1=0.1$ for the others. Good agreements are found for all the macroscopic properties. We analyze the mixture shear stress predicted by two methods across various mole fractions $\chi_1$ and Knudsen numbers $\text{Kn}_1$ (Table \ref{tab:1DCouetteFlow_Pxy}). Note that the actual plate's velocities vary significantly for mixtures with different mass ratios and mole fractions, and the boundary velocity is supersonic or even hypersonic for the heavier species, but subsonic for lighter gas in most cases (exclude the one for Mixture 1 with $\chi_1=0.9$). Nevertheless, the maximum relative error of shear stress in all the considered cases remains below $3.2\%$, and occurs at a moderate degree of rarefaction. It should be noted that, the moderate rarefaction level extends to a wider range of Knudsen numbers for a mixture with larger mass difference, because of the presence of multiscale species $\text{Kn}$ and relaxation times of inter-species collisions. Importantly, the data demonstrates a consistent level of accuracy for our model, regardless of variations in the mass ratio of the gas species.
 

\section{Numerical results for two-dimensional problems}\label{sec:2D_cases}

In this section, the kinetic model is further applied to solve two-dimensional mixture flows, that is, a supersonic mixture passing a circular cylinder and a gas mixture flowing through a nozzle. Also, the results are compared to the DSMC simulations to evaluate the accuracy of our kinetic model. 

The discretized velocity method is applied to solve the model equation that is reduced to be quasi-two-dimensional in velocity space using \eqref{eq:quasi_two_dimensional}. The spatial domains simulated in all the following cases are discretized by structured quadrilateral mesh with refinement near the surfaces, and the grid size is maintained smaller than the local mean free path of the gas molecules, to ensure a reliable solution of the kinetic model equation. The finite volume method is used in the numerical implementation, where the convention fluxes are evaluated implicitly by the LU-SGS technique, and the collision terms are calculated with the Venkata limiter.

\subsection{Supersonic flow around a circular cylinder}

\begin{figure}[t]
	\centering
	\sidesubfloat[]{\includegraphics[scale=0.3,clip=true]{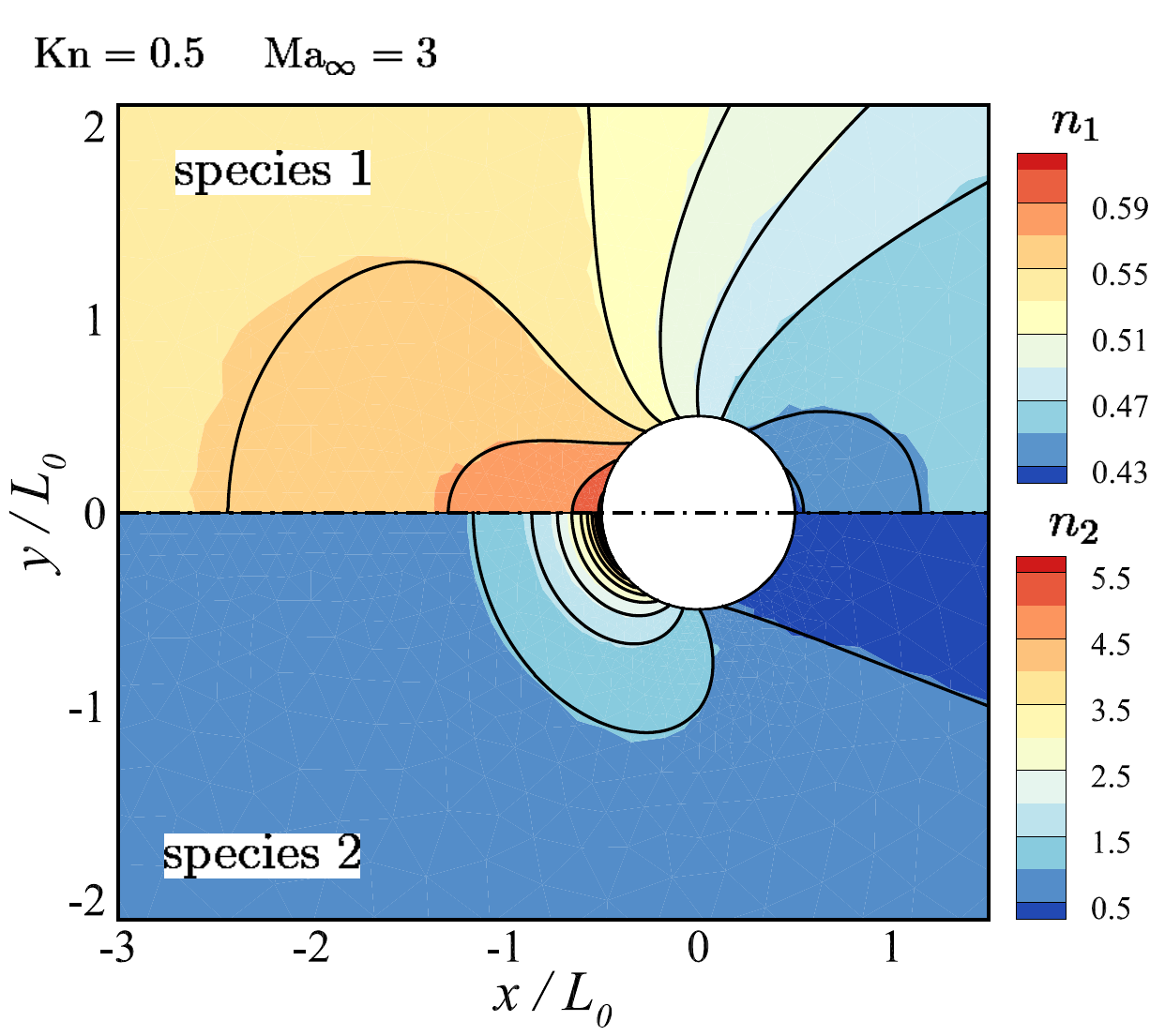}\label{fig:2DShockWave_field_Mix2:n}}   
    \sidesubfloat[]{\includegraphics[scale=0.3,clip=true]{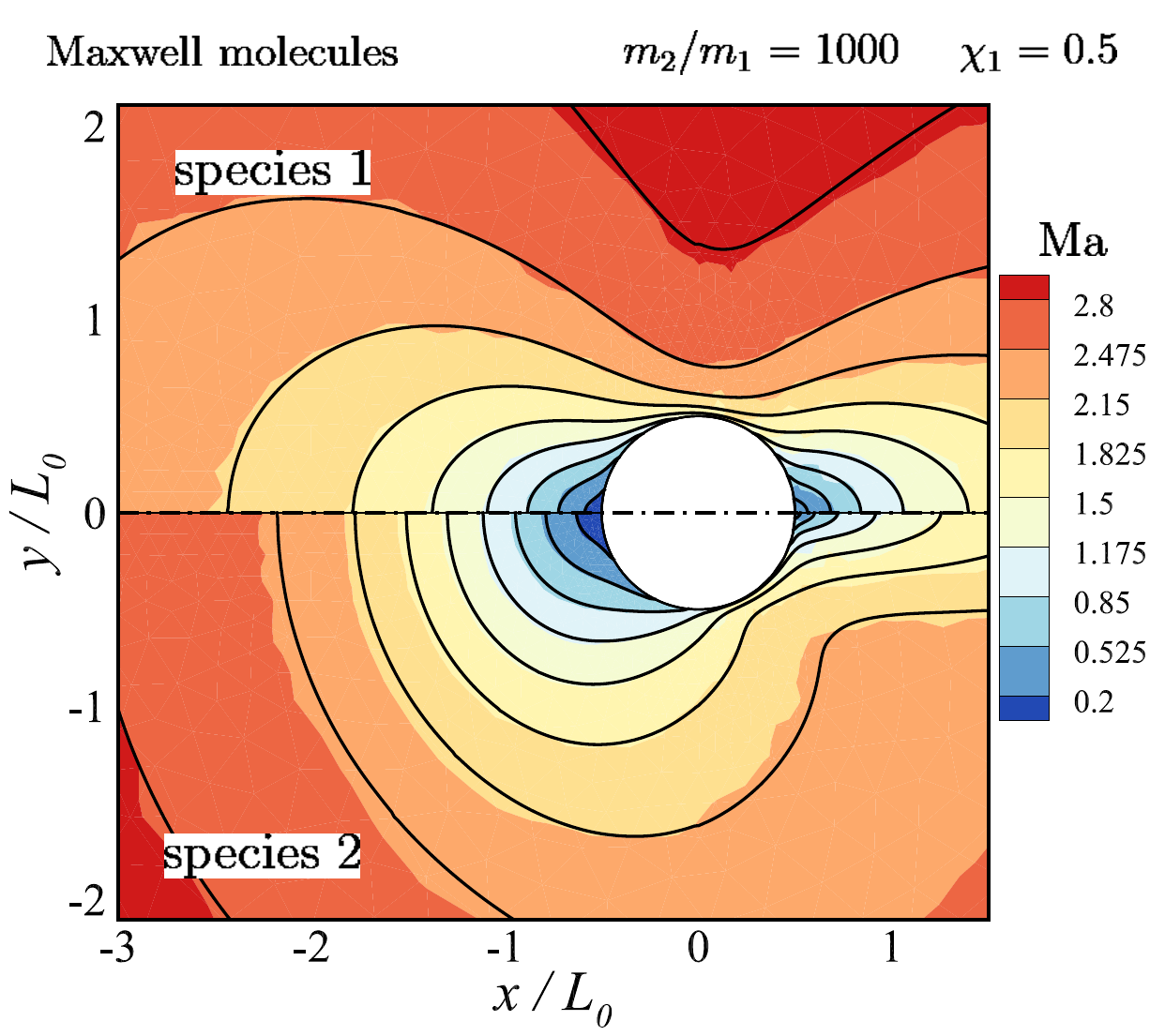}\label{fig:2DShockWave_field_Mix2:u}}
	\caption{Comparisons of the dimensionless (a) number density and (b) flow velocity between the results of the kinetic model (black lines) and DSMC method (background contours) for a supersonic mixture flow passing a cylinder, when the mole fraction of the freestream is $\chi_1=0.5$, $\text{Ma}_{\infty}=3$ and $\text{Kn}_{1}=0.5$. The binary mixture consists of Maxwell molecules with a mass ratio $m_2/m_1=1000$ and a diameter ratio $d_2/d_1=1$.}
	\label{fig:2DShockWave_field}
\end{figure}

\begin{figure}[t]
	\centering
	\sidesubfloat[]{\includegraphics[scale=0.19,clip=true]{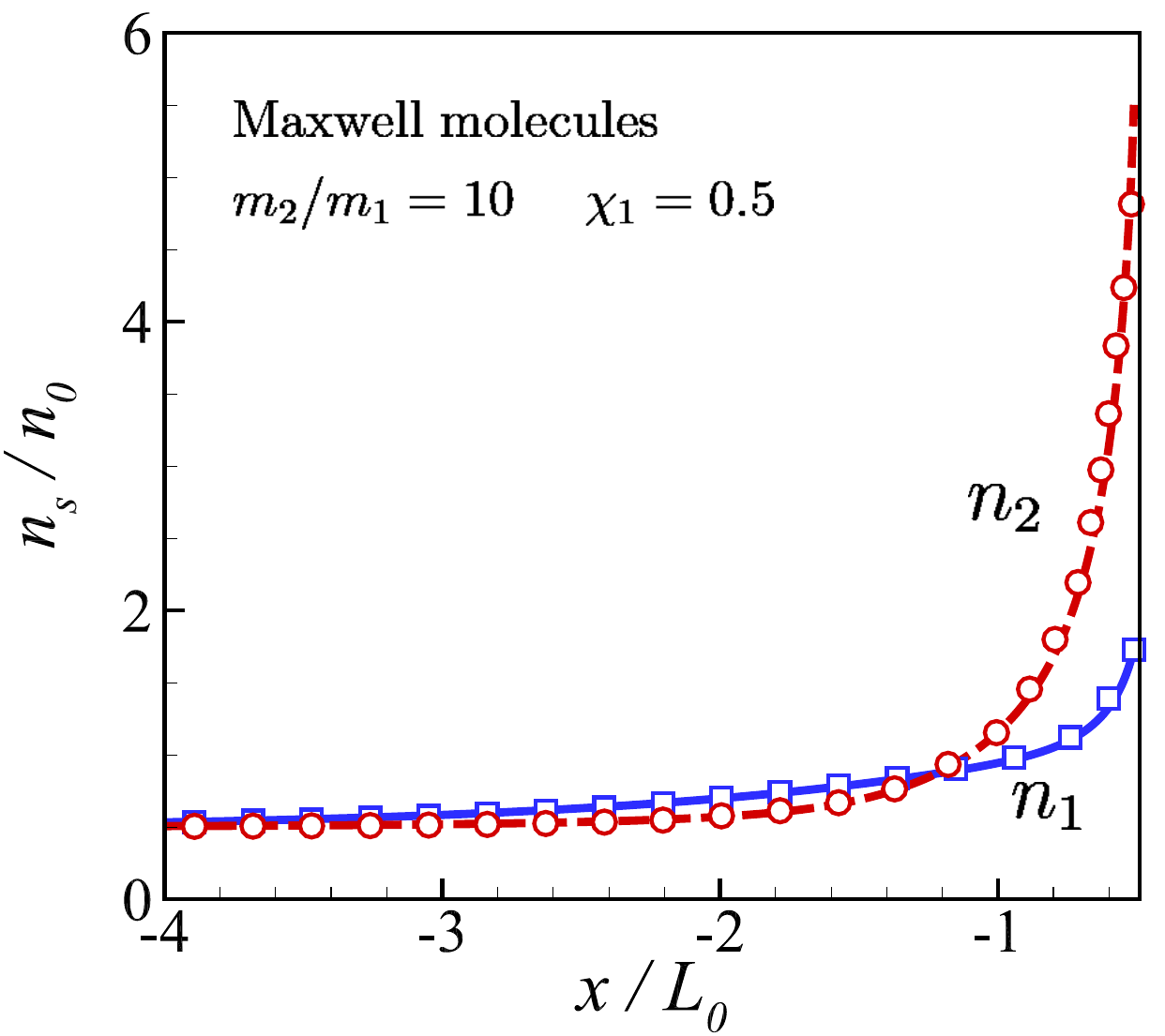}\label{fig:2DShockWave_center_Mix1:n}}   
    \sidesubfloat[]{\includegraphics[scale=0.19,clip=true]{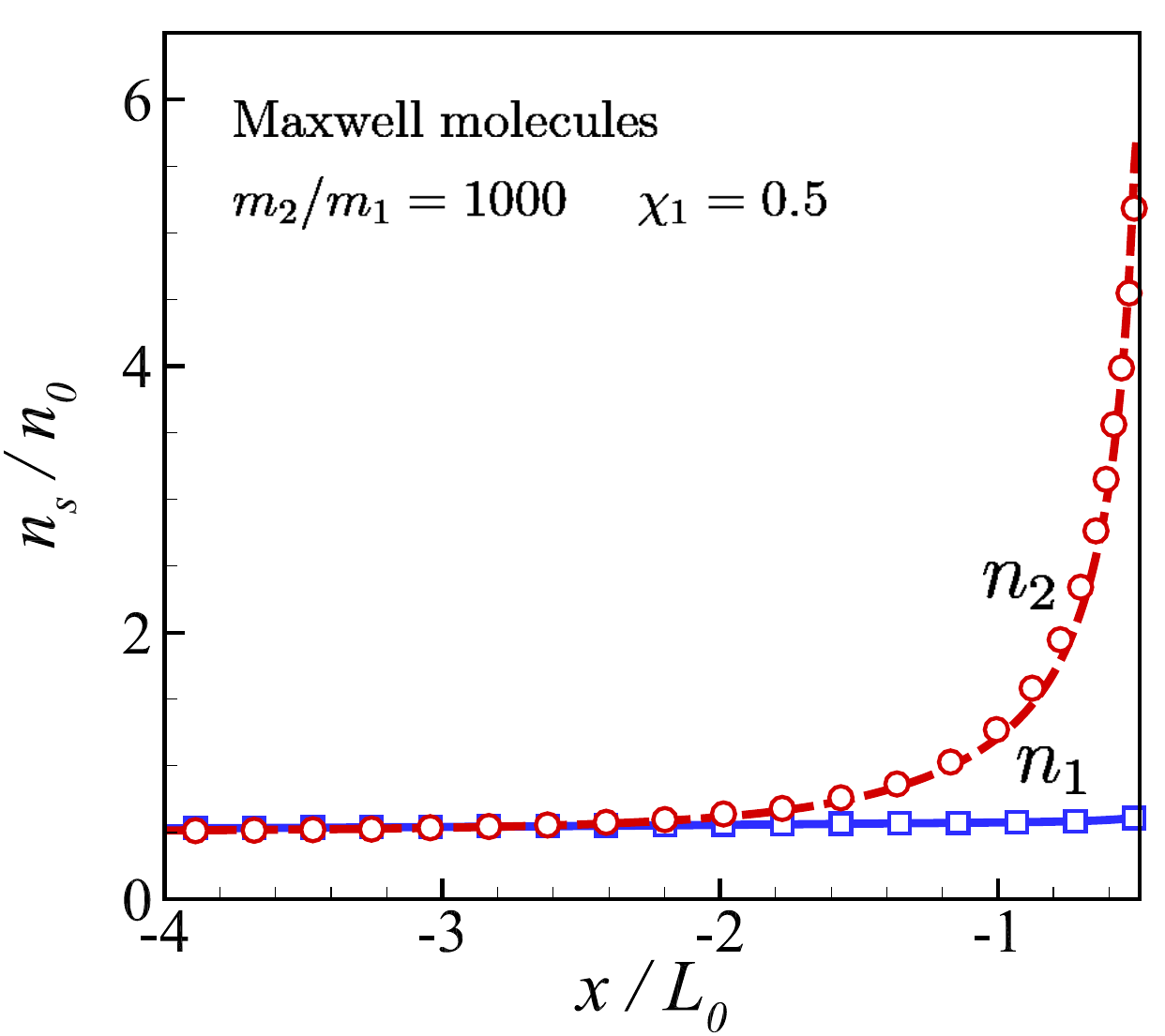}\label{fig:2DShockWave_center_Mix2:n}}
    \sidesubfloat[]{\includegraphics[scale=0.19,clip=true]{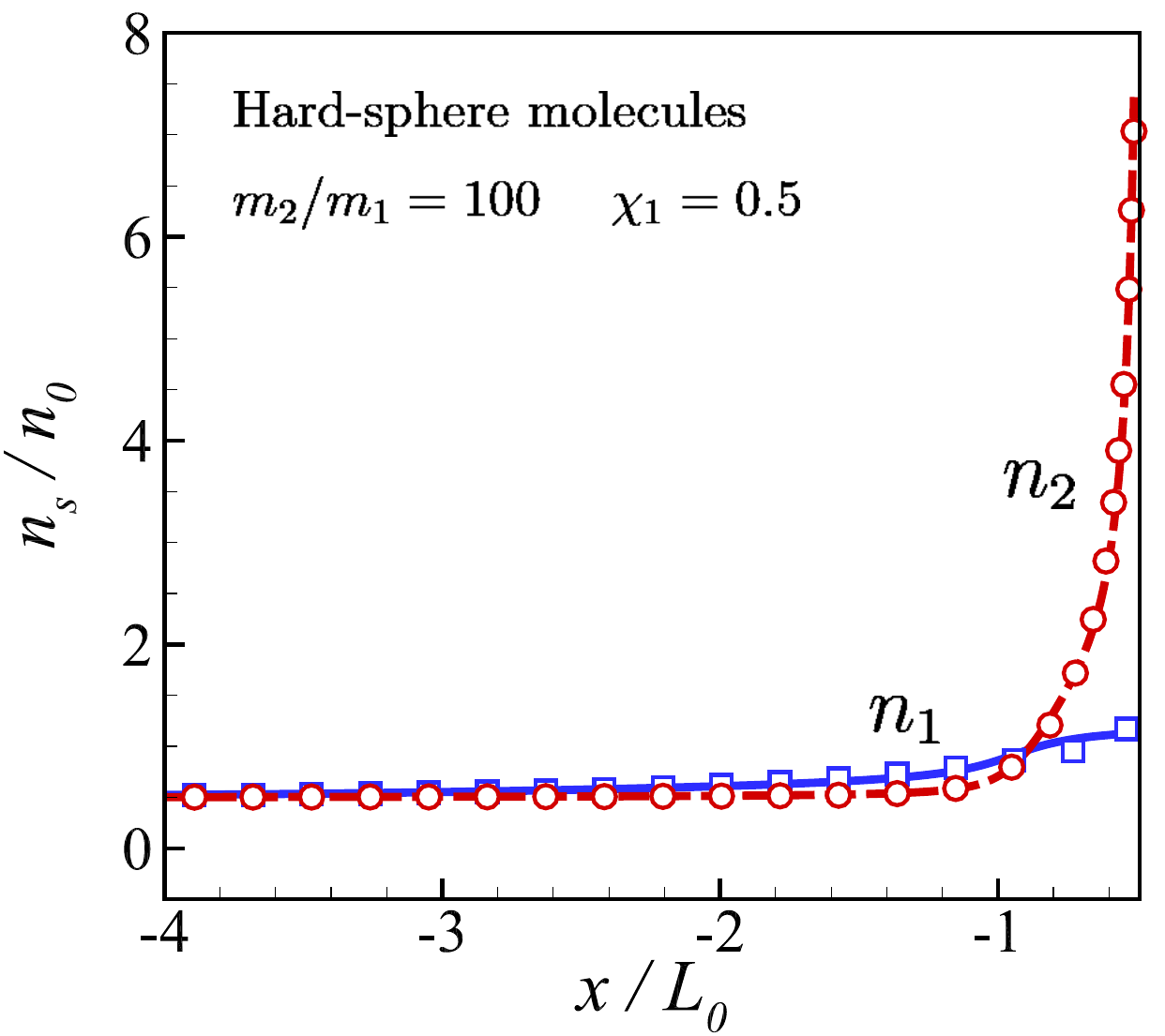}\label{fig:2DShockWave_center_Mix3:n}} \\
    \sidesubfloat[]{\includegraphics[scale=0.19,clip=true]{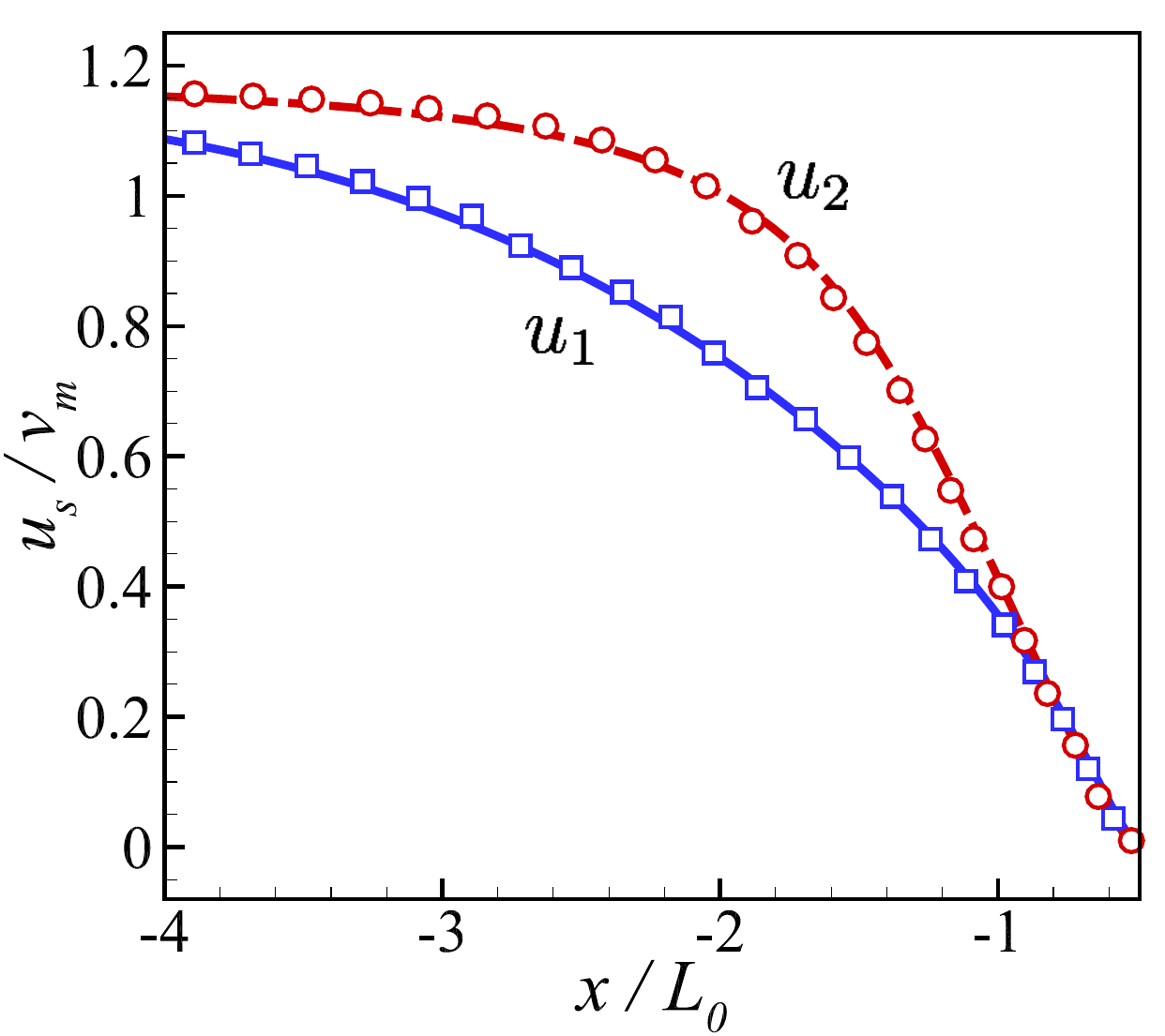}\label{fig:2DShockWave_center_Mix1:u}}   
    \sidesubfloat[]{\includegraphics[scale=0.19,clip=true]{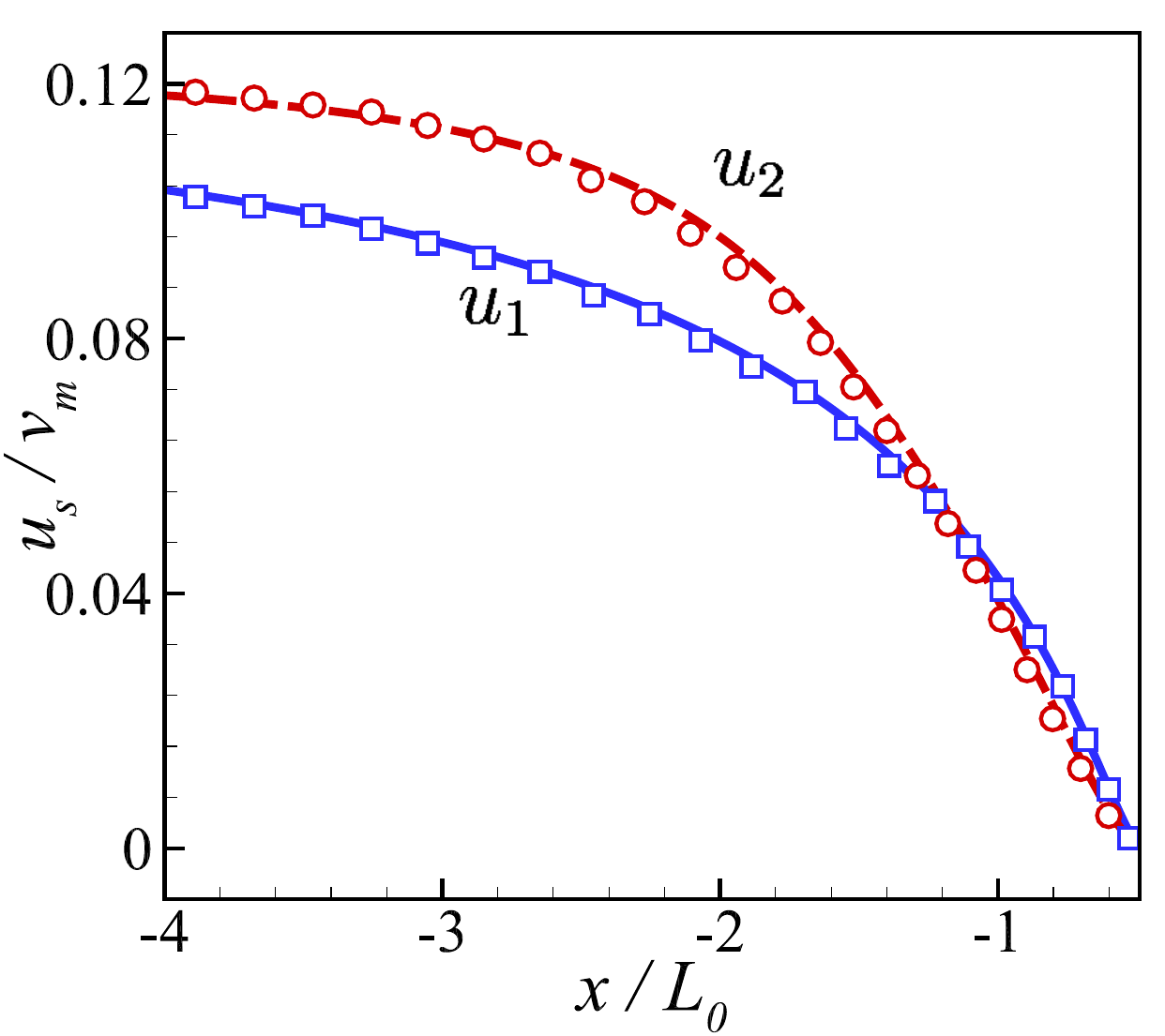}\label{fig:2DShockWave_center_Mix2:u}}
    \sidesubfloat[]{\includegraphics[scale=0.19,clip=true]{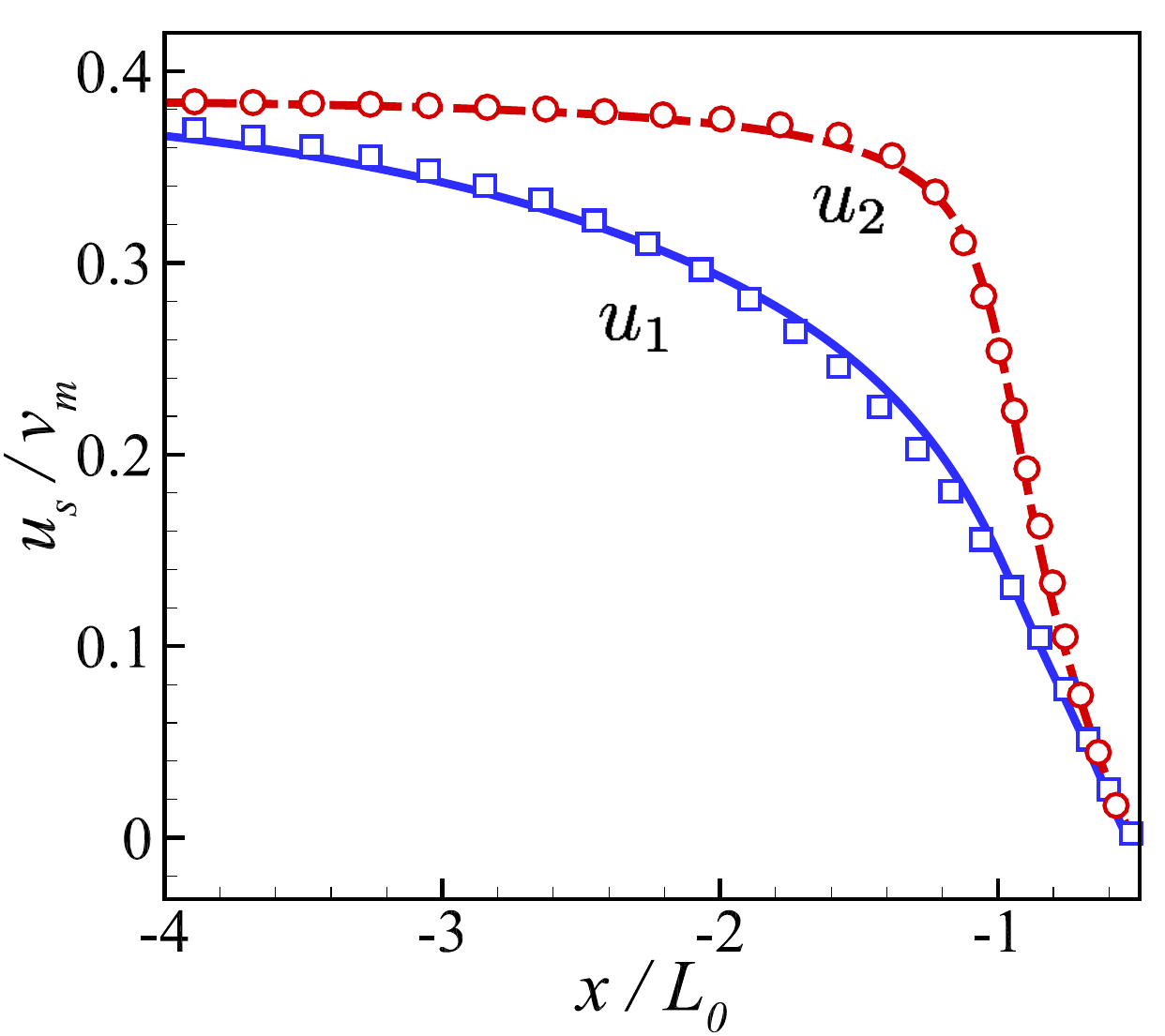}\label{fig:2DShockWave_center_Mix3:u}} \\
    \sidesubfloat[]{\includegraphics[scale=0.19,clip=true]{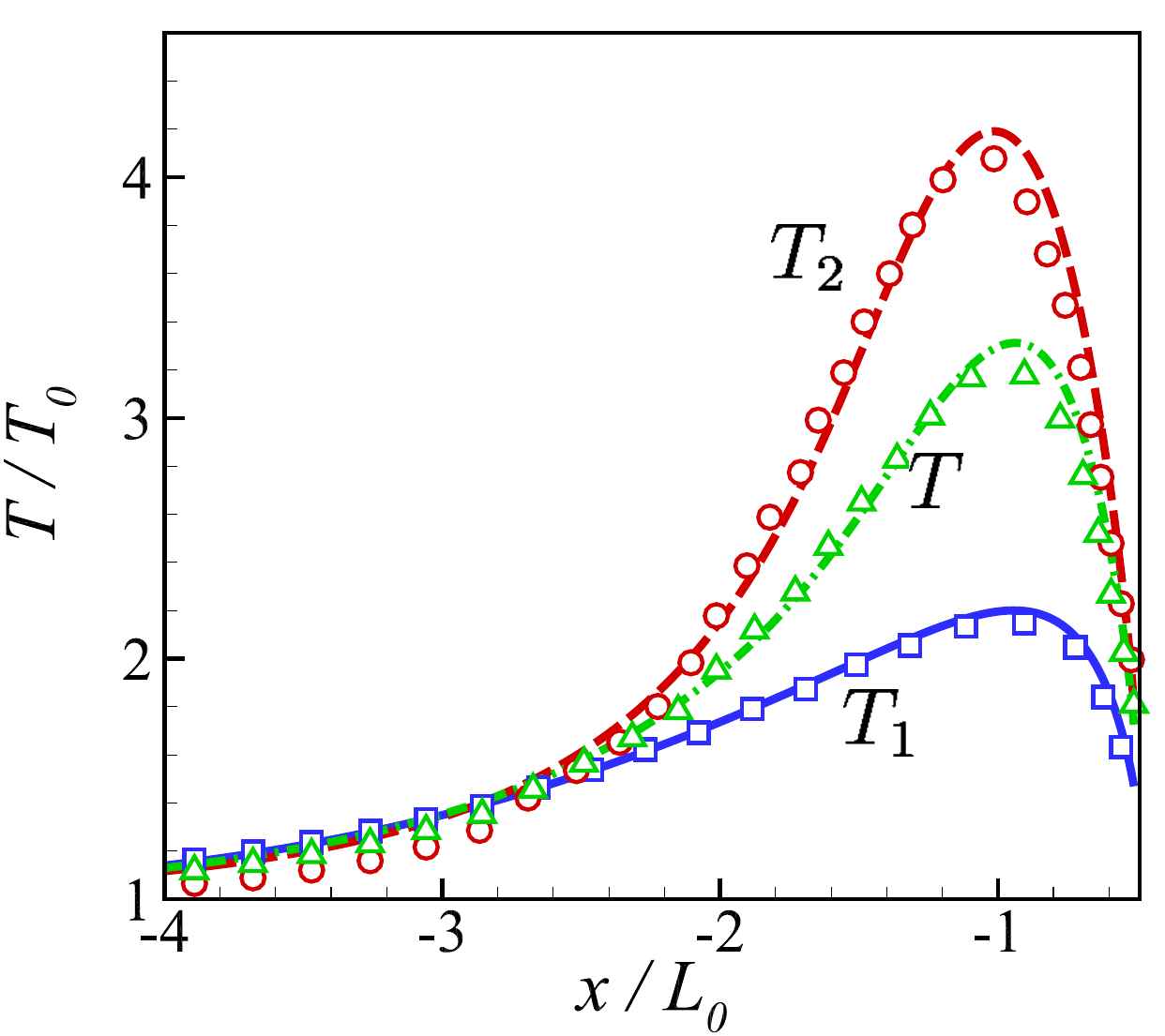}\label{fig:2DShockWave_center_Mix1:T}}   
    \sidesubfloat[]{\includegraphics[scale=0.19,clip=true]{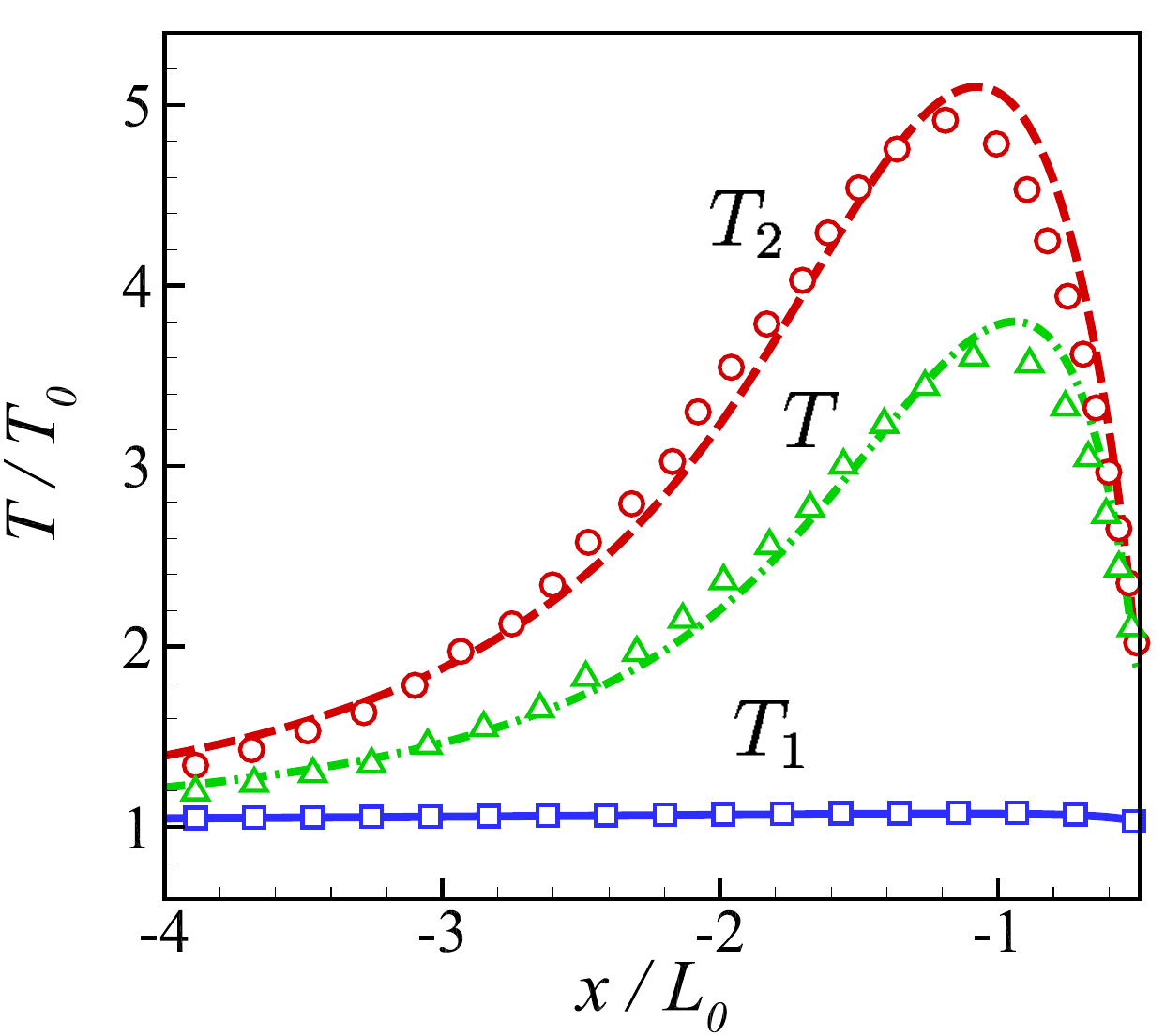}\label{fig:2DShockWave_center_Mix2:T}}
    \sidesubfloat[]{\includegraphics[scale=0.19,clip=true]{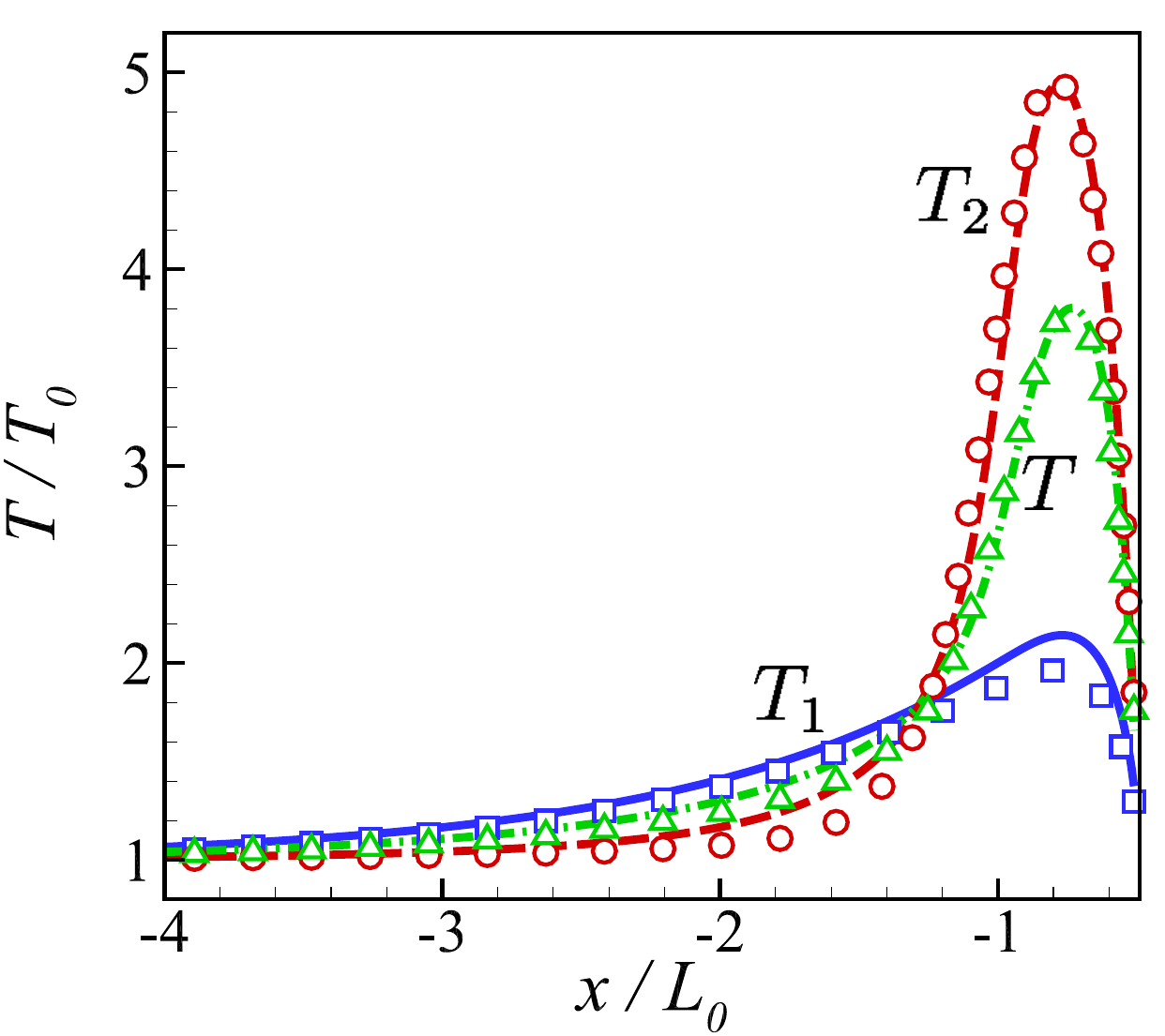}\label{fig:2DShockWave_center_Mix3:T}} \\
    \sidesubfloat[]{\includegraphics[scale=0.19,clip=true]{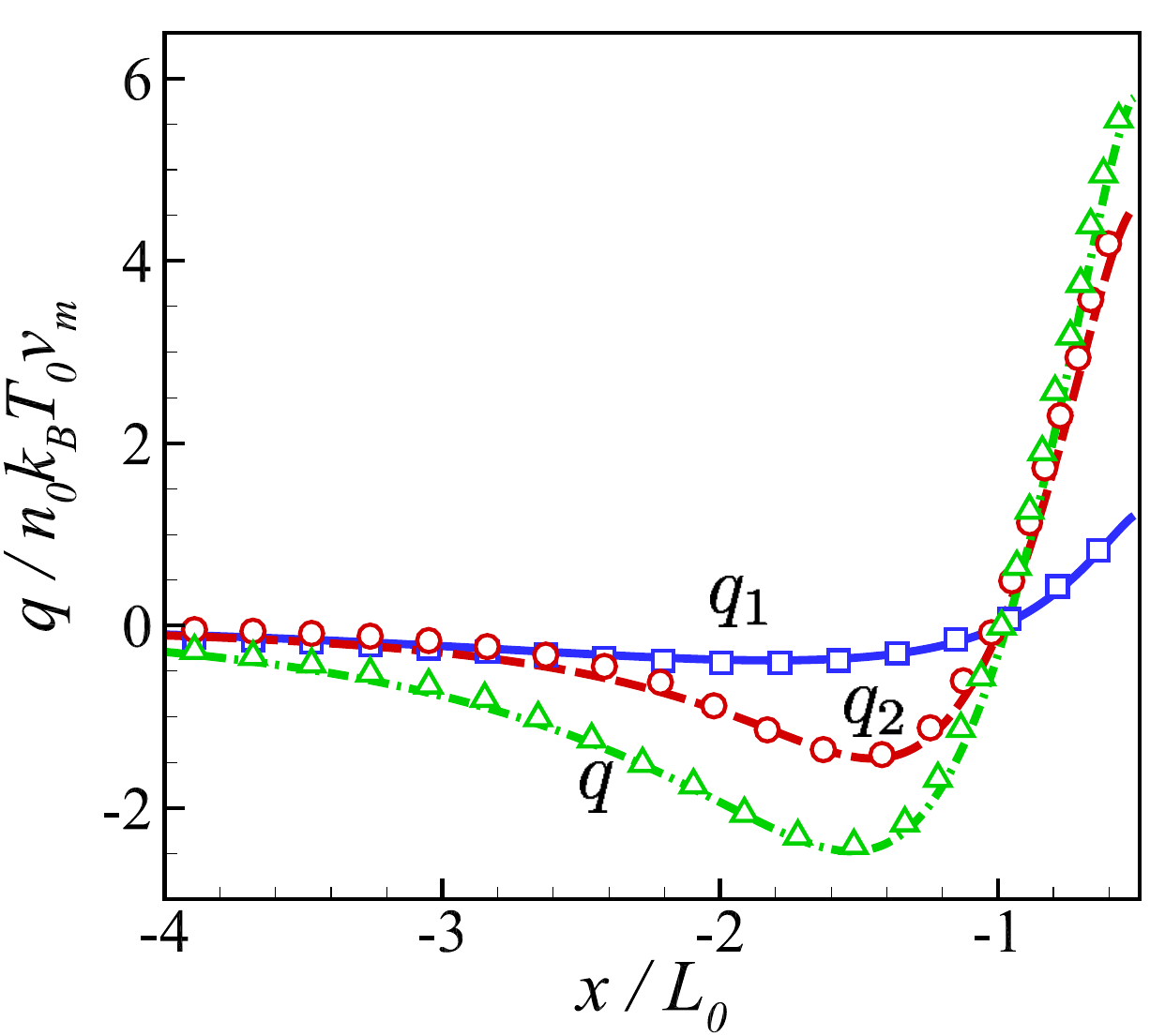}\label{fig:2DShockWave_center_Mix1:q}}   
    \sidesubfloat[]{\includegraphics[scale=0.19,clip=true]{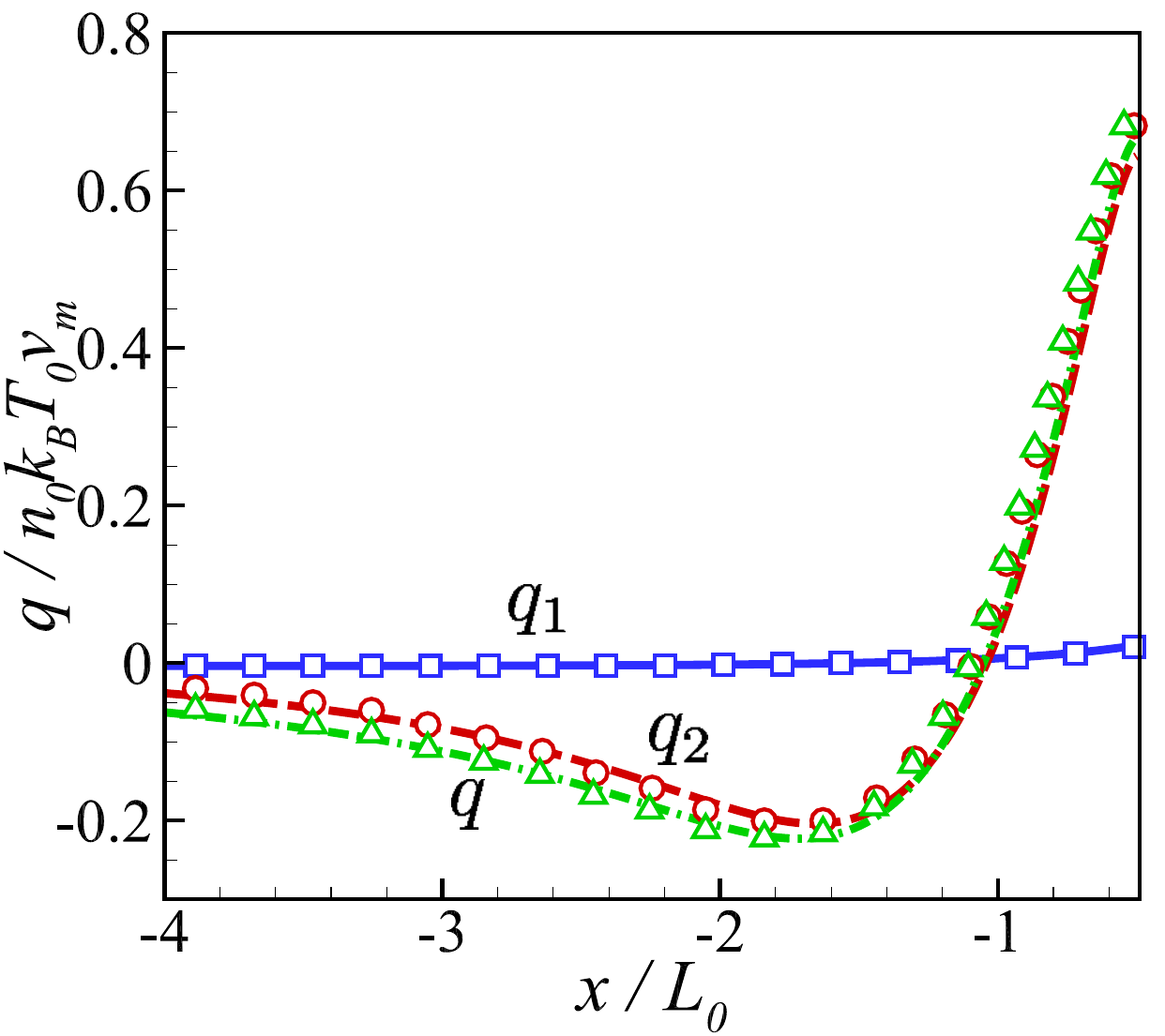}\label{fig:2DShockWave_center_Mix2:q}}
    \sidesubfloat[]{\includegraphics[scale=0.19,clip=true]{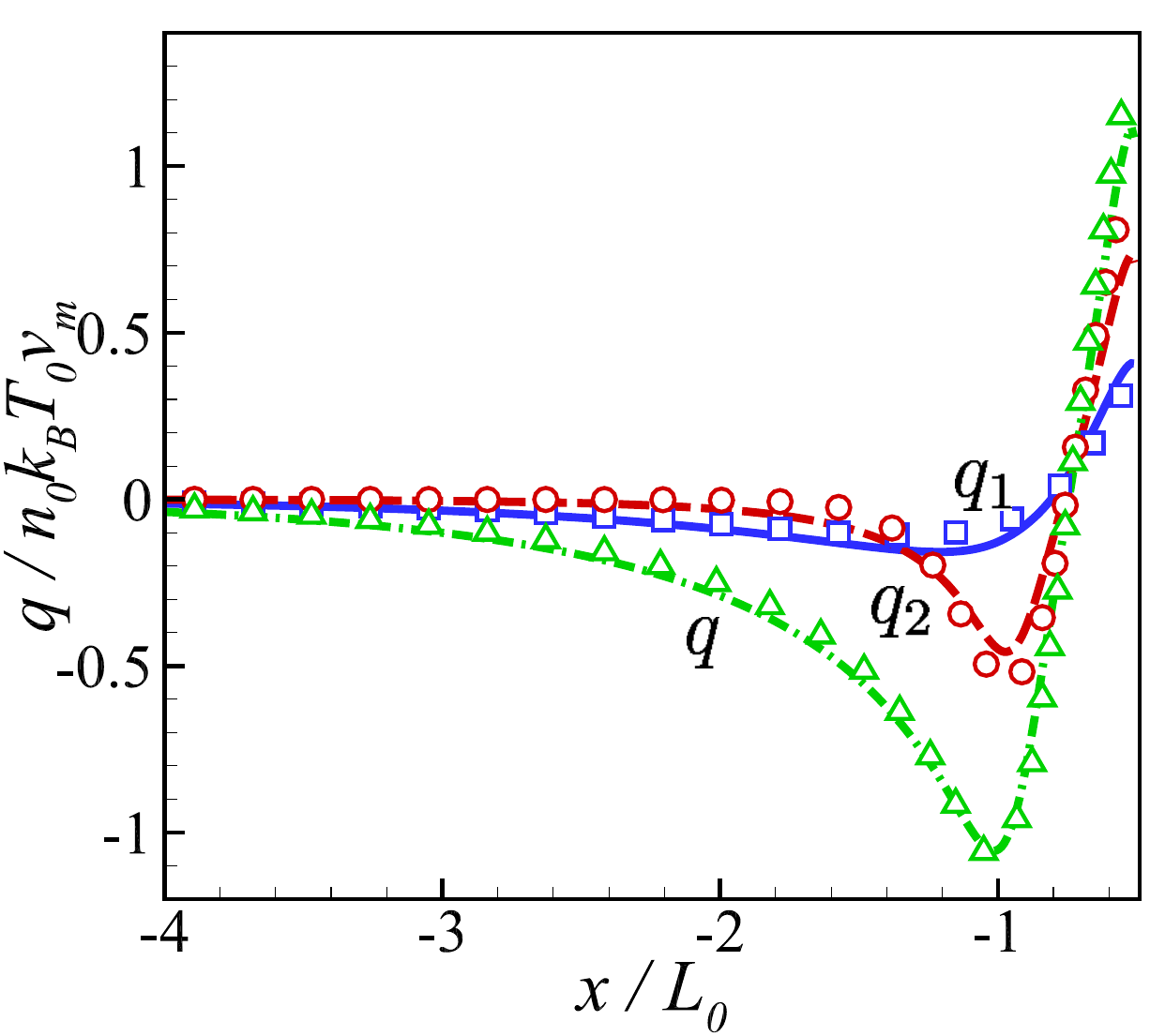}\label{fig:2DShockWave_center_Mix3:q}} \\
    \includegraphics[scale=0.22,clip=true]{Figures/legend_1D.png}
	\caption{Comparisons of the dimensionless (a-c) number density, (d-f) flow velocity, (g-i) temperature and (j-l) heat flux along the windward side stagnation line between the results of the kinetic model and DSMC for a supersonic mixture flow passing a cylinder, when the mole fraction of the freestream is $\chi_1=0.5$, $\text{Ma}_{\infty}=3$ and $\text{Kn}_{1}=0.5$. The first, second and third column corresponds to Mixture 1, 2 and 3, respectively.}
	\label{fig:2DShockWave_center}
\end{figure}

\begin{figure}[t]
	\centering
	\sidesubfloat[]{\includegraphics[scale=0.19,clip=true]{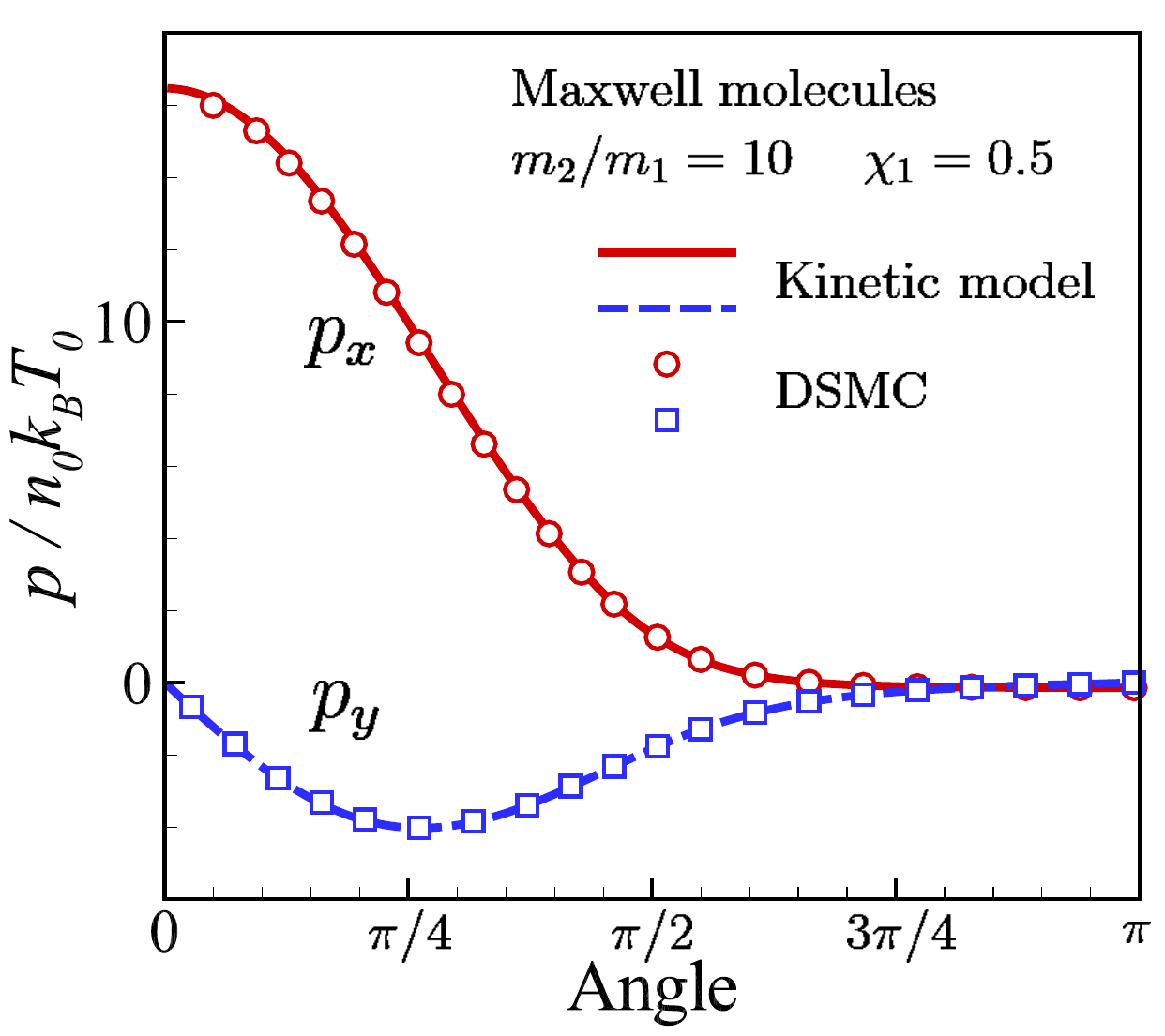}\label{fig:2DShockWave_surface_Mix1:p}}   
    \sidesubfloat[]{\includegraphics[scale=0.19,clip=true]{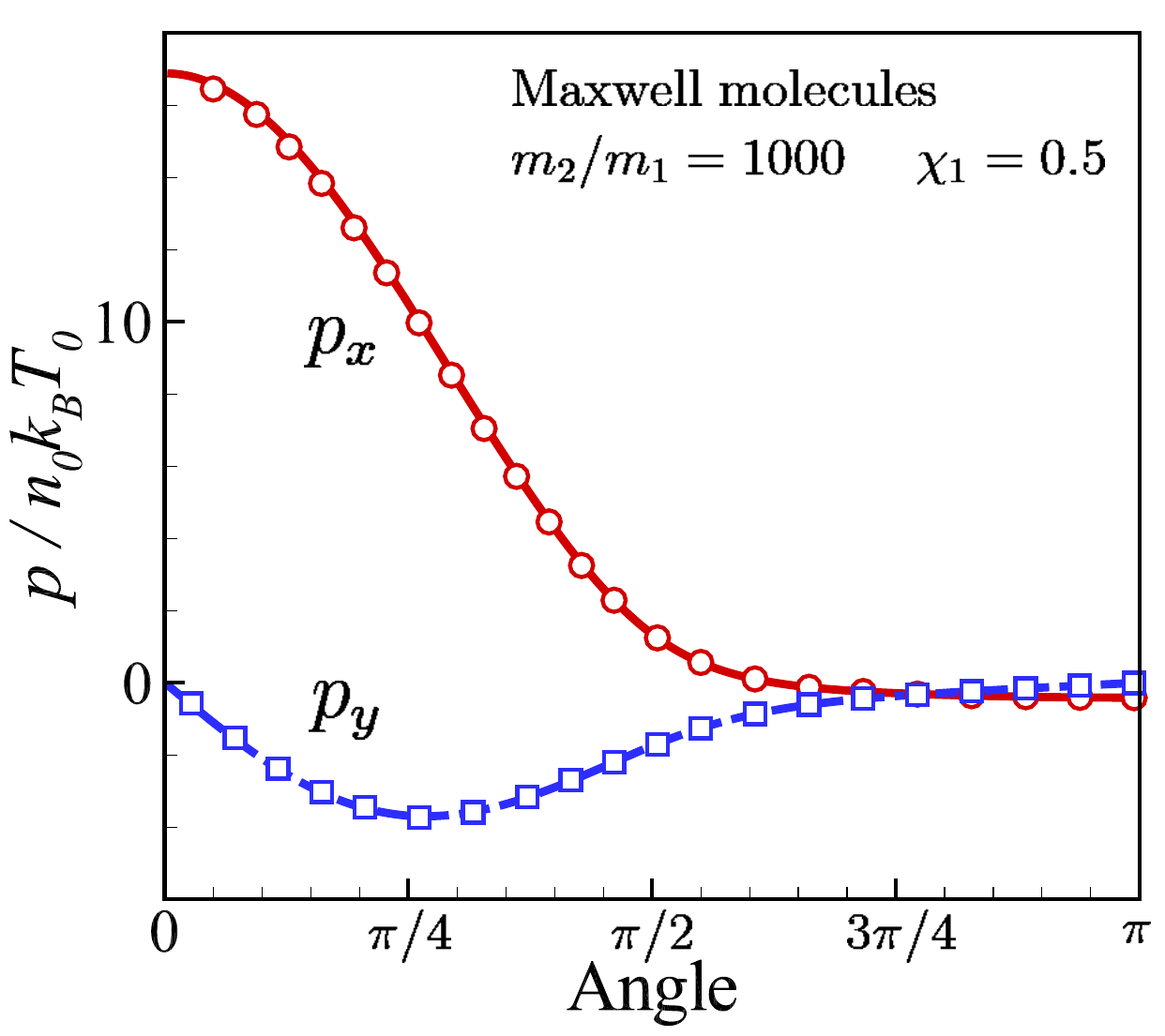}\label{fig:2DShockWave_surface_Mix2:p}}
    \sidesubfloat[]{\includegraphics[scale=0.19,clip=true]{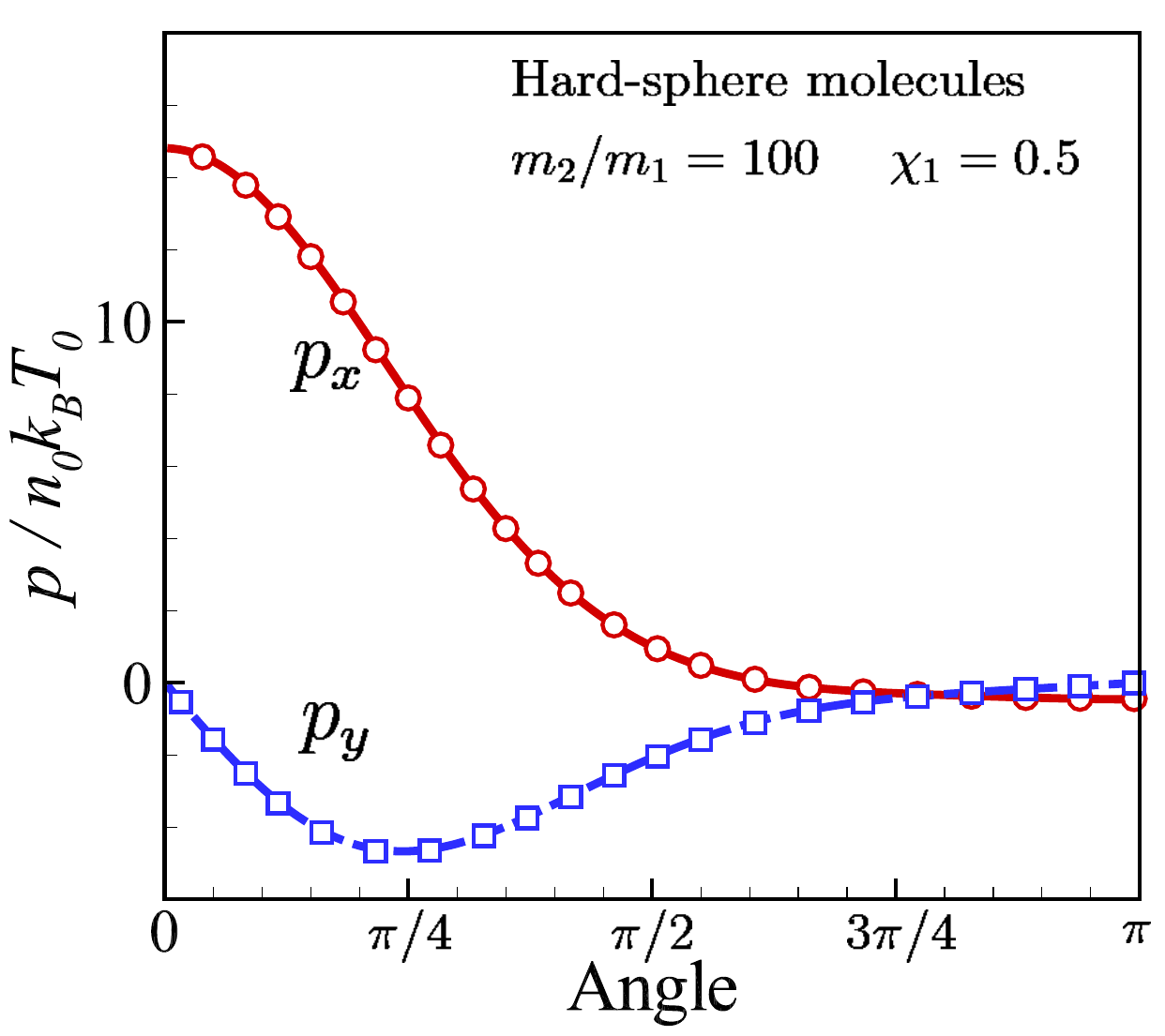}\label{fig:2DShockWave_surface_Mix3:p}} \\
    \sidesubfloat[]{\includegraphics[scale=0.19,clip=true]{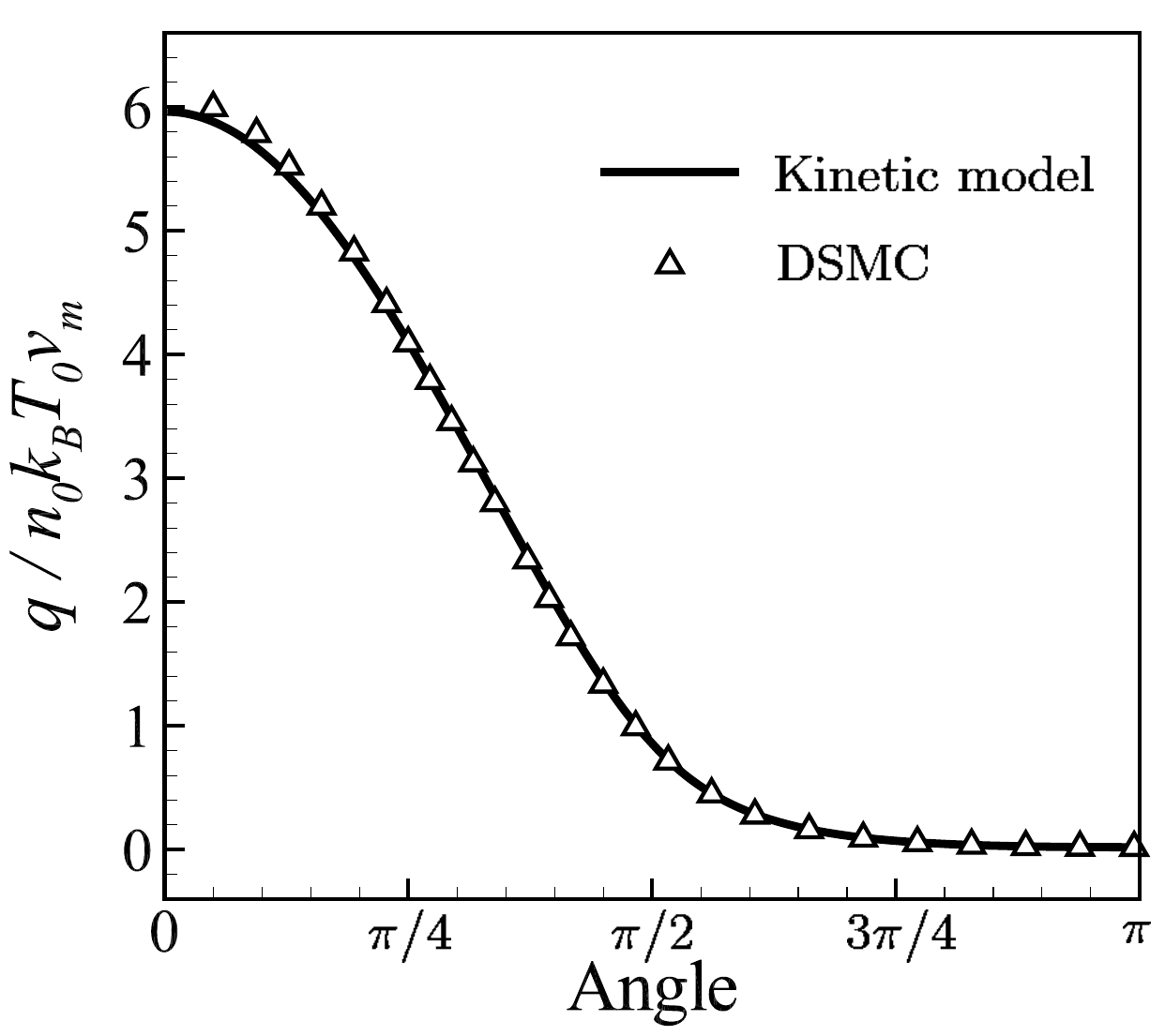}\label{fig:2DShockWave_surface_Mix1:q}}   
    \sidesubfloat[]{\includegraphics[scale=0.19,clip=true]{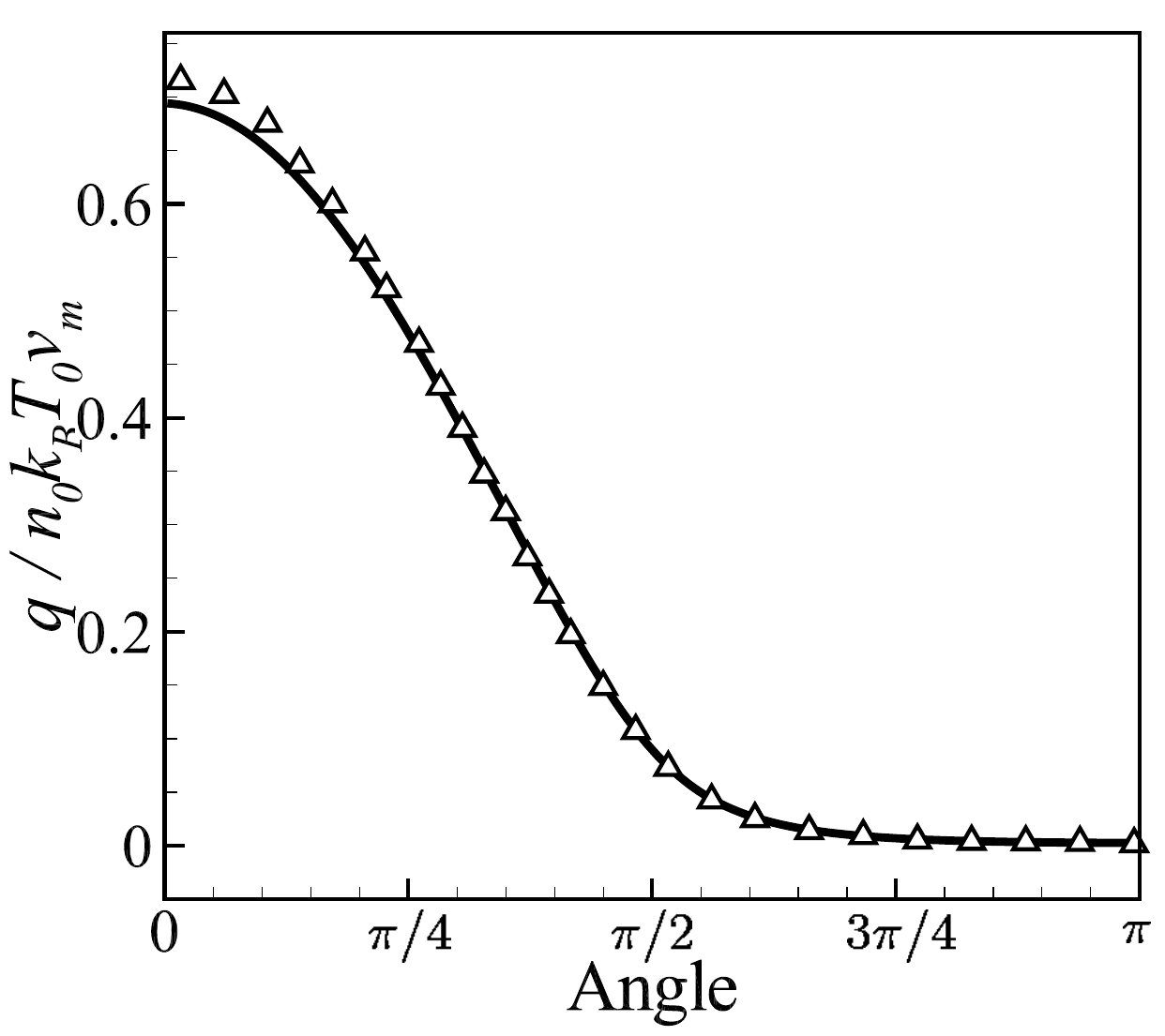}\label{fig:2DShockWave_surface_Mix2:q}}
    \sidesubfloat[]{\includegraphics[scale=0.19,clip=true]{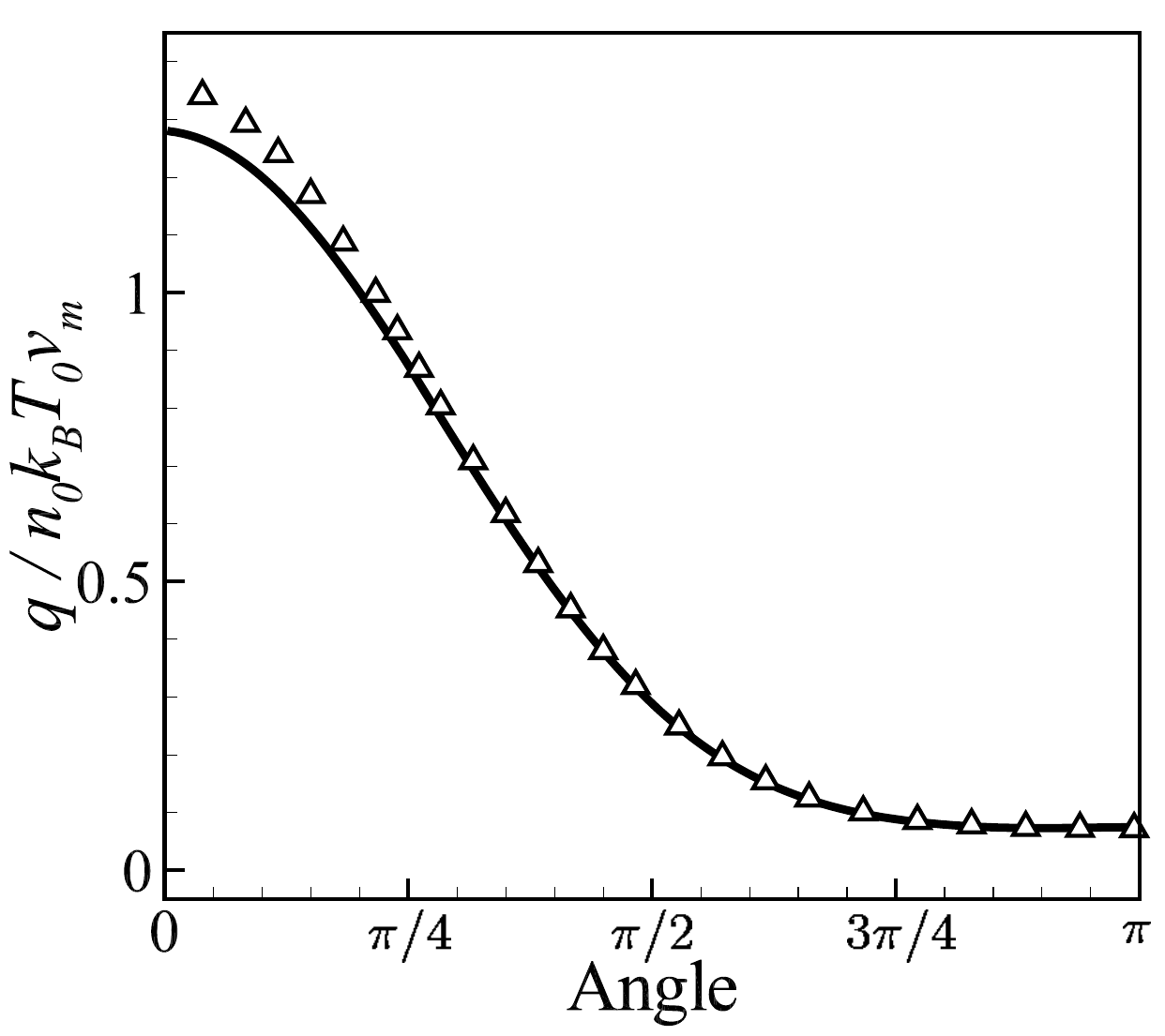}\label{fig:2DShockWave_surface_Mix3:q}} 
	\caption{Comparisons of the dimensionless (a-c) pressure and (d-f) heat flux along the surface of the cylinder between the results of kinetic model and DSMC for the mixtures, when the mole fraction of the freestream is $\chi_1=0.5$, $\text{Ma}_{\infty}=3$ and $\text{Kn}_{1}=0.5$. The angle is measured from the windward to the leeward side. The first, second and third column corresponds to Mixture 1, 2 and 3, respectively.}
	\label{fig:2DShockWave_surface}
\end{figure}

We consider the supersonic gas flow with density $n_0$ at $\text{Ma}_{\infty}=3$ passing a cylinder with diameter $L_0$. The temperatures of both the freestream and isothermal surfaces of the cylinder are maintained at $T_0$. The Knudsen number of the lighter species of the freestream is $\text{Kn}_{1}=0.5$. Based on the diameter ratio of mixtures' components, the Knudsen number of the heavier species in Mixture 1 and 2 is $\text{Kn}_{2}=0.5$ as well, while that in Mixture 3 of hard-sphere molecules is $\text{Kn}_{2}=0.125$. Besides, the mole fraction considered in this problem is $\chi_1=0.5$ for all the mixtures. The simulations are conducted only in the upper half-domain $[-L,L]\times[0,L]$ due to symmetry, with $L/L_0=6,12,10$ for Mixture 1, 2 and 3, respectively.

The detailed flow fields about species number density and flow velocity of the surrounding gas are presented in figure \ref{fig:2DShockWave_field} for the gas mixture with a mass ratio of 1000, where the result given by the kinetic model matches the DSMC data well. Figure \ref{fig:2DShockWave_center} compares kinetic model results of the windward side number density, flow velocity, temperature and heat flux along the stagnation line with those solved by the DSMC method, and the overall agreement is very good. 

Despite having the same freestream flow velocity that exceeds the sound speed of the mixture, individual species in a gas mixture with disparate mass experience vastly different flow characteristics. This is due to their distinct species Mach numbers $\text{Ma}_{\infty,s}$, which are defined based on their own individual sound speeds $\sqrt{5k_BT_0/3m_{s}}$. For instance, $\text{Ma}_{\infty,1}=0.13$ and $\text{Ma}_{\infty,2}=4.24$ for the mixture shown in figure \ref{fig:2DShockWave_field}. Therefore, the lighter component forms a subsonic flow field and exhibits significantly less compression compared to the heavier species. As shown by the density profiles in figure \ref{fig:2DShockWave_center} for the mixture with a mass ratio of 1000, the number density of the heavier gas gets nearly 10 times that of the lighter one near the stagnation point.

The overall properties of the shock are mainly determined by the heavier species, because of its higher number density and molecular mass, as long as the components' mole fractions in the freestream are comparable. However, it is found that, as the mass ratio changes from 10 to 1000 for Mixture 1 and 2, the thickness of the shock wave and the peak values of the mixture temperature increase only slightly (10\%), while the thinner shock structure of Mixture 3 composed of hard-sphere molecules arises from the lower Knudsen number of its heavier component. Note that the actual freestream velocities of the mixtures vary significantly due to the distinct average mixture mass, though $\text{Ma}_{\infty}$ keeps constant.

Figure \ref{fig:2DShockWave_surface} shows the pressure and heat flux along the surface of the cylinder. The pressure predicted by the kinetic model matches the DSMC results very well, while the heat fluxes given by the two methods have a 6\% relative difference around the windward side stagnation region for Mixture 3 consisting of hard-sphere molecules. Interestingly, the forces acting on the object are found to be very close in the three types of mixtures, despite the disparity in average mixture mass, and even the intermolecular potential. Particularly, for the flow of Mixture 1 and 2, the aerodynamic forces are nearly the same hence roughly independent of the mass ratio, which is the only different dimensionless variable in the two flows. Also, the force is found to be insensitive to the intermolecular potential and molecular diameter ratio (Mixture 3). On the other hand, the values of heat flux on the surface are not only inversely scaled by the square root of the mass ratio, but also significantly influenced by the intermolecular potential and molecular diameter ratio.

\subsection{Nozzle flow}

\begin{figure}[t]
	\centering
	\sidesubfloat[]{\includegraphics[scale=0.3,clip=true]{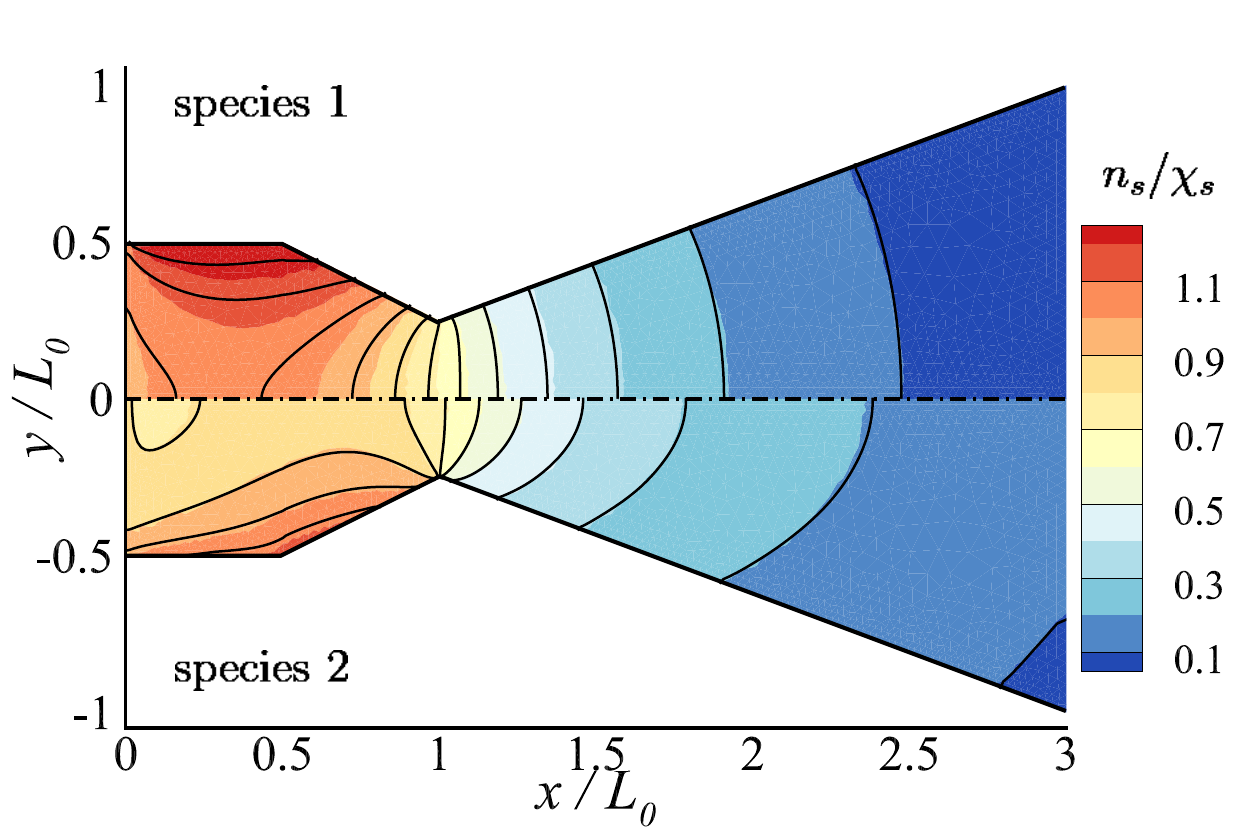}\label{fig:2DNozzle_field:Mix2_n_field}}   
    \sidesubfloat[]{\includegraphics[scale=0.3,clip=true]{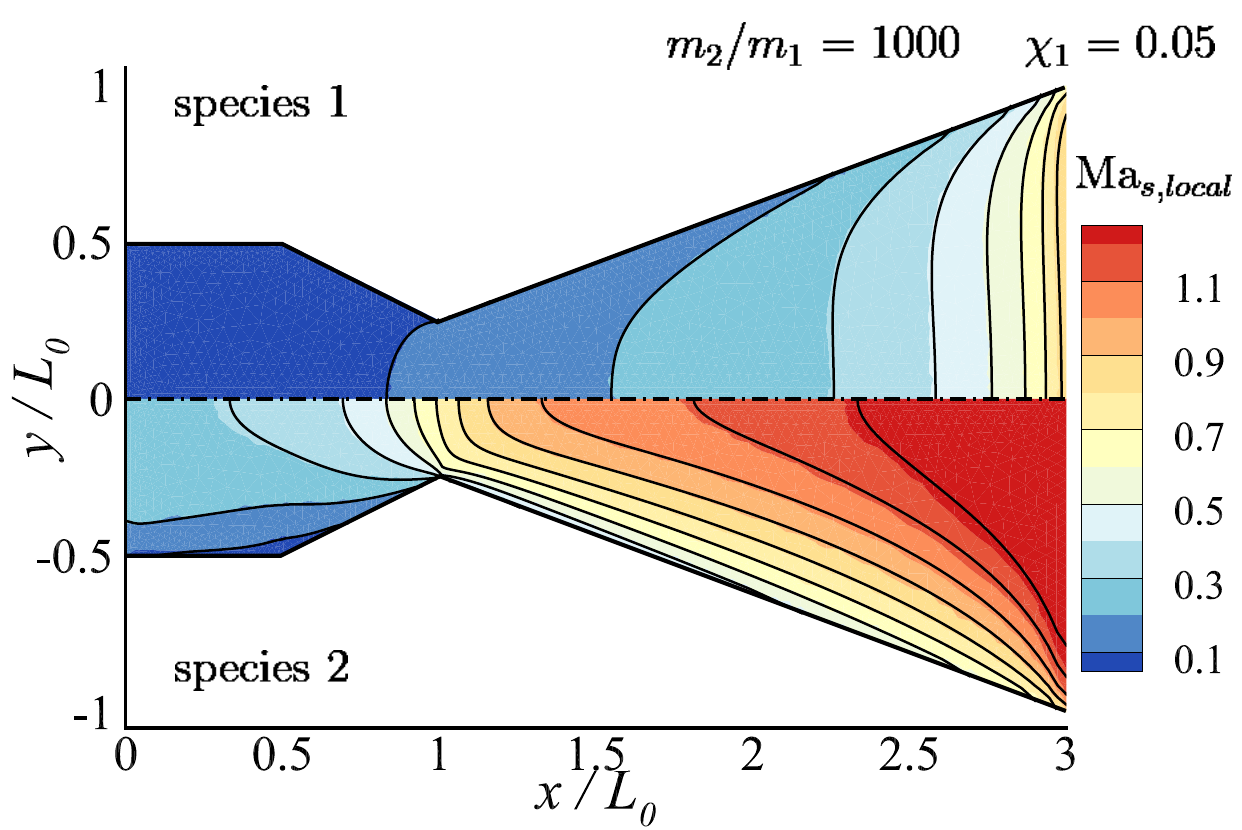}\label{fig:2DNozzle_field:Mix2_Ma_field}}  \\  
    \sidesubfloat[]{\includegraphics[scale=0.3,clip=true]{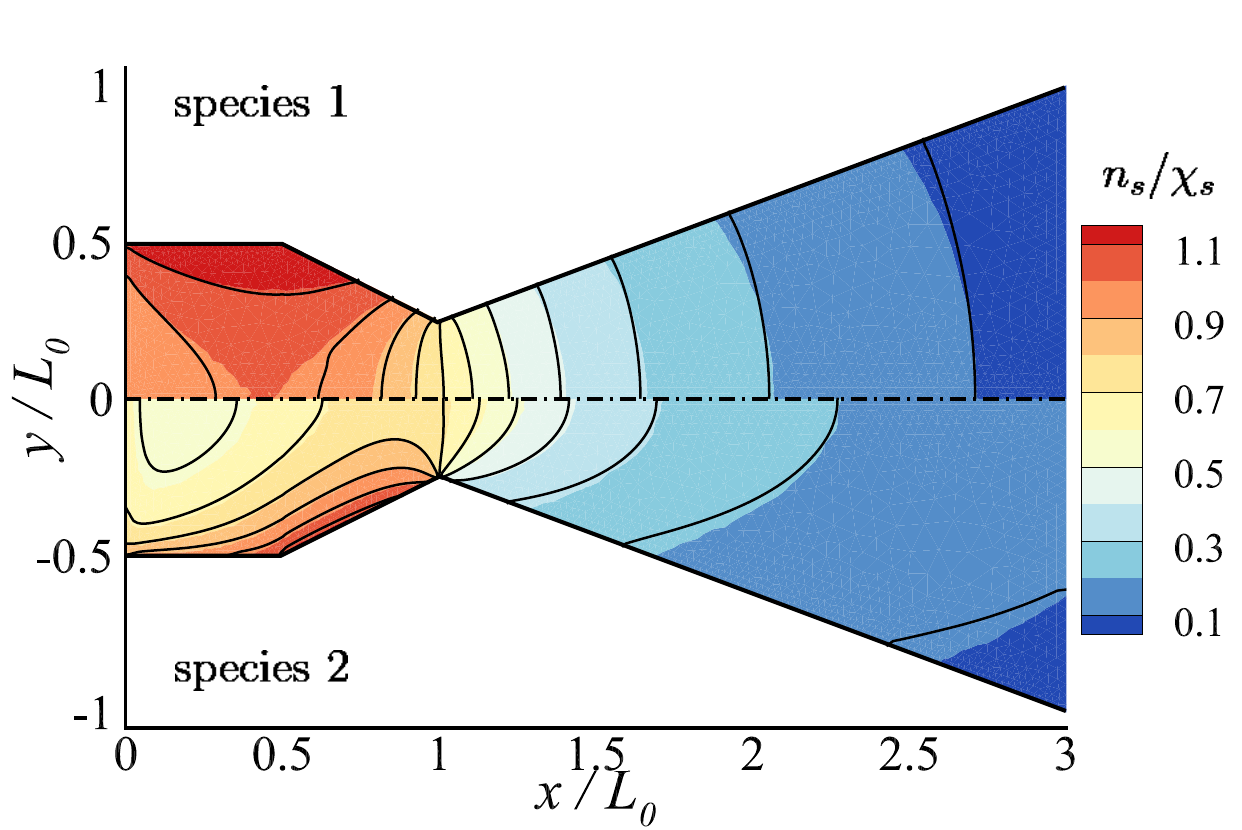}\label{fig:2DNozzle_field:Mix3_n_field}}   
    \sidesubfloat[]{\includegraphics[scale=0.3,clip=true]{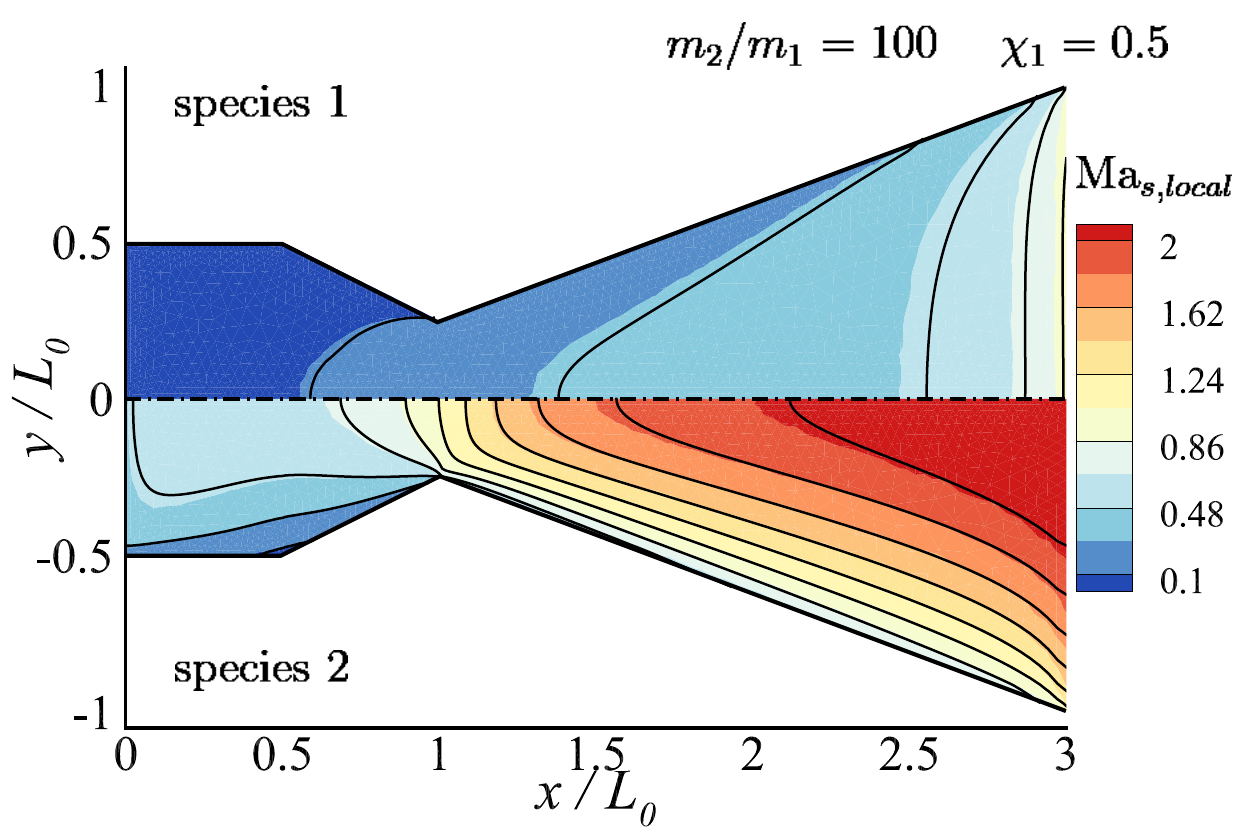}\label{fig:2DNozzle_field:Mix3_Ma_field}}  
	\caption{The dimensionless (a,c) number density and (b,d) local Mach number solved by the kinetic model (black lines) and DSMC method (background contours) for a gas mixture flowing through a nozzle, when $\text{Kn}_{1}=0.1$ at the inlet. The inlet mole fraction is $\chi_1=0.05$ for Mixture 2 (first row) and $\chi_1=0.5$ for Mixture 3 (second row).}
	\label{fig:2DNozzle_field}
\end{figure}

\begin{figure}[t]
	\centering
    \sidesubfloat[]{\includegraphics[scale=0.19,clip=true]{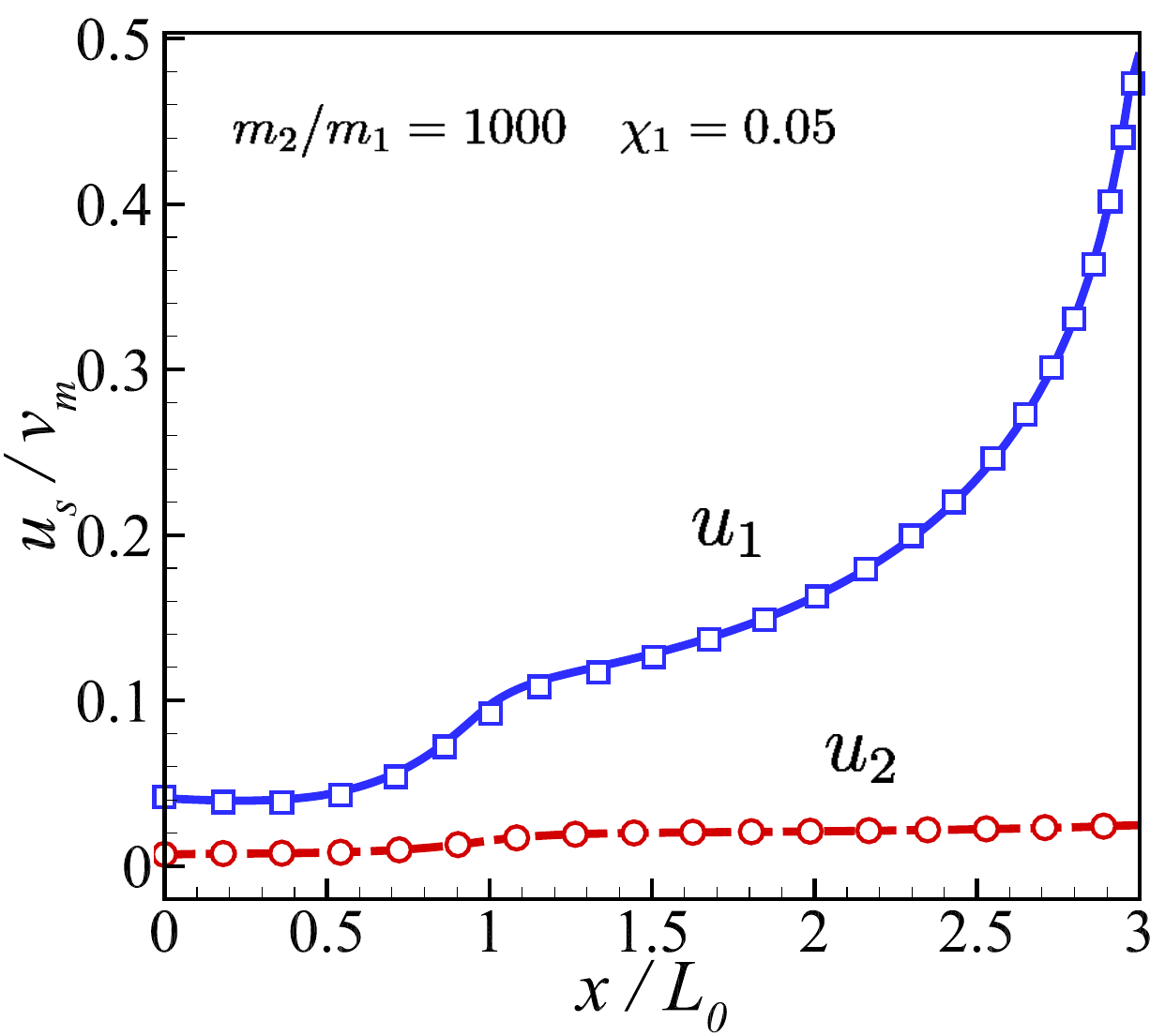}\label{fig:2DNozzle_Mix2:u_center}} 
	\sidesubfloat[]{\includegraphics[scale=0.19,clip=true]{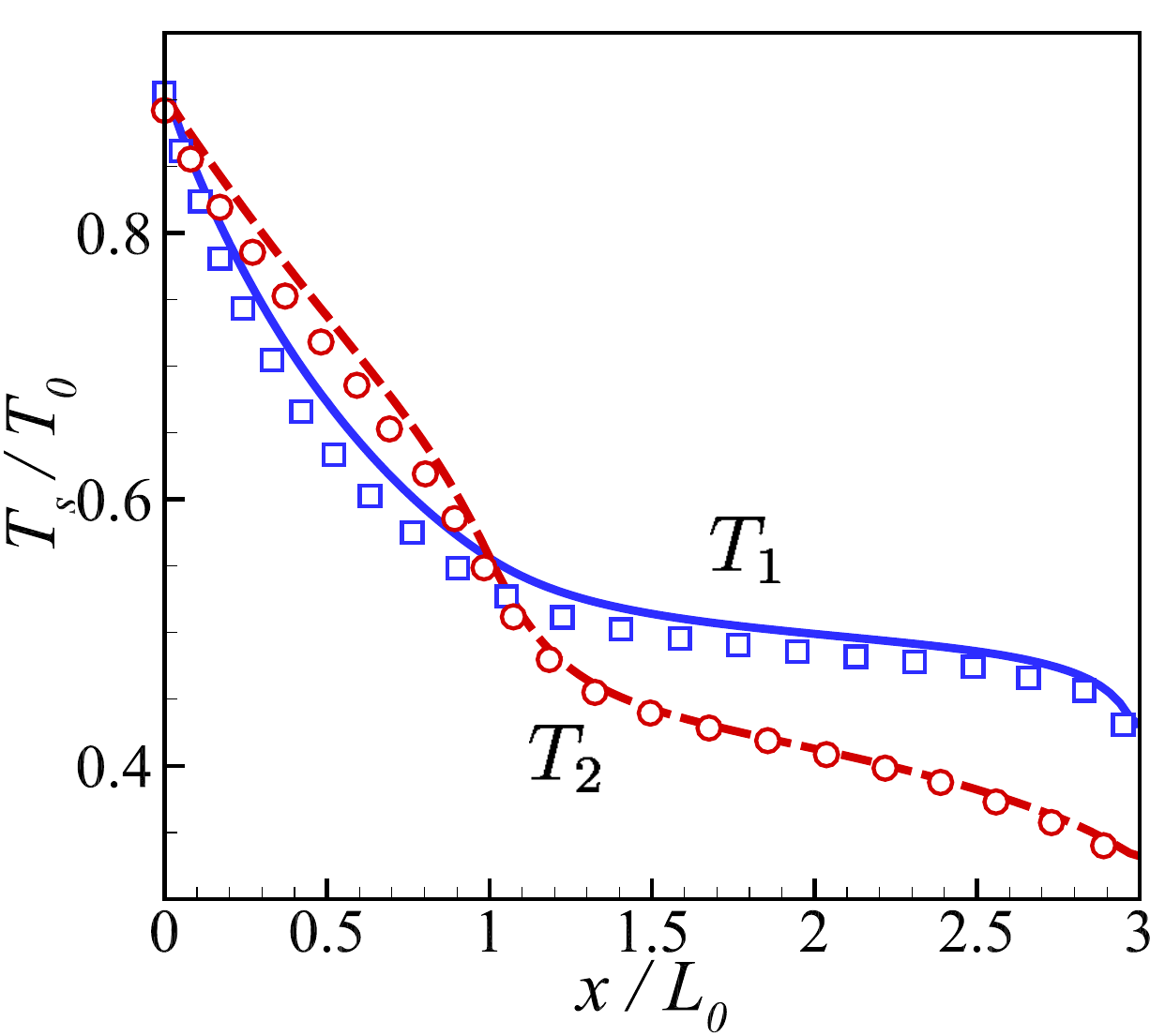}\label{fig:2DNozzle_Mix2:T_center}} 
    \sidesubfloat[]{\includegraphics[scale=0.19,clip=true]{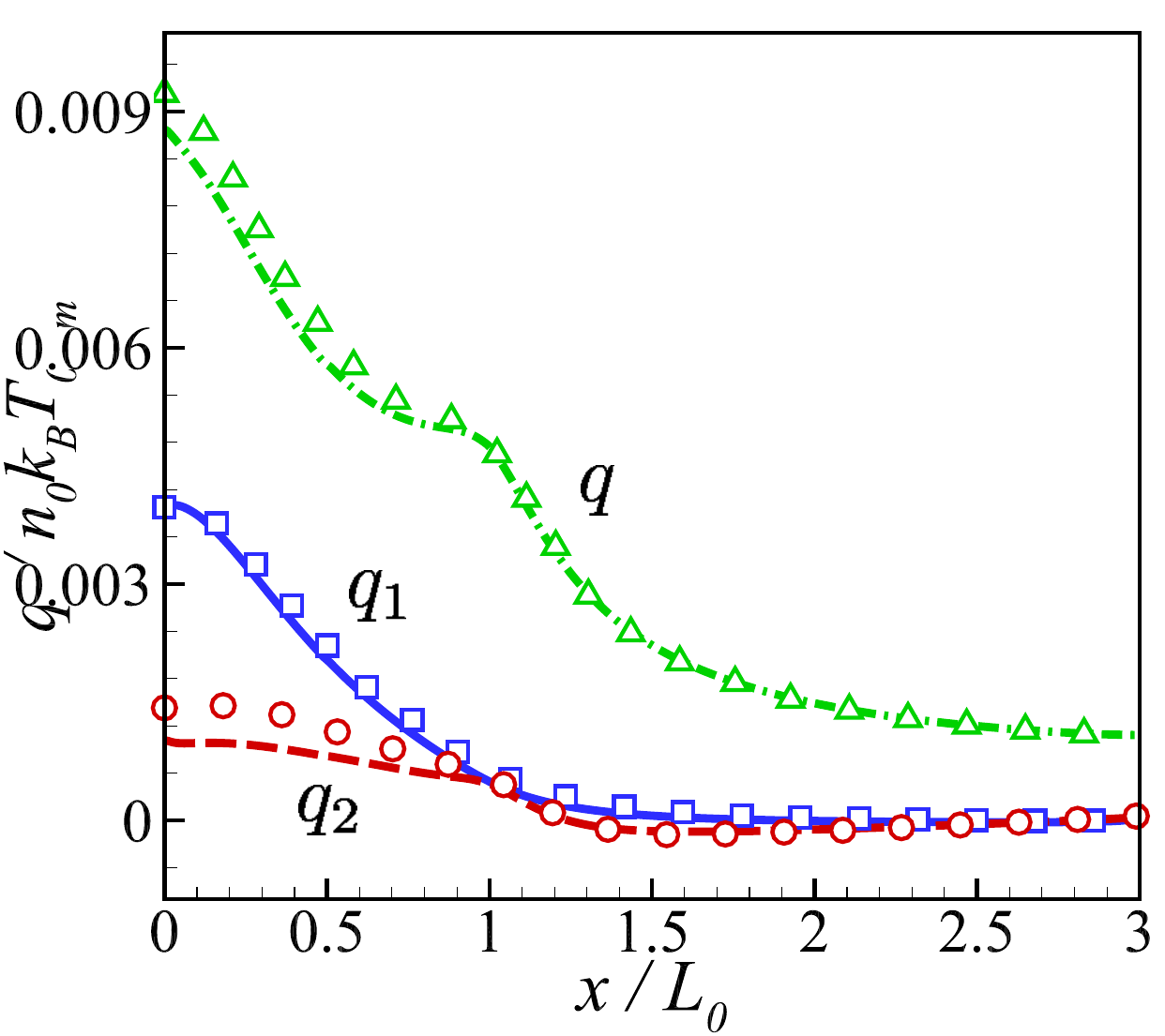}\label{fig:2DNozzle_Mix2:q_center}} \\
    \sidesubfloat[]{\includegraphics[scale=0.19,clip=true]{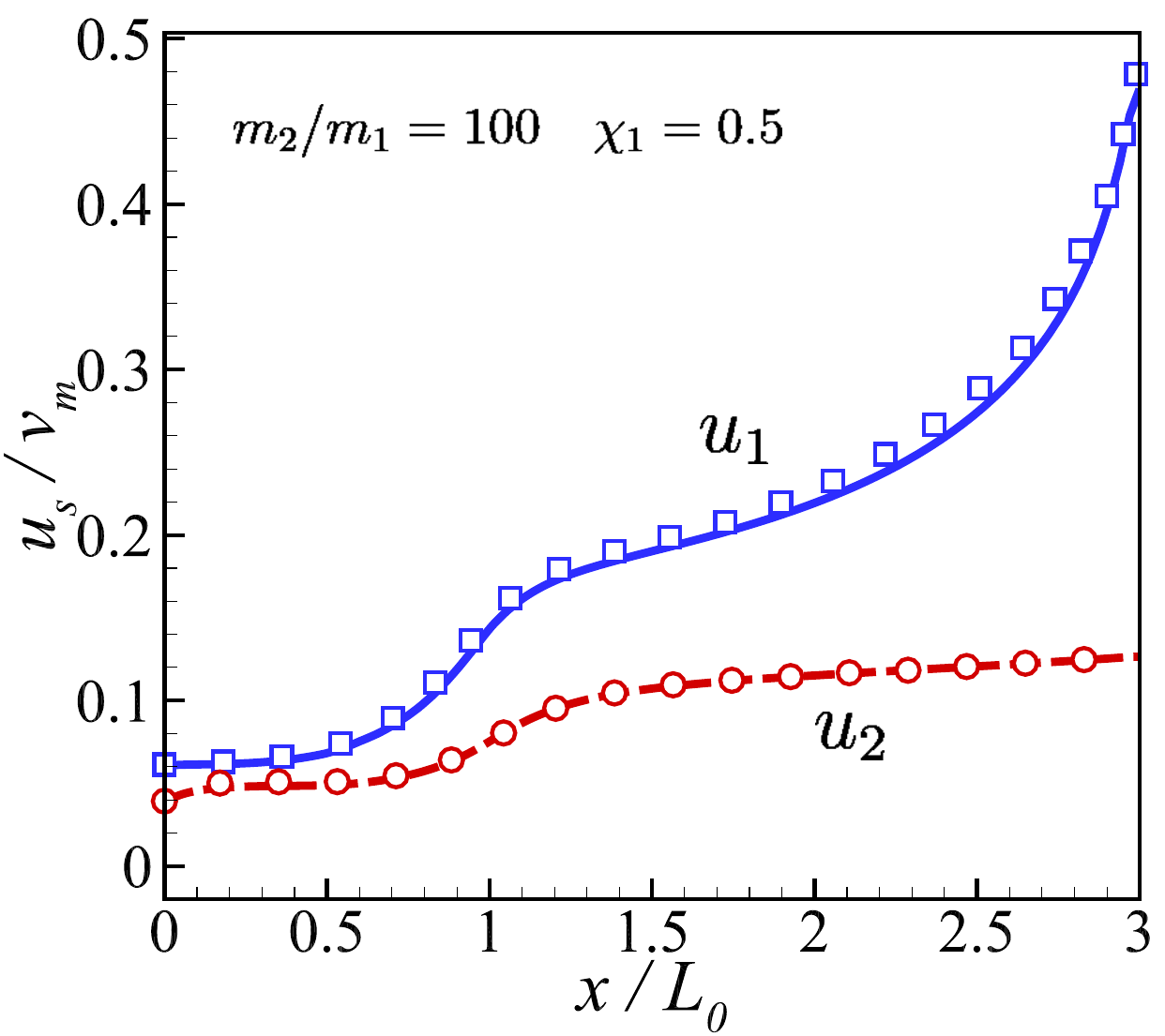}\label{fig:2DNozzle_Mix3:u_center}} 
    \sidesubfloat[]{\includegraphics[scale=0.19,clip=true]{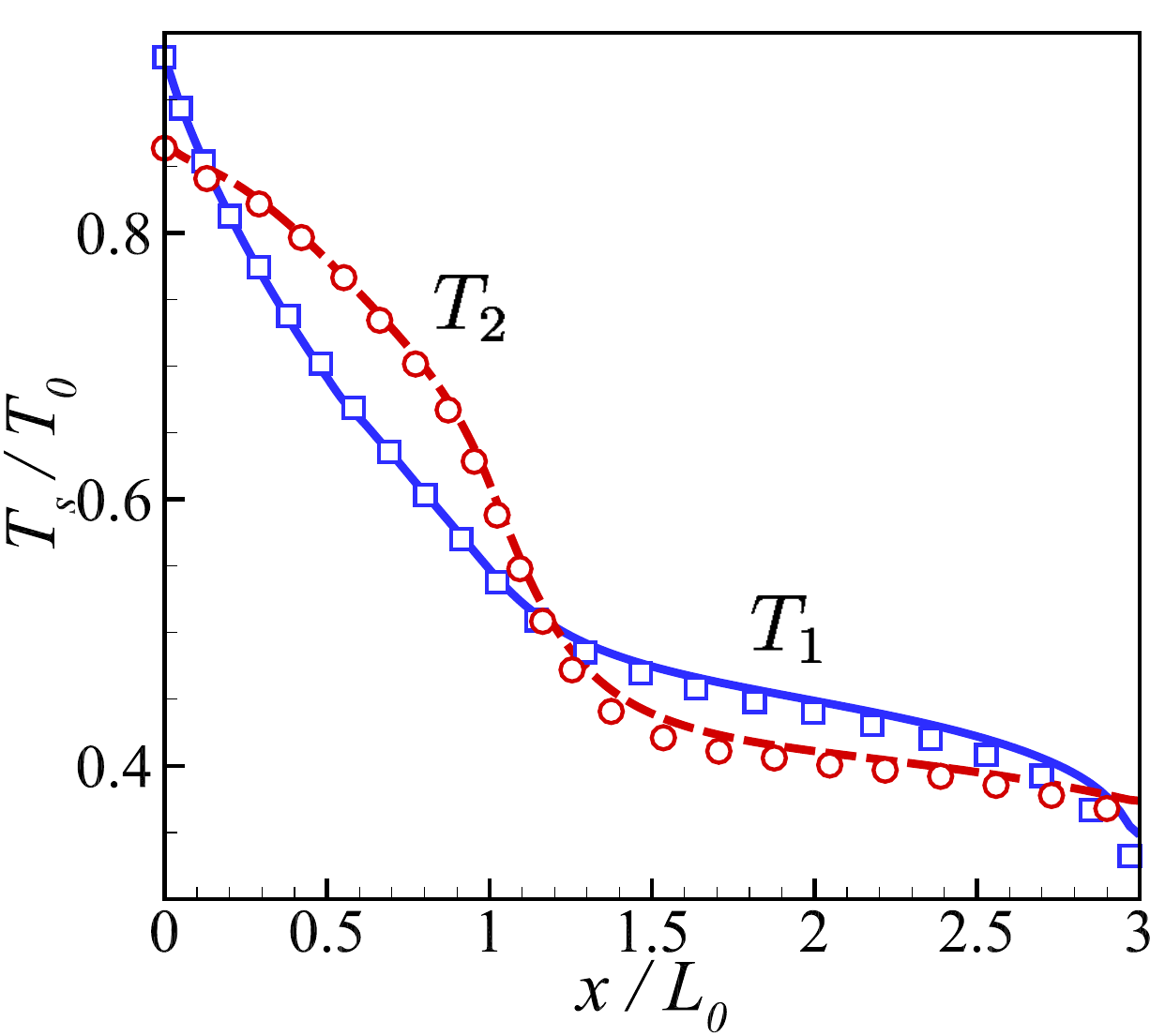}\label{fig:2DNozzle_Mix3:T_center}} 
    \sidesubfloat[]{\includegraphics[scale=0.19,clip=true]{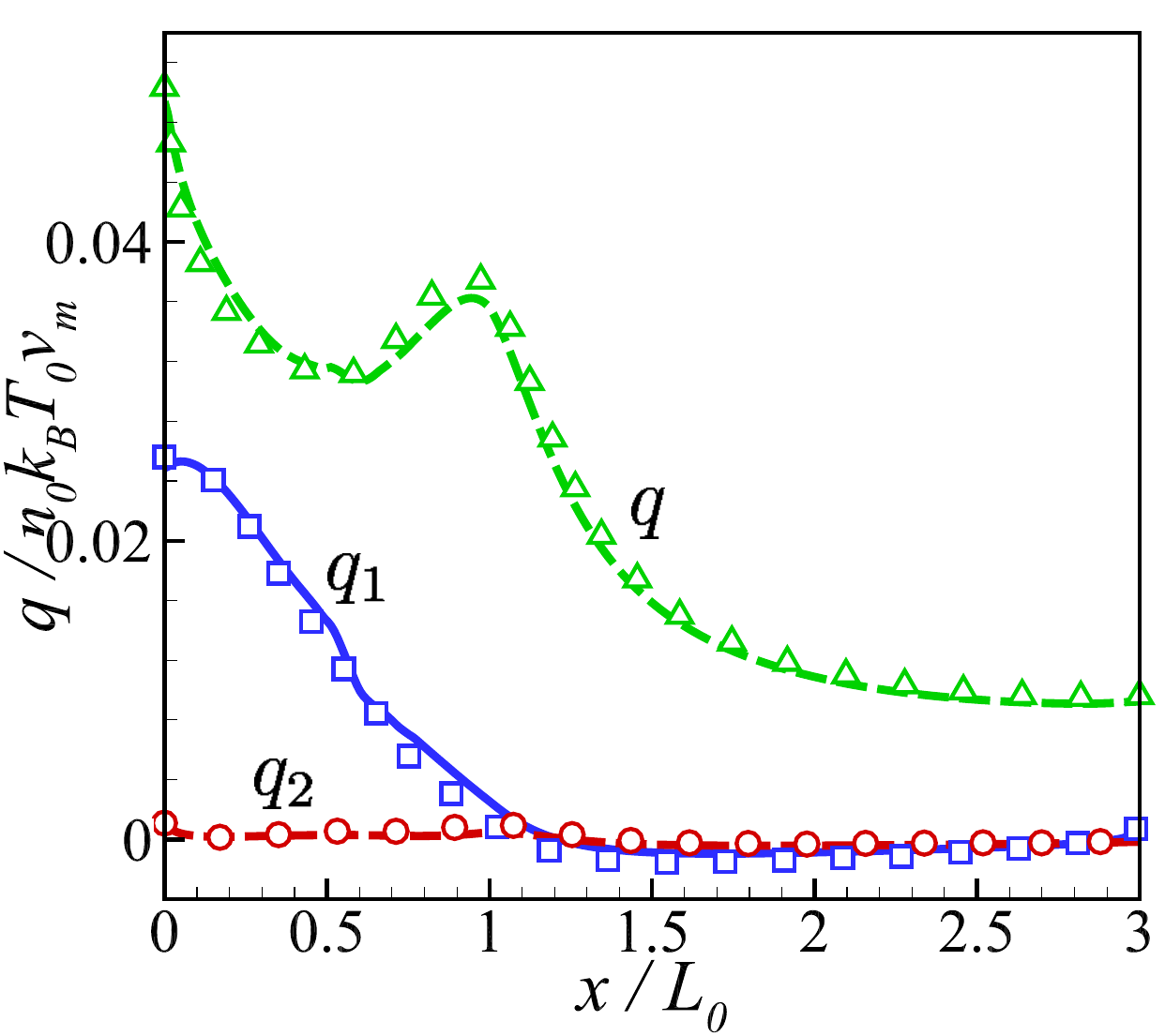}\label{fig:2DNozzle_Mix3:q_center}} \\
    \includegraphics[scale=0.22,clip=true]{Figures/legend_1D.png}
	\caption{Comparisons of the dimensionless (a,d) flow velocity, (b,e) temperature and (c,f) heat flux along the center line between the results of the kinetic model and DSMC for a gas mixture flowing through a nozzle, when $\text{Kn}_{1}=0.1$ at the inlet. The inlet mole fraction is $\chi_1=0.05$ for Mixture 2 (first row) and $\chi_1=0.5$ for Mixture 3 (second row).}
	\label{fig:2DNozzle_center}
\end{figure}

\begin{figure}[t]
	\centering
    \sidesubfloat[]{\includegraphics[scale=0.19,clip=true]{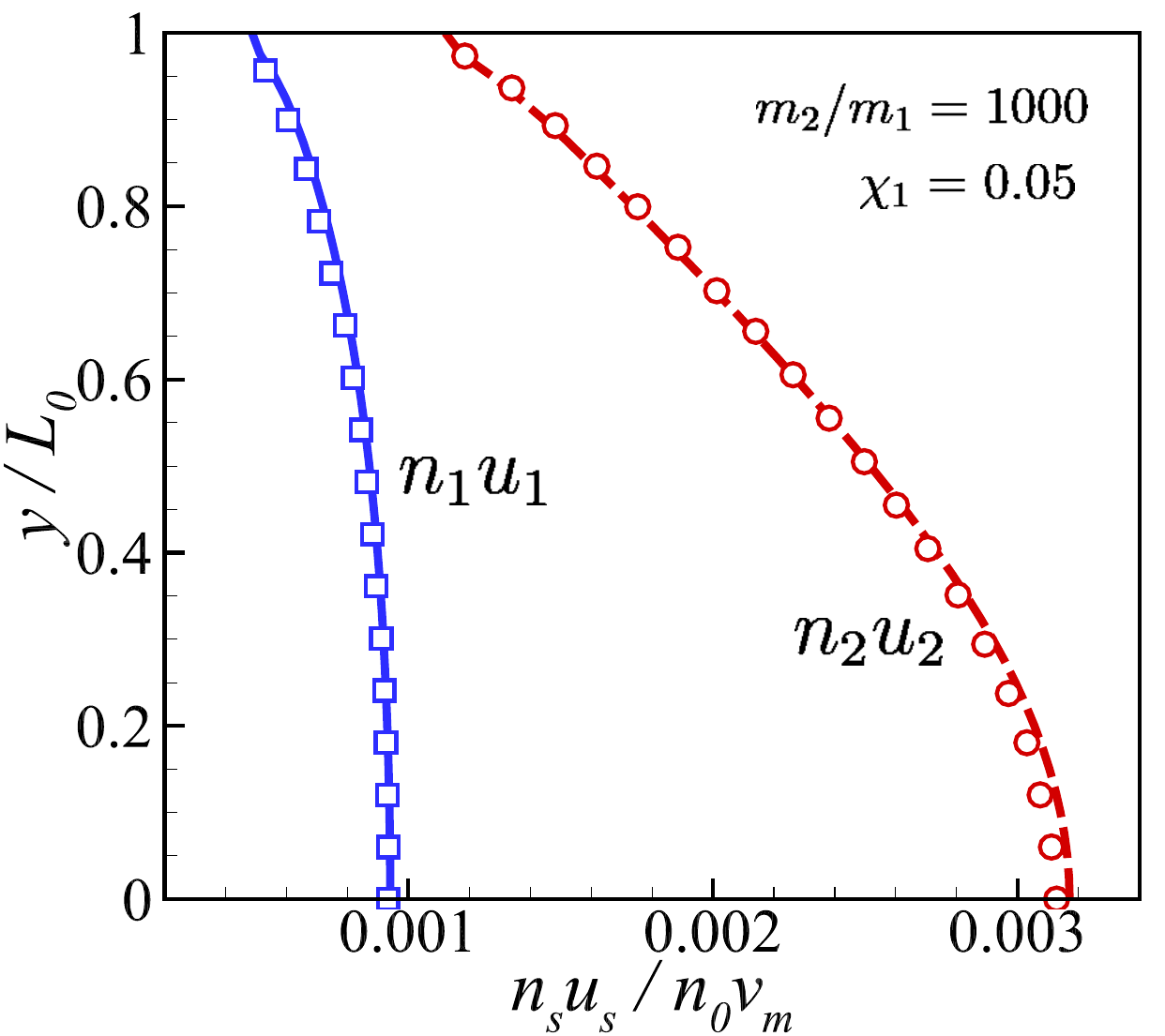}\label{fig:2DNozzle_Mix2:nu_outlet}} 
    \sidesubfloat[]{\includegraphics[scale=0.19,clip=true]{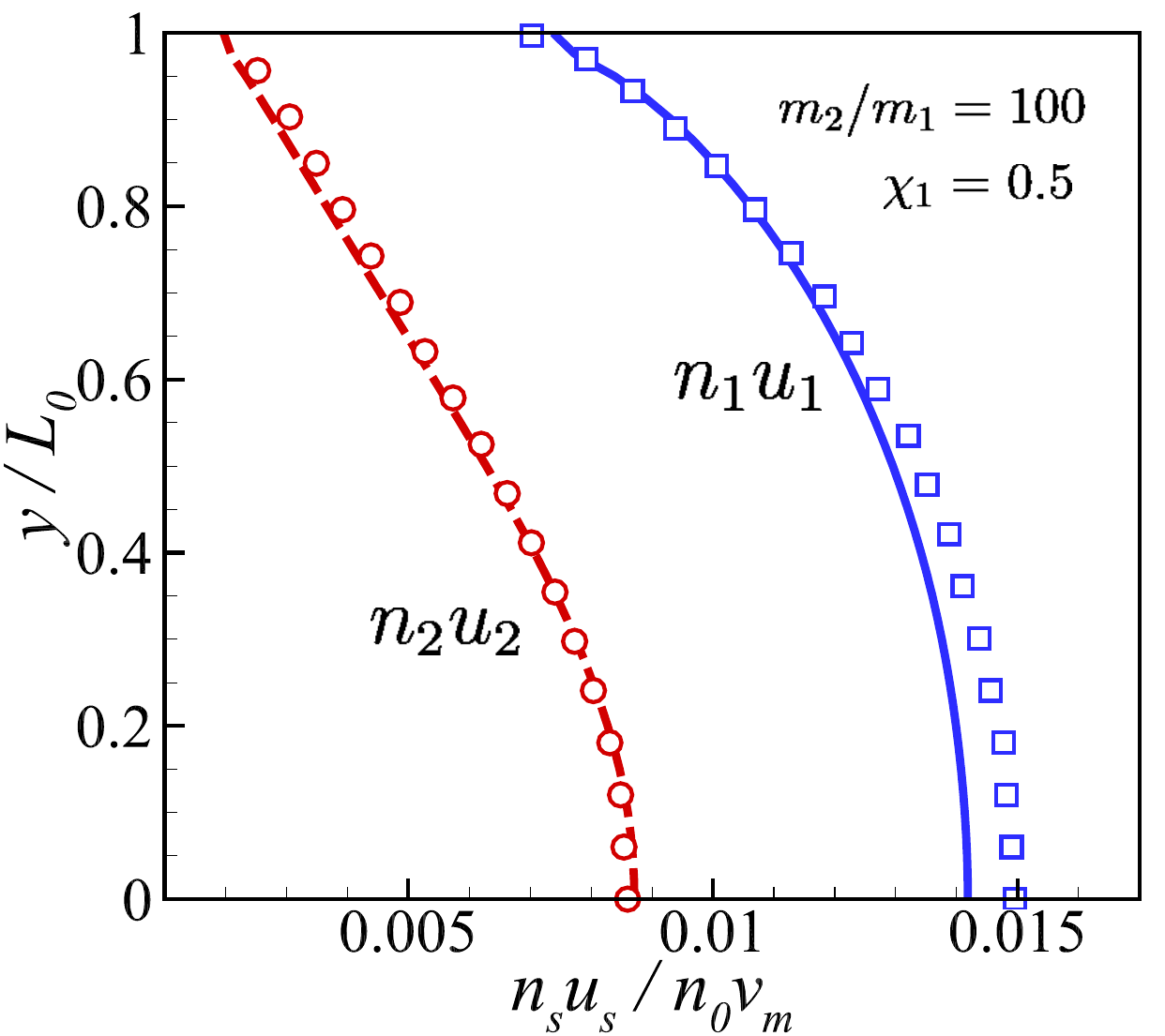}\label{fig:2DNozzle_Mix3:nu_outlet}} 
	\caption{Comparisons of the dimensionless number flow rate of each species passing through the outlet of the nozzle between the results of the kinetic model (lines) and DSMC (symbols). The results of Mixture 2 and 3 are given in (a) and (b), respectively.}
	\label{fig:2DNozzle_outlet}
\end{figure}

In this case, the kinetic model is applied to simulate a two-dimensional rarefied gas mixture flowing through a nozzle into the vacuum. The structure of the nozzle is shown in figure \ref{fig:2DNozzle_field}, which has a straight channel with width $L_0$, a converging section shrinking the width to $L_0/2$ at the throat, and a diverging section. At the inlet of the nozzle ($x=0$), the flow was assumed to be considerably subsonic $\text{Ma}_{in}=0.05$ with Maxwellian velocity distributions at temperature $T_0$. The gas molecules are reflected on the cold walls of the nozzle ($T_w=T_0/2$) with complete thermal accommodation, and then go through the outlet ($x=3L_0$) into the vacuum. The Knudsen number, defined in terms of the gas properties at the inlet, is $\text{Kn}_1=0.1$ for lighter species for all the mixtures.

We consider a very small proportion of lighter molecules ($\chi_1=0.05$) mixed with heavier ones for Mixture 2, and a half-half mix ($\chi_1=0.5$) of hard-sphere molecules for Mixture 3. Figure \ref{fig:2DNozzle_field} shows the species number density and local Mach number distribution in the nozzle solved by the kinetic model and DSMC. Also, Figure \ref{fig:2DNozzle_center} shows the density, temperature and heat flux along the center line of the nozzle. The good accuracy of our model equation is demonstrated. 

It is found that, when a gas mixture with disparate mass flows through a nozzle, the density and temperatures of each component do not experience remarkable separation across the nozzle. However, the species velocities at the outlet become noticeably different. As illustrated in figure \ref{fig:2DNozzle_Mix2:u_center} and \ref{fig:2DNozzle_Mix3:u_center}, Mixture 2, having a mass ratio of 1000, exhibits an outlet velocity for the lighter species that is 20 times higher than the velocity of the heavier one; similarly, Mixture 3 shows a velocity ratio of 3.7 between its lighter and heavier components. Consequently, the significant diffusion velocities make a noticeable contribution to the total heat flux of the mixture, particularly in the diverging section of the nozzle, where the heat flux due to conductance becomes negligible. On the other hand, the two components achieve close values of species Mach number $\text{Ma}_{s,local}$ calculated using their respective local sound speeds, as plotted in figure \ref{fig:2DNozzle_field:Mix2_Ma_field} and \ref{fig:2DNozzle_field:Mix3_Ma_field}, which also indicates similar degrees of compression of the disparate species flowing through the nozzle. While the relatively higher $\text{Ma}_{s,local}$ of the heavier components primarily arises from inter-species collisions with lighter molecules, which accelerate the heavier ones. 

We also calculate the flow rate of individual components passing through the nozzle outlet, as plotted in figure \ref{fig:2DNozzle_outlet}. The number flow rate is found to be comparable for the mixture components, while the heavier species will hence dominate the mass flow rate due to its much higher molecular mass. Compared with DSMC results, the mass flow rate predicted by the kinetic model yields less than 1.7\% relative difference for all the species in Mixture 2, and 3.2\% relative difference for those in Mixture 3.

\section{Conclusions}\label{sec:conclusion}

In summary, we have proposed a new kinetic model for monatomic gas mixtures, which can describe the dynamics of rarefied gas flow with disparate molecular mass, and reduce to the Shakhov model for single-species gas when the components are mechanically identical and the diffuse velocity vanishes. The tunable parameters in the kinetic model are uniquely determined by the transport properties of the gas mixture, and thus the shear viscosity, thermal conductivity, diffusion coefficient and thermal diffusion coefficients can be recovered by the model equation in the continuum limit.

The accuracy of the proposed models has been assessed by comparing with DSMC simulations for various binary gas mixture flows in representative problems, including the one-dimensional Fourier flow, Couette flow and normal shock waves, as well as the two-dimensional supersonic flow passing a cylinder and nozzle flow into a vacuum. A wide range of mass ratios, species concentrations, and different intermolecular potentials have been considered. The kinetic model demonstrates its high accuracy not only for predicting the average mixture properties but also for capturing the flow fields of the individual components.

The proposed kinetic model benefits from the following features that enable its applicability in the accurate modelling of a rarefied gas mixture with disparate mass:
\begin{enumerate}
	\item The model equation is constructed using a sum of relaxation operators imitating each type of binary collision individually, thus it can correctly capture the multiscale relaxation rates inherent in different collision processes. We reveal the dependence of the inter-species relaxation timescales on mass difference, which differ for the lighter and heavier components by orders of magnitude in a mixture with disparate mass. 
	\item All the transport coefficients can be recovered simultaneously by the kinetic model in the continuum limit. Particularly, the model goes beyond previous BGK-type models by incorporating the thermal diffusion effect for mixtures composed of non-Maxwell gases. The concentration variation is correctly captured by our model in simulations of the Fourier flow of hard-sphere gas mixtures. Therefore, the kinetic model can be applied to investigate the gas separation processes driven by temperature gradients, such as those occurring in a Knudsen compressor \citep{Takata2007EJMB}. 
	\item More importantly, the model equation exhibits minimal loss of accuracy when the mass ratio between components increases from 10 to 1000, when compared to the original Boltzmann equation. This remarkable consistency suggests the accuracy of the kinetic model for the gas mixtures with even larger mass disparities, and particularly, its potential extension to the kinetic modeling of plasmas.
\end{enumerate}

Gas mixtures with disparate mass possess substantial velocity and temperature non-equilibrium due to significant slow inter-species relaxations, thus forming unique flow characteristics, which are most evident in supersonic mixture flows:
\begin{enumerate}
    \item Previous observations of temperature overshoots and two-stage structures in normal shock waves are found at low concentrations of heavier gas in mixtures with moderate mass ratios. However, mixtures with significant mass disparity exhibit these phenomena under different conditions. Specifically, the temperature overshoot occurs when the heavier molecules are present in equal and even lower amounts compared to the lighter molecules. Additionally, the shock wave displays a reversed two-stage structure, characterized by a smooth and expansive upstream region followed by a steep change in properties on the downstream side.
    \item The supersonic mixture around an object leads to the coexistence of a subsonic lighter gas flow and a super/hypersonic heavier gas flow, hence posing a dramatic temperature difference between the components. Interestingly, the aerodynamic force acting on the cylinder is found to be independent of the mass ratio and insensitive to the intermolecular potential, while the heat transfer to the cylinder can be significantly affected by these factors.
\end{enumerate}

Last but not least, a gas mixture with disparate mass usually exhibits multiscale features both spatially and temporally, as well as significant concentration differences between its components. These characteristics bring unaffordable computational costs when solving the Boltzmann equation or conducting DSMC simulations in such mixture flows. On the other hand, the deterministic numerical methods with multiscale schemes that solve the kinetic equations have shown their excellent performance in multiscale problems. For example, the general synthetic iterative scheme developed in recent years can asymptotically preserve the Navier-Stokes equation in the continuum limit (thus removing the constraint on the spatial cell size), and find the steady-state solution of a kinetic equation within dozens of iterations at any Knudsen number \citep{Su2020JCP,Su2020SIAM,Liu2024JCP}. Therefore, with the computationally tractable kinetic model proposed here and the multiscale numerical methods, the gas mixture flows with disparate mass can be accurately and efficiently solved and found promising engineering applications, e.g., the particle exhaust system in nuclear fusion device \citep{Tantos2024NuclFusion} and the gas dynamics locker in EUV lithography \citep{Teng2023JCProd}. Moreover, with our experience in the kinetic modeling of single-species with internal degrees of freedom \citep{LeiJFM2015,Li2021JFM,Li2023JFM}, the kinetic models for multi-species gas mixtures with internal degrees of freedom are expected to be established in the near future. 


\section*{Acknowledgements} 
This work is supported by the National Natural Science Foundation of China under grants No. 12172162 and No. 12202177.

\section*{Declaration of interests} 
The authors report no conflict of interest.

\appendix{

\section{Chapman-Enskog analysis of kinetic model equation}\label{app:A}

The transport coefficients and macroscopic equations given by our kinetic model \eqref{eq:kinetic_equation} in continuum limit can be obtained by the Chapman-Enskog method \citep{CE}, where the distribution functions $f_s$ are expansions in the form of an infinite series,
\begin{equation}\label{eq:expansions_f}
	\begin{aligned}[b]
		f_s = f_s^{(0)}+\varepsilon f_s^{(1)} +\varepsilon^2 f_s^{(2)} +\cdots, \quad s=1,2.
	\end{aligned}
\end{equation} 
The conserved macroscopic properties $n_s,~\bm{u},~T$ remain unexpanded and thus are determined only by $f_s^{(0)}$, while the other quantities $h$, including macroscopic variables and auxiliary properties, are also expanded as,
\begin{equation}\label{eq:expansions_h}
	\begin{aligned}[b]
		h = h^{(0)}+\varepsilon h^{(1)} +\varepsilon^2 h^{(2)} +\cdots. 
	\end{aligned}
\end{equation}

Substituting the expansions into the model equation \eqref{eq:kinetic_equation}, the zero-order distribution functions $f_s^{(0)}$ are given by the solution of the kinetic equations,
\begin{equation}\label{eq:Js0}
	\begin{aligned}[b]
		\mathcal{D}^{(0)}f_1 &= {\frac{1}{\tau_{11}^{(0)}}\left(g_{11}^{(0)}-f_1^{(0)}\right)} + {\frac{1}{\tau_{12}^{(0)}}\left(g_{12}^{(0)}-f_1^{(0)}\right)}, \\
		\mathcal{D}^{(0)}f_2 &= {\frac{1}{\tau_{21}^{(0)}}\left(g_{21}^{(0)}-f_2^{(0)}\right)} + {\frac{1}{\tau_{22}^{(0)}}\left(g_{22}^{(0)}-f_2^{(0)}\right)},
	\end{aligned}
\end{equation} 
where $\mathcal{D}^{(0)}f_s=0$, and the reference distribution functions $g_{sr}~(s,r=1,2)$, which depend on auxiliary properties, are expanded around the Maxwellian distribution of the conserved macroscopic variables $n_s,~\bm{u},~T$,
\begin{equation}\label{eq:expansions_g}
	\begin{aligned}[b]
		g_{sr}^{(0)} &= {n}_{s}\left(\frac{m_s}{2\pi k_B{T}}\right)^{3/2}\exp\left(-\frac{m_sc^2}{2k_B{T}}\right), \quad s,r=1,2,\\
		g_{sr}^{(1)} &= g_{sr}^{(0)}\left[\frac{m_s\hat{\bm{u}}_{sr}^{(1)}\cdot\bm{c}}{k_BT} + \frac{\hat T_{sr}^{(1)}}{T}\left(\frac{m_sc^2}{2k_B{T}}-\frac{3}{2}\right) + \frac{2m_s\hat{\bm{q}}_{sr}^{(1)}\cdot\bm{c}}{5{n}_{s}k_B^2{T}^2}\left(\frac{m_sc^2}{2k_B{T}}-\frac{5}{2}\right)\right], \quad s,r=1,2,
	\end{aligned}
\end{equation}
where $\bm{c}=\bm{v}-\bm{u}$ is the peculiar velocity with respect to the mixture velocity $\bm{u}$. Therefore, the first approximation to $f_s=f_s^{(0)}$ gives the local equilibrium state of each species with the common flow velocity $\bm{u}$ and temperature $T$,
\begin{equation}\label{eq:f0}
	\begin{aligned}[b]
		f_s^{(0)} = n_s\left(\frac{m_s}{2\pi k_BT}\right)^{3/2}\exp{\left(-\frac{m_s\left(\bm{v}-{\bm{u}}\right)^2}{2k_BT}\right)}, \quad s=1,2.
	\end{aligned}
\end{equation} 
Then, the zero-order macroscopic properties can be obtained by taking respective moments of $f_s^{(0)}$,
\begin{equation}\label{eq:us_P_q_T_0th}
	\begin{aligned}[b]
		\bm{u}_s^{(0)} = \bm{u}, \quad T_s^{(0)} = 0, \quad \bm{P}^{(0)} = nk_BT\bm{\mathrm{I}}, \quad \bm{q}^{(0)} =0, \quad s=1,2,
	\end{aligned}
\end{equation} 
where $\bm{\mathrm{I}}$ is the identity matrix. Meanwhile, the zero-order auxiliary parameters are also obtained from equation \eqref{eq:auxiliary_u_T} as $\hat{\bm{u}}_{12}^{(0)}=\hat{\bm{u}}_{21}^{(0)}=\bm{u},~\hat{T}_{12}^{(0)}=\hat{T}_{21}^{(0)}=T$.

To the second approximation of the distribution function $f_s=f_s^{(0)}+\varepsilon f_s^{(1)}$, the first-order correction $f_s^{(1)}$ is solved from the kinetic equations,
\begin{equation}\label{eq:Js1}
	\begin{aligned}[b]
		\mathcal{D}^{(1)}f_1 &= {\frac{1}{\tau_{11}^{(0)}}\left(g_{11}^{(1)}-f_1^{(1)}\right)} + {\frac{1}{\tau_{12}^{(0)}}\left(g_{12}^{(1)}-f_1^{(1)}\right)}, \\
		\mathcal{D}^{(1)}f_2 &= {\frac{1}{\tau_{21}^{(0)}}\left(g_{21}^{(1)}-f_2^{(1)}\right)} + {\frac{1}{\tau_{22}^{(0)}}\left(g_{22}^{(1)}-f_2^{(1)}\right)},
	\end{aligned}
\end{equation}
where $\mathcal{D}^{(1)}f_s$ can be explicitly evaluated as,
\begin{equation}\label{eq:D1f}
	\begin{aligned}[b]
		\mathcal{D}^{(1)}f_s &= \frac{\partial{f_s^{(0)}}}{\partial{t}}+\bm{v} \cdot \frac{\partial{f_s^{(0)}}}{\partial{\bm{x}}}+\bm{a}_s \cdot \frac{\partial{f_s^{(0)}}}{\partial{\bm{v}}} \\
	&= f_s^{(0)}\left[ \left(\frac{m_sc^{2}}{2k_BT}-\frac{5}{2}\right)\bm{c}\cdot\nabla\ln{T} + \frac{n}{n_s}\bm{d}_{sr}\cdot\bm{c} + \frac{m_s}{k_BT}\left(\bm{c}\bm{c}-\frac{1}{3}c^2\bm{\mathrm{I}}\right):\nabla\bm{u} \right],\\ 
	&s=1,2,~r\ne s.
	\end{aligned}
\end{equation}
Here, $\bm{d}_{sr}~(s\ne r)$ represents the diffusive driving force
\begin{equation}\label{eq:ds}
	\begin{aligned}[b]
		\bm{d}_{12} = -\bm{d}_{21} = \frac{\rho_1\rho_2}{\rho p}\left[\left(\frac{\nabla p_1}{\rho_1}-\frac{\nabla p_2}{\rho_2}\right)-\left(\bm{a}_1-\bm{a}_2\right)\right].
	\end{aligned}
\end{equation}
Therefore, the first-order correction of the distribution function is obtained,
\begin{equation}\label{eq:f1_A}
	\begin{aligned}[b]
		f_s^{(1)} = \frac{1}{\tau_{s1}^{(0)}+\tau_{s2}^{(0)}}\left(\tau_{s2}^{(0)}g_{s1}^{(1)} + \tau_{s1}^{(0)}g_{s2}^{(1)} - \tau_{s1}^{(0)}\tau_{s2}^{(0)}\mathcal{D}^{(1)}f_s\right), \quad s=1,2.
	\end{aligned}
\end{equation}

Substituting the second approximation to $f_s$ into the definition of diffusion velocity, stress tensor and heat flux, the transport terms as functions of the gradients of the macroscopic properties can be calculated according to \eqref{eq:species_macroscopic_variables_f} and \eqref{eq:mixture_macroscopic_variables_f}.

The first-order correction of the species velocity $\bm{u}_s^{(1)}$ is,
\begin{equation}\label{eq:us_1st}
	\begin{aligned}[b]
		\bm{u}_1^{(1)} =&~ \frac{1}{n_1}\int{\bm{v}f_1^{(1)}}\mathrm{d}\bm{v} 
		= \frac{\tau_{12}^{(0)}}{\tau_{11}^{(0)}+\tau_{12}^{(0)}}\bm{u}_{1}^{(1)} + \frac{\tau_{11}^{(0)}}{\tau_{11}^{(0)}+\tau_{12}^{(0)}}\hat{\bm{u}}_{12}^{(1)} - \frac{\tau_{11}^{(0)}\tau_{12}^{(0)}}{\tau_{11}^{(0)}+\tau_{12}^{(0)}}\frac{p}{\rho_1}\bm{d}_{12}, \\
		\bm{u}_2^{(1)} =&~ \frac{1}{n_2}\int{\bm{v}f_2^{(1)}}\mathrm{d}\bm{v} 
		= \frac{\tau_{21}^{(0)}}{\tau_{22}^{(0)}+\tau_{21}^{(0)}}\bm{u}_{2}^{(1)} + \frac{\tau_{22}^{(0)}}{\tau_{21}^{(0)}+\tau_{21}^{(0)}}\hat{\bm{u}}_{21}^{(1)} + \frac{\tau_{22}^{(0)}\tau_{21}^{(0)}}{\tau_{22}^{(0)}+\tau_{21}^{(0)}}\frac{p}{\rho_2}\bm{d}_{12}, 
	\end{aligned}
\end{equation} 
where $\hat{\bm{u}}_{12}$ and $\hat{\bm{u}}_{21}$ are given by \eqref{eq:auxiliary_u_T} and \eqref{eq:auxiliary_X_Y}. Considering that the mixture velocity $\bm{u}$ is unexpanded, which gives a constraint $\rho_1\bm{u}_1^{(1)}+\rho_2\bm{u}_2^{(1)}=0$, the first-order species velocities are obtained,
\begin{equation}\label{eq:u1_u2_1st}
	\begin{aligned}[b]
		\bm{u}_1^{(1)} &= -\frac{\rho_1\tau_{21}^{(0)}+\rho_2\tau_{12}^{(0)}}{a_{12}}\frac{p}{\rho\rho_1}\bm{d}_{12}-\frac{2b_{12}\rho_2}{a_{12}\rho}\nabla{\ln T}, \\
		\bm{u}_2^{(1)} &= -\frac{\rho_1\tau_{21}^{(0)}+\rho_2\tau_{12}^{(0)}}{a_{12}}\frac{p}{\rho\rho_2}\bm{d}_{21}-\frac{2b_{21}\rho_1}{a_{12}\rho}\nabla{\ln T}.
	\end{aligned}
\end{equation} 

Similarly, the first-order correction of the species temperature $T_s^{(1)}$ is calculated based on \eqref{eq:species_macroscopic_variables_f} and the auxiliary properties \eqref{eq:auxiliary_u_T} and \eqref{eq:auxiliary_X_Y},
\begin{equation}\label{eq:Ts_0th_1st}
	\begin{aligned}[b]
		T+\varepsilon T_s^{(1)} =&~ \frac{2}{3n_sk_B}\int{\frac{1}{2}m_s(\bm{v}-\bm{u}_s)^2\left(f_s^{(0)}+\varepsilon f_s^{(1)}\right)}\mathrm{d}\bm{v} \\
		=&~\left(T+\varepsilon T_s^{(1)}\right) + \varepsilon \frac{\tau_{ss}^{(0)}}{\tau_{ss}^{(0)}+\tau_{sr}^{(0)}}\frac{n_r\tau_{sr}^{(0)}}{n_s\tau_{rs}^{(0)}+n_r\tau_{sr}^{(0)}}c_{sr}\left(T_r^{(1)}-T_s^{(1)}\right) + O\left(\varepsilon^2\right), \quad s\ne r. 
	\end{aligned}
\end{equation} 
Since the mixture temperature $T$ is unexpanded, the constraint $n_1T_1^{(1)}+n_2T_2^{(1)}=0$ needs to be satisfied by ignoring the higher-order terms. It is found that the first-order correction of the species temperature vanishes
\begin{equation}\label{eq:Ts_1st}
	\begin{aligned}[b]
		T_1^{(1)}=T_2^{(1)}=0. 
	\end{aligned}
\end{equation} 

The first-order correction of the mixture stress tensor $\bm{P}^{(1)}$ is calculated based on \eqref{eq:mixture_macroscopic_variables_f},
\begin{equation}\label{eq:P_1st}
	\begin{aligned}[b]
		\bm{P}^{(1)} =&~ \int{m_1\bm{c}\bm{c}f_1^{(1)}}\mathrm{d}\bm{v} + \int{m_2\bm{c}\bm{c}f_2^{(1)}}\mathrm{d}\bm{v}\\
		=&~ -k_BT\left(\frac{\tau_{11}^{(0)}\tau_{12}^{(0)}}{\tau_{11}^{(0)}+\tau_{12}^{(0)}}n_1+\frac{\tau_{22}^{(0)}\tau_{21}^{(0)}}{\tau_{22}^{(0)}+\tau_{21}^{(0)}}n_2\right)\left(\nabla\bm{u}+\nabla\bm{u}^{\mathrm{T}}-\frac{2}{3}\nabla\cdot\bm{u}\bm{\mathrm{I}}\right).
	\end{aligned}
\end{equation} 

The first-order correction of the species heat fluxes $\bm{q}_s^{(1)}$ and $\bm{q}_{sr}^{(1)}$ are,
\begin{equation}\label{eq:qs_qsr_1st}
	\begin{aligned}[b]
		\bm{q}_s^{(1)} =&~ \frac{1}{\varepsilon}\int{\frac{1}{2}m_s(\bm{v}-\bm{u}_s)^2(\bm{v}-\bm{u}_s)\left(f_s^{(0)}+\varepsilon f_s^{(1)}\right)}\mathrm{d}\bm{v} \\
		=&~ \frac{\tau_{sr}^{(0)}}{\tau_{ss}^{(0)}+\tau_{sr}^{(0)}}\hat{\bm{q}}_{ss}^{(1)} + \frac{\tau_{ss}^{(0)}}{\tau_{ss}^{(0)}+\tau_{sr}^{(0)}}\hat{\bm{q}}_{sr}^{(1)} 
		- \frac{\tau_{ss}^{(0)}\tau_{sr}^{(0)}}{\tau_{ss}^{(0)}+\tau_{sr}^{(0)}}\frac{5k_BT}{2m_s}n_1k_B\nabla T + O\left(\varepsilon\right), \\
		\bm{q}_{sr}^{(1)} =&~ \frac{1}{\varepsilon}\int{\frac{1}{2}m_s(\bm{v}-\hat{\bm{u}}_{sr})^2(\bm{v}-\hat{\bm{u}}_{sr})\left(f_s^{(0)}+\varepsilon f_s^{(1)}\right)}\mathrm{d}\bm{v} \\
		=&~ \frac{\tau_{sr}^{(0)}}{\tau_{ss}^{(0)}+\tau_{sr}^{(0)}}\hat{\bm{q}}_{ss}^{(1)} + \frac{\tau_{ss}^{(0)}}{\tau_{ss}^{(0)}+\tau_{sr}^{(0)}}\hat{\bm{q}}_{sr}^{(1)} 
		-\tau_{sr}^{(0)}\frac{5k_BT}{2m_s}p\bm{d}_{sr}- \frac{\tau_{ss}^{(0)}\tau_{sr}^{(0)}}{\tau_{ss}^{(0)}\tau_{sr}^{(0)}}\frac{5k_BT}{2m_s}n_1k_B\nabla T + O\left(\varepsilon\right), \quad s\ne r,
	\end{aligned}
\end{equation} 
where the auxiliary heat fluxes $\hat{\bm{q}}_{ss},~\hat{\bm{q}}_{sr}$ are constructed as \eqref{eq:auxiliary_q}, and thus we have,
\begin{equation}\label{eq:qs_1st}
	\begin{aligned}[b]
		\bm{q}_s^{(1)} &= -\frac{\tau_{ss}^{(0)}\tau_{sr}^{(0)}}{\text{Pr}_{sr}\tau_{ss}^{(0)}+\text{Pr}_{ss}\tau_{sr}^{(0)}}\frac{5k_BT}{2m_s}\left(\gamma_{sr}p\bm{d}_{sr}+n_sk_B\nabla T\right), \quad s\ne r. 
	\end{aligned}
\end{equation} 
Then the first-order correction of the mixture heat flux is:
\begin{equation}\label{eq:q_1st}
	\begin{aligned}[b]
		\bm{q}^{(1)} =&~ \int{\frac{1}{2}m_1c^2\bm{c}f_1^{(1)}}\mathrm{d}\bm{v} + \int{\frac{1}{2}m_2c^2\bm{c}f_2^{(1)}}\mathrm{d}\bm{v}\\
		=&~ \bm{q}_1^{(1)}+\bm{q}_2^{(1)}+ \frac{5}{2}k_BT\left(n_1\bm{u}_1^{(1)}+n_2\bm{u}_2^{(1)}\right) + O\left(\varepsilon\right), \\
		=&~ \frac{5}{2}k_BT\left(n_1\bm{u}_1^{(1)}+n_2\bm{u}_2^{(1)}\right) \\
		&- \left(\frac{n_1}{m_1}\frac{\tau_{11}^{(0)}\tau_{12}^{(0)}}{\text{Pr}_{12}\tau_{11}^{(0)}+\text{Pr}_{11}\tau_{12}^{(0)}}+\frac{n_2}{m_2}\frac{\tau_{22}^{(0)}\tau_{21}^{(0)}}{\text{Pr}_{21}\tau_{22}^{(0)}+\text{Pr}_{22}\tau_{21}^{(0)}}\right)\frac{5k_BT}{2}k_B\nabla T \\
		& +\gamma_{12}\left(\frac{1}{m_1}\frac{\tau_{11}^{(0)}\tau_{12}^{(0)}}{\text{Pr}_{12}\tau_{11}^{(0)}+\text{Pr}_{11}\tau_{12}^{(0)}}+\frac{1}{m_2}\frac{\tau_{22}^{(0)}\tau_{21}^{(0)}}{\text{Pr}_{21}\tau_{22}^{(0)}+\text{Pr}_{22}\tau_{21}^{(0)}}\right)\frac{5k_BT}{2} \times \\
		&\frac{\rho_1\rho_2}{\rho_1\tau_{21}^{(0)}+\rho_2\tau_{12}^{(0)}}\left[a_{12}\left({\bm{u}}_{1}^{(1)}-{\bm{u}}_{2}^{(1)}\right)+2b_{12}\nabla{\ln T}\right] + O\left(\varepsilon\right). 
	\end{aligned}
\end{equation}

}

\bibliographystyle{jfm}
\bibliography{bibnew}

\begin{thebibliography}{77}
\expandafter\ifx\csname natexlab\endcsname\relax\def\natexlab#1{#1}\fi
\def\au#1{#1} \def\ed#1{#1} \def\yr#1{#1}\def\at#1{#1}\def\jt#1{\textit{#1}}
  \def\bt#1{#1}\def\bvol#1{\textbf{#1}} \def\vol#1{#1} \def\pg#1{#1}
  \def\publ#1{#1}\def\arxiv#1{#1}\def\org#1{#1}\def\st#1{\textit{#1}}

\bibitem[Agrawal {\em et~al.\/}(2020)Agrawal, Singh \&
  Ansumali]{Agrawal2020JFM}
{\sc \au{Agrawal, S.}, \au{Singh, S.~K.} \& \au{Ansumali, S.}} \yr{2020}
  \at{{Fokker--Planck} model for binary mixtures}.  \jt{Journal of Fluid
  Mechanics}  \bvol{899},  \pg{A25}.

\bibitem[Alves {\em et~al.\/}(2018)Alves, Bogaerts, Guerra \&
  Turner]{Alves2018PSST}
{\sc \au{Alves, L.~L.}, \au{Bogaerts, A.}, \au{Guerra, V.} \& \au{Turner,
  M.~M.}} \yr{2018}  \at{Foundations of modelling of nonequilibrium
  low-temperature plasmas}.  \jt{Plasma Sources Science and Technology}
  \bvol{27}~(2).

\bibitem[Andries {\em et~al.\/}(2002)Andries, Aoki \& Perthame]{Andries2002JSP}
{\sc \au{Andries, P.}, \au{Aoki, K} \& \au{Perthame, B.}} \yr{2002}  \at{A
  consistent {BGK}-type model for gas mixtures}.  \jt{Journal of Statistical
  Physics}  \bvol{106}~(516),  \pg{993--1018}.

\bibitem[Andries {\em et~al.\/}(2000)Andries, Le~Tallec, Perlat \&
  Perthame]{andries2000gaussian}
{\sc \au{Andries, P.}, \au{Le~Tallec, P.}, \au{Perlat, J.} \& \au{Perthame,
  B.}} \yr{2000}  \at{The {Gaussian-BGK} model of {Boltzmann} equation with
  small {Prandtl} number}.  \jt{European Journal of Mechanics-B/Fluids}
  \bvol{19}~(6),  \pg{813--830}.

\bibitem[Bellan(2006)]{Bellan2006}
{\sc \au{Bellan, P.~M.}} \yr{2006} {\em Fundamentals of Plasma Physics\/}.
  \publ{Cambridge University Press}.

\bibitem[Bhatnagar {\em et~al.\/}(1954)Bhatnagar, Gross \&
  Krook]{Bhatnagar1954}
{\sc \au{Bhatnagar, P.~L.}, \au{Gross, E.~P.} \& \au{Krook, M.}} \yr{1954}
  \at{A model for collision processes in gases. {I}. {S}mall amplitude
  processes in charged and neutral one-component systems}.  \jt{Physical
  review}  \bvol{94},  \pg{511--525}.

\bibitem[Bird(1968)]{Bird1968JFM}
{\sc \au{Bird, G.~A.}} \yr{1968}  \at{The structure of normal shockwaves in a
  binary gas mixture}.  \jt{Journal of Fluid Mechanics}  \bvol{31}~(4),
  \pg{657--668}.

\bibitem[Bird(1994)]{Bird1994}
{\sc \au{Bird, G.~A.}} \yr{1994} {\em Molecular Gas Dynamics and the Direct
  Simulation of Gas Flows\/}.  \publ{Oxford University Press Inc, New York:
  Oxford Science Publications}.

\bibitem[Bisi {\em et~al.\/}(2022)Bisi, Boscheri, Dimarco, Groppi \&
  Martal{\`o}]{Bisi2022AMC}
{\sc \au{Bisi, M.}, \au{Boscheri, W.}, \au{Dimarco, G.}, \au{Groppi, M.} \&
  \au{Martal{\`o}, G.}} \yr{2022}  \at{A new mixed {Boltzmann-BGK} model for
  mixtures solved with an {IMEX} finite volume scheme on unstructured meshes}.
  \jt{Applied Mathematics and Computation}  \bvol{433},  \pg{127416}.

\bibitem[Bisi {\em et~al.\/}(2018)Bisi, Monaco \& Soares]{Bisi2018JPA}
{\sc \au{Bisi, M.}, \au{Monaco, R.} \& \au{Soares, A.~J.}} \yr{2018}  \at{A
  {BGK} model for reactive mixtures of polyatomic gases with continuous
  internal energy}.  \jt{Journal of Physics A: Mathematical and Theoretical}
  \bvol{51}~(12),  \pg{125501}.

\bibitem[Bisi \& Travaglini(2020)]{Bisi2020PhysicaA}
{\sc \au{Bisi, M.} \& \au{Travaglini, R.}} \yr{2020}  \at{A {BGK} model for
  mixtures of monoatomic and polyatomic gases with discrete internal energy}.
  \jt{Physica A: Statistical Mechanics and its Applications}  \bvol{547},
  \pg{124441}.

\bibitem[Bobylev {\em et~al.\/}(2018)Bobylev, Bisi, Groppi, Spiga \&
  Potapenko]{Bobylev2018KRM}
{\sc \au{Bobylev, A.}, \au{Bisi, M.}, \au{Groppi, M.}, \au{Spiga, G.} \&
  \au{Potapenko, I.}} \yr{2018}  \at{A general consistent {BGK} model for gas
  mixtures}.  \jt{Kinetic and Related Models}  \bvol{11}~(6),  \pg{1377--1393}.

\bibitem[Boyd(1996)]{Boyd1996JTHT}
{\sc \au{Boyd, D.}} \yr{1996}  \at{Conservative species weighting scheme for
  the direct simulationmonte carlo method}.  \jt{Journal of Thermophysics and
  Heat Transfer}  \bvol{10},  \pg{579--585}.

\bibitem[Brull(2015)]{Brull2015CMS}
{\sc \au{Brull, S.}} \yr{2015}  \at{An ellipsoidal statistical model for gas
  mixtures}.  \jt{Communications in Mathematical Sciences}  \bvol{13}~(1),
  \pg{1--13}.

\bibitem[Brull {\em et~al.\/}(2012)Brull, Pavan \& Schneider]{Brull2012EJMB}
{\sc \au{Brull, S.}, \au{Pavan, V.} \& \au{Schneider, J.}} \yr{2012}
  \at{Derivation of a {BGK} model for mixtures}.  \jt{European Journal of
  Mechanics - B/Fluids}  \bvol{33},  \pg{74--86}.

\bibitem[Brun(2012)]{Brun2012High}
{\sc \au{Brun, Raymond}} \yr{2012} {\em High Temperature Phenomena in Shock
  Waves\/}.  \publ{Berlin Heidelberg: Springer-Verlag}.

\bibitem[Chapman(1958)]{Chapman1958PPS}
{\sc \au{Chapman, S.}} \yr{1958}  \at{Thermal diffusion in ionized gases}.
  \jt{Proceedings of the Physical Society}  \bvol{72}~(3),  \pg{353--362}.

\bibitem[Chapman \& Cowling(1970)]{CE}
{\sc \au{Chapman, S.} \& \au{Cowling, T.~G.}} \yr{1970} {\em {The Mathematical
  Theory of Non-uniform Gases}\/}.  \publ{Cambridge University Press}.

\bibitem[Chen {\em et~al.\/}(2015)Chen, Xu \& Cai]{ChenAAMM2015}
{\sc \au{Chen, S.}, \au{Xu, K.} \& \au{Cai, Q.}} \yr{2015}  \at{A comparison
  and unification of ellipsoidal statistical and {Shakhov BGK} models}.
  \jt{Advances in Applied Mathematics and Mechanics}  \bvol{7}~(2),
  \pg{245--266}.

\bibitem[Erdman {\em et~al.\/}(1993)Erdman, Zipf, Hewlett, Loda, Collins, Levin
  \& Candler]{Erdman1993JTHT}
{\sc \au{Erdman, P.}, \au{Zipf, E.}, \au{Hewlett, C.}, \au{Loda, R.},
  \au{Collins, R.~J.}, \au{Levin, D.~A.} \& \au{Candler, G.~V.}} \yr{1993}
  \at{Flight measurements of low velocity bow shock ultraviolet radiation}.
  \jt{Journal of Thermophysics and Heat Transfer}  \bvol{7}~(1),  \pg{37--42}.

\bibitem[Farbar \& Boyd(2010)]{Farbar2010PoF}
{\sc \au{Farbar, E.} \& \au{Boyd, D.}} \yr{2010}  \at{Modeling of the plasma
  generated in a rarefied hypersonic shock layer}.  \jt{Physics of Fluid}
  \bvol{22},  \pg{106101}.

\bibitem[Fei {\em et~al.\/}(2020)Fei, Liu, Liu \& Zhang]{Fei2020AIAA}
{\sc \au{Fei, F.}, \au{Liu, H.}, \au{Liu, Z.} \& \au{Zhang, J.}} \yr{2020}
  \at{A benchmark study of kinetic models for shock waves}.  \jt{AIAA Journal}
  \bvol{58}~(6),  \pg{2596--2608}.

\bibitem[Garz{\'o} {\em et~al.\/}(1989)Garz{\'o}, Santos \&
  Brey]{Garzo1989PoFA}
{\sc \au{Garz{\'o}, V.}, \au{Santos, A.} \& \au{Brey, J.~J.}} \yr{1989}  \at{A
  kinetic model for a multicomponent gas}.  \jt{Physics of Fluids A: Fluid
  Dynamics}  \bvol{1}~(2),  \pg{380--383}.

\bibitem[Gmurczyk {\em et~al.\/}(1979)Gmurczyk, Tarczynski \&
  Walenta]{Gmurczyk1979RGD}
{\sc \au{Gmurczyk, A.~S.}, \au{Tarczynski, M.} \& \au{Walenta, Z.~A.}}
  \yr{1979}  \at{Shock wave structure in the binary mixtures of gases with
  disparate molecular masses}.  \jt{Rarefied gas dynamics; Proceedings of the
  11th International Symposium}  \bvol{1},  \pg{333--341}.

\bibitem[Goldman \& Sirovich(1969)]{Goldman1969JFM}
{\sc \au{Goldman, E.} \& \au{Sirovich, L.}} \yr{1969}  \at{The structure of
  shock-waves in gas mixtures}.  \jt{Journal of Fluid Mechanics}
  \bvol{35}~(3),  \pg{575--597}.

\bibitem[Grad(1960)]{Grad1960}
{\sc \au{Grad, H.}} \yr{1960}  \at{Theory of rarefied gases}.  \jt{Rarefied gas
  dynamics; Proceedings of the 1st International Symposium}  \pg{pp. 100--138}.

\bibitem[Greene(1973)]{Greene1973PoF}
{\sc \au{Greene, J.~M.}} \yr{1973}  \at{Improved {Bhatnagar-Gross-Krook} model
  of electron-ion collisions}.  \jt{The Physics of Fluids}  \bvol{16}~(11),
  \pg{2022--2023}.

\bibitem[Groppi {\em et~al.\/}(2011)Groppi, Monica \& Spiga]{Groppi2011EPL}
{\sc \au{Groppi, M.}, \au{Monica, S.} \& \au{Spiga, G.}} \yr{2011}  \at{A
  kinetic ellipsoidal {BGK} model for a binary gas mixture}.  \jt{EPL
  (Europhysics Letters)}  \bvol{96}~(6),  \pg{64002}.

\bibitem[Groppi \& Spiga(2004)]{Groppi2004PoF}
{\sc \au{Groppi, M.} \& \au{Spiga, G.}} \yr{2004}  \at{A
  {Bhatnagar--Gross--Krook}-type approach for chemically reacting gas
  mixtures}.  \jt{Physics of Fluids}  \bvol{16}~(12),  \pg{4273--4284}.

\bibitem[Haack {\em et~al.\/}(2017)Haack, Hauck \& Murillo]{Haack2017JSP}
{\sc \au{Haack, J.~R.}, \au{Hauck, C.~D.} \& \au{Murillo, M.~S.}} \yr{2017}
  \at{A conservative, entropic multispecies {BGK} model}.  \jt{Journal of
  Statistical Physics}  \bvol{168}~(4),  \pg{826--856}.

\bibitem[Hamel(1965)]{Hamel1965PoF}
{\sc \au{Hamel, B.~B.}} \yr{1965}  \at{Kinetic model for binary gas mixtures}.
  \jt{The Physics of Fluids}  \bvol{8}~(3),  \pg{418--425}.

\bibitem[Hirschfelder {\em et~al.\/}(1954)Hirschfelder, Curtiss \&
  Bird]{Hirschfelder1954}
{\sc \au{Hirschfelder, J.O.}, \au{Curtiss, C.F.} \& \au{Bird, R.~B.}} \yr{1954}
  {\em Molecular Theory of Gases and Liquids\/}.  \publ{John Wiley \& Sons}.

\bibitem[Ho {\em et~al.\/}(2016)Ho, Wu, Graur, Zhang \& Reese]{Ho2016IJHMT}
{\sc \au{Ho, M.~T.}, \au{Wu, L.}, \au{Graur, I.}, \au{Zhang, Y.} \& \au{Reese,
  J.~M.}} \yr{2016}  \at{Comparative study of the boltzmann and mccormack
  equations for couette and fourier flows of binary gaseous mixtures}.
  \jt{International Journal of Heat and Mass Transfer}  \bvol{96},
  \pg{29--41}.

\bibitem[Holway(1966)]{Holway1966}
{\sc \au{Holway, L.~H.}} \yr{1966}  \at{New statistical models for kinetic
  theory: methods of construction}.  \jt{The Physics of Fluids}  \bvol{9},
  \pg{1658--1673}.

\bibitem[Klingenberg {\em et~al.\/}(2017)Klingenberg, Pirner \&
  Puppo]{Klingenberg2017KRM}
{\sc \au{Klingenberg, C.}, \au{Pirner, M.} \& \au{Puppo, G.}} \yr{2017}  \at{A
  consistent kinetic model for a two-component mixture with an application to
  plasma}.  \jt{Kinetic and Related Models}  \bvol{10}~(2),  \pg{445--465}.

\bibitem[Kosuge(2009)]{Kosuge2009EJMB}
{\sc \au{Kosuge, S.}} \yr{2009}  \at{Model {Boltzmann} equation for gas
  mixtures: Construction and numerical comparison}.  \jt{European Journal of
  Mechanics - B/Fluids}  \bvol{28}~(1),  \pg{170--184}.

\bibitem[Kosuge {\em et~al.\/}({2001})Kosuge, Aoki \& Takata]{Kosuge2001EJMB}
{\sc \au{Kosuge, S.}, \au{Aoki, K.} \& \au{Takata, S.}} \yr{{2001}}
  \at{{Shock-wave structure for a binary gas mixture: finite-difference
  analysis of the Boltzmann equation for hard-sphere molecules}}.  \jt{European
  Journal of Mechanics - B/Fluids}  \bvol{{20}}~({1}),  \pg{{87--126}}.

\bibitem[Koura \& Matsumoto(1991)]{Koura1991PoFA}
{\sc \au{Koura, K.} \& \au{Matsumoto, H.}} \yr{1991}  \at{Variable soft sphere
  molecular model for inverse-power-law or {Lennard-Jones} potential}.
  \jt{Physics of Fluids A: Fluid Dynamics}  \bvol{3}~(10),  \pg{2459--2465}.

\bibitem[Li {\em et~al.\/}(2023)Li, Zeng, Huang \& Wu]{Li2023JFM}
{\sc \au{Li, Q.}, \au{Zeng, J.}, \au{Huang, Z.} \& \au{Wu, L.}} \yr{2023}
  \at{Kinetic modelling of rarefied gas flows with radiation}.  \jt{Journal of
  Fluid Mechanics}  \bvol{965},  \pg{A13}.

\bibitem[Li {\em et~al.\/}(2021)Li, Zeng, Su \& Wu]{Li2021JFM}
{\sc \au{Li, Q.}, \au{Zeng, J.}, \au{Su, W.} \& \au{Wu, L.}} \yr{2021}
  \at{Uncertainty quantification in rarefied dynamics of molecular gas: rate
  effect of thermal relaxation}.  \jt{Journal of Fluid Mechanics}  \bvol{917},
  \pg{A58}.

\bibitem[Liu {\em et~al.\/}(2020)Liu, Zhu \& Xu]{Liu2020JCP}
{\sc \au{Liu, C.}, \au{Zhu, Y.} \& \au{Xu, K.}} \yr{2020}  \at{{Unified
  gas-kinetic wave-particle methods I: Continuum and rarefied gas flow}}.
  \jt{Journal of Computational Physics}  \bvol{401},  \pg{108977}.

\bibitem[Liu {\em et~al.\/}(2024)Liu, Zhang, Zeng \& Wu]{Liu2024JCP}
{\sc \au{Liu, W.}, \au{Zhang, Y.~B.}, \au{Zeng, J.~N.} \& \au{Wu, L.}}
  \yr{2024}  \at{{Further acceleration of multiscale simulation of rarefied gas
  flow via a generalized boundary treatment}}.  \jt{Journal of Computational
  Physics}  \bvol{503},  \pg{112830}.

\bibitem[Mathiaud {\em et~al.\/}(2022)Mathiaud, Mieussens \&
  Pfeiffer]{Mathiaud2022EjM}
{\sc \au{Mathiaud, J.}, \au{Mieussens, L.} \& \au{Pfeiffer, M.}} \yr{2022}
  \at{{An ES-BGK model for diatomic gases with correct relaxation rates for
  internal energies}}.  \jt{European Journal of Mechanics - B/Fluids}
  \bvol{96},  \pg{65--77}.

\bibitem[McCormack(1973)]{McCormack1973PoF}
{\sc \au{McCormack, F.~J.}} \yr{1973}  \at{Construction of linearized kinetic
  models for gaseous mixtures and molecular gases}.  \jt{The Physics of Fluids}
   \bvol{16}~(12),  \pg{2095--2105}.

\bibitem[Morse(1964)]{Morse1964PoF}
{\sc \au{Morse, T.~F.}} \yr{1964}  \at{Kinetic model equations for a gas
  mixture}.  \jt{The Physics of Fluids}  \bvol{7}~(12),  \pg{2012--2013}.

\bibitem[Nagnibeda \& Kustova(2009)]{Nagnibeda2009}
{\sc \au{Nagnibeda, E.} \& \au{Kustova, E.}} \yr{2009} {\em Non-Equilibrium
  Reacting Gas Flows: Kinetic Theory of Transport and Relaxation Processes\/}.
  \publ{Berlin Heidelberg: Springer-Verlag}.

\bibitem[Park {\em et~al.\/}(2024)Park, Kim, Pfeiffer \& Jun]{Park2024PoF}
{\sc \au{Park, W.}, \au{Kim, S.}, \au{Pfeiffer, M.} \& \au{Jun, E.}} \yr{2024}
  \at{Evaluation of stochastic particle {Bhatnagar--Gross--Krook} methods with
  a focus on velocity distribution function}.  \jt{Physics of Fluids}
  \bvol{36}~(2).

\bibitem[Pfeiffer {\em et~al.\/}(2022)Pfeiffer, Garmirian \&
  Gorji]{Pfeiffer2022PRE}
{\sc \au{Pfeiffer, M.}, \au{Garmirian, F.} \& \au{Gorji, M.~H.}} \yr{2022}
  \at{{Exponential Bhatnagar-Gross-Krook integrator for multiscale
  particle-based kinetic simulations}}.  \jt{Physical Review E}  \bvol{106},
  \pg{025303}.

\bibitem[Pfeiffer {\em et~al.\/}(2021)Pfeiffer, Mirza \&
  Nizenkov]{Pfeiffer2021PoF}
{\sc \au{Pfeiffer, M.}, \au{Mirza, A.} \& \au{Nizenkov, P.}} \yr{2021}
  \at{Multi-species modeling in the particle-based ellipsoidal statistical
  {Bhatnagar--Gross--Krook} method for monatomic gas species}.  \jt{Physics of
  Fluids}  \bvol{33}~(3),  \pg{036106}.

\bibitem[Pirner(2021)]{Pirner2021Fluids}
{\sc \au{Pirner, M.}} \yr{2021}  \at{A review on {BGK} models for gas mixtures
  of mono and polyatomic molecules}.  \jt{Fluids}  \bvol{6}~(11),  \pg{393}.

\bibitem[Plimpton {\em et~al.\/}(2019)Plimpton, Moore, Borner, Stagg, Koehler,
  Torczynski \& Gallis]{SPARTA}
{\sc \au{Plimpton, S.~J.}, \au{Moore, S.~G.}, \au{Borner, A.}, \au{Stagg,
  A.~K.}, \au{Koehler, T.~P.}, \au{Torczynski, J.~R.} \& \au{Gallis, M.~A.}}
  \yr{2019}  \at{{Direct Simulation Monte Carlo} on petaflop supercomputers and
  beyond}.  \jt{Physics of Fluids}  \bvol{31}~(8),  \pg{086101}.

\bibitem[Ruyev {\em et~al.\/}(2002)Ruyev, Fedorov \& Fomin]{Ruyev2002JAMTP}
{\sc \au{Ruyev, G.~A.}, \au{Fedorov, A.~V.} \& \au{Fomin, V.~M.}} \yr{2002}
  \at{Special features of the shock-wave structure in mixtures of gases with
  disparate molecular masses}.  \jt{Journal of Applied Mechanics and Technical
  Physics}  \bvol{43}~(4),  \pg{529--537}.

\bibitem[Sawant {\em et~al.\/}(2020)Sawant, Dorschner \& Karlin]{Sawant2020JFM}
{\sc \au{Sawant, N.}, \au{Dorschner, B.} \& \au{Karlin, I.~V.}} \yr{2020}
  \at{Consistent lattice {Boltzmann} model for multicomponent mixtures}.
  \jt{Journal of Fluid Mechanics}  \bvol{909},  \pg{A1}.

\bibitem[Schmidt {\em et~al.\/}(1984)Schmidt, Seiler \&
  W{\"o}rner]{Schmidt1984JFM}
{\sc \au{Schmidt, B.}, \au{Seiler, F.} \& \au{W{\"o}rner, M.}} \yr{1984}
  \at{Shock structure near a wall in pure inert gas and in binary inert-gas
  mixtures}.  \jt{Journal of Fluid Mechanics}  \bvol{143},  \pg{305--326}.

\bibitem[Shakhov(1968{\natexlab{{\em a\/}}})]{Shakhov1968}
{\sc \au{Shakhov, E.~M.}} \yr{1968{\natexlab{{\em a\/}}}}  \at{Approximate
  kinetic equations in rarefied gas theory}.  \jt{Fluid Dynamics}  \bvol{3},
  \pg{112--115}.

\bibitem[Shakhov(1968{\natexlab{{\em b\/}}})]{Shakhov_S}
{\sc \au{Shakhov, E.~M.}} \yr{1968{\natexlab{{\em b\/}}}}  \at{{Generalization
  of the Krook kinetic relaxation equation}}.  \jt{Fluid Dynamics}
  \bvol{3}~(5),  \pg{95--96}.

\bibitem[Sharipov(2024)]{Sharipov2024IJHMT}
{\sc \au{Sharipov, F.}} \yr{2024}  \at{Ab initio modelling of transport
  phenomena in multi-component mixtures of rarefied gases}.  \jt{International
  Journal of Heat and Mass Transfer}  \bvol{220},  \pg{124906}.

\bibitem[Sharipov \& Dias(2018)]{Sharipov2018EJMB}
{\sc \au{Sharipov, F.} \& \au{Dias, F.~C.}} \yr{2018}  \at{Structure of planar
  shock waves in gaseous mixtures based on ab initio direct simulation}.
  \jt{European Journal of Mechanics - B/Fluids}  \bvol{72},  \pg{251--263}.

\bibitem[Stephani {\em et~al.\/}(2012)Stephani, Goldstein \&
  Varghese]{Stephani2012PoF}
{\sc \au{Stephani, K.~A.}, \au{Goldstein, D.~B.} \& \au{Varghese, P.~L.}}
  \yr{2012}  \at{Consistent treatment of transport properties for five-species
  air direct simulation {Monte Carlo/Navier-Stokes} applications}.  \jt{Physics
  of Fluids}  \bvol{24}~(7).

\bibitem[Su {\em et~al.\/}(2020{\natexlab{{\em a\/}}})Su, Zhu, Wang, Zhang \&
  Wu]{Su2020JCP}
{\sc \au{Su, W.}, \au{Zhu, L.}, \au{Wang, P.}, \au{Zhang, Y.} \& \au{Wu, L.}}
  \yr{2020{\natexlab{{\em a\/}}}}  \at{Can we find steady-state solutions to
  multiscale rarefied gas flows within dozens of iterations?}  \jt{Journal of
  Computational Physics}  \bvol{407},  \pg{109245}.

\bibitem[Su {\em et~al.\/}(2020{\natexlab{{\em b\/}}})Su, Zhu \&
  Wu]{Su2020SIAM}
{\sc \au{Su, W.}, \au{Zhu, L.~H.} \& \au{Wu, L.}} \yr{2020{\natexlab{{\em
  b\/}}}}  \at{Fast convergence and asymptotic preserving of the general
  synthetic iterative scheme}.  \jt{SIAM Journal on Scientific Computing}
  \bvol{42}~(6),  \pg{B1517--B1540}.

\bibitem[Takata {\em et~al.\/}(2007)Takata, Sugimoto \& Kosuge]{Takata2007EJMB}
{\sc \au{Takata, S.}, \au{Sugimoto, H.} \& \au{Kosuge, S.}} \yr{2007}  \at{Gas
  separation by means of the {Knudsen} compressor}.  \jt{European Journal of
  Mechanics - B/Fluids}  \bvol{26}~(2),  \pg{155--181}.

\bibitem[Tantos \& Valougeorgis(2018)]{Tantos2018IJHMT}
{\sc \au{Tantos, C.} \& \au{Valougeorgis, D.}} \yr{2018}  \at{Conductive heat
  transfer in rarefied binary gas mixtures confined between parallel plates
  based on kinetic modeling}.  \jt{International Journal of Heat and Mass
  Transfer}  \bvol{117},  \pg{846--860}.

\bibitem[Tantos {\em et~al.\/}(2021)Tantos, Varoutis \& Day]{Tantos2021PoF}
{\sc \au{Tantos, C.}, \au{Varoutis, S.} \& \au{Day, C.}} \yr{2021}  \at{Heat
  transfer in binary polyatomic gas mixtures over the whole range of the gas
  rarefaction based on kinetic deterministic modeling}.  \jt{Physics of Fluids}
   \bvol{33}~(2),  \pg{022004}.

\bibitem[Tantos {\em et~al.\/}(2024)Tantos, Waroutis, Hauer, Day \&
  Innocente]{Tantos2024NuclFusion}
{\sc \au{Tantos, C.}, \au{Waroutis, S.}, \au{Hauer, V.}, \au{Day, C.} \&
  \au{Innocente, P.}} \yr{2024}  \at{{3D numerical study of netural gas
  dynamics in the DTT particle exhaust using the DSMC method}}.  \jt{Nuclear
  Fusion}  \bvol{64},  \pg{016019}.

\bibitem[Teng {\em et~al.\/}(2023)Teng, Hao, Liu, Bian, Xie \&
  Liu]{Teng2023JCProd}
{\sc \au{Teng, S.}, \au{Hao, M.}, \au{Liu, J.~X.}, \au{Bian, X.}, \au{Xie,
  Y.~H.} \& \au{Liu, K.}} \yr{2023}  \at{Pollutant inhibition in an extreme
  ultraviolet lighography machine by dynamic gas lock}.  \jt{Journal of Cleaner
  Production}  \bvol{430},  \pg{139664}.

\bibitem[Tipton {\em et~al.\/}(2009{\natexlab{{\em a\/}}})Tipton, Tompson \&
  Loyalka]{Tipton2009EJMB}
{\sc \au{Tipton, E.~L.}, \au{Tompson, R.~V.} \& \au{Loyalka, S.~K.}}
  \yr{2009{\natexlab{{\em a\/}}}}  \at{{Chapman--Enskog solutions to arbitrary
  order in Sonine polynomials II: Viscosity in a binary, rigid-sphere, gas
  mixture}}.  \jt{European Journal of Mechanics - B/Fluids}  \bvol{28}~(3),
  \pg{335--352}.

\bibitem[Tipton {\em et~al.\/}(2009{\natexlab{{\em b\/}}})Tipton, Tompson \&
  Loyalka]{Tipton2009EJMB_2}
{\sc \au{Tipton, E.~L.}, \au{Tompson, R.~V.} \& \au{Loyalka, S.~K.}}
  \yr{2009{\natexlab{{\em b\/}}}}  \at{{Chapman--Enskog solutions to arbitrary
  order in Sonine polynomials III: Diffusion, thermal diffusion, and thermal
  conductivity in a binary, rigid-sphere, gas mixture}}.  \jt{European Journal
  of Mechanics - B/Fluids}  \bvol{28}~(3),  \pg{353--386}.

\bibitem[Todorova \& Steijl(2019)]{Todorova2019EJMB}
{\sc \au{Todorova, B.~N.} \& \au{Steijl, R.}} \yr{2019}  \at{Derivation and
  numerical comparison of {Shakhov} and ellipsoidal statistical kinetic models
  for a monoatomic gas mixture}.  \jt{European Journal of Mechanics - B/Fluids}
   \bvol{76},  \pg{390--402}.

\bibitem[Todorova {\em et~al.\/}(2020)Todorova, White \&
  Steijl]{Todorova2020AIP}
{\sc \au{Todorova, B.~N.}, \au{White, C.} \& \au{Steijl, R.}} \yr{2020}
  \at{Modeling of nitrogen and oxygen gas mixture with a novel diatomic kinetic
  model}.  \jt{AIP Advances}  \bvol{10}~(9),  \pg{095218}.

\bibitem[Wagner(1992)]{wagner_consist}
{\sc \au{Wagner, W.}} \yr{1992}  \at{A convergence proof for {B}ird's direct
  simulation {M}onte {C}arlo method for the {B}oltzmann equation}.  \jt{Journal
  of Statistical Physics}  \bvol{66},  \pg{1011--1044}.

\bibitem[Wilke(1950)]{Wilke1950JCP}
{\sc \au{Wilke, C.~R.}} \yr{1950}  \at{A viscosity equation for gas mixtures}.
  \jt{The Journal of Chemical Physics}  \bvol{18}~(4),  \pg{517--519}.

\bibitem[Wu {\em et~al.\/}(2015{\natexlab{{\em a\/}}})Wu, White, Scanlon, Reese
  \& Zhang]{LeiJFM2015}
{\sc \au{Wu, L.}, \au{White, C.}, \au{Scanlon, T.~J.}, \au{Reese, J.~M.} \&
  \au{Zhang, Y.~H.}} \yr{2015{\natexlab{{\em a\/}}}}  \at{A kinetic model of
  the {Boltzmann equation for} non-vibrating polyatomic gases}.  \jt{Journal of
  Fluid Mechanics}  \bvol{763},  \pg{24--50}.

\bibitem[Wu {\em et~al.\/}(2015{\natexlab{{\em b\/}}})Wu, Zhang, Reese \&
  Zhang]{Wu2015JCP}
{\sc \au{Wu, L.}, \au{Zhang, J.}, \au{Reese, J.~M.} \& \au{Zhang, Y.}}
  \yr{2015{\natexlab{{\em b\/}}}}  \at{A fast spectral method for the
  {Boltzmann} equation for monatomic gas mixtures}.  \jt{Journal of
  Computational Physics}  \bvol{298},  \pg{602--621}.

\bibitem[Yuan \& Wu(2022)]{Yuan2022JFM}
{\sc \au{Yuan, R.~F.} \& \au{Wu, L.}} \yr{2022}  \at{Capturing the influence of
  intermolecular potential in rarefied gas flows by a kinetic model with
  velocity-dependent collision frequency}.  \jt{Journal of Fluid Mechanics}
  \bvol{942},  \pg{A13}.

\bibitem[Zeng {\em et~al.\/}(2022)Zeng, Li \& Wu]{Zeng2022AAS}
{\sc \au{Zeng, J.}, \au{Li, Q.} \& \au{Wu, L.}} \yr{2022}  \at{Kinetic modeling
  of rarefied molecular gas dynamics {(in Chinese)}}.  \jt{Acta Aerodynamica
  Sinica}  \bvol{40}~(2),  \pg{1--30}.

\bibitem[Zeng {\em et~al.\/}(2023)Zeng, Su \& Wu]{Zeng2023CiCP}
{\sc \au{Zeng, J.}, \au{Su, W.} \& \au{Wu, L.}} \yr{2023}  \at{General
  synthetic iterative scheme for unsteady rarefied gas flows}.
  \jt{Communications in Computational Physics}  \bvol{34}~(1),  \pg{173--207}.

\end{thebibliography}

\end{document}